\documentclass[pra,aps,reprint,twocolumn,superscriptaddress,10pt,nofootinbib,longbibliography,floatfix]{revtex4-2}

\usepackage[utf8]{inputenc}
\usepackage[T1]{fontenc}
\usepackage[english]{babel}
\usepackage{graphicx,epsfig}
\graphicspath{{images/}}
\usepackage{color}
\usepackage[usenames,dvipsnames]{xcolor}
\usepackage{rotating}

\usepackage{longtable}
\usepackage{graphicx}
\usepackage{multirow}
\usepackage{makecell}

\usepackage{physics}
\usepackage{amsmath,amssymb,amsthm}
\usepackage{bbm,dsfont} 
\usepackage{stmaryrd}
\usepackage{enumitem}
\usepackage[caption=false]{subfig}
\usepackage[normalem]{ulem}

\usepackage{mathtools}
\usepackage{cancel}

\definecolor{linkcolor}{rgb}{0,0,0.6}	
\usepackage[colorlinks=true,pdfstartview=FitV,linkcolor=linkcolor,citecolor=linkcolor,urlcolor=linkcolor,hyperindex=true,hyperfigures=false]{hyperref}

\usepackage{numprint}

\usepackage[bb=boondox]{mathalfa}

\usepackage{upgreek}

\newcommand{\kket}[1]{{|#1 \rangle \!\rangle}}
\newcommand{\bbra}[1]{{\langle \!\langle #1 |}}
\renewcommand{\ket}[1]{{|#1 \rangle}}
\renewcommand{\bra}[1]{{\langle #1 |}}
\renewcommand{\ketbra}[2]{{|#1 \rangle \!\langle #2|}}
\def\kketbra#1#2{\mathinner{|{#1}\rangle\!\rangle\!\langle\!\langle{#2}|}}

\newcommand{\id}{\mathbbm{1}}

\newcommand{\HS}{\mathcal{H}}
\renewcommand\L{\mathcal{L}}

\renewcommand\L{\mathcal{L}}
\newcommand{\M}{\mathcal{M}}
\newcommand{\N}{\mathcal{N}}
\newcommand{\K}{\mathcal{K}}
\newcommand{\bigpi}{\scalebox{1.6}{$\pi$}}
\newcommand{\A}{\mathcal{A}}

\usepackage{array}
\usepackage{ragged2e}

\newcolumntype{J}[1]{>{\justifying\arraybackslash}p{#1}}

\makeatletter 
    
\renewcommand\onecolumngrid{%
\do@columngrid{one}{\@ne}%
\def\set@footnotewidth{\onecolumngrid}%
\def\footnoterule{\kern-6pt\hrule width 1.5in\kern6pt}%
}

\renewcommand\twocolumngrid{%
        \def\footnoterule{
        \dimen@\skip\footins\divide\dimen@\thr@@
        \kern-\dimen@\hrule width.5in\kern\dimen@}
        \do@columngrid{mlt}{\tw@}
}%

\newcommand{\AppendixTOC}{
    \begingroup
    \renewcommand{\contentsname}{Appendix Table of Contents}
    \tableofcontents
    \endgroup
}

\makeatother

\begin{document}

\title{Correlations and quantum circuits with dynamical causal order}

\author{Raphaël Mothe}
\affiliation{Univ.\ Grenoble Alpes, CNRS, Grenoble INP\footnote{Institute of Engineering Univ.\ Grenoble Alpes}, Institut N\'eel, 38000 Grenoble, France}
\affiliation{Univ.\ Grenoble Alpes, Inria, 38000 Grenoble, France}
\affiliation{Naturwissenschaftlich-Technische Fakultät, Universität Siegen, Walter-Flex-Straße 3, 57068 Siegen, Germany}

\author{Alastair A.\ Abbott}
\affiliation{Univ.\ Grenoble Alpes, Inria, 38000 Grenoble, France}

\author{Cyril Branciard}
\affiliation{Univ.\ Grenoble Alpes, CNRS, Grenoble INP\footnote{Institute of Engineering Univ.\ Grenoble Alpes}, Institut N\'eel, 38000 Grenoble, France}

\date{November 23, 2026} 


\begin{abstract}
Requiring that the causal structure between different parties is well-defined imposes constraints on the correlations they can establish, which define so-called causal correlations. Some of these are known to have a ``dynamical'' causal order in the sense that their causal structure is not fixed \emph{a priori} but is instead established on the fly, with for instance the causal order between future parties depending on some choice of action of parties in the past. Here we identify a new way that the causal order between the parties can be dynamical: with at least four parties, there can be some dynamical order which can nevertheless not be influenced by the actions of past parties. This leads us to introduce an intermediate class of correlations with what we call non-influenceable causal order, in between the set of correlations with static (non-dynamical) causal order and the set of general causal correlations. We then define analogous classes of quantum processes, considering recently introduced classes of quantum circuits with classical or quantum control of causal order---the latter being the largest class within the process matrix formalism known to have a clear interpretation in terms of coherent superpositions of causal orders. This allows us to formalise precisely in which sense certain quantum processes can have both indefinite and dynamical causal order.

\end{abstract}


\maketitle


\addtocontents{toc}{\string\iffalse}

\section{Introduction}

The study of the causal structure of the physical world has led to many rich and fascinating research directions. 
On the one hand, the assumption of particular causal structures places constraints on observations that may be challenged, for example, by quantum mechanics~\cite{wood15} or gravity~\cite{hardy05}, while on the other hand, the possibility of causally indefinite processes opens intriguing possibilities, both foundational~\cite{oreshkov12} and practical~\cite{chiribella12,araujo14,Rozema24,Costa26} in nature.

One approach has been to investigate how the assumption of a well-defined causal structure between different events constrains the correlations that can be established between them~\cite{oreshkov12,Baumeler14_2,Araujo15,branciard15,Abbott16,Baumeler16,Tselentis23a}. 
Formally, such correlations can be studied through the notion of \emph{causal correlations} obtainable by $N$ parties receiving classical inputs and producing classical outputs within a well-defined causal structure.
In the simplest nontrivial case with $N=2$ parties $A$ and $B$, the correlations are said to be causal if they can be established with the two parties acting one after the other in a fixed causal order (that may be chosen probabilistically for each run of the experiment), either ``$A$ before $B$'' or ``$B$ before $A$''~\cite{oreshkov12}.
However, it was quickly realised that, as soon as one considers a third party, the assumption of a well-defined causal structure between the parties no longer imposes that the correlations established are necessarily compatible with a probabilistic fixed causal order: the correlations may follow a \emph{dynamical} causal order that is not predetermined, or fixed \emph{a priori}, but instead established on the fly as the parties act one after another~\cite{hardy05,BaumelerWolf,Oreshkov16,Abbott16}.
Causal correlations with dynamical causal order can be found in situations where the order between ``future'' parties depends on the actions of some ``past'' parties.
While some recent works have studied correlations with dynamical, or ``non-static'', causal order in specific scenarios~\cite{Tselentis23,Tselentis24,Baumann24}, little effort has so far been made to characterise more directly the notion of correlations with dynamical causal order as opposed to ones with non-dynamical causal order, or to understand their properties.

At the same time, the causal structure of quantum processes has been explored, in particular within the process matrix formalism~\cite{oreshkov12}. 
By describing how different parties' actions, which are locally causal and consistent with quantum mechanics, can be related without assuming any \emph{a priori} well-defined global causal structure, this formalism distinguishes between processes that are compatible with a definite causal order from those that are not, the latter having an \emph{indefinite causal order}. 
Indeed, the set of quantum processes compatible with a well-defined causal structure, the so-called \emph{causally separable processes}, has been well studied~\cite{oreshkov12,Araujo15,wechs19} and, just as for causal correlations, it was observed that one must take into account processes with dynamical causal order in the definition of causally separable processes for $N\ge 3$ parties~\cite{Oreshkov16,wechs19}.
For instance, one can create quantum processes in which the operation implemented by a party $A$ can influence, or even determine, the order between two later parties, $B$ and $C$, in the causal future of $A$; in any given run, however, a well-defined causal order can nonetheless be identified, albeit dynamically chosen. 
The notion of causal separability also, in particular, allows one to precisely argue that certain processes are causally indefinite, or \emph{causally nonseparable}~\cite{oreshkov12}.
The \emph{quantum switch}~\cite{Chiribella13} is a notable example of such a process, in which the two possible causal orders between two parties $A$ and $B$, namely ``$A$ before $B$'' and ``$B$ before $A$'', are superposed in a coherent manner with the help of a control qubit, leading to a so-called \emph{causal superposition}. 
In the quantum switch and its natural generalisations~\cite{Colnaghi12,araujo14}, the parties never intervene on the causal order itself, and 
thus, despite being causally nonseparable, the quantum switch does not intuitively exhibit a dynamical causal order.
More recently, examples of processes that intuitively combine dynamical and indefinite causal order have been presented within the larger class of \emph{quantum circuits with quantum control of causal order} (QC-QCs)~\cite{wechs21}.
These examples have highlighted, however, that the concept of dynamical causal orders has largely been overlooked, except for its consideration in the definition of causally separable processes.
Until now, no precise definition of what it means for a process to have (or not) a dynamical causal order has been proposed, hindering attempts to argue precisely, for example, that the quantum switch does not feature any dynamical causal order while other QC-QCs may do. 
Moreover, the idea of combining dynamical and indefinite causal order finds motivations in a wide range of research lines, from foundational problems~\cite{hardy05}---as a theory of quantum gravity is likely to involve a dynamical causal structure because of general relativity theory, with an indefinite causal structure because of quantum theory---to more applied problems---as it could offer new forms of advantages for quantum information processing over standard quantum circuits. 

In this work, we propose precise definitions of (non)-dynamical causal structures, both at the level of causal correlations and quantum processes, and study in detail these notions of dynamical causal orders. 
Exploiting a characterisation of causal correlations in which the causal order appears explicitly as an additional (unobserved) random variable, we identify a novel way in which the causal order between different parties that establish causal correlations can be dynamical: for $N\ge 4$ parties, there can be some dynamical order which can nevertheless not be influenced by the actions of past parties. 
We thereby observe that the standard understanding of dynamical causal orders---as parties in the past influencing the order of those in the future~\cite{BaumelerWolf,Abbott16,Oreshkov16,Tselentis23,Tselentis24,Baumann24}---is incomplete.
In addition to the already known classes of causal correlations, namely mixtures of fixed causal order correlations (which correspond precisely to those that have non-dynamical causal order), and general causal correlations (which allow for any kind of dynamicality), this allows us to define the class of causal correlations with ``non-influenceable orders'', and a weaker form of it, the class of causal correlations with ``non-influenceable coarse-grained orders''. Using a fourpartite causal game, we show that (for $N\ge 4$) these classes form a strict hierarchy of correlations.

Inspired by these classes of causal correlations, we next look to define analogous classes at the level of quantum processes. 
To this end, we restrict ourselves to subclasses of quantum circuits with classical or quantum control of causal order (QC-CCs or QC-QCs), as these have clear physical interpretations as generalised circuits in which a classical or quantum system is used to control, at each step, which party should act~\cite{wechs21}. (This is in contrast to some processes beyond QC-QCs, for which physical realisability remains an open question~\cite{oreshkov12,araujo17}.)
QC-QCs in particular include causally indefinite processes such as the quantum switch, but nevertheless can only produce causal correlations (potentially with dynamical causal order). 
We first study the class of QC-CCs, which are causally separable processes, allowing us to study dynamicality independently from causal indefiniteness.
We consider QC-CCs with non-dynamical causal order as precisely those that are convex mixtures of quantum circuits with fixed causal order (i.e., standard quantum circuits), a class that we denote \textup{\textsf{QC-convFO}}. 
Just as for causal correlations, we find that a more subtle type of ``non-influenceable'' dynamical causal order also exists for QC-CCs.
We formalise this as a condition for the internal control system of the QC-CC to be independent of the quantum operations chosen by previous parties, and we show how the corresponding class, \textup{\textsf{QC-NICC}}, can be explicitly characterised.  
Turning our attention to QC-QCs, for which notions of (non)-dynamicality are less intuitive to define, we then describe the class of QC-QCs with non-dynamical causal order, which allow for the quantum control system to create superpositions of causal orders, but with the weight of each order being fixed from the start, and not altered dynamically---a class that we call \textup{\textsf{QC-supFO}}. 
The quantum switch (with a fixed control system) and its generalisations are shown to be contained in this class.
By contrast, the existence of causally nonseparable quantum circuits beyond this class, in $\textup{\textsf{QC-QC}}\backslash\textup{\textsf{QC-supFO}}$, allows us to formalise precisely in which sense certain quantum processes can have both indefinite and dynamical causal order.
Finally, we formulate an analogous condition for QC-QCs to have non-influenceable quantum control of causal order, defining the class \textup{\textsf{QC-NIQC}}, and providing nontrivial examples of QC-QCs with dynamical but non-influenceable causal order. 

In Table~\ref{tab:summary} we provide a summary to help the reader navigate this manuscript. It includes a list of the four classes of correlations and the six classes of quantum circuits discussed in the main text, the abbreviations they are referred to by, as well as short descriptions and references to their definitions or characterisations, and to key examples.

\renewcommand{\arraystretch}{1.3}

\begin{table*}[h]
\caption{\label{tab:summary}Summary of the abbreviations and full names of the four classes of correlations and six classes of quantum circuits considered throughout this work, along with references to the respective definitions or characterisations, and key examples. The inclusion relations between the classes of correlations, the classes of quantum circuits with classical control and the classes of quantum circuits with quantum control are depicted in the Abbreviation column; see also Figs.~\ref{fig:polytopes}, \ref{fig:QCCC-classes} and~\ref{fig:QCQC-classes}. \\}
\begin{ruledtabular}
\begin{tabular}{c|>{\centering\arraybackslash}m{2cm}|>{\centering\arraybackslash}p{6cm}|p{9cm}}
 & Abbreviation & \centering Full name & \multicolumn{1}{c}{Description} \\
\hline\hline
\multirow{12.5}{*}{\rotatebox{90}{Sec.~\ref{sec:corr_causal}: causal correlations}} 
& \raisebox{-3mm}{convFO} & \raisebox{-3mm}{\parbox{6cm}{\centering convex mixtures of causal correlations\\ with fixed order}} & Causal correlations with non-dynamical causal order, i.e.\ where the full order can be fixed (with some probability) in advance; see Eq.~\eqref{eq:def_convFO_corr_2} for the definition and below Eq.~\eqref{eq:bnd_I3_FO} for an example. \\[-2mm]
& $\cap$ & \\[-2mm]
& \raisebox{-8mm}{NIO} & \raisebox{-7mm}{\parbox{6cm}{\centering
causal correlations with\\
non-influenceable orders}} & Causal correlations with possibly dynamical
but non-influenceable causal order, i.e.\ where the causal order cannot be fixed from the start, but the parties' choices of inputs still cannot influence the order between subsequent parties; see Eq.~\eqref{eq:def_NIO} for the definition and Eq.~\eqref{eq:saturate_NIO}  for an example.\\[-2mm]
& $\cap$ & \\[-2mm]
& \raisebox{-3mm}{$\text{NIO}^\prime$} & \raisebox{-3mm}{\parbox{6cm}{\centering causal correlations with \\ non-influenceable coarse-grained orders}} & Causal correlations with possibly dynamical but non-influenceable coarse-grained causal order; see Eq.~\eqref{eq:def_NIO'} for the definition and Eq.~\eqref{eq:saturate_NIO'} for an example.\\[-2mm]
& $\cap$ & \\[-2mm]
& \raisebox{-3mm}{$\text{causal}$} &  \raisebox{-3mm}{\parbox{6cm}{\centering causal correlations}} & General case of causal correlations, with possibly dynamical and influenceable causal order; see Eq.~\eqref{eq:def_causal_p} for the definition and Eq.~\eqref{eq:correl_I4_1} for an example.\\
\hline \hline
\multirow{23.5}{*}{\rotatebox{90}{Sec.~\ref{sec:quantum_circuits}: quantum circuits}} 
& \raisebox{-5mm}{QC-convFO} & \raisebox{-5mm}{\parbox{6cm}{\centering convex mixtures of quantum circuits \\ with fixed causal order}} & Quantum circuits with non-dynamical classically controlled causal order, where the state of the control system can determine the full order from the start; see Eq.~\eqref{eq:charact_W_QC_convFO_decomp} for a characterisation and Eq.~\eqref{eq:QC-convFO-example} for an example.\\[-2mm]
& $\cap$ & \\[-2mm]
& \raisebox{-9mm}{QC-NICC} & \raisebox{-9mm}{\parbox{6cm}{\centering quantum circuits with non-influenceable \\ classical control of causal order}} & Quantum circuits with possibly dynamical but non-influenceable classically controlled causal order, where the state of the control system cannot determine the full order from the start but also cannot be influenced by the parties' operations; see Eq.~\eqref{eq:cstr_NICC} for a characterisation and Eq.~\eqref{eq:QC-NICC-example} for an example, with its circuit representation given in Fig.~\ref{fig:QC-NICC-example}. \\[-2mm]
& $\cap$ & \\[-2mm]
& \raisebox{-5mm}{QC-CC} & \raisebox{-5mm}{\parbox{6cm}{\centering quantum circuits with \\ classical control of causal order}} & General case of quantum circuits with possibly dynamical and influenceable classically controlled causal order; see Fig.~\ref{fig:QCCC_graph} for their general circuit representation, Eq.~\eqref{eq:charact_W_QCCC_decomp} for a characterisation, and Eq.~\eqref{eq:CS-example} for an example.\\ \cline{2-4}
& \raisebox{-5.5mm}{QC-supFO} & \raisebox{-5.5mm}{\parbox{6cm}{\centering quantum circuits with\\ superposition of fixed causal orders}} & Quantum circuits with non-dynamical coherently controlled causal order, where the state of the control system can (coherently) encode the full order from the start; see Eq.~\eqref{eq:charact_W_QC_SupFO_decomp} for a characterisation and Eq.~\eqref{eq:QC-supFO-example} for an example.\\[-2mm]
& $\cap$ & \\[-2mm]
& \raisebox{-9mm}{QC-NIQC} & \raisebox{-9mm}{\parbox{6cm}{\centering quantum circuits with non-influenceable\\ quantum control of causal order}} & Quantum circuits with possibly dynamical but non-influenceable coherently controlled causal order, where the state of the control system cannot encode the full order from the start but also cannot be influenced by the parties' operations; see Eq.~\eqref{eq:cstr_NIQC} for a characterisation and Eq.~\eqref{eq:QC-NIQC-example} for an example, with its circuit representation given in Fig.~\ref{fig:QC-NIQC-example}.\\[-2mm]
& $\cap$ & \\[-2mm]
& \raisebox{-5.5mm}{QC-QC} & \raisebox{-5.5mm}{\parbox{6cm}{\centering quantum circuits with\\ quantum control of causal order}} & General case of quantum circuits with possibly dynamical and influenceable coherently controlled causal order; see Fig.~\ref{fig:QCQC_graph} for their general circuit representation, Eq.~\eqref{eq:charact_W_QCQC_decomp} for a characterisation, and Eq.~\eqref{eq:QC-QC-example} for an example. 
\end{tabular}
\end{ruledtabular}
\end{table*}

The rest of this paper is organised as follows. 
In Sec.~\ref{sec:corr_causal} we introduce four classes of causal correlations: the class of mixtures of correlations with fixed causal orders, two classes of correlations with possibly dynamical but non-influenceable order, and the class of general causal correlations. We present a causal game that allows us to show a strict separation between these different classes. 
In Sec.~\ref{sec:quantum_circuits} we introduce the corresponding classes of quantum circuits with classical and quantum control of causal order, and present examples of processes in each class. 
In Sec.~\ref{sec:QCQC_correlations} we investigate the ability for processes in the different classes just introduced to saturate or violate inequalities that bound the different sets of causal correlations. 
Then, in Sec.~\ref{sec:conclusion}, we conclude with a discussion of our results and some related open questions.

\section{Correlations with dynamical causal order}
\label{sec:corr_causal}

We will be interested throughout this paper in different scenarios, given by a finite number $N \ge 1$ of parties $A_k$ ($k\in \mathcal{N} \coloneqq \{1,\ldots,N\}$) that each chooses, or receives from an external referee, a classical input $x_k$ and produces a classical output $a_k$. 
Both the input and output belong to finite sets (not necessarily identical), which may differ for each party. 
When convenient (for a small enough, fixed value of $N$), we will sometimes equivalently refer to the parties by $A_1,A_2,A_3,\dots$ or $A,B,C,\dots$, and analogously denote their inputs and outputs $x_1,x_2,x_3,\dots$ and $a_1,a_2,a_3,\dots$, or $x,y,z,\dots$ and $a,b,c,\dots$. 
An ordered set of $n$ parties, where $A_{k_1}$ is first and $A_{k_n}$ is last, will be denoted as $(k_1,\ldots,k_n)$, while the unordered set composed of the same $n$ parties will be denoted as $\K_n=\{k_1,\ldots,k_n\}$. Both notations implicitly assume that all $k_i$'s are different (with the notation $\K_n$ implying it contains $n$ elements); furthermore, we simply identify the parties $A_k$ with their label $k$.

Due to a common cause prior to the parties' actions, or because of signalling between the parties, their outputs might be correlated, conditioned on their inputs. 
We are thus interested in the correlations established by the different parties that are described using the conditional probability distribution $p(\vec a | \vec x)$, where $\vec x =(x_1,\ldots,x_N)$ and $\vec a =(a_1,\ldots,a_N)$ are the vectors of inputs and outputs. 
For the sake of conciseness, we will denote the vector of inputs and outputs of the set of parties $\K_n=\{k_1,\ldots,k_n\}$ by $\vec x_{\K_n}$ or $\vec x_{k_1,\ldots,k_n} = (x_{k_1},\ldots,x_{k_n})$ and $\vec a_{\K_n}$ or $\vec a_{k_1,\ldots,k_n} = (a_{k_1},\ldots,a_{k_n})$.
We will study here the correlations obtained when imposing that the causal structure between the parties is well-defined.

\subsection{Multipartite causal correlations}

The causal structure between parties constrains the correlations they can establish; indeed, a party cannot signal to any party in its causal past. This observation forms the basis of the definition of causal correlations, which are those compatible with a well-defined causal structure.

To start with, let us consider the case of $N=2$ parties, the $N=1$ case being rather trivial. 
Assuming a well-defined causal order between $A$ and $B$ implies that either $A$ causally precedes $B$ (which we will denote $A\prec B$), $B$ causally precedes $A$ (which we will denote $B\prec A$), or one may be in a probabilistic mixture of the two possibilities.
The case that $A$ and $B$ are causally independent can be included in either of the above cases, as the observed correlations will be non-signalling in both directions and thus compatible with both $A\prec B$ and $B\prec A$.
Thus, we say that the causal structure between the two parties is well-defined, or equivalently, that the probability distribution $p(a,b|x,y)$ is a bipartite causal correlation, if and only if
\begin{align}
    p(a,b|x,y) = q \,p_{A \prec B}(a,b|x,y) + (1-q)\,p_{B \prec A}(a,b|x,y), \label{eq:causal_N_2}
\end{align}
where $q \in [0,1]$, and $p_{A \prec B},p_{B \prec A}$ are probability distributions whose marginal distributions satisfy the one-way-signalling constraints $p_{A \prec B}(a|x,y)=p_{A \prec B}(a|x)$ and $p_{B \prec A}(b|x,y)=p_{B \prec A}(b|y)$~\cite{oreshkov12}.

In the general multipartite scenario with $N$ parties, we say that a correlation $p_\pi(\vec a|\vec x)$ is compatible with a fixed causal order (encoded as a permutation of the parties) $\pi = (k_1,\ldots,k_N)$ if and only if%
\footnote{More formally, that the specified probability does not depend on $\vec x_{\mathcal{N}\backslash\{k_1,\ldots,k_n\}}$ means here that $\forall\, \vec a_{k_1,\ldots,k_n}$, $\forall\, \vec x_{k_1,\ldots,k_n}$, $\forall\, \vec x_{\mathcal{N}\backslash\{k_1,\ldots,k_n\}}$, $\forall\, \vec x_{\mathcal{N}\backslash\{k_1,\ldots,k_n\}}'$, $p_\pi\big(\vec a_{k_1,\ldots,k_n}\big|\vec x_{k_1,\ldots,k_n}, \vec x_{\mathcal{N}\backslash\{k_1,\ldots,k_n\}}\big) = p_\pi\big(\vec a_{k_1,\ldots,k_n}\big|\vec x_{k_1,\ldots,k_n}, \vec x_{\mathcal{N}\backslash\{k_1,\ldots,k_n\}}'\big)$.}
\begin{align}
    & \forall\, n = 1,\ldots,N-1, \notag \\[1mm]
    & \ \ p_\pi(\vec a_{k_1,\ldots,k_n}|\vec x) \text{ does not depend on } \vec x_{\mathcal{N}\backslash\{k_1,\ldots,k_n\}}, \label{eq:def_convFO_corr}
\end{align}
where $p_\pi(\vec a_{k_1,\ldots,k_n}|\vec x) \coloneqq \sum_{\vec a_{\mathcal{N}\backslash\{k_1,\ldots,k_n\}}} p_\pi(\vec a|\vec x)$. As (for any fixed number of parties, and of inputs and outputs for each party) the set of correlations compatible with a fixed causal order $\pi$ is defined by a finite number of linear constraints on a bounded probability space, it defines a convex polytope~\cite{ziegler95}. The convex hull of all the polytopes of correlations compatible with a given fixed causal order $\pi$ is therefore also a convex polytope, that we call $\mathcal{P}_{\text{convFO}}$. It comprises any probabilistic mixture of correlations, each of which is compatible with a fixed causal order $\pi$---i.e., \textit{convex mixtures of causal correlations with fixed order}, of the form
\begin{align}
    p(\vec a|\vec x) = \sum_\pi q_\pi \,p_\pi(\vec a|\vec x),
    \label{eq:def_convFO_corr_2}
\end{align}
with $q_\pi \geq 0$, $\sum_\pi q_\pi = 1$.
Note that in the bipartite scenario, $\mathcal{P}_{\textup{convFO}}$ corresponds precisely to the set of causal correlations.

As soon as one considers scenarios with more than two parties, however, there are correlations that can be understood as being compatible with a well-defined causal structure between the parties, and yet do not correspond to probabilistic mixtures of correlations with fixed causal order.
As proposed in~\cite{hardy05} and formalised in~\cite{BaumelerWolf,Oreshkov16,Abbott16}, fundamentally new, dynamical properties appear from the tripartite case onwards. These are usually referred to as dynamical causal orders, in the sense that the causal order of future parties is not fixed from the start, but is established on the fly (``dynamically'') as the correlations are constructed, i.e., as the inputs of the parties are chosen and their outputs are obtained.
For instance, the input choice of a first party, say $A$, may determine the causal order between two subsequent parties, $B$ and $C$.
Intuitively, we say that a causal structure is well-defined if, after the parties have acted, one can interpret them as having acted in a strict order, where a given party can only affect the order between the parties that have not already acted, as well as their outputs (together with the given party's own output). 
More precisely, we say that a correlation $p(\vec a|\vec x)$ between $N$ parties has a well-defined causal structure if one can identify a hidden variable $\pi$ encoding the causal order, and a joint distribution $p(\pi,\vec a|\vec x)$, such that the probability to observe $n$ parties following the order $(k_1,\ldots,k_{n})$ with given outputs $\vec a_{k_1,\ldots,k_n}$, and the choice of the next party $A_{k_{n+1}}$ to act, cannot be influenced by the choice of inputs made by the $N-n$ parties that remain to act. 
Formally, a distribution $p(\vec a|\vec x)$ is thus a \textit{causal correlation} if and only if
\begin{align}
    \exists\, p(\pi,\vec a|\vec x) \text{ s.t. } & p(\vec a|\vec x) = \sum_\pi p(\pi,\vec a|\vec x) \notag \\
    \text{and} \ & \forall\, n = 0,\ldots,N-1, \ \forall\, (k_1,\ldots,k_n,k_{n+1}), \notag \\
    & p\big((k_1,\ldots,k_n,k_{n+1}),\vec a_{k_1,\ldots,k_n}\big|\vec x\big) \notag \\
    & \ \ \text{does not depend on } \vec x_{\mathcal{N}\backslash\{k_1,\ldots,k_n\}},
    \label{eq:def_causal_p}
\end{align}
where the marginal distributions are defined according to
\begin{align}
    &p\big((k_1,\ldots,k_{n+1}),\cdot\big|\cdot\big) \coloneqq \sum_{\substack{\pi\text{ s.t. }\pi(1)=k_1,\ldots \\ \qquad \ldots,\pi(n+1)=k_{n+1}}} \!\!p(\pi,\cdot|\cdot) \notag \\
    &= \!\!\sum_{(k_{n+2},\ldots,k_N)} \!\!\!p\big(\pi=(k_1,\ldots,k_{n+1},k_{n+2},\ldots,k_N),\cdot\big|\cdot\big),
\end{align}
and  $p(\cdot,\vec a_{k_1,\ldots,k_n}|\cdot) \coloneqq \sum_{\vec a_{\mathcal{N}\backslash\{k_1,\ldots,k_n\}}} p(\cdot,\vec a|\cdot)$. 

From the general requirement that the causal structure between the parties should be well defined, Refs.~\cite{Oreshkov16,Abbott16} proposed two equivalent definitions of causal correlations $p(\vec a|\vec x)$. The definition of~\cite{Oreshkov16} proposes a decomposition of a causal correlation in terms of components compatible with a strict partial order, while the one of~\cite{Abbott16} proposes a recursive characterisation of causal correlations.
Our definition in Eq.~\eqref{eq:def_causal_p} turns out to also be equivalent to the ones proposed in~\cite{Oreshkov16,Abbott16}, as we prove in Appendix~\ref{app:def_causal_correl} (where we also provide two more possible characterisations of causal correlations). 
Our approach based on the compatibility with a total ordering of the parties will prove useful to investigate the notion of causal correlations with dynamical causal order, and somewhat simpler to deal with than with strict partial orders (as considered in~\cite{Oreshkov16}).

As any convex combination of causal correlations is a causal correlation, the set of causal correlations is convex. As noted in~\cite{Oreshkov16, branciard15}, this set (for any given scenario, i.e.\ for any given number of parties, and of inputs and outputs for each party) forms a convex polytope ${\cal P}_\textup{causal}$, called the causal polytope, that clearly contains ${\cal P}_\textup{convFO}$. Its vertices correspond to deterministic strategies, i.e., correlations for which the probabilities $p(\vec a|\vec x)$ (as well as $p(\pi,\vec a|\vec x)$ in their causal decomposition) only take the values 0 and 1. In principle one can list all such vertices, and use standard techniques to then enumerate the facets of the causal polytope~\cite{ziegler95,branciard15,Abbott16}, which define so-called causal inequalities---although this becomes computationally quite demanding when the numbers of parties, of inputs and outputs go beyond the simplest cases.

An especially simple scenario of particular interest, which we will focus on in this paper, is the so-called ``lazy'' scenario, in which each party $A_k$ has a binary input $x_k$ (we will take $x_k=0,1$), a binary output for one of the two inputs ($a_k=0,1$ if $x_k=1$), and a single fixed output for the other input ($a_k=0$ if $x_k=0$). The bipartite and tripartite causal polytopes were fully characterised in the lazy scenario in Refs.~\cite{branciard15,Abbott16} (while for the more usual scenario with binary inputs and outputs, this could only be done for $N=2$~\cite{branciard15}). It was found in particular that in the bipartite lazy scenario, the causal polytope has only one type of nontrivial facet inequality,%
\footnote{More specifically, in that scenario ${\cal P}_\textup{causal}$ has 4 nontrivial facets: the other ones are obtained by flipping the outputs $a$ and/or $b$ when $x$ and/or $y=1$. Additionally, it has 8 trivial facets of the form $p(a,b|x,y)\ge 0$.}
which can be interpreted as a bound on the maximal probability of success in a causal game, the \textit{Lazy Guess Your Neighbour's Input} (LGYNI) game~\cite{branciard15}. 
The goal of this game is for $A$ and $B$ to output the input bit of the other party, but only when their input is 1; otherwise, the lazy scenario imposes that they must output 0. In short, $A$ and $B$ win the game if they both output $a=b=xy$. If one assumes the input bits to be uniformly distributed, the causal bound on the success probability for the LGYNI game is\footnote{Here, and throughout this work, we use the short-hand notation that implicitly assumes uniform inputs, so that $p(a=b=xy) = \sum_{x,y,a,b=0,1}p(x,y)\delta_{a,xy}\delta_{b,xy}p(a,b|x,y)=\frac{1}{4}\sum_{x,y,a,b=0,1}\delta_{a,xy}\delta_{b,xy}p(a,b|x,y)$, where $\delta$ denotes the Kronecker delta.}
\begin{equation}
    p_{\textup{LGYNI}} \coloneqq p(a=b=xy) \underset{\text{causal}}{\leq} \frac{3}{4}.
    \label{eq:LGYNI}
\end{equation}
This bound captures the intuition that if the causal structure between $A$ and $B$ is well-defined, then the party that acts last can guess the input of the first party with certainty, while the party that comes first cannot do better than output a random guess for the input of the second party. Given that the latter is asked to guess correctly that input in half of the cases (when receiving input 1), then overall they have at least a $\frac12\times\frac12=\frac14$ probability of losing at the LGYNI game.

\subsection{Static vs.\ dynamical order}

As introduced above, for $N \ge 3$ parties, some causal correlations are not just convex mixtures of fixed causal orders. The correlations with ``static'', or ``non-dynamical'' causal order, are those that are convex mixtures of fixed causal orders, and that define ${\cal P}_\textup{convFO}$.

To illustrate the difference between static and dynamical causal order, we consider the following tripartite scenario. One of the three parties (here party $A$) has a binary input but no output (or fixed output taking a single value), and the two other parties are in the lazy scenario.  
Assuming uniform inputs, i.e.\ $p(x,y,z)=\frac18$, we introduce the expression
\begin{align}
    I_3 \coloneqq \ & p\big(x=0, c=yz\big) + p\big(x=1, b=yz\big). \label{eq:def_I3}
\end{align}
This equation can be interpreted as the success probability of a causal game where, depending on the input $x=0$ or 1 of $A$, either $C$ should guess the input of $B$ (when they get the input $z=1$), or vice-versa (when $y=1$).

As it turns out, one obtains the following bound on $I_3$, assuming that $p(b,c|x,y,z)$ should be a convex mixture of fixed causal order correlations:
\begin{align}
    I_3 \underset{\text{convFO}}{\leq} \frac{7}{8}. \label{eq:bnd_I3_FO}
\end{align}
This bound can be understood intuitively in a similar way to that of Eq.~\eqref{eq:LGYNI} above. A formal proof is given in Appendix~\ref{app:simplest_N3}, where we also note that Eq.~\eqref{eq:bnd_I3_FO} is actually a facet inequality of the polytope ${\cal P}_\textup{convFO}$. The inequality \eqref{eq:bnd_I3_FO} is thus tight, and it is  saturated, e.g., by $p(b,c|x,y,z)=\delta_{b,0}\,\delta_{c,yz}$, which is compatible with the causal order $A \prec B \prec C$.

For general causal correlations however, we have the larger (trivial) bound
\begin{align}
    I_3 \underset{\text{causal}}{\leq} 1,
\end{align}
which can be reached, e.g., by
\begin{align}
    p(b,c|x,y,z)=\left\{\begin{array}{ll}
    \delta_{b,0}\,\delta_{c,yz} & \text{if}\ x=0 \\
    \delta_{c,0}\,\delta_{b,yz} & \text{if}\ x=1 
\end{array}\right.. \label{eq:pbc_dynamical}
\end{align}
This correlation is indeed causal, as it satisfies the characterisation of Eq.~\eqref{eq:def_causal_p}, with a decomposition given by
\begin{align}
    p(\pi,b,c|x,y,z)=\left\{\begin{array}{ll}
    \delta_{\pi,(A,B,C)}\,\delta_{b,0}\,\delta_{c,yz} & \text{if}\ x=0 \\
    \delta_{\pi,(A,C,B)}\,\delta_{c,0}\,\delta_{b,yz} & \text{if}\ x=1 
\end{array}\right.. 
\end{align}
Hence, there is clearly a gap between convex mixtures of causal orders and general causal correlations. Causal correlations that are not convex mixtures of fixed orders are said to have \emph{dynamical} causal order.

Note that in the specific example above,
\begin{align}
    p(\pi|x,y,z)=\left\{\begin{array}{ll}
    \delta_{\pi,(A,B,C)} & \text{if}\ x=0 \\
    \delta_{\pi,(A,C,B)} & \text{if}\ x=1 
\end{array}\right..
\end{align}
 Hence, here the causal order is determined, or ``influenced'' by the input $x$. As we shall see, for $N\ge 4$ parties there can also be dynamical causal order with no such influence.

\subsection{Dynamical but non-influenceable orders}
\label{subsec:NIO}

We now study in more detail causal correlations with dynamical causal order. Inspired by the last comment above, we formalise the notion of causal correlations with (non\nobreakdash-)influenceable causal order in order to compare it to the notion of causal correlations with (non\nobreakdash-)dynamical causal order.

Intuitively, we say that a causal correlation has a non-influenceable causal order between the parties if none of the choices of inputs for the parties can influence the causal order between them. 
This can be understood as requiring that the hidden variable $\pi$ describing the causal order that unfolds in causal correlations (cf.\ Eq.~\eqref{eq:def_causal_p}) is independent of the inputs $\vec x$.
Formally, we thus define \emph{causal correlations with non-influenceable orders} as those that admit a causal decomposition in terms of $p(\pi,\vec a|\vec x)$ as in Eq.~\eqref{eq:def_causal_p}, which further satisfies
\begin{align}
    \forall\, \pi, \ p(\pi|\vec x)\ \text{does not depend on } \vec x, \label{eq:def_NIO}
\end{align}
where $p(\pi|\vec x) \coloneqq \sum_{\vec a} p(\pi,\vec a|\vec x)$. We emphasise that this definition does not forbid, in general, the existence of another causal decomposition with $p'(\pi|\vec x) \neq p'(\pi|\vec x')$ for some $\vec x, \vec x', \pi$. 
We note also that one could equivalently impose a similar condition starting from a causal decomposition as defined in the language of~\cite{Oreshkov16} (in terms of strict partial orders rather than total orderings), see Appendix~\ref{app:subsec:equiv_def_NIO}.

\begin{figure}[t]
\centering
 \includegraphics[width=0.8\columnwidth]{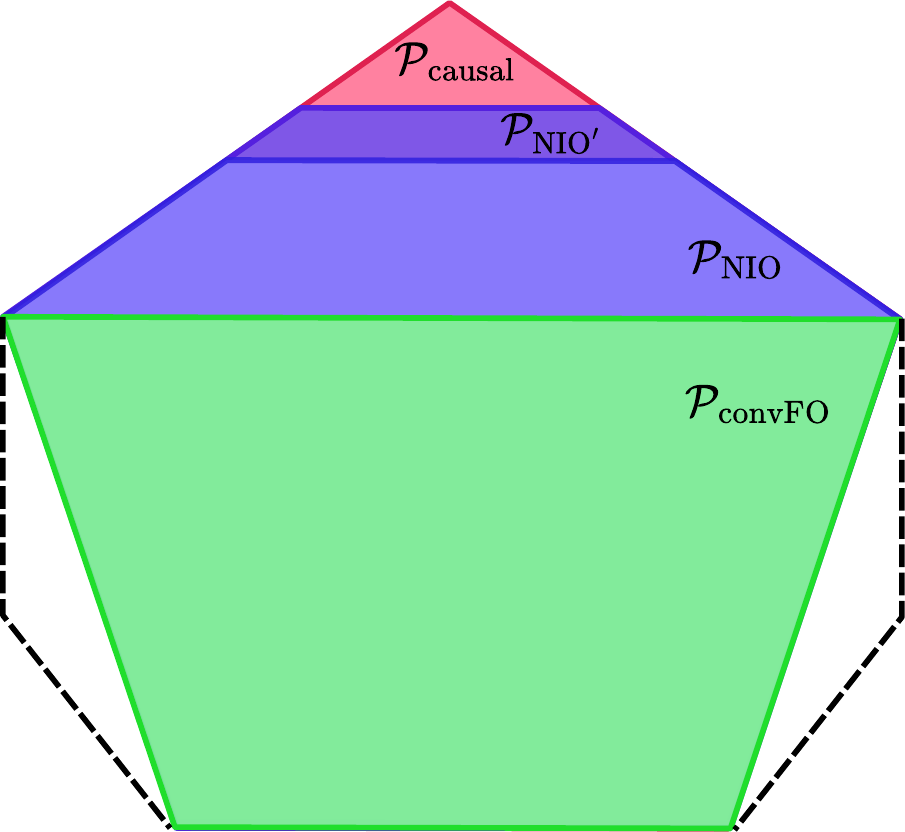}
\caption{Illustration of the inclusion relations between the polytopes: ${\cal P}_\textup{convFO} \subset {\cal P}_\textup{NIO} \subset {\cal P}_{\textup{NIO}'} \subset {\cal P}_\textup{causal}$. In general (for $N\ge 4$) the inclusions are strict; for $N\le 3$ however, ${\cal P}_\textup{NIO}$ and ${\cal P}_{\textup{NIO}'}$ reduce to ${\cal P}_\textup{convFO}$; and for $N \le 2$, they all coincide. The polytopes are also all (in general strictly) included in the polytope of all correlations (all valid probability distributions), illustrated by the dashed lines.}
\label{fig:polytopes}
\end{figure}

The set of correlations $p(\pi,\vec a|\vec x)$ that satisfy Eq.~\eqref{eq:def_causal_p} is a bounded probability space that is linearly constrained, and thus defines a convex polytope. 
The set of causal correlations with non-influenceable orders can then be obtained as a projection of the latter polytope (projecting out the unobserved variable $\pi$), and hence also defines a convex polytope. 
This polytope, which we call ${\cal P}_\textup{NIO}$, is included in the causal polytope ${\cal P}_\textup{causal}$ and moreover clearly contains the polytope ${\cal P}_\textup{convFO}$.
The relation between these sets is shown in Fig.~\ref{fig:polytopes}; we will see in the following section that the inclusions are indeed generally strict.
Characterising explicitly the vertices and facet inequalities of the polytope ${\cal P}_\textup{NIO}$ is somewhat more complicated than for ${\cal P}_\textup{convFO}$ or ${\cal P}_\textup{causal}$. While for these latter polytopes the vertices are deterministic and can be enumerated relatively directly, this is not the case for the vertices of ${\cal P}_\textup{NIO}$. 
Instead, the polytope of correlations $p(\pi,\vec a|\vec x)$ involving also $\pi$ is defined by Eqs.~\eqref{eq:def_causal_p} and~\eqref{eq:def_NIO} along with nonnegativity and normalisation.
Obtaining an explicit characterisation of ${\cal P}_\textup{NIO}$ is thus computationally costly. One can either first transform this higher-dimensional polytope into its vertex representation before projecting out $\pi$ to obtain the vertices of ${\cal P}_\textup{NIO}$, which are in general not deterministic, or adopt variable elimination methods in order to compute directly facet inequalities bounding ${\cal P}_\textup{NIO}$~\cite{williams86}.
Nonetheless, one can still test whether a given correlation is inside the polytope or not by using linear programming to search explicitly for a causal decomposition satisfying Eq.~\eqref{eq:def_NIO}.

Before moving on to comparing the different sets of correlations introduced so far, let us introduce one more, which will prove relevant when analysing the correlations produced by certain quantum circuits with quantum control of causal order, see Sec.~\ref{sec:QCQC_all}. As we will recall, QC-QCs involve a control system which, at any given time, only encodes the unordered set $\K_{n-1}$ of parties that have already acted, together with the label $k_n$ of the party that acts at that given time. Looking back at causal correlations and at their decomposition in terms of $p(\pi,\vec a|\vec x)$, this suggests it may be relevant in that context to consider the probabilities that the first $n-1$ parties are those in $\K_{n-1}$, and that the $n^\text{th}$ party is $k_n$, i.e.\ the ``coarse-grained'' probabilities $p((\mathcal{K}_{n-1},k_n)|\vec x) \coloneqq \sum_{(k_1,\ldots,k_{n-1}) \in \mathcal{K}_{n-1}} p((k_1,\ldots,k_{n-1},k_n)|\vec x)$. In a similar way to how we introduced correlations with non-influenceable order, let us then define \emph{causal correlations with non-influenceable coarse-grained orders} as those that admit a causal decomposition in terms of $p(\pi,\vec a|\vec x)$ as in Eq.~\eqref{eq:def_causal_p}, which further satisfies%
\footnote{Notice that one could have equivalently defined correlations with non-influenceable order by imposing that for any $n=1,\ldots,N$, $p((k_1,\ldots,k_n)|\vec x)$ does not depend on $\vec x$---which would be slightly closer to Eq.~\eqref{eq:def_NIO'}. In that case however, considering $n=N$ was enough; for correlations with non-influenceable coarse-grained orders on the other hand, imposing the corresponding  constraint only for $n=N$ would in general define a larger set of correlations.}
\begin{align}
    & \forall\, \mathcal{K}_{n-1} \subsetneq \mathcal{N}, \forall\, k_n \in \mathcal{N}\backslash \mathcal{K}_{n-1}, \notag \\
    & \qquad p\big((\mathcal{K}_{n-1},k_n)\big|\vec x\big)\ \text{does not depend on } \vec x. \label{eq:def_NIO'}
\end{align}
As for the set of correlations with non-influenceable orders, the set of causal correlations with non-influenceable coarse-grained orders forms a convex polytope, which we denote $\mathcal{P}_{\textup{NIO}'}$, that includes $\mathcal{P}_{\textup{NIO}}$ and is included in $\mathcal{P}_{\textup{causal}}$. Characterising this polytope in terms of vertices (which, in general, are not deterministic) and facets is of a similar difficulty as for $\mathcal{P}_{\textup{NIO}}$.

It turns out that for $N \le 3$, correlations with non-influenceable orders, as well as correlations with non-influenceable coarse-grained orders, are nothing but convex mixtures of fixed orders, i.e.\ ${\cal P}_\textup{NIO}$  and ${\cal P}_{\textup{NIO}'}$ reduce to ${\cal P}_\textup{convFO}$ (see the proof in Appendix~\ref{app:NIO_convFO_N3}). 
Thus, the notions of dynamicality and influenceability coincide in that case. 
However, we will see that as soon as $N\ge 4$, these form genuinely distinct classes of correlations.

\subsection{Separation between the four polytopes of correlations}

To show that the classes of correlations introduced above indeed form a strict hierarchy, let us consider the fourpartite case with all parties in the lazy scenario, and let us introduce
\begin{align}
    I_4 \coloneqq \ & p\big(a=b=xy , d=zt\big) \notag \\
    & + p\big(\neg(a=b=xy) , c=zt\big). \label{eq:def_I4}
\end{align}
This expression can be understood as the success probability of a new causal game involving all four parties, defined as follows: $A$ and $B$ play the LGYNI game, as presented before Eq.~\eqref{eq:LGYNI}; depending on whether they win or lose the LGYNI game, either $D$ must guess (when receiving the input 1) the input of $C$, or vice-versa.

One can readily see the connection with the expression of $I_3$, Eq.~\eqref{eq:def_I3}: instead of looking at how the direction of signalling between the last two parties correlates with the input of the first party, as in $I_3$, here we look at how that direction of signalling correlates with the success or failure in the LGYNI game between the first two parties---which, if these are causally connected, cannot be predetermined since the LGYNI cannot be won, nor lost,%
\footnote{Notice that the LGYNI game is always won when the inputs are $x=y=0$, so $p_{\textup{LGYNI}}\ge\frac14$ always holds.}
with certainty.

As we show in Appendix~\ref{app:proofs_bounds}, the bounds on $I_4$ are different for the different sets of correlations introduced above, namely:%
\footnote{We note that neither Eq.~\eqref{eq:bnd_I4_FO}, nor Eq.~\eqref{eq:bnd_I4_causal}, are facets of the corresponding polytopes: we find that the vertices that saturate them span affine subspaces of dimensions 55 and 33, respectively, while the dimension of the total correlation space in the 4-partite lazy scenario, and of the four polytopes under consideration here, is 65. We also do not expect Eqs.~\eqref{eq:bnd_I4_NIO} and~\eqref{eq:bnd_I4_NIO'} to define facet inequalities, although we could not check this because of the difficulty of enumerating the vertices of $\mathcal{P}_{\textup{NIO}}$ and $\mathcal{P}_{\textup{NIO}'}$. \\
In Appendix~\ref{app:violations} we present an inequality that defines a facet of $\mathcal{P}_{\textup{convFO}}$. Obtaining more facet inequalities for the various polytopes---possibly inequalities that can be given nice interpretations in terms of causal games---is left for future work.}
\begin{itemize}
    \item For convex mixtures of fixed orders: 
        \begin{align}
            I_4 \underset{\text{convFO}}{\leq} \frac{15}{16}. \label{eq:bnd_I4_FO}
        \end{align}
        This bound is reached for instance by the correlation (compatible with any fixed order where $C \prec D$) $p(a,b,c,d|x,y,z,t) = \delta_{a,0}\,\delta_{b,0}\,\delta_{c,0}\,\delta_{d,zt}$.
    \item For non-influenceable orders:
        \begin{align}
            I_4 \underset{\text{NIO}}{\leq} \frac{31}{32}. \label{eq:bnd_I4_NIO}
        \end{align}
        This bound is reached for instance by the correlation presented in Eq.~\eqref{eq:saturate_NIO} of Appendix~\ref{app:proof_NIO_bound}.
    \item For non-influenceable coarse-grained orders:
        \begin{align}
            I_4 \underset{\text{NIO}'}{\leq} \frac{47}{48}. \label{eq:bnd_I4_NIO'}
        \end{align}
        This bound is reached for instance by the correlation presented in Eq.~\eqref{eq:saturate_NIO'} of Appendix~\ref{app:proof_NIO'_bound}.
       
    \item For general causal correlations, one has the trivial bound:
        \begin{align}
            I_4 \underset{\text{causal}}{\leq} 1. \label{eq:bnd_I4_causal}
        \end{align}
        This bound is reached for instance by
        \begin{align}
            \hspace{-2mm} p(a,b,c,d|x,y,z,t) = \delta_{a,0}\,\delta_{b,0} \left\{\begin{array}{ll}
    \!\delta_{c,0}\,\delta_{d,zt} & \text{if}\ xy=0 \\
    \!\delta_{d,0}\,\delta_{c,zt} & \text{if}\ xy=1
\end{array}\right.\!\!, \label{eq:correl_I4_1}
        \end{align}
        where the order between $C$ and $D$ depends on the product of inputs $xy$.
\end{itemize}

The fact that the bounds above are different implies that there is a clear separation between the different types of correlations and the corresponding polytopes, as depicted in Fig.~\ref{fig:polytopes}:
\begin{align}
    \mathcal{P}_{\textup{convFO}} \ \subset \ \mathcal{P}_{\textup{NIO}} \ \subset \  \mathcal{P}_{\textup{NIO}'} \ \subset \  \mathcal{P}_{\textup{causal}},
    \label{eq:hierarchy_polytopes}
\end{align}
where the inclusions are in general all strict for $N \ge 4$.
In particular, we see that correlations with static causal order (in $\mathcal{P}_{\textup{convFO}}$) are only a restricted case of correlations with non-influenceable order: one can find correlations with non-influenceable order that are not just convex mixtures of correlations with fixed orders. In other words, there exist correlations with \emph{dynamical but non-influenceable order}, i.e.\ dynamicality and influenceability are inequivalent notions in general, despite the fact that these notions have largely been equated in the literature~\cite{Abbott16,Oreshkov16,Tselentis23,Tselentis24,Baumann24}.

\section{Quantum circuits with dynamical causal order}
\label{sec:quantum_circuits}

Now that we have clarified the definition of correlations with dynamical causal order and explored some of their features, let us turn to the quantum processes that may generate such correlations, and see how one may characterise the dynamicality of these processes directly. More specifically, we will focus on quantum circuits, although with a less constrained definition than the usual one: we allow for quantum circuits that do not necessarily have a fixed, or even well-defined causal order. Two important, physically relevant classes of such circuits were introduced in Ref.~\cite{wechs21}: \emph{quantum circuits with classical} or \emph{with quantum control of causal order} (QC-CCs or QC-QCs). The latter includes in particular processes that intuitively feature both indefinite and dynamical causal order, although no clear definition of dynamicality was given so far, that would formalise this intuition. Below we will remedy this by looking at whether the classical or quantum systems used to control the causal order can be fixed from the start, or can be influenced by the choice of the parties' operations.

QC-CCs and QC-QCs are conveniently described within the framework of process matrices~\cite{oreshkov12}; let us start by recalling this formalism.

\subsection{The process matrix formalism}

The process matrix formalism was introduced in Ref.~\cite{oreshkov12} in order to characterise correlations that could be obtained from quantum processes, without imposing \emph{a priori} that the operations giving rise to these correlations are applied according to any specific, well-defined causal structure. By relaxing the assumption of a well-defined global causal order between the different operations, this formalism allows one to go beyond standard, causally ordered quantum circuits, and describe processes with potentially indefinite causal order. 

More formally, one considers a situation with $N$ parties $A_k$ applying some quantum operations, described as quantum instruments~\cite{davies70}, i.e.\ as collections $\{\mathcal{M}^{A_k}_{a_k|x_k}\}_{a_k}$ of completely positive (CP) maps $\mathcal{M}^{A_k}_{a_k|x_k}: \mathcal{L}(\mathcal{H}^{A_k^I}) \to \mathcal{L}(\mathcal{H}^{A_k^O})$ from party $A_k$'s input Hilbert space $\mathcal{H}^{A_k^I}$ to their output Hilbert space $\mathcal{H}^{A_k^O}$, with $\mathcal{L}(\mathcal{H}^{X})$ denoting the space of linear operators acting on $\mathcal{H}^X$ (and restricting ourselves, throughout this paper, to finite-dimensional Hilbert spaces).
Here $x_k$ labels the instrument, and thus describes the choice of a ``setting'' for $A_k$'s operation. Each map $\mathcal{M}^{A_k}_{a_k|x_k}$ in the instrument is associated with a classical output $a_k$, so that for any given (trace 1) input state $\rho\in\mathcal{L}(\mathcal{H}^{A_k^I})$, the probability for that map to be applied and for that output to be produced is $\Tr[\mathcal{M}^{A_k}_{a_k|x_k}(\rho)]$. To ensure that these probabilities sum up to 1, the sum over $a_k$ of all CP maps $\mathcal{M}^{A_k}_{a_k|x_k}$ in a given instrument must be trace-preserving~(TP).

A convenient way to represent such quantum operations, that the process matrix formalism relies on, is through the Choi–Jamiołkowski isomorphism~\cite{jamiolkowski72,Choi}. For any given map $\mathcal{M}: \mathcal{L}(\mathcal{H}^{A^I}) \to \mathcal{L}(\mathcal{H}^{A^O})$, one can define its Choi matrix $M$ as%
\footnote{We often use superscripts on operators and vectors to indicate the spaces they belong to. These may be dropped when the situation is clear enough.}
\begin{align}
    M^{A^{IO}} \coloneqq \sum_{i,j} \ketbra{i}{j}^{A^I}\otimes \mathcal{M}\big(\ketbra{i}{j}^{A^I}\big) \ \in \mathcal{L}\big(\mathcal{H}^{A^{IO}}\big), \label{eq:Choi_def}
\end{align}
where $\{\ket{i}^X\}_i$ generically denotes a fixed orthonormal basis---the ``computational basis''---of $\mathcal{H}^{X}$, and where we use the short-hand notations $\ketbra{i}{j}^X \coloneqq \ket{i}^{X}\bra{j}^X$, $\mathcal{H}^{XY}\coloneqq\mathcal{H}^{X} \otimes \mathcal{H}^{Y}$ and $\mathcal{H}^{A^{IO}} \coloneqq \mathcal{H}^{A^IA^O}$.
With this, a quantum instrument $\{\mathcal{M}^{A_k}_{a_k|x_k}\}_{a_k}$ can equivalently be described by the set of Choi matrices $\{M^{A_k}_{a_k|x_k}\}_{a_k}$, with each $M^{A_k}_{a_k|x_k}\in\L(\HS^{A_k^{IO}})$. The complete positivity of each $\mathcal{M}^{A_k}_{a_k|x_k}$ translates to the positive semidefiniteness of their Choi matrices, $M^{A_k}_{a_k|x_k}\ge 0$, while the trace-preserving property of their sum (for each $x_k$) translates to the property that $\sum_{a_k} \Tr_{A_k^O}M^{A_k}_{a_k|x_k} = \id^{A_k^I}$ (where $\Tr_X$ denotes the partial trace over $\HS^X$, and $\id^Y$ denotes the identity operator on $\HS^Y$).

Ref.~\cite{oreshkov12} then considered the question of what are the most general correlations---i.e., the most general conditional probability distributions $p(\vec a|\vec x)$ of obtaining the outputs $\vec a=(a_1,\ldots,a_N)$ for a choice of settings (of instruments) $\vec x=(x_1,\ldots,x_N)$---that would be consistent with the quantum description of the $N$ parties' local operations, but would not require \emph{a priori} that these operations are applied in any well-defined order. Imposing that the probabilities $p(\vec a|\vec x)$ only depend on the local maps $\mathcal{M}^{A_k}_{a_k|x_k}$ (rather than on the full instruments $\{\mathcal{M}^{A_k}_{a_k|x_k}\}_{a_k}$) and that these probabilities are linear in the local maps (for consistency with probabilistic mixtures and coarse-graining of operations, assuming that each of them is applied once and only once), it was shown that they can most generally be written in the following form---via the ``generalised Born rule'':
\begin{align}
    p(\vec a|\vec x) = \Tr [\Big(\bigotimes_k M^{A_k}_{a_k|x_k}\Big)^{\!T} \, W],
    \label{eq:Born_rule}
\end{align}
where $W \in \mathcal{L}(\mathcal{H}^{A_{\mathcal{N}}^{IO}})$ (with $\mathcal{H}^{A_\mathcal{N}^{IO}}\coloneqq\bigotimes_{k\in\mathcal{N}}\mathcal{H}^{A_k^{IO}}$) is the so-called ``process matrix''. 
In order to only generate nonnegative and normalised probabilities through Eq.~\eqref{eq:Born_rule}, for any actions of the parties (even when the latter share auxiliary, possibly entangled systems and their local operations are allowed to also act on these), the process matrix must be positive semidefinite (PSD) and must satisfy certain linear constraints~\cite{oreshkov12,Araujo15}, which we recall in Appendix~\ref{app:subsec_validity_cstr}. If that is the case, we say that the process matrix is \emph{``valid''}.
 
The process matrix framework has also been extended to describe how $N$ quantum operations may be combined so as to generate a new ``global'' quantum map (rather than probabilities, i.e.\ scalar values), from some ``past'' Hilbert space $\HS^P$ to some ``future'' Hilbert space $\HS^F$---again without assuming \emph{a priori} that the operations are composed in any well-defined order~\cite{araujo17}. In that case the process matrix lives in $\mathcal{L}(\mathcal{H}^{PA_{\mathcal{N}}^{IO}F}) = \mathcal{L}(\HS^P\otimes\mathcal{H}^{A_{\mathcal{N}}^{IO}}\otimes\HS^F)$ and acts as a ``supermap'': a map that transforms quantum maps to a new quantum map~\cite{Chiribella08supermaps}. Specifically, the Choi matrix of the resulting global map is given through a further generalisation of the Born rule above:
\begin{align}
    M_{\vec a|\vec x}^{PF} = \Tr_{A_\N^{IO}} \!\left[\Big[\id^{\!P}\!\!\otimes\!\Big(\!\bigotimes_k M^{A_k}_{a_k|x_k}\Big)^{\!T} \!\!\otimes\!\id^{\!F} \Big] W\right] \in \L(\HS^{PF})
    \label{eq:Born_rule_PF}
\end{align}
(which indeed recovers Eq.~\eqref{eq:Born_rule} when $\HS^P, \HS^F$ are ``trivial'', i.e.\ 1-dimensional spaces). When the local input maps (possibly extended to auxiliary systems) are CP, then the resulting global map is required to also be CP. When the local input maps are TP (e.g.\ considering the sum of the parties' instrument elements), then the resulting global map is required to also be TP. This implies that the process matrix must still be PSD and satisfy certain linear constraints (also recalled in Appendix~\ref{app:subsec_validity_cstr}), for it to be \emph{valid}.%
\footnote{We recall however that the two versions of the process matrix formalism, with ``closed'' or ``open'' past and future spaces $\HS^P, \HS^F$, are in fact equivalent~\cite{araujo17,wechs21}. Indeed, the past Hilbert space $\HS^P$ (the future Hilbert space $\HS^F$, respectively) can be considered as the output space (the input space, resp.) of an additional party with no input space (no output space, resp.). \label{ftn:equiv_PF}}

By just requiring that only valid probabilities, or only valid CP(TP) maps are obtained through Eq.~\eqref{eq:Born_rule} or~\eqref{eq:Born_rule_PF}, the ``top-down'' approach of the framework allows for quite general forms of process matrices, including some for which no clear physical interpretation is known~\cite{oreshkov12,Baumeler16}. A more constructive, bottom-up approach was taken in Ref.~\cite{wechs21} to describe (using the same formalism) certain processes with a clearer physical understanding. This led to the introduction of the subclasses of quantum circuits with classical or quantum control of causal order. Interestingly these subclasses can be characterised in terms of linear constraints on PSD matrices, as we will recall below.

\subsection{Quantum Circuits with Classical Control of causal order}
\label{sec:subQC-CC}

Let us start with the case of quantum circuits with classical control of causal order (QC-CCs). We will first recall their general form and then introduce two subclasses so as to characterise the dynamicality of their causal order.

\subsubsection{The general \textup{\textsf{QC-CC}} class}

\begin{figure*}[t]
\centering
\includegraphics[width=2\columnwidth]{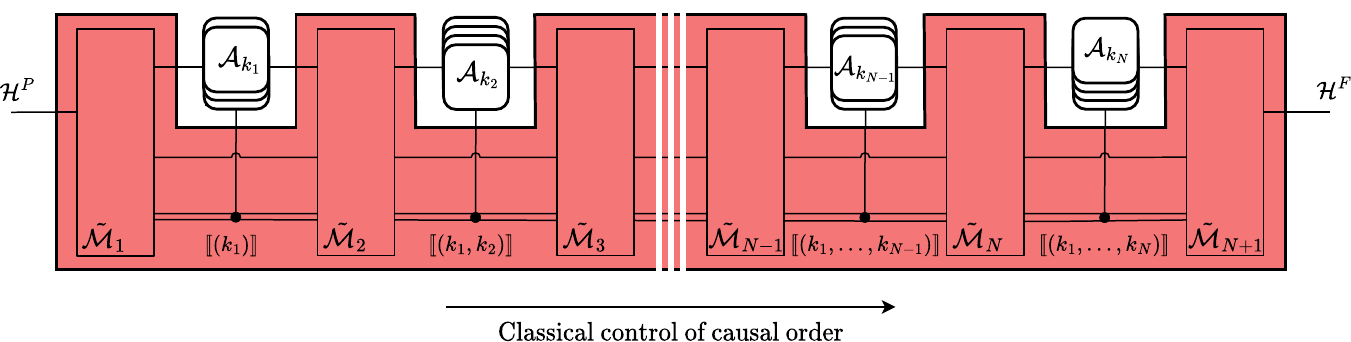}
\caption{Graphical representation of a quantum circuit with classical control (QC-CC) of causal order (a simplified version of Fig.~9 from Ref.~\cite{wechs21}, also reproduced with more details as Fig.~\ref{fig:QCCC_graph_full} in Appendix~\ref{app:subsec_QCCC_classes}).
The QC-CC is defined as the composition of the ``internal'' operations $\tilde{\mathcal{M}}_1,\ldots,\tilde{\mathcal{M}}_{N+1}$ (shown in the red shaded area), leaving ``open slots'' for the parties' ``external'' operations $\A_k$---or the CP maps $\mathcal{M}^{A_k}_{a_k|x_k}$, as we denote them in the main text, for a choice of setting $x_k$ and associated with a classical output $a_k$ (not shown on the figure).
Each internal operation $\tilde\M_n$ (for $n\le N$) determines which party $A_{k_{n}}$ should act next; it transmits a target system from the previous party $A_{k_{n-1}}$ to the next party $A_{k_{n}}$ (or from the global past space $\HS^P$ to $A_{k_1}$ for $\tilde\M_1$), potentially along with an auxiliary system (mid-height line on the figure), and updates the state of the classical control system (bottom double-stroke line) into the state $[\![(k_1,\ldots,k_{n-1},k_{n})]\!]\coloneqq \ketbra{(k_1,\ldots,k_{n-1},k_n)}{(k_1,\ldots,k_{n-1},k_n)}$, which keeps track of the ordered list  of all external operations used so far. The last internal operation $\tilde\M_{N+1}$ just outputs a state in the global future space $\HS^F$ (when the latter is nontrivial). According to this description, the causal order between the external operations is thus established on the fly or dynamically, as each internal operation is applied.}
\label{fig:QCCC_graph}
\end{figure*}

QC-CCs were introduced in~\cite{wechs21}, in an attempt to construct the most general class of processes with a well-defined causal order. The causal order of a QC-CC is not necessarily predetermined: it is in general established on the fly, and can hence be dynamical. More precisely, the action of a QC-CC can be understood as a protocol that alternatively applies ``internal'' and ``external'' operations, the latter corresponding to the quantum operations chosen by the different parties $A_k$. As depicted in Fig.~\ref{fig:QCCC_graph}, a first internal operation $\tilde{\mathcal{M}}_1$ acts on the system entering the circuit in the ``past'' Hilbert space $\HS^P$ (if any: the past system could also be trivial) and determines which party $A_{k_1}$ will apply its external operation first. Then each subsequent internal operation $\tilde{\mathcal{M}}_n$ determines (potentially, probabilistically) which party $A_{k_n}$ will be the next to apply its external operation, conditioned on which previous external operations have already been applied (to ensure that each of them is applied once and only once), and in which order.
This conditioning can be physically encoded in a control system that acts classically, with its state of the form $[\![(k_1,\ldots,k_{n})]\!]\coloneqq \ketbra{(k_1,\ldots,k_n)}{(k_1,\ldots,k_n)}$ being updated to  $[\![(k_1,\ldots,k_{n},k_{n+1})]\!]$ as the next party $A_{k_{n+1}}$ is chosen---thereby dynamically storing the causal order of the parties as it gets established.
The QC-CC under consideration is then defined as the composition of the internal maps $\tilde{\mathcal{M}}_1,\ldots,\tilde{\mathcal{M}}_{N+1}$, leaving ``open slots'' for the parties' external operations; see Appendix~\ref{app:QC-CC} and Ref.~\cite{wechs21} for more details on this construction. Notice that different internal maps may define the same QC-CC (the same ``supermap'') after being composed; different choices of internal maps can be thought of as defining different ``implementations'' of a given QC-CC.

We denote by \textup{\textsf{QC-CC}} the set of process matrices that describe QC-CCs.
As shown in~\cite{wechs21}, this set can be characterised as follows: $W \in \mathcal{L}(\mathcal{H}^{PA_\mathcal{N}^{IO}F})$ is in \textup{\textsf{QC-CC}} iff $W \ge 0$, and for all $n = 1, \ldots, N$ and all $(k_1,\ldots,k_n)$, there exist PSD matrices $W_{(k_1,\ldots,k_N,F)} \in {\cal L}(\HS^{PA_{{\cal N}}^{IO} F})$ and $W_{(k_1,\ldots,k_n)} \in {\cal L}(\HS^{PA_{\{k_1,\ldots,k_{n-1}\}}^{IO} A_{k_n}^I})$ such that
\begin{align}
    \left\{
    \begin{array}{l}
        W = \sum_{(k_1,\ldots,k_N)} W_{(k_1,\ldots,k_{N},F)}, \\[3mm]
        \forall \, (k_1, \ldots, k_N), \ \Tr_F W_{(k_1,\ldots,k_N,F)} = W_{(k_1,\ldots,k_N)} \otimes \id^{A_{k_N}^O}, \\[3mm]
        \forall \, n = 1, \ldots, N-1, \ \forall \, (k_1, \ldots, k_n), \\[1mm]
        \quad \sum_{k_{n+1}} \Tr_{A_{k_{n+1}}^I} \! W_{(k_1,\ldots,k_n,k_{n+1})} = W_{(k_1,\ldots,k_n)} \otimes \id^{A_{k_n}^O}, \\[3mm]
        \sum_{k_1 \in {\cal N}} \Tr_{A_{k_1}^I} W_{(k_1)} = \id^P.
    \end{array}
    \right. \label{eq:charact_W_QCCC_decomp}
\end{align}

As these are linear and affine constraints on PSD matrices, the existence of such a decomposition for a matrix $W$ can be (for small enough dimensions) efficiently and reliably tested via semidefinite programming (SDP)~\cite{Skrzypczyk2023}.
Furthermore, given the various matrices $W_{(k_1,\ldots,k_N,F)}$ and $W_{(k_1,\ldots,k_n)}$ that appear in the decomposition above, one can explicitly reconstruct (in a non-unique manner) the internal operations that, composed together, give the QC-CC. These same two remarks apply to all the classes that will be defined below; see Appendix~\ref{app:sec_classes} and \cite{wechs21} for more details.

Since QC-CCs are processes with a well-defined causal order, they clearly only produce causal correlations in $\mathcal{P}_{\textup{causal}}$ through the generalised Born rule of Eq.~\eqref{eq:Born_rule}~\cite{wechs21}, as we illustrate in Fig.~\ref{fig:QCCC-classes}. In Appendix~\ref{app:corr_QC-CC} we provide an explicit proof of this, which uses the definition of causal correlations given in Eq.~\eqref{eq:def_causal_p}, and uses the fact that each $W_{(k_1,\ldots,k_N,F)}$ and $W_{(k_1,\ldots,k_n)}$ in the decomposition of Eq.~\eqref{eq:charact_W_QCCC_decomp} relates to the causal order $\pi=(k_1,\ldots,k_N)$ or to the order $(k_1,\ldots,k_n)$ of the first $n$ parties, respectively.

\begin{figure}[t]
\centering
\includegraphics[width=\columnwidth]{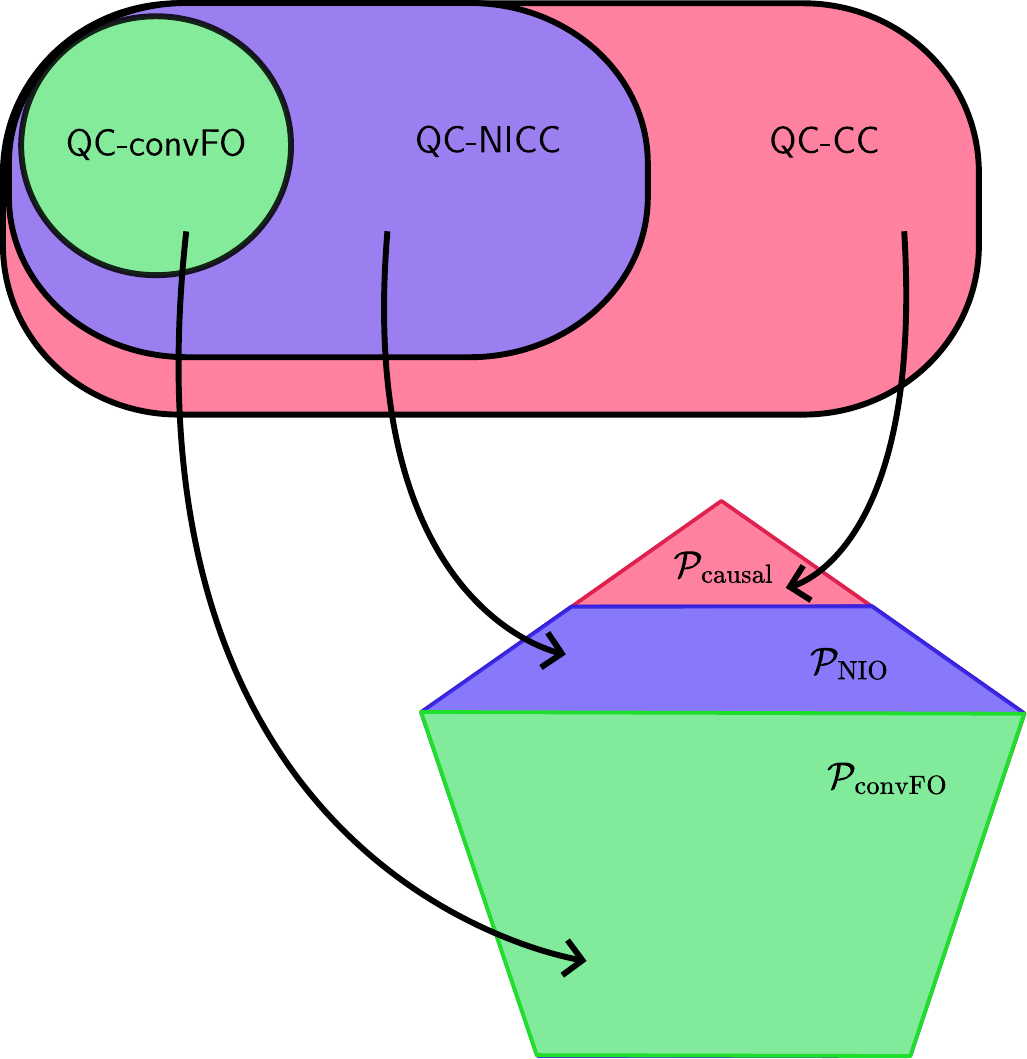}
\caption{The upper part of the figure depicts the inclusion relations between the different (sub)classes of QC-CCs considered in this paper. For $N\ge 4$ (and nontrivial input and output spaces for the different parties) the inclusions are strict; for $N\le3$, \textup{\textsf{QC-NICC}} reduces to \textup{\textsf{QC-convFO}}. \\
The arrows directed to the lower part of the figure indicate the kind of causal correlations (defined through the generalised Born rule, Eq.~\eqref{eq:Born_rule}) that process matrices in each class can produce: process matrices in \textup{\textsf{QC-convFO}} only give correlations in ${\cal P}_\textup{convFO}$; those in \textup{\textsf{QC-NICC}} only give correlations in ${\cal P}_\textup{NIO}$; those in \textup{\textsf{QC-CC}} give causal correlations in ${\cal P}_\textup{causal}$.}
\label{fig:QCCC-classes}
\end{figure}

The connection between the structure of the control system of a QC-CC, which in general is in the intermediate state $[\![(k_1,\ldots,k_n)]\!]\coloneqq \ketbra{(k_1,\ldots,k_n)}{(k_1,\ldots,k_n)}$, and the definition of causal correlations given in Eq.~\eqref{eq:def_causal_p}, which involves the corresponding intermediate probability to obtain the order $(k_1,\ldots,k_{n(+1)})$, suggests that one could define subclasses of $\textup{\textsf{QC-CC}}$ inspired by the classes of correlations introduced above. Indeed, we are going to consider particular QC-CCs by constraining their control systems, first to be fixed from the beginning of the process, analogously to correlations with fixed causal order, and then by requiring their evolution to be unaffected by the choice of the parties' operations, analogously to correlations with non-influenceable causal order.

\subsubsection{The \textup{\textsf{QC-convFO}} subclass}

We thus move on to consider the simplest class of quantum circuits, also called ``quantum channels with memory''~\cite{Kretschmann05}, ``quantum strategies''~\cite{Gutoski06} or ``quantum combs''~\cite{Chiribella08,Chiribella09}, for which the causal order between the external operations is fixed. A quantum circuit with fixed causal order can be seen as a particular QC-CC, where the control system is determined from the beginning, and therefore not dynamical. Furthermore, as in the case of correlations, we consider statistical mixtures of quantum circuits with fixed order as having a static, non-dynamical order.

We thus define QC-CCs with non-dynamical causal order to be those that can be written as \emph{convex mixtures of quantum circuits with fixed causal order} (QC-convFOs), and we denote by \textup{\textsf{QC-convFO}} the set of process matrices describing them.
Extending in a rather straightforward way the characterisation of quantum channels with memory, quantum strategies or quantum combs~\cite{Gutoski06,Chiribella08,Chiribella09} to convex mixtures thereof, one easily obtains the following process matrix characterisation (see Appendix~\ref{app:QC-convFO}): $W \in \mathcal{L}(\mathcal{H}^{PA_\mathcal{N}^{IO}F})$ is in \textup{\textsf{QC-convFO}} iff $W \ge 0$, and for all $(k_1,\ldots,k_N)$ and all $n = 1, \ldots, N$, there exist PSD matrices $W_{(k_1,\ldots,k_N,F)} \in {\cal L}(\HS^{PA_{{\cal N}}^{IO} F})$ and $W_{(k_1,\ldots,k_n)}^{(k_{n+1},\ldots,k_N)} \in {\cal L}(\HS^{PA_{\{k_1,\ldots,k_{n-1}\}}^{IO} A_{k_n}^I})$ such that
\begin{align}
    \left\{
    \begin{array}{l}
        W = \sum_{(k_1,\ldots,k_N)} W_{(k_1,\ldots,k_N,F)}, \\[3mm]
        \forall \, (k_1,\ldots,k_N), \Tr_F W_{(k_1,\ldots,k_N,F)} = W_{(k_1,\ldots,k_N)}^\emptyset \otimes \id^{A_{k_N}^O}, \\[3mm]
        \forall \, n=1,\ldots,N-1, \ \forall \, (k_1,\ldots,k_n,k_{n+1},k_{n+2},\ldots,k_N), \\[1mm]
        \hspace{8mm} \Tr_{A_{k_{n+1}}^I} W_{(k_1,\ldots,k_{n+1})}^{(k_{n+2},\ldots,k_N)} = W_{(k_1,\ldots,k_{n})}^{(k_{n+1},\ldots,k_N)} \otimes \id^{A_{k_n}^O}, \\[3mm]
        \forall \, \pi=(k_1,k_2,\ldots,k_N), \ \Tr_{A_{k_1}^I} W_{(k_1)}^{(k_{2},\ldots,k_N)} = q_\pi \id^P,\\[3mm]
        \hspace{45mm} q_\pi\ge 0, \, \sum_\pi q_\pi = 1.
    \end{array}
    \right. \label{eq:charact_W_QC_convFO_decomp}
\end{align}

Note that each matrix $W_{(k_1,\ldots,k_N,F)}$ and $W_{(k_1,\ldots,k_n)}^{(k_{n+1},\ldots,k_N)}$ relates to a specific fixed order $(k_1,\ldots,k_n,k_{n+1},\ldots,k_N)$ in the convex mixture. 
Each $W_{(k_1,\ldots,k_n)}^{(k_{n+1},\ldots,k_N)}$ describes the reduced (valid, up to normalisation) process matrix for the first $n$ parties, for that specific order. The notation we use keeps track explicitly of the full order, by displaying those first $n$ parties in the subscript, and the (ordered) list of remaining parties in the superscript.

It can be seen that as soon as $N\ge 2$ with a nontrivial $\HS^P$, or as $N\ge 3$ if $\HS^P$ is trivial (see Appendix~\ref{app:subsubsec_QCCC_inclusions}), \textup{\textsf{QC-convFO}} is in general (at least, when the different parties have nontrivial input/output spaces) a strict subclass of \textup{\textsf{QC-CC}}. Furthermore, as QC-convFOs are convex mixtures of processes compatible with a fixed causal order, it is rather straightforward to see that the correlations they create are convex mixtures of correlations compatible with a fixed causal order, in $\mathcal{P}_{\textup{convFO}}$. See Fig.~\ref{fig:QCCC-classes} for an illustration.

\subsubsection{The \textup{\textsf{QC-NICC}} subclass}

\label{sec:QC-NICC}

In analogy to the notion of causal correlations with non-influenceable orders introduced in the previous section, we now define \emph{quantum circuits with non-influenceable classical control of causal order} (QC-NICCs).

Recall that in the case of correlations, we imposed that the probability distribution of the underlying variable $\pi$ describing the causal order should not depend on the choice of the parties' settings, see Eq.~\eqref{eq:def_NIO}. Given how, in a QC-CC, the causal order is encoded in the classical control system, we similarly impose here that (at least for some implementation of the QC-NICC, i.e.\ for some decomposition in terms of internal operations) the state of the control system should not depend on what TP operations the parties implement---nor on the initial state that may be fed into the circuit, as one may consider it to be influenced by some other party in the global past. More formally:
\begin{align*}
  \parbox{0.85\linewidth}{%
    \emph{QC-NICCs are the QC-CCs for which there exists an implementation such that, at any intermediate time step, the state of the control system is independent of the choice of previously applied external operations and of the potential initial state preparation in the global past.}%
  }
\end{align*}

We denote by \textup{\textsf{QC-NICC}} the set of process matrices that describe QC-NICCs. We show in Appendix~\ref{app:def_QC-NICC} that the above constraint implies the following characterisation: $W \in \mathcal{L}(\mathcal{H}^{PA_\mathcal{N}^{IO}F})$ is in \textup{\textsf{QC-NICC}} if and only if it has a \textup{\textsf{QC-CC}} decomposition in terms of PSD matrices $W_{(k_1,\ldots,k_N,F)}$ and $W_{(k_1,\ldots,k_n)}$ as in Eq.~\eqref{eq:charact_W_QCCC_decomp}, with the additional constraint that
\begin{align}
    & \begin{array}{r}
        \text{all} \ W_{(k_1,\ldots,k_n)} \text{'s are valid process matrices} \\
        \text{(up to normalisation)}.
    \end{array} \label{eq:cstr_NICC}
\end{align}
We note that it suffices to impose that all $W_{(k_1,\ldots,k_N)}$'s are valid process matrices (up to normalisation), or that all $W_{(k_1,\ldots,k_N,F)}$'s are valid process matrices (up to normalisation), to imply (through Eq.~\eqref{eq:charact_W_QCCC_decomp}) the validity of all $W_{(k_1,\ldots,k_n)}$'s. 
We also emphasise that Eq.~\eqref{eq:cstr_NICC} does not imply that the $W_{(k_1,\ldots,k_n)}$'s must be compatible with the causal order $(k_1,\ldots,k_n)$, otherwise \textup{\textsf{QC-NICC}} would reduce to \textup{\textsf{QC-convFO}}; see the example below for an illustration.

As we show in Appendix~\ref{app:subsubsec_QCCC_inclusions}, for $N\le 3$ \textup{\textsf{QC-NICC}} reduces to \textup{\textsf{QC-convFO}}. For $N\ge 4$ however, we obtain a new class in between
\textup{\textsf{QC-convFO}} and \textup{\textsf{QC-CC}}, as depicted in Fig.~\ref{fig:QCCC-classes}:
\begin{align}
    \text{\textup{\textsf{QC-convFO}}} \ \subset \ \text{\textup{\textsf{QC-NICC}}} \ \subset \ \text{\textup{\textsf{QC-CC}}}. \label{eq:inclusions_QC-CCs}
\end{align}
The situation here is somewhat similar to the case of correlations, where we observed that ${\cal P}_\textup{NIO}$ reduces to ${\cal P}_\textup{convFO}$ for $N \le 3$, and ${\cal P}_\textup{NIO}$ becomes a strictly larger polytope than ${\cal P}_\textup{convFO}$ for $N \ge 4$.

As the analogy between correlations with non-influenceable orders and $\textup{\textsf{QC-NICC}}$ may have suggested, we also find that correlations generated by process matrices in $\textup{\textsf{QC-NICC}}$ are in $\mathcal{P}_{\textup{NIO}}$, as illustrated in Fig.~\ref{fig:QCCC-classes}. See Appendix~\ref{app:corr_QC-NICC} for the proof.

\subsubsection{Examples}
\label{subsubsec:examples_QCCCs}

We will now provide examples of process matrices that belong to each of the three classes with classical control of causal order introduced above.%
\footnote{For each of the three examples below, we provide in Appendix~\ref{app:decomp_ex_QCCCs} an explicit decomposition, which matches the characterisation of the class that we claim they belong to. \\
On the other hand, our claims that a given process matrix does \emph{not} belong to a given class, i.e.\ that it does not admit a given decomposition, can be verified using SDP (for some choice of the arbitrary states $\ket{\psi}$ appearing below, and of the dimension of the unspecified spaces---we typically consider qubits). One may also construct explicit ``witnesses'' to prove that a process is not in a given class, using the technique of causal witnesses~\cite{Araujo15,branciard16}. \label{ftn:check_SDP}}
Before that, let us introduce an important building block that will be used in all our examples: for any two isomorphic spaces $\mathcal{H}^X$ and $\mathcal{H}^Y$ whose computational bases $\{\ket{i}^X\}_i$ and $\{\ket{i}^Y\}_i$ are in one-to-one correspondence, we define the unnormalised maximally entangled state $\kket{\id}^{XY}\coloneqq \sum_i \ket{i}^X\otimes \ket{i}^Y$, such that $\kketbra{\id}{\id}^{XY}$ is the Choi matrix of an identity channel between the two spaces.

With this,
\begin{align}
    W_{\textup{convFO}} =& \, {\textstyle\frac12}\ketbra{\psi}{\psi}^{A^I} \otimes \kketbra{\id}{\id}^{A^OB^I}\otimes \id^{B^O} \notag \\
    & + {\textstyle\frac12}\ketbra{\psi}{\psi}^{B^I} \otimes \kketbra{\id}{\id}^{B^OA^I}\otimes \id^{A^O},
    \label{eq:QC-convFO-example}
\end{align}
where (here and in all our examples below) $\ket{\psi}$ denotes any fixed state received by one of the parties, is clearly an example of a process matrix in \textup{\textsf{QC-convFO}} (for $N=2$, with trivial spaces $\HS^P, \HS^F$). It corresponds indeed to a convex mixture of two identity channels, one from $A$ to $B$ and one from $B$ to $A$ (and with the first party receiving the initial state $\ket{\psi}$). 

An example of a process with dynamical but non-influenceable classical control of causal order in $\textup{\textsf{QC-NICC}} \setminus \textup{\textsf{QC-convFO}}$ (for $N=4$, still with trivial $\HS^P, \HS^F$) is given by the process matrix
\begin{align}
    W_{\textup{NICC}} =& \, {\textstyle\frac12} W_{+\alpha} \otimes \ketbra{\psi}{\psi}^{C^I} \otimes \kketbra{\id}{\id}^{C^O D^I} \otimes \id^{D^{O}} \notag \\
    & + {\textstyle\frac12} W_{-\alpha}\!\otimes \ketbra{\psi}{\psi}^{D^I} \!\otimes \kketbra{\id}{\id}^{D^O C^I} \!\!\otimes\!\id^{C^{O}}\!, \label{eq:QC-NICC-example}
\end{align}
where $W_{\pm \alpha} \coloneqq \frac{1}{4} (\id^{\otimes 4} \pm \alpha (Z Z Z \id + Z \id X X)) \in \mathcal{L}(\mathcal{H}^{{A^{IO}B^{IO}}})$, with $\alpha = \frac{1}{\sqrt{2}}$ here, $X$ and $Z$ (together with $Y$, which will be used later on) being the Pauli matrices, and with implicit tensor products. The matrix $W_{\pm\alpha}$ is a bipartite causally nonseparable process matrix (with qubit input/output systems for both parties), and it was shown in~\cite{branciard15} that $W_{+\alpha}$ can produce correlations that violate  the LGYNI causal inequality, Eq.~\eqref{eq:LGYNI} (so can $W_{-\alpha}$, with different instruments).

\begin{figure}[t]
\includegraphics[width=1\columnwidth]{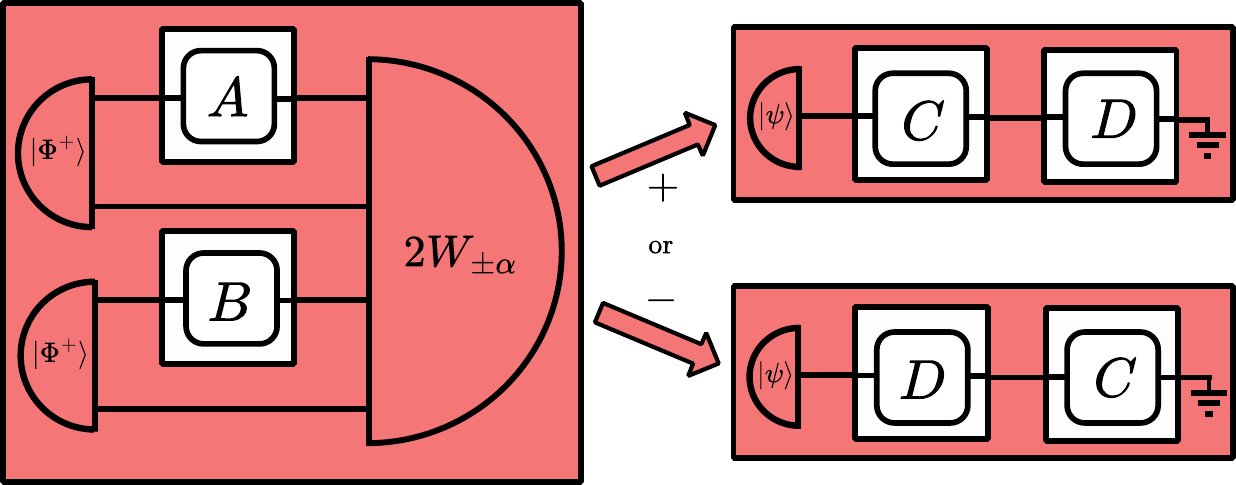}
\centering
\caption{A circuit-like representation of $W_{\textup{NICC}}$. $A$ and $B$ are each initially given half of a maximally entangled pair of qubits in the state $\ket{\Phi^+}~=~\frac{1}{\sqrt{2}}(\ket{0,0} + \ket{1,1})$. The global output state is then measured by the projective measurement $\{2W_{+\alpha},2W_{-\alpha}\}$. A new system initialised in the state $\ket{\psi}$ is then sent to $C$ and subsequently forwarded to $D$, or sent to $D$ and forwarded to $C$, with the order depending on the outcome of the previous measurement. The two branches of the circuit containing $C$ and $D$ are exclusive: on each run only one ``happens'', so that $C$ and $D$ (like $A$ and $B$) are each applied once and only once.}
\label{fig:QC-NICC-example}
\end{figure}

From an explicit decomposition of $W_{\textup{NICC}}$ as in Eq.~\eqref{eq:charact_W_QCCC_decomp}, in terms of matrices $W_{(k_1,\ldots,k_n)}$ (see Eq.~\eqref{app:dec_WNICC} of Appendix~\ref{app:decomp_ex_QCCCs}), one can construct internal operations $\tilde\M_n$ which, composed together, give the process matrix $W_{\textup{NICC}}$ (as in the general description of QC-CCs above). This allows us to give the circuit-like representation of $W_{\textup{NICC}}$ depicted in Fig.~\ref{fig:QC-NICC-example}.
This implementation is compatible with the ``coarse-grained'' causal order $\{A,B\} \prec \{C,D\}$, i.e.\ with both $A$ and $B$ acting before $C$ and before $D$, without assuming \emph{a priori} any particular causal relation between respectively $A$ and $B$, and between $C$ and $D$.%
\footnote{As it turns out, $W_{\textup{NICC}}$ can also be interpreted as a QC-CC compatible with $\{C,D\} \prec \{A,B\}$, however at the cost of losing the non-influenceable property, as it cannot be interpreted as a QC-NICC compatible with $\{C,D\} \prec \{A,B\}$.}
Here, the parties $A$ and $B$ are  both given half of a maximally entangled pair of qubits and can in fact act in parallel. Then, $\{2W_{+\alpha},2W_{-\alpha}\}$ defines a projective measurement performed on the output spaces of $A$ and $B$, along with the auxiliary spaces containing the other halves of the maximally entangled states, whose outcome controls classically the causal order between $C$ and $D$: $C\prec D$ for outcome $+$, and $D\prec C$ for outcome $-$. The probability to obtain $+$ or $-$ is always $1/2$ for each measurement outcome, independently of the CPTP maps performed by $A$ and $B$; hence, the latter cannot influence the causal order between $C$ and $D$ and the process has non-influenceable control of causal order. Yet, the process has dynamical causal order in the sense that the causal order is not and cannot be fixed from the start: $A$ and $B$ have to first perform their operations before the measurement is performed and its outcome controls the causal order between $C$ and $D$.
This is fundamentally related to the fact that $W_{\pm \alpha}$ are causally nonseparable process matrices. Indeed, if we were to define a similar process with $W_{\pm \alpha}$ being causally separable processes, then one would be able to find decompositions of the resulting process matrix of the QC-convFO form of Eq.~\eqref{eq:charact_W_QC_convFO_decomp}. This is \emph{not} the case here. Note that while the terms $W_{(A,B,C(,D))}$ and $W_{(A,B,D(,C))}$ appearing in the QC-NICC decomposition of $W_{\textup{NICC}}$ (e.g., $W_{(A,B,C)} = \frac12 W_{+\alpha} \otimes \ketbra{\psi}{\psi}^{C^I}$; see Eq.~\eqref{app:dec_WNICC} of Appendix~\ref{app:decomp_ex_QCCCs}) are valid process matrices (up to normalisation), these are not compatible with the corresponding orders $(A,B,C(,D))$ and $(A,B,D(,C))$---see our comment below Eq.~\eqref{eq:cstr_NICC}.

Finally, the ``classical switch''~\cite{Chiribella13}, where a (2-dimensional) state prepared in the past Hilbert space $\HS^P$ classically controls the causal order between two parties $A$ and $B$ acting on a target system, is an example of a process with dynamical and influenceable classical control of causal order in $\textup{\textsf{QC-CC}}\setminus\text{\textup{\textsf{QC-NICC}}}$ (for $N=2$, with a trivial $\HS^F$). Its process matrix is defined as
\begin{align}
    W_{\textup{CS}} =&\, \ketbra{0}{0}^P \otimes \ketbra{\psi}{\psi}^{A^I} \otimes \kketbra{\id}{\id}^{A^OB^I} \otimes \id^{B^O} \notag \\
    & +\ketbra{1}{1}^P\otimes \ketbra{\psi}{\psi}^{B^I} \otimes \kketbra{\id}{\id}^{B^OA^I} \otimes \id^{A^O}.
    \label{eq:CS-example}
\end{align}
The first term relates to the order $A\prec B$, while the second term relates to the order $B\prec A$. These two terms individually are not valid process matrices; rather than being in a convex mixture (as were the two terms in $W_{\textup{convFO}}$, Eq.~\eqref{eq:QC-convFO-example}), here their contribution is controlled by the state prepared in $\HS^P$. The first term indeed contributes when that state is $\ket{0}^P$, while the second term contributes when it is $\ket{1}^P$---hence the influenceability of the control by the initial state preparation in $\HS^P$.%
\footnote{By considering $\HS^P$ to be the output space of an additional party (that has no input space, see Footnote~\ref{ftn:equiv_PF}), then one would equivalently say that it is the choice of operation (the state preparation) of that new party that influences the order between $A$ and $B$.}

\subsection{Quantum Circuits with Quantum Control of causal order}
\label{sec:QCQC_all}

Let us now turn our attention to the second main class of quantum circuits introduced in Ref.~\cite{wechs21}: circuits with quantum, rather than classical, control of causal order. QC-QCs allow for indefinite causal orders; nevertheless, we will see that one can still characterise some form of dynamicality for the quantum control.

\subsubsection{The general \textup{\textsf{QC-QC}} class}

\begin{figure*}[t]
\centering
\includegraphics[width=2\columnwidth]{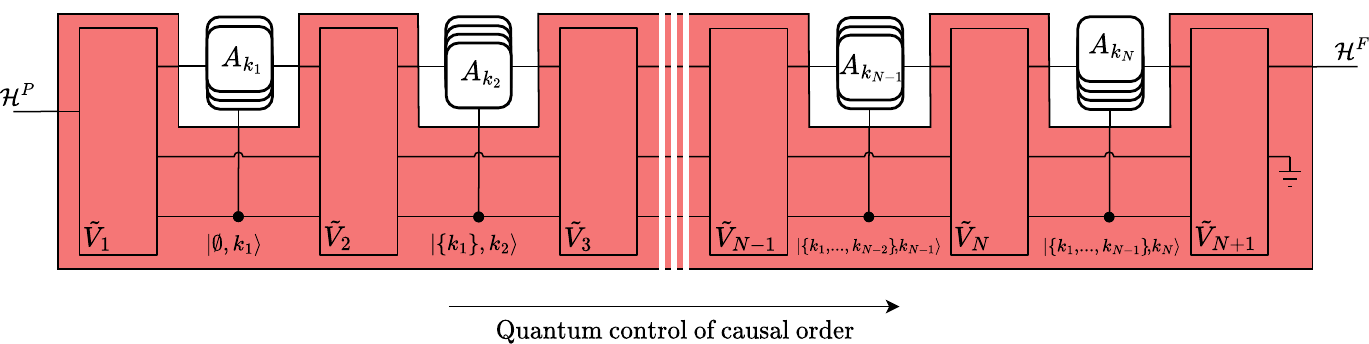}
\caption{Graphical representation of a quantum circuit with quantum control (QC-QC) of causal order (a simplified version of Fig.~10 from Ref.~\cite{wechs21}, also reproduced with more details as Fig.~\ref{fig:QCQC_graph_full} in Appendix~\ref{app:subsec_QCQC_classes}). The QC-QC is defined as the composition of the ``internal'' operations $\tilde{V}_1,\ldots,\tilde{V}_{N+1}$ (in the red shaded area), leaving ``open slots'' for the parties' ``external'' operations $A_k$. The construction is similar to that of QC-CCs (see Fig.~\ref{fig:QCCC_graph}), except that here the control system (the now single-stroke bottom line) acts in a quantum manner, and can coherently control different causal orders, in a superposition. Furthermore, the state of the control system, in the form $\ket{\mathcal{K}_{n-1},k_{n}}$, only keeps track of the unordered set $\K_{n-1}$ of parties that have already acted, together with the next party $A_{k_n}$ to come. The causal order between the external operations can again be seen as being established on the fly or dynamically, as each internal operation is applied.}
\label{fig:QCQC_graph}
\end{figure*}

QC-QCs are built in analogy to QC-CCs, with now a quantum (coherent) control of causal order. As for a QC-CC, the action of a QC-QC (depicted in Fig.~\ref{fig:QCQC_graph}) can be understood as a protocol that alternatively applies ``internal'' and ``external'' operations, the latter corresponding to the quantum operations realised by the different parties $A_k$. Each internal operation $\tilde{V}_n$ produces the system that coherently controls which party $A_{k_n}$ will be the next to apply its external operation, and which keeps track of which previous external operations have already been applied (to ensure that each is applied once and only once). The state of the control system is now of the form $\ket{\mathcal{K}_{n-1},k_{n}}$, where $\mathcal{K}_{n-1}$ describes the set of $n-1$ parties that already applied their external operation, while $k_n$ describes the choice of the next party that applies its external operation. Contrarily to QC-CCs, only the unordered set of past operations is encoded in the control system: for the branch in which the next party to act is chosen to be $A_{k_{n+1}}$, the control state is updated from $\ket{\mathcal{K}_{n-1},k_{n}}$ to $\ket{\mathcal{K}_{n-1}\cup\{k_n\},k_{n+1}}$, thereby forgetting which was the previous party $A_{k_n}$. This eventually makes it possible for different causal orders corresponding to the same unordered set to interfere, once the information about the causal order is erased. The QC-QC under consideration is then defined as the composition of the internal operations $\tilde{V}_1,\ldots,\tilde{V}_{N+1}$ (possibly tracing out some system at the end), leaving ``open slots'' for the parties' external operations; see Appendix~\ref{app:introQCQC} and Ref.~\cite{wechs21} for more details on this construction. Again, different internal operations may define the same QC-QC after being composed.

We denote by \textup{\textsf{QC-QC}} the set of process matrices that describe QC-QCs. As shown in~\cite{wechs21}, this set admits the following characterisation:
$W \in \mathcal{L}(\mathcal{H}^{PA_\mathcal{N}^{IO}F})$ is in \textup{\textsf{QC-QC}} iff $W \ge 0$, and for all $n=1,\ldots, N$, for all strict subsets ${\cal K}_{n-1}$ of ${\cal N}$ (with $|\K_{n-1}|=n-1$), and all $k_n \in {\cal N}\backslash{\cal K}_{n-1}$, there exist PSD matrices $W_{({\cal K}_{n-1},k_n)} \in {\cal L}(\HS^{PA_{{\cal K}_{n-1}}^{IO} A_{k_n}^I})$ such that
\begin{align}
    \left\{
    \begin{array}{l}
        \Tr_F W = \sum_{k_N \in {\cal N}} W_{({\cal N} \backslash \{k_N\},k_N)}\otimes \id^{A_{k_N}^O}, \\[3mm]
        \forall \, \emptyset \subsetneq {\cal K}_n \subsetneq {\cal N}, \ \sum_{k_{n+1} \in {\cal N} \backslash {\cal K}_n} \Tr_{A_{k_{n+1}}^I} W_{({\cal K}_n,k_{n+1})} \\
        \hspace{25mm} = \sum_{k_n \in {\cal K}_n} W_{({\cal K}_n \backslash \{k_n\},k_n)}\otimes \id^{A_{k_n}^O}, \\[3mm]
        \sum_{k_1 \in {\cal N}} \Tr_{A_{k_1}^I} W_{(\emptyset,k_1)} = \id^P.
    \end{array}
    \right. \label{eq:charact_W_QCQC_decomp}
\end{align}

The \textup{\textsf{QC-QC}} set contains \textup{\textsf{QC-CC}} as a generally strict subset (for $N\ge 2$ if $\HS^F$ is nontrivial, or for $N\ge 3$ if $\HS^F$ is trivial). Beyond this, it also contains circuits with genuinely quantum control of causal order, which may feature indefinite causal order.  Despite this, QC-QCs are known to create only causal correlations, in $\mathcal{P}_{\textup{causal}}$~\cite{wechs21,Purves21}; see Fig.~\ref{fig:QCQC-classes} for an illustration. We provide in Appendix~\ref{app:corr_QC-QC} a new formulation of this proof that is more directly related to the definition of causal correlations given in Eq.~\eqref{eq:def_causal_p}, and to an alternative form given in Appendix~\ref{app:more_causal_charact}.

\begin{figure}[hbtp]
\centering
\includegraphics[width=\columnwidth]{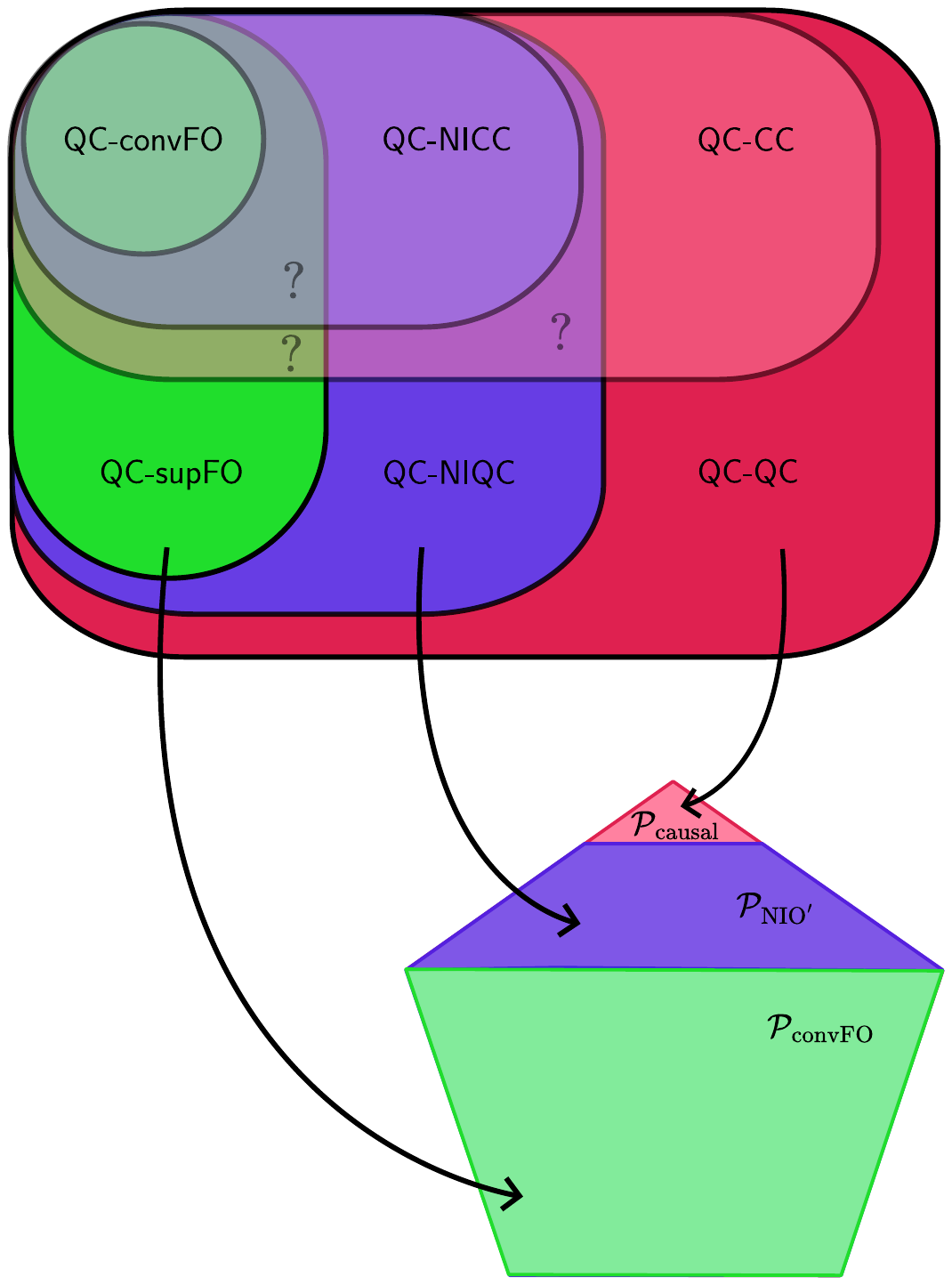}
\caption{The upper part of the figure depicts the inclusion relations between the different subclasses of QC-QCs considered in this paper (including the subclasses of QC-CCs, already presented in Fig.~\ref{fig:QCCC-classes}). For $N\ge 4$ (and nontrivial input and output spaces for the different parties) the inclusions between \textup{\textsf{QC-supFO}}, \textup{\textsf{QC-NIQC}} and \textup{\textsf{QC-QC}} are strict; for $N\le3$, \textup{\textsf{QC-NIQC}} reduces to \textup{\textsf{QC-supFO}}. It remains an open question whether the intersections with question marks are empty or not. \\
The arrows directed to the lower part of the figure indicate the kind of causal correlations that process matrices in each (sub)class of QC-QCs can produce: process matrices in \textup{\textsf{QC-supFO}} only give correlations in ${\cal P}_\textup{convFO}$; those in \textup{\textsf{QC-NIQC}} only give correlations in ${\cal P}_{\textup{NIO}'}$; those in \textup{\textsf{QC-QC}} give causal correlations in ${\cal P}_\textup{causal}$.}
\label{fig:QCQC-classes}
\end{figure}

We will now see how to characterise the dynamicality of causal orders within QC-QCs, in analogy with what we did for QC-CCs, by referring to the state of the control system.

\subsubsection{The \textup{\textsf{QC-supFO}} subclass}
\label{sec:QC-supFO}

One may first wonder what is the most general way to coherently control the causal order within QC-QCs, without any sort of dynamicality. In some sense, one wishes to look for the largest class that generalises the quantum switch with a fixed initial control state, keeping this ``static'' form. 

To define such QC-QCs with non-dynamical orders, we propose a similar construction to that of general QC-QCs, except that we now impose that the total causal order should be encoded in the control system from the very beginning, and that the ``weight'' of each order should not be affected by the internal operations. More specifically, referring to Fig.~\ref{fig:QCQC_graph}: we take the basis states of the control system between the internal operations $\tilde V_1$ and $\tilde V_2$ to be of the form $\ket{\emptyset,k_1; (k_2,\ldots,k_N)}$ (instead of just $\ket{\emptyset,k_1}$), each of them coming with a weight that has a fixed value $q_{\pi=(k_1,k_2,\ldots,k_N)}$ (which does not depend, in particular, on the initial state fed into the circuit in the global past space $\HS^P$). Subsequent internal operations then update the control system from states of the form $\ket{\K_{n-1},k_n; (k_{n+1},\ldots,k_N)}$ to states of the form $\ket{\K_{n-1}\cup\{k_n\},k_{n+1}; (k_{n+2},\ldots,k_N)}$ (instead of $\ket{\K_{n-1},k_n}$ to $\ket{\K_{n-1}\cup\{k_n\},k_{n+1}}$). We keep here the feature that the ``past'' order is erased: the control system only keeps track of the unordered set of past parties. On the other hand, the order of remaining parties is left unchanged, and we impose that the internal operations cannot change it, nor can they change the weight of each of the ``future'' orders $(k_{n+2},\ldots,k_N)$: see  Appendix~\ref{app:def_QC-supFO} for more details on this construction. 

We call circuits that abide by this description \emph{quantum circuits with superposition of fixed causal orders} (QC-supFO), and we denote by \textup{\textsf{QC-supFO}} the set of process matrices that describe them. As we show in Appendix~\ref{app:def_QC-supFO}, one obtains the following characterisation: $W \in \mathcal{L}(\mathcal{H}^{PA_\mathcal{N}^{IO}F})$ is in \textup{\textsf{QC-supFO}} iff $W \ge 0$, and for all $n = 1, \ldots, N$, all $\K_{n-1}\subsetneq \N$ and all $(k_n,k_{n+1},\ldots,k_N)\not\in\K_{n-1}$, there exist PSD matrices $W_{({\cal K}_{n-1},k_n)}^{(k_{n+1},\ldots,k_N)} \in {\cal L}(\HS^{PA_{{\cal K}_{n-1}}^{IO} A_{k_n}^I})$ such that
\begin{align}
    \left\{
    \begin{array}{l}
        \Tr_F W = \sum_{k_N \in {\cal N}} W_{({\cal N} \backslash \{k_N\},k_N)}^{\emptyset}\otimes \id^{A_{k_N}^O}, \\[3mm]
        \forall \, \emptyset \subsetneq {\cal K}_n \subsetneq {\cal N}, \, \forall \, (k_{n+1},\ldots,k_N) \in {\cal N}\backslash{\cal K}_n, \\[1mm]
        \ \ \Tr_{A_{k_{n+1}}^I} \!\!W_{({\cal K}_n,k_{n+1})}^{(k_{n+2},\ldots,k_N)} \!=\! \sum_{k_n \in {\cal K}_n} \!W_{\!({\cal K}_n \backslash \{k_n\},k_n)}^{(k_{n+1},\ldots,k_N)}\!\otimes\! \id^{\!A_{k_n}^O}, \\[3mm]
        \forall \, \pi=(k_1,\ldots,k_N) \in {\cal N}, \ \Tr_{A_{k_1}^I} W_{(\emptyset,k_1)}^{(k_2,\ldots,k_N)} = q_\pi \,\id^P, \\[3mm]
        \hspace{45mm} q_\pi\ge 0, \, \sum_\pi q_\pi = 1.
    \end{array}
    \right. \label{eq:charact_W_QC_SupFO_decomp}
\end{align}
This \textup{\textsf{QC-supFO}} class is the quantum-controlled analogue of the classically and non-dynamically-controlled \textup{\textsf{QC-convFO}}, in the same way as QC-QCs are the quantum-controlled analogue of QC-CCs. The characterisations and the notations we use reflect this: Eq.~\eqref{eq:charact_W_QC_SupFO_decomp} above involves matrices $W_{({\cal K}_{n-1},k_n)}^{(k_{n+1},\ldots,k_N)}$, while the characterisation of \textup{\textsf{QC-convFO}} in Eq.~\eqref{eq:charact_W_QC_convFO_decomp} involved matrices $W_{(k_1,\ldots,k_n)}^{(k_{n+1},\ldots,k_N)}$; in both cases, the subscripts include the first $n$ parties of a given order (forgetting about the order of the first $n-1$ parties, in the QC-QC or QC-supFO case), while the superscript (not present in the general QC-CC and QC-QC decompositions) display the ordered list of remaining parties. We note also that just like the $W_{(k_1,\ldots,k_n)}^{(k_{n+1},\ldots,k_N)}$'s, the matrices $W_{({\cal K}_{n-1},k_n)}^{(k_{n+1},\ldots,k_N)}$ here can be seen to be valid process matrices (up to normalisation).
Comparing the two classes, one has $\textup{\textsf{QC-convFO}} \subset \textup{\textsf{QC-supFO}}$, the inclusion being strict (in general) for $N\ge 2$ if $\HS^F$ is nontrivial, or for $N\ge 3$ if $\HS^F$ is trivial.

It is interesting to observe, also, that the characterisation of QC-supFOs above displays some form of symmetry with that of general QC-CCs, Eq.~\eqref{eq:charact_W_QCCC_decomp}: for QC-CCs the total causal order is not encoded in the decomposition at the ``lower levels'' (for small $n$), but it appears in the matrices involved in the decomposition as we go ``up'' (to larger $n$) in the decomposition. For QC-supFOs it is the opposite: the causal order is encoded in the decomposition at the lower levels, but gets erased as we go up in the decomposition. 
This symmetry can also be seen in the equations: the sums appear on the left-hand side of the constraints in Eq.~\eqref{eq:charact_W_QCCC_decomp}, and on the right-hand side of the constraints in Eq.~\eqref{eq:charact_W_QC_SupFO_decomp} (notice that for QC-QCs they appear on both sides). In Appendix~\ref{app:subsec:graphical_rep} we propose a graphical representation of the SDP constraints for our different classes, which allows one, in particular, to clearly visualise the symmetry between the decompositions of QC-CCs and of QC-supFOs.

Comparing it to the full set of QC-QCs, it can clearly be seen that \textup{\textsf{QC-supFO}} is a subclass of \textup{\textsf{QC-QC}}: $\textup{\textsf{QC-supFO}} \subset \textup{\textsf{QC-QC}}$, the inclusion being in general strict for $N\ge2$ with a nontrivial $\HS^P$, or for $N\ge3$ with a trivial $\HS^P$ (see Appendix~\ref{app:subsubsec_QCQC_inclusions}; the situation is thus similar to that of $\textup{\textsf{QC-convFO}}$ vs $\textup{\textsf{QC-CC}}$). It contains on the other hand the subclass of QC-convFOs: $\textup{\textsf{QC-convFO}} \subset \textup{\textsf{QC-supFO}}$, the inclusion being in general strict for $N\ge2$ with a nontrivial $\HS^F$, or for $N\ge3$ with a trivial $\HS^F$ (see also Appendix~\ref{app:subsubsec_QCQC_inclusions}; here the situation is similar to that of $\textup{\textsf{QC-CC}}$ vs $\textup{\textsf{QC-QC}}$).
Furthermore, building on the proof that QC-QCs produce causal correlations, we show in Appendix~\ref{app:corr_QC-supFO} that QC-supFOs produce correlations in $\mathcal{P}_{\textup{convFO}}$, which is consistent with the claim that these processes feature no dynamical causal order; see Fig.~\ref{fig:QCQC-classes}.

The definition and characterisation of QC-supFOs allows us to now properly formalise the claim that some processes can have \emph{both indefinite and dynamical causal order}. Indeed, this can be said of any QC-QC that is not a QC-CC,%
\footnote{We recall that it remains an open question, whether for \mbox{$N\ge 4$}, all causally separable processes are QC-CCs~\cite{wechs19,wechs21}. Here we overlook this subtlety for the case of quantum circuits, and consider all those that are not QC-CCs to have indefinite causal order.}
but also not a QC-supFO---i.e., whose process matrix is in $\textup{\textsf{QC-QC}}\setminus(\textup{\textsf{QC-CC}}\cup\textup{\textsf{QC-supFO}})$. Explicit examples will be given below.

To finish with the presentation of QC-supFOs, let us mention that another somewhat related class was introduced in Ref.~\cite{Liu23} (just called \textup{\textsf{Sup}} there), which may have also looked like a natural candidate to characterise processes with non-dynamical superpositions of causal orders. The idea was to impose that if one traces out the global future system in $\HS^F$, the reduced circuit should end up in just a convex mixture of fixed-order processes (in \textup{\textsf{QC-convFO}})---as is indeed the case for the ``static'' quantum switch. However, we argue that such a constraint is too restrictive. It also has some undesirable features, e.g.\ that if several parties have no output Hilbert space and could be considered to be in the global future, then the definition crucially depends on whose Hilbert space is called $\HS^F$.%
\footnote{In contrast, it can be verified that in such a case, the classes we introduce here do not depend on whose space is called $\HS^F$; similarly if several parties have no input Hilbert space and could be considered to be in the global past, our classes do not depend on whose space is called $\HS^P$: see Appendix~\ref{app:subsec:P/F-label indep}.}
We discuss this in more details in Appendix~\ref{app:discuss_Sup}. In fact, we show that by generalising the \textup{\textsf{Sup}} class in an adequate way, so as precisely to avoid the unwanted features, one naturally obtains the same characterisation as in Eq.~\eqref{eq:charact_W_QC_SupFO_decomp}. This alternative perspective lends further support to the claim that our \textup{\textsf{QC-supFO}} class appropriately characterises quantum circuits with non-dynamical quantum control of causal order.

\subsubsection{The \textup{\textsf{QC-NIQC}} subclass}
\label{sec:QC-NIQC}

Let us finally introduce one last class: the quantum-controlled counterpart of QC-NICCs, namely \emph{quantum circuits with non-influenceable quantum control of causal order} (QC-NIQCs).

Recall that for QC-NICCs, although the (classical) state of the control could be dynamical (if it cannot be fixed in advance), we imposed that it should not depend on the choice of parties' TP operations (it must be non-influenceable). One may be tempted to impose the same condition, now for the quantum control system. However, this would not quite correspond to what we want. To see this, consider the ``static'' quantum switch, where a control qubit, initialised in the fixed state $\ket{+}=\frac{1}{\sqrt{2}}(\ket{0} + \ket{1})$, coherently controls the order between two operations applied on a target system~\cite{Chiribella13} (see also Eq.~\eqref{eq:QC-supFO-example} below for its process matrix description). This example is clearly a QC-supFO, so we do not want to say that its quantum control features any form of dynamicality. However, the reduced state of the control system \emph{does} depend on the two external operations: e.g.\ the output state of the control qubit is $\ket{+}$ if (the Kraus operators of) the two external operations commute, or $\ket{-}=\frac{1}{\sqrt{2}}(\ket{0} - \ket{1})$ if these anti-commute---this is in fact at the very heart of the best-known application of the quantum switch~\cite{chiribella12}.
What \emph{does not} depend on the external operations in this example are the \emph{classical weights} corresponding to each fixed order in the quantum-controlled superposition of orders---that is, the diagonal elements of the state of the control system, when written in the basis that defines the controllisation (the $\{\ket{0},\ket{1}\}$ basis here).

This is precisely what we shall impose here to define QC-NIQCs, extending it to the general form of the control system as defined in QC-QCs:
\begin{align*}
  \parbox{0.85\linewidth}{%
    \emph{QC-NIQCs are the QC-QCs for which there exists an implementation such that, at any intermediate time step, the classical weights of each term in the basis $\{\ket{\K_{n-1},k_n}^{C_n}\}_{\K_{n-1},k_n}$ that defines the controllisation---i.e.\ the diagonal elements of the state of the control system in that basis---are independent of the choice of previously applied external operations and of the potential initial state preparation in the global past.}%
  }
\end{align*}

We denote by \textup{\textsf{QC-NIQC}} the set of process matrices that describe QC-NIQCs. We show in Appendix~\ref{app:def_QC-NIQC} that the above constraint implies the following characterisation: $W \in \mathcal{L}(\mathcal{H}^{PA_\mathcal{N}^{IO}F})$ is in \textup{\textsf{QC-NIQC}} if and only if $W\ge 0$ and it has a \textup{\textsf{QC-QC}} decomposition in terms of PSD matrices $W_{({\cal K}_{n-1},k_n)}$ as in Eq.~\eqref{eq:charact_W_QCQC_decomp}, with the additional constraint that
\begin{align}
    & \begin{array}{r}
        \text{all} \ W_{({\cal K}_{n-1},k_n)} \text{'s are valid process matrices} \\
        \text{(up to normalisation).}
    \end{array} \label{eq:cstr_NIQC}
\end{align}
This additional constraint is thus quite analogous to the one we obtained for QC-NICCs, Eq.~\eqref{eq:cstr_NICC}.
Note however that, contrary to QC-NICCs, it is not sufficient here to verify the validity for only the ``highest'' terms $W_{({\cal K}_{N-1},k_N)}$ to ensure that Eq.~\eqref{eq:cstr_NIQC} is satisfied.

We show in Appendix~\ref{app:subsubsec_QCQC_inclusions} that for $N\le 3$, \text{\textup{\textsf{QC-NIQC}}} coincides with \text{\textup{\textsf{QC-supFO}}}, but for $N\ge 4$ it defines a class in between \textup{\textsf{QC-supFO}} and \textup{\textsf{QC-QC}}, as depicted in Fig.~\ref{fig:QCQC-classes}:
\begin{align}
    \text{\textup{\textsf{QC-supFO}}} \ \subset \ \text{\textup{\textsf{QC-NIQC}}} \ \subset \ \text{\textup{\textsf{QC-QC}}}. \label{eq:inclusions_QC-QCs}
\end{align}
It can also easily be seen that $\text{\textup{\textsf{QC-NICC}}} \subset \text{\textup{\textsf{QC-NIQC}}}$ (the inclusion being again generally strict for $N\ge2$ if $\HS^F$ is nontrivial, or for $N\ge3$ if $\HS^F$ is trivial, as in the previous cases of \textup{\textsf{QC-CC}} vs \textup{\textsf{QC-QC}}, and of \textup{\textsf{QC-convFO}} vs \textup{\textsf{QC-supFO}}).

Considering now the correlations generated by QC-NIQCs, one can show (see Appendix~\ref{app:corr_QC-NIQC}) that these are necessarily in the polytope $\mathcal{P}_{\textup{NIO}'}$ of causal correlations with ``non-influenceable coarse-grained order'', which we pre-emptively introduced in Sec.~\ref{subsec:NIO}. This can be understood by recalling that the control state in a QC-QC only encodes, at any given time step, the ``coarse-grained'' order $(\K_{n-1},k_n)$ (with $\K_{n-1}$ containing all ``past'' parties, grouped together). When imposing the non-influenceability of (the diagonal elements of) the control state, it is then natural to expect that the distribution of $p\big((\K_{n-1},k_n)\big|\vec x\big)$ obtained from a causal decomposition of some induced correlation would not depend on the choice of external operations.

One may then wonder whether the stronger statement holds, that the non-coarse-grained probabilities $p\big((k_1,\ldots,k_n)\big|\vec x\big)$ also do not depend on the external operations, i.e.\ that the correlations created by QC-NIQCs are in the smaller polytope $\mathcal{P}_{\textup{NIO}}$. We leave this as an open question: we could not show this in general, but neither did we find any QC-NIQC correlation in $\mathcal{P}_{\textup{NIO}'}\backslash\mathcal{P}_{\textup{NIO}}$ (see Sec.~\ref{sec:QCQC_correlations}).

\subsubsection{Examples}
\label{subsubsec:examples_QCQCs}

We now provide examples of process matrices that belong to each of the
three classes with quantum control of causal order introduced above.%
\footnote{For each of the three examples detailed below, we provide in Appendix~\ref{app:decomp_ex_QCQCs} an explicit decomposition, which matches the characterisation of the corresponding class. \\
Again, our claims that a given process matrix does \emph{not} belong to a given class can be verified through SDP (see Footnote~\ref{ftn:check_SDP}).}

First, the quantum switch with a fixed initial state of the control system~\cite{Chiribella13}---the ``static'', or ``non-dynamical'' quantum switch (already referred to at the beginning of Sec.~\ref{sec:QC-NIQC})---is an example of a process in $\text{\textup{\textsf{QC-supFO}}}\setminus \text{\textup{\textsf{QC-convFO}}}$ for $N=2$ parties (with a trivial global past space $\HS^P$ but a nontrivial future space $\HS^F$). Considering that the future Hilbert space $\HS^F = \HS^{F_\text{t}}\otimes\HS^{F_\text{c}}$ receives the target (in $\HS^{F_\text{t}}$) and control (in $\HS^{F_\text{c}}$) systems, its process matrix is defined as
\begin{align}
& W_{\textup{supFO}}=\ketbra{w_\textup{supFO}}{w_\textup{supFO}} \ \ \text{with} \notag \\[2mm]
& \ket{w_\textup{supFO}} = {\textstyle\frac{1}{\sqrt{2}}} \ket{\psi}^{A^I} \otimes \kket{\id}^{A^OB^I} \otimes 
 \kket{\id}^{B^OF_\text{t}} \otimes \ket{0}^{F_\text{c}} \notag \\
    & \hspace{17mm} + {\textstyle\frac{1}{\sqrt{2}}}\ket{\psi}^{B^I} \otimes \kket{\id}^{B^OA^I} \otimes \kket{\id}^{A^OF_\text{t}} \otimes \ket{1}^{F_\text{c}}.
    \label{eq:QC-supFO-example}
\end{align}
This process corresponds to a coherent superposition of two branches where $A \prec B$ or $B \prec A$, with the target system being sent in each branch to the first party and forwarded through identity channels to the second party and then to $\HS^{F_\text{t}}$, and with a control qubit initially in the fixed state $\ket{+}=\frac{1}{\sqrt{2}}(\ket{0} + \ket{1})$ and finally sent to $\HS^{F_\text{c}}$.

\begin{figure}
\includegraphics[width=1\columnwidth]{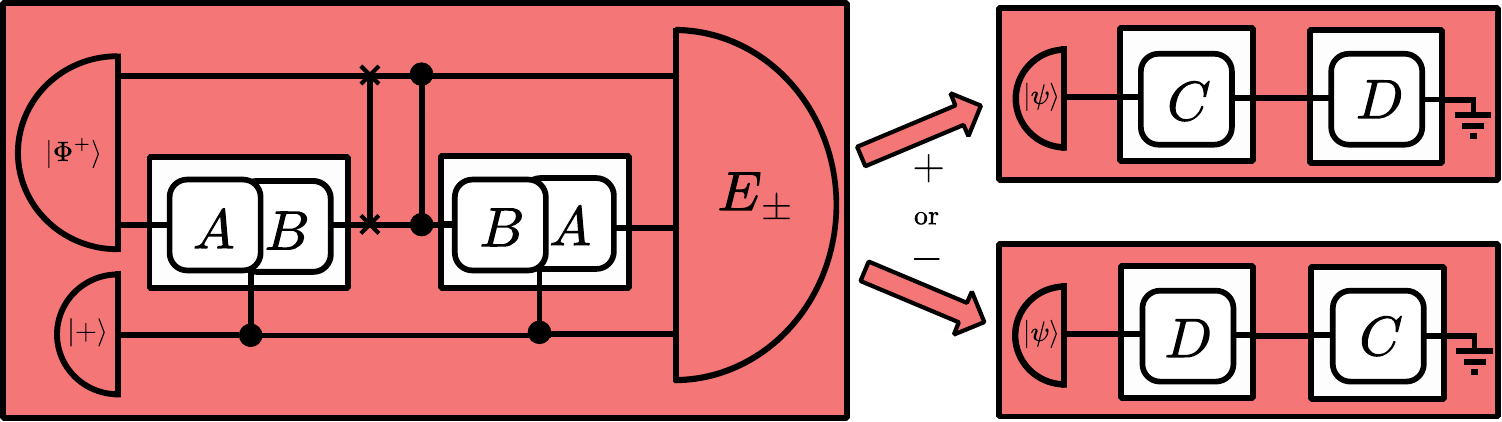}
\caption{Circuit-like representation of $W_{\textup{NIQC}}$. The causal order between $A$ and $B$ is coherently controlled (illustrated by the overlap between gates $A$ and $B$) by a control qubit prepared in the state $\ket{+}=\frac{1}{\sqrt{2}}(\ket{0} + \ket{1})$. The first party to act (be it $A$ or $B$) is given half of the state $\ket{\Phi^+}=\frac{1}{\sqrt{2}}(\ket{0,0}+\ket{1,1})$. An internal unitary, composed of a Swap and a Controlled-$Z$ gates, is applied between the first and the second operations in both branches of the causal superposition between $A$ and $B$. The output state coming from the output Hilbert space of the second operation, the auxiliary space and the control Hilbert spaces are then measured by the POVM defined by $\{E_+,E_-\}$ (see main text). 
As in the implementation of $W_{\textup{NICC}}$ shown on Fig.~\ref{fig:QC-NICC-example}, a new system in the state $\ket{\psi}$ is then, according to the measurement result, either sent to $C$ then forwarded to $D$, or sent to $D$ then forwarded to $C$---the two corresponding branches of the circuit being exclusive.}
\label{fig:QC-NIQC-example}
\end{figure}

Our second example here is of a process with dynamical but non-influenceable quantum control of causal order, i.e.\ in $\text{\textup{\textsf{QC-NIQC}}}\setminus(\text{\textup{\textsf{QC-supFO}}}\cup\text{\textup{\textsf{QC-NICC}}})$, for the $N=4$ case with trivial spaces $\HS^P, \HS^F$. Let us first define the ket vectors
\begin{align}
    \ket{\Phi_{00}} & = {\textstyle\frac{1}{\sqrt{2}}}(\ket{0,0}+\ket{1,1}), & \ket{\Phi_{11}} & = {\textstyle\frac{1}{\sqrt{2}}}(\ket{0,0}-\ket{1,1}), \notag \\
    \ket{\Phi_{01}} & = {\textstyle\frac{1}{\sqrt{2}}}(\ket{0,0}+i\ket{1,1}), & \ket{\Phi_{10}} & = {\textstyle\frac{1}{\sqrt{2}}}(\ket{0,0}-i\ket{1,1}),
\end{align}
and the matrices
\begin{align}
    W_0 & = {\textstyle\sum_{a,b}}\ \ketbra{\Phi_{ab}}{\Phi_{ab}}^{A^IB^I}\otimes\ketbra{a,b}{a,b}^{A^OB^O}, \notag \\
    \quad W_1 &= {\textstyle\sum_{a,b}}\ \ketbra{\Phi_{ba}}{\Phi_{ba}}^{A^IB^I}\otimes\ketbra{a,b}{a,b}^{A^OB^O}, \label{eq:W0/W1}
\end{align}
which can both be shown to be bipartite causally nonseparable process matrices in $\mathcal{L}(\mathcal{H}^{A^{IO}B^{IO}})$.
We then define $W_{\textup{NIQC}}$ as:
\begin{align}
    W_{\textup{NIQC}} =& \, {\textstyle\frac12} W_0\otimes \ketbra{\psi}{\psi}^{C^I} \otimes \kketbra{\id}{\id}^{C^OD^I} \otimes \id^{D^O} \notag \\
    & + {\textstyle\frac12}W_1\otimes \ketbra{\psi}{\psi}^{D^I} \!\otimes \kketbra{\id}{\id}^{D^OC^I} \!\otimes \id^{C^O}. \label{eq:QC-NIQC-example}
\end{align}
As for any QC-QC, one can find an implementation of this process by constructing internal operations from its QC-QC decomposition, through a procedure detailed in Appendix~\ref{app:def_QC-NIQC}~\cite{wechs21}.
The process matrix $W_{\textup{NIQC}}$ has a similar structure to $W_{\textup{NICC}}$, and one can also find an implementation compatible with the coarse-grained causal order $\{A,B\} \prec \{C,D\}$, which is graphically represented in Fig.~\ref{fig:QC-NIQC-example}. However, in contrast to $W_{\textup{NICC}}$, the order between $A$ and $B$ is now coherently controlled using a control qubit initialised in the state $\ket{+}$. Then, $\{E_+,E_-\}$, with $E_\pm = \frac{1}{2}[\id^{\otimes 3} \pm \frac{1}{2}(XX+YY)Y]$ and implicit tensor products, defines a POVM (no longer a projective measurement) performed on the output spaces of $A$ and $B$, along with the auxiliary and control spaces, whose outcome controls classically the causal order between $C$ and $D$ ($C \prec D$ for outcome $+$ and $D \prec C$ for outcome $-$). One can explain why $W_{\textup{NIQC}}$ has dynamical but non-influenceable causal order with the same reasoning as for $W_{\textup{NICC}}$: dynamicality relates to the fact  that $W_{0/1}$ are causally nonseparable, while non-influenceability is due to the fact that the probabilities for the outcomes $\pm$ of the POVM $\{E_+,E_-\}$ are independent of the operations performed by $A$ and $B$ (these probabilities are $\frac12$ for both outcomes). The only different qualitative feature between these two processes is the coherent control of causal order between $A$ and $B$ in $W_{\textup{NIQC}}$. This raises questions concerning other potential features that would differentiate \text{\textup{\textsf{QC-NICC}}} from \text{\textup{\textsf{QC-NIQC}}}; we will comment on this in the outlook below.

Next, the ``dynamical quantum switch'' (as opposed to the static one), where the qubit that coherently controls the causal order between two parties $A$ and $B$ is not fixed \emph{a priori}, but is left to be prepared (in any possible state) in the past Hilbert space $\HS^P$, is an example of a process with influenceable quantum control of causal order in $\textup{\textsf{QC-QC}}\setminus(\textup{\textsf{QC-NIQC}} \cup \textup{\textsf{QC-CC}})$ (for $N=2$, with now both $\HS^P$ and $\HS^F$ being nontrivial). Considering again that the target and control systems are ultimately sent to the future Hilbert space $\HS^F = \HS^{F_\text{t}}\otimes\HS^{F_\text{c}}$, its process matrix is defined here as
\begin{align}
& W_{\textup{QS}}=\ketbra{w_\textup{QS}}{w_\textup{QS}} \ \ \text{with} \notag \\[2mm]
& \ket{w_\textup{QS}} = \ket{0}^P \otimes \ket{\psi}^{A^I} \otimes \kket{\id}^{A^OB^I} \otimes 
 \kket{\id}^{B^OF_\text{t}}  \otimes 
 \ket{0}^{F_\text{c}} \notag \\
    & \hspace{12mm} + \ket{1}^P \otimes 
 \ket{\psi}^{B^I} \otimes \kket{\id}^{B^OA^I} \otimes \kket{\id}^{A^OF_\text{t}} \otimes 
 \ket{1}^{F_\text{c}}.
    \label{eq:QC-QC-example}
\end{align} 
Compared to $\ket{w_\textup{supFO}}$ in Eq.~\eqref{eq:QC-supFO-example}, here the past is ``open'', so that the input state of the control system in $\HS^P$ can influence the weight of the two branches, corresponding to $A\prec B$ and $B\prec A$.

The other new explicit example of a QC-QC presented in~\cite{wechs21} (dubbed the ``Grenoble process'' in Ref.~\cite{vanrietvelde22}) is also an example of a process in $\textup{\textsf{QC-QC}}\setminus(\textup{\textsf{QC-NIQC}}\cup\textup{\textsf{QC-CC}})$, for $N=3$. The ``double quantum switch'' from Ref.~\cite{Salzger23} is yet another example in the same class, now for $N=4$. This process is obtained by composing two quantum switches, where the control qubit of the first quantum switch (between $A$ and $B$, controlled in the $\{\ket{0},\ket{1}\}$ basis) is transmitted in order to control the second quantum switch (between $C$ and $D$), in the complementary $\{\ket{+},\ket{-}\}$ basis. 
Since the projection onto the $\{\ket{\pm}\}$ basis of the control qubit after the first quantum switch depends on the operations performed by $A$ and $B$ (e.g., it is $\ket{+}$ if those operations commute, or $\ket{-}$ if they anti-commute, as already discussed~\cite{chiribella12}), then $A$ and $B$ can indeed influence the causal order between $C$ and $D$.

\section{Correlations with dynamical causal order from quantum circuits}
\label{sec:QCQC_correlations}

We have already clarified, for each class of quantum circuits introduced in the previous section, the type of correlations they may generate---i.e., in which of the polytopes introduced in Sec.~\ref{sec:corr_causal} their correlations can be found (see Figs.~\ref{fig:QCCC-classes} and \ref{fig:QCQC-classes}). One may now wonder whether the circuits from a given class can reach \emph{all} correlations in the corresponding polytopes (and if not, whether some further restrictions can be evidenced) and whether one can exhibit correlations that are outside of a smaller polytope.
This can be done, in particular, by considering the quantity $I_4$ from Eq.~\eqref{eq:def_I4}---which we introduced precisely to separate our four polytopes of correlations---and by looking at the maximal value that processes in a given class, with appropriate quantum instruments for the different parties, can reach.

As one may have expected, it can be shown that any correlation in ${\cal P}_{\text{convFO}}$ can be obtained from a quantum (and even a classical) circuit in \textup{\textsf{QC-convFO}}---hence also from circuits in larger classes, in particular in \textup{\textsf{QC-supFO}}---and that any correlation in ${\cal P}_{\text{causal}}$ can be obtained from a quantum (and even a classical) circuit in \textup{\textsf{QC-CC}}---hence also from circuits in \textup{\textsf{QC-QC}}; see Appendix~\ref{app:obtaining_any_correl}.
Thus, quantum circuits in \textup{\textsf{QC-convFO}} and \textup{\textsf{QC-supFO}} (which both only generate correlations in ${\cal P}_{\text{convFO}}$) saturate the convFO bound of $I_4 \leq_{\text{convFO}} \frac{15}{16}$; while quantum circuits in \textup{\textsf{QC-CC}} and \textup{\textsf{QC-QC}} can reach the  trivial upper bound of $I_4 \leq_{\text{causal}} 1$, thereby violating the bounds of Eqs.~\eqref{eq:bnd_I4_FO}--\eqref{eq:bnd_I4_NIO'} for ${\cal P}_{\text{convFO}}$, ${\cal P}_{\text{NIO}}$ and ${\cal P}_{\text{NIO}'}$. In Appendix~\ref{app:obtaining_any_correl} we present explicit process matrices, together with choices of instruments, that can be used to reach these different bounds.

The situation is more subtle for the circuits with non-influenceable classical or quantum control of causal order. Recall that we showed that circuits in \textup{\textsf{QC-NICC}} and \textup{\textsf{QC-NIQC}} only generate correlations in ${\cal P}_{\text{NIO}}$ and ${\cal P}_{\text{NIO}'}$, respectively.
It is however less clear \emph{a priori} whether the whole polytopes can be obtained.
Indeed while the correlation with non-influenceable order that reaches the bound $I_4 \leq_{\text{NIO}} \frac{31}{32}$ (Eq.~\eqref{eq:saturate_NIO} in Appendix~\ref{app:proof_NIO_bound}), for instance, can be obtained naturally by a QC-CC and appropriate instruments, that QC-CC is not a QC-NICC: other choices of instruments give correlations with influenceable order (see again Appendix~\ref{app:obtaining_any_correl}).

To investigate this question further, we exploited the techniques of Ref.~\cite{liu24}, which provide means to compute process matrix upper bounds on causal inequalities via SDPs.
In particular, for so-called ``single-trigger inequalities'', where for each party $A_k$ there is only a single input $x_k=\xi_k$ (the ``trigger'') for which the inequality depends non-trivially on $a_k$, Ref.~\cite{liu24} shows that one can assume, without loss of generality, that the parties use some ``canonical'' instruments, and thereby the process matrix bound of the inequality can be computed exactly as an SDP.
It is easy to see that in the lazy scenario, all inequalities are single-trigger (with triggers $\xi_k=1$ for each party $A_k$).
Moreover, one can readily verify that the results of Ref.~\cite{liu24} are applicable not only to correlations from valid process matrices, but also to all the classes of quantum circuits we consider in this work.%
\footnote{Indeed, when optimising the bound of single-trigger inequalities over a class of process matrices, one can restrict without loss of generality to the canonical instruments of Ref.~\cite{liu24} as long as that class is closed under the composition with ``local one-slot combs'' at each slot---i.e., for any $W$ in that given class, for any party $A_k$, $W*C_k$ remains in the class for any one-slot quantum comb $C_k$ (with local past $P_k$ and future $F_k$ that are plugged into $A_k^I$ and $A_k^O$, respectively, and whose slot becomes $A_k$'s new slot) representing local pre- and post-processing accompanied by a quantum memory.}
This allows us to calculate numerically the maximal value of $I_4$ over the classes \textup{\textsf{QC-NICC}} and \textup{\textsf{QC-NIQC}} as a single SDP with the parties' instruments fixed to the canonical instruments.
In the lazy scenario, these instruments act on qubit inputs and output two qubits to the process, requiring process matrices of dimension $(2\times 4)^4 \times (2\times 4)^4=4096 \times 4096$ to be considered.
Despite the size of the optimisation problem, we were able to determine that the maximal values of $I_4$ are $0.9482$ for \textup{\textsf{QC-NICC}}, and $0.9527$ for \textup{\textsf{QC-NIQC}}, up to precision of the SDP solver~\cite{scs}.
These values can in fact be obtained (up to the same numerical precision) with simple qubit instruments (the same ones both for \textup{\textsf{QC-NICC}} and \textup{\textsf{QC-NIQC}}) and processes compatible with the coarse-grained causal orders $A\prec B\prec \{C,D\}$ (for QC-NICCs) and $\{A,B\} \prec \{C,D\}$ (for QC-NIQCs), with a similar general structure as in Figs.~\ref{fig:QC-NICC-example} and~\ref{fig:QC-NIQC-example}.
The precise form of the processes obtaining these values, however, is somewhat more complicated than the examples $W_{\text{NICC}}$ (Eq.~\eqref{eq:QC-NICC-example}) and $W_{\text{NIQC}}$ (Eq.~\eqref{eq:QC-NIQC-example}) we introduced, so we do not reproduce them here.
Further details of the instruments and processes obtaining these violations are given in Appendix~\ref{app:violations}.

\begin{figure}[t]
\centering
\includegraphics[width=1\columnwidth]{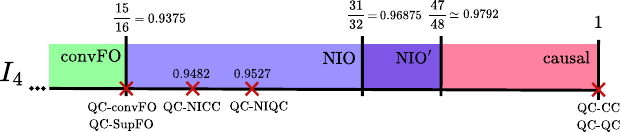}
\caption{Different bounds on the quantity $I_4$ defined in Eq.~\eqref{eq:def_I4}, attainable by circuits in the different classes under consideration. QC-convFOs and QC-supFOs can only generate correlations in ${\cal P}_{\text{convFO}}$ and saturate the corresponding bound $I_4 \leq_{\text{convFO}} \frac{15}{16}$ from Eq.~\eqref{eq:bnd_I4_FO}. Similarly, general QC-CCs and QC-QCs can only generate correlations in ${\cal P}_{\text{causal}}$ and saturate the corresponding (trivial) bound $I_4 \leq_{\text{causal}} 1$ from Eq.~\eqref{eq:bnd_I4_causal}. On the other hand, the optimal values obtainable with QC-NICCs and QC-NIQCS do not reach the corresponding bounds for ${\cal P}_{\text{NIO}}$ and ${\cal P}_{\text{NIO}'}$, $I_4 \leq_{\text{NIO}} \frac{31}{32}$ and $I_4 \leq_{\text{NIO}'} \frac{47}{48}$ from Eqs.~\eqref{eq:bnd_I4_NIO}--\eqref{eq:bnd_I4_NIO'}.}
\label{fig:bounds_I4}
\end{figure}

The optimal values of $I_4$ for these two classes both violate the convFO bound $I_4 \leq_{\text{convFO}} \frac{15}{16} = 0.9375$.
However, they are far from saturating the NIO and NIO$^\prime$ bounds $I_4 \leq_{\text{NIO}} \frac{31}{32} = 0.96875$ and $I_4 \leq_{\text{NIO}'} \frac{47}{48} \simeq 0.9792$; see Fig.~\ref{fig:bounds_I4}.
As a result, the classes \textup{\textsf{QC-NICC}} and \textup{\textsf{QC-NIQC}} cannot cover the whole polytopes ${\cal P}_{\text{NIO}}$ and ${\cal P}_{\text{NIO}'}$. 
Understanding precisely how these correlations are further restricted remains an open problem.

A related open question is whether QC-NIQCs can give correlations outside of ${\cal P}_{\text{NIO}}$, or if for some reason their correlations are also constrained to be in that smaller polytope. Indeed, while we found that QC-NIQCs cannot violate the NIO bound for $I_4$, it could still be possible to witness a violation of the NIO bound for different quantities than $I_4$. However, since we were unable to characterise the polytope $\mathcal{P}_{\textup{NIO}}$ in terms of its vertices or facets due to its high-dimensionality (even in the simplest nontrivial scenario), we have no easy method to obtain other inequalities that bound these correlations, and hence study their potential violation. 

Adopting a complementary perspective, one may also wonder whether one could define any other subclass of \textup{\textsf{QC-NIQC}}, larger than \textup{\textsf{QC-NICC}}, for which one can prove that the correlations produced are in ${\cal P}_{\textup{NIO}}$. We formally define such a class in Appendix~\ref{app:new_class}, even if its definition is somewhat ad hoc and lacking---for now, at least---a clear physical motivation.

Finally, we considered whether our explicit examples of circuits with dynamical but non-influenceable control of causal order, $W_{\text{NICC}}\in \textup{\textsf{QC-NICC}} \setminus \textup{\textsf{QC-convFO}}$ and $W_{\text{NIQC}}\in\text{\textup{\textsf{QC-NIQC}}}\setminus(\text{\textup{\textsf{QC-supFO}}}\cup\text{\textup{\textsf{QC-NICC}}})$ from Eqs.~\eqref{eq:QC-NICC-example} and~\eqref{eq:QC-NIQC-example}, could provide violations of some inequality for non-dynamical causal orders. Although we did not find violations of Eq.~\eqref{eq:bnd_I4_FO}, we were able to find another inequality that bounds $\mathcal{P}_{\textup{convFO}}$ and can be violated by $W_{\text{NICC}}$. We did not find such a violation with $W_{\text{NIQC}}$, but we found another similar process in $\text{\textup{\textsf{QC-NIQC}}}\setminus(\text{\textup{\textsf{QC-supFO}}}\cup\text{\textup{\textsf{QC-NICC}}})$ that also provides an analytical violation of the same inequality as for $W_{\text{NICC}}$; see Appendix~\ref{app:violations}.

\section{Discussion and Outlook}
\label{sec:conclusion}

As soon as $N\ge 3$ parties establish correlations in a well-defined causal structure, there may be some dynamical causal order, which cannot be explained in terms of a (probabilistic) fixed causal order between the parties. 
Previously, all the examples of causal structures with dynamical causal order have been understood as the influence of past parties on the causal order of future parties. In this work, we have critically reassessed this assertion and shown it to be incomplete. 
We identified a new way for the causal order to be dynamical without any influence of the past parties on the causal order of the future parties, and which arises as soon as $N\ge 4$.
We termed this dynamical but non-influenceable causal order, and we showed that the set of causal correlations with non-influenceable order forms a subpolytope ${\cal P}_\textup{NIO}$ of the causal polytope ${\cal P}_\textup{causal}$, that reduces to the convex mixture of fixed orders polytope ${\cal P}_\textup{convFO}$ for $N \le 3$. 
Yet, for $N \ge 4$, ${\cal P}_\textup{NIO}$ is strictly larger than ${\cal P}_\textup{convFO}$, and there exist causal correlations with dynamical but non-influenceable order. We also proposed a relaxed form of non-influenceable causal order, namely causal correlations with non-influenceable coarse-grained orders. 

We then built on the formalism of quantum circuits with classical and quantum control of causal order~\cite{wechs21} to define various classes of processes inspired by the different classes of causal correlations presented (and indeed generating correlations in the corresponding classes, cf.\ Fig.~\ref{fig:QCQC-classes}). We are thus able to clearly distinguish between the classes of quantum circuits with non-dynamical, dynamical but non-influenceable and dynamical and influenceable classical control of causal order (\textup{\textsf{QC-convFO}}, \textup{\textsf{QC-NICC}} and \textup{\textsf{QC-CC}}), and analogously in the case of quantum circuits with quantum control of causal order (with the classes \textup{\textsf{QC-supFO}}, \textup{\textsf{QC-NIQC}} and \textup{\textsf{QC-QC}} respectively). We summarise the inclusion relations between these classes as follows (see also Fig.~\ref{fig:QCQC-classes}):
\begin{align}
    \begin{array}{ccccc}
        \text{\textup{\textsf{QC-convFO}}} & \subset & \text{\textup{\textsf{QC-NICC}}} & \subset & \text{\textup{\textsf{QC-CC}}} \\[1mm]
        \rotatebox[origin=c]{270}{$\subset$} &  & \rotatebox[origin=c]{270}{$\subset$} &  & \rotatebox[origin=c]{270}{$\subset$} \\[1mm]
        \text{\textup{\textsf{QC-supFO}}} & \subset & \text{\textup{\textsf{QC-NIQC}}} & \subset & \text{\textup{\textsf{QC-QC}}}
    \end{array}\ . \label{eq:inclusion_classes}
\end{align}
In general, for $N\ge 4$, the inclusions are all strict, while for $N\le 3$, \text{\textup{\textsf{QC-NICC}}} coincides with \text{\textup{\textsf{QC-convFO}}} and \text{\textup{\textsf{QC-NIQC}}} coincides with \text{\textup{\textsf{QC-supFO}}}. 
Furthermore, the inclusions between \textup{\textsf{QC-NICC}} and \textup{\textsf{QC-CC}}, and between \textup{\textsf{QC-NIQC}} and \textup{\textsf{QC-QC}} are in general strict as soon as $N\ge 2$ if $\HS^P$ is nontrivial (or $N\ge 3$ if $\HS^P$ is trivial); the inclusions between the QC-CC (sub)-classes on the top row and the corresponding QC-QC (sub)-classes on the bottom row are in general strict as soon as $N\ge 2$ if $\HS^F$ is nontrivial (or $N\ge 3$ if $\HS^F$ is trivial).\\

Below we list a number of follow-up questions, open problems and perspectives.

\textit{Inclusion relations between the classes.}
From the inclusion relations defined in Eq.~\eqref{eq:inclusion_classes}, one may wonder whether there remain nontrivial and interesting intersections between the classes, as denoted by the question marks in Fig.~\ref{fig:QCQC-classes}, or whether it is the case that (some of) these intersections are empty, so that $\text{\textup{\textsf{QC-supFO}}} \cap \text{\textup{\textsf{QC-NICC}}} = \text{\textup{\textsf{QC-convFO}}}$, $\text{\textup{\textsf{QC-supFO}}} \cap \text{\textup{\textsf{QC-CC}}} = \text{\textup{\textsf{QC-convFO}}}$, and/or that $\text{\textup{\textsf{QC-NIQC}}} \cap \text{\textup{\textsf{QC-CC}}} = \text{\textup{\textsf{QC-NICC}}}$? 

\textit{Graphical representation of the classes.}
All the classes of quantum circuits considered in this work were characterised in terms of SDP constraints on positive semidefinite matrices. We propose in Appendix~\ref{app:subsec:graphical_rep} a graphical representation of these constraints which highlights the structural differences between the considered classes. Considering new graphical structures could inspire the definition of new classes of circuits. It would also be interesting to explore the connection between this graphical representation with other graphical understandings of the causal structure that underlies process matrices. One can for instance find similar structures in the graph that represents the SDP constraints for the \textup{\textsf{QC-QC}} class (see Fig.~\ref{fig:graph_QCQC} in Appendix~\ref{app:subsec:graphical_rep}) as in the so-called ``branch graph''~\cite{vanrietvelde22} that one can build from a possible generic routed quantum circuit representation of QC-QCs~\cite{grothus25}.

\textit{New examples of QC-NICCs/QC-NIQCs with qualitatively different structures}.
We exhibited in particular two examples of quantum circuits with dynamical but non-influenceable classical or quantum control of causal order, characterised by their process matrices $W_{\textup{NICC}}$ and $W_{\textup{NIQC}}$ (Eqs.~\eqref{eq:QC-NICC-example} and~\eqref{eq:QC-NIQC-example}). 
Admittedly, for these two examples the dynamical behaviour has a strong classical flavour: a measurement is performed after the action of $A$ and $B$, whose classical outcome controls the order between $C$ and $D$. (This is also the case for the additional examples in Appendix~\ref{app:violations}.) One may wonder whether there exist other examples of quantum circuits with non-influenceable causal order, where the dynamical behaviour has a ``more quantum'' flavour. How this question could be formalised, and whether a scenario with more parties involved would be required---so that for instance a quantum system that coherently controls the causal order between $C$ and $D$ is sent to a future party---is left for further work.
 
\textit{Reconciling dynamicality and influenceability.}
A corollary of the existence of dynamical but non-influenceable causal structures is that the two notions of dynamicality and influenceability are in general (for $N \geq 4$) inequivalent, both at the level of causal correlations and of quantum circuits with classical or quantum control of causal order. An interesting follow-up question would be to clarify under which conditions dynamicality and influenceability are found to be equivalent notions. For example, one may wonder whether dynamicality without influenceability may still be found in ``pure'' QC-QCs~\cite{araujo17}, that is, QC-QCs that transform unitary maps into a unitary map from a global past Hilbert space to a global future Hilbert space. Indeed, we note that the property of dynamicality without influenceability of the processes $W_{\textup{NICC}}$ and $W_{\textup{NIQC}}$ presented above crucially relies on the maximally entangled pair of qubits initially prepared. If one purifies these processes (in the sense of~\cite{araujo17}), the system initially prepared will then depend on the action of the global past, in turn allowing the parties' actions to---in general---influence the causal order. 
Along this line, an alternative path worth exploring further is to study whether the possibility of causal structures with dynamical but non-influenceable causal order can be understood as some sort of fine-tuning that prevents the parties from influencing the causal order at the statistical level, even if such an influence is in principle possible at the level of the underlying causal structure.

\textit{Refining the characterisation of dynamical causal order.}
More generally, this work also opens several follow-up questions to better understand causal structures between $N$ parties. One first idea could be to refine the analysis performed in this work in order to define ``partially dynamical'' causal structures, so as to provide a clearer characterisation of which parties contribute to the dynamical behaviour of a process and which do not, allowing one to quantify some notion of ``depth'' of dynamicality. 
Another interesting perspective is to look whether any fundamentally new feature could occur if one studies scenarios with $N\ge 5$ parties, in the same way that the tripartite scenario gives rise to dynamical properties, and the fourpartite scenario gives rise to dynamical but non-influenceable causal order. 

\textit{Formalising dynamicality beyond QC-QCs.}
An important question that remains open is whether one could formalise a notion of dynamical causal order beyond the class of QC-QCs we restricted ourselves to here, as there are processes---such as the Lugano (or ``AF/BW'') process~\cite{Baumeler14_2,Baumeler16}---which cannot be expressed as QC-QCs and whose dynamicality remains ambiguous. This would be rather challenging as the formalisation of the notion of dynamical causal order presented in this present work crucially relies on the notion of a quantum control system that dictates the causal order between the parties. Yet, no clear physical interpretation of process matrices that go beyond \textup{\textsf{QC-QC}} has been proposed so far, which makes it unclear what the notion of dynamical causal order could mean in general. Nevertheless, a potential track to tackle this problem could be to consider the framework of routed quantum circuits~\cite{vanrietvelde21,vanrietvelde22}, that can model process matrices beyond QC-QCs, and in which a graphical criterion to detect the dynamical causal order in the ``Grenoble process'' was hinted at, in terms of ``bifurcation choices" of the parties~\cite{vanrietvelde22}. One could first try to see whether this graphical criterion captures the same notion of dynamical causal order for QC-QCs as ours (or perhaps of influenceable order) and then generalise it to processes beyond QC-QCs---including processes that may violate causal inequalities.

\textit{Computational advantages from dynamical causal order.}
From a more practical point of view, the study of indefinite causal order has aroused some particular interest for the potential information processing advantages that they offer (see \cite{Rozema24,Costa26} for extensive reviews of these). 
Whether dynamical causal order, both classically or coherently controlled, can provide such kind of advantages thus arises as a natural question.
One may wonder, for instance, whether the advantage of QC-CCs over different fixed-order processes obtained in Ref.~\cite{bavaresco24} (in simulating the quantum switch with two queries to $A$'s operation and one to $B$'s operation) or the advantage of QC-QCs over a particular class of quantum circuits with non-dynamical control of causal order (the \textup{\textsf{Sup}} class from Ref.~\cite{Liu23}, which turns out to be strictly included in \text{\textup{\textsf{QC-supFO}}}) for a tripartite task in quantum metrology could be ascribed to dynamicality.
In both cases, it seems the answer is no: QC-convFOs turn out to obtain the same performance as QC-CCs in the task of~\cite{bavaresco24}, while QC-supFOs perform as well as QC-QCs in the task of~\cite{mothe23}, for the channels considered there.
Further work thus remains to be done to show whether dynamical causal order, both in the presence and absence of indefinite causal order, is indeed a useful resource in quantum information. 
In this context, we note that the characterisations of the various classes of processes proposed in this work, in terms of linear constraints on positive semidefinite matrices, provide a readily usable way to compare the performances of quantum circuits with dynamical control of causal order with those of quantum circuits with non-dynamical causal order, for any task that could be solved or optimised through semidefinite programming techniques.

\begin{acknowledgments}
\vspace{-2mm}
We thank Maarten Grothus, Augustin Vanrietvelde and V.\ Vilasini for helpful discussions and comments on this work.
This research was funded in part by l’Agence Nationale de la Recherche (ANR) projects ANR-15-IDEX-02 and ANR-22-CE47-0012, and the PEPR integrated project EPiQ ANR-22-PETQ-0007 as part of Plan France 2030. RM
acknowledges the support from the Alexander von Humboldt Foundation. 
\end{acknowledgments}



%

\clearpage
\onecolumngrid
\appendix

\section*{Appendices}

\vspace{5mm}

\addtocontents{toc}{\string\fi}

\AppendixTOC

\clearpage

\section{Multipartite causal correlations: equivalence between the different definitions}
\label{app:def_causal_correl}

\subsection{Abbott \emph{et al.}'s definition}

In Ref.~\cite{Abbott16} Abbott \emph{et al.}~proposed the following recursive definition for causal correlations:
an $N$-partite probability distribution $p(\vec a|\vec x)$ is causal if and only if $N=1$, or it has a decomposition as
\begin{align}
    p(\vec a|\vec x) = \sum_{k_1} q_{k_1} \, p_{k_1}(a_{k_1}|x_{k_1}) \, p_{\mathcal{N}\backslash k_1,x_{k_1},a_{k_1}}(\vec a_{\mathcal{N}\backslash k_1}|\vec x_{\mathcal{N}\backslash k_1}), 
        \label{eq:def_causal_p_recursive}
\end{align}
with $q_{k_1}\ge 0$, $\sum_{k_1}q_{k_1}=1$, where for each $k_1\in\mathcal{N}$, $p_{k_1}(a_{k_1}|x_{k_1})$ is a probability distribution for the single party $A_{k_1}$, and where for each $k_1,x_{k_1},a_{k_1}$, $p_{\mathcal{N}\backslash k_1,x_{k_1},a_{k_1}}(\vec a_{\mathcal{N}\backslash k_1}|\vec x_{\mathcal{N}\backslash k_1})$ is a conditional probability distribution for the remaining $N-1$ parties in $\mathcal{N}\backslash k_1$,%
\footnote{We use the short-hand notation $\mathcal{N}\backslash k_1=\mathcal{N}\backslash\{k_1\}$ (or more generally, $\K_n\backslash k_n=\K_n\backslash\{k_n\}$).}
which is itself causal (according to the present definition).

\bigskip

Our definition in Eq.~\eqref{eq:def_causal_p} can be seen as an ``unravelling'' of this definition. Let us indeed prove by induction that the two are equivalent.
This is clearly the case for $N=1$ (they are both trivial); let us then assume they are equivalent for $N-1 \ (\ge 1)$:

\begin{itemize}

    \item \underline{Proof that \eqref{eq:def_causal_p_recursive} $\Rightarrow$ \eqref{eq:def_causal_p}:} Assume Eq.~\eqref{eq:def_causal_p_recursive}. Since $p_{\mathcal{N}\backslash k_1,x_{k_1},a_{k_1}}(\vec a_{\mathcal{N}\backslash k_1}|\vec x_{\mathcal{N}\backslash k_1})$ is required to be an $(N-1)$-partite causal distribution according to Abbott \emph{et al.}'s definition, then by the induction hypothesis, it is also causal according to our definition in the main text, i.e., it has a decomposition of the form of Eq.~\eqref{eq:def_causal_p}, with permutations $\pi'=(k_2,\ldots,k_N)$ of $\mathcal{N}\backslash k_1$:
    \begin{align}
        & p_{\mathcal{N}\backslash k_1,x_{k_1},a_{k_1}}(\vec a_{\mathcal{N}\backslash k_1}|\vec x_{\mathcal{N}\backslash k_1}) = \sum_{\pi'=(k_2,\ldots,k_N)} p_{\mathcal{N}\backslash k_1,x_{k_1},a_{k_1}}\big(\pi'=(k_2,\ldots,k_N),\vec a_{\mathcal{N}\backslash k_1}\big|\vec x_{\mathcal{N}\backslash k_1}\big) \notag \\
        & \text{s.t.} \ \forall\, n = 1,\ldots,N-1, \ \forall\, (k_2,\ldots,k_n,k_{n+1}), \notag \\
        & \hspace{5mm} p_{\mathcal{N}\backslash k_1,x_{k_1},a_{k_1}}\big((k_2,\ldots,k_n,k_{n+1}),\vec a_{k_2,\ldots,k_n}\big|\vec x_{\mathcal{N}\backslash k_1}\big) \text{ does not depend on } \vec x_{\mathcal{N}\backslash k_1\backslash\{k_2,\ldots,k_n\}}.  \label{eq:def_rec_proof2}
    \end{align}
    Defining $p\big((k_1,\ldots,k_N),\vec a\big|\vec x\big) \coloneqq q_{k_1} \, p_{k_1}(a_{k_1}|x_{k_1}) \, p_{\mathcal{N}\backslash k_1,x_{k_1},a_{k_1}}\big((k_2,\ldots,k_N),\vec a_{\mathcal{N}\backslash k_1}\big|\vec x_{\mathcal{N}\backslash k_1}\big)$, the combination of Eq.~\eqref{eq:def_causal_p_recursive} with Eq.~\eqref{eq:def_rec_proof2} then readily provides the decomposition
    \begin{align}
        p(\vec a|\vec x) & = \sum_{k_1} q_{k_1} \, p_{k_1}(a_{k_1}|x_{k_1}) \, \sum_{\pi'=(k_2,\ldots,k_N)} p_{\mathcal{N}\backslash k_1,x_{k_1},a_{k_1}}\big(\pi'=(k_2,\ldots,k_N),\vec a_{\mathcal{N}\backslash k_1}\big|\vec x_{\mathcal{N}\backslash k_1}\big), \notag \\
        & = \sum_{\pi=(k_1,\ldots,k_N)} p\big(\pi=(k_1,\ldots,k_N),\vec a|\vec x\big),
    \end{align}
    in terms of a valid probability distribution $p\big(\pi,\vec a|\vec x\big)$ which can be shown to satisfy Eq.~\eqref{eq:def_causal_p}: indeed, for any $n = 1,\ldots,N-1$ and any $(k_1,\ldots,k_n,k_{n+1})$, 
    \begin{align}
        p\big((k_1,\ldots,k_n,k_{n+1}),\vec a_{k_1,\ldots,k_n}\big|\vec x\big) = q_{k_1} \, p_{k_1}(a_{k_1}|x_{k_1}) \, p_{\mathcal{N}\backslash k_1,x_{k_1},a_{k_1}}\big((k_2,\ldots,k_n,k_{n+1}),\vec a_{k_2,\ldots,k_n}\big|\vec x_{\mathcal{N}\backslash k_1}\big)
    \end{align}
    does not, according to Eq.~\eqref{eq:def_rec_proof2} above, depend on $\vec x_{\mathcal{N}\backslash\{k_1,\ldots,k_n\}}$; while for $n = 0$ and any $k_1$, $p\big((k_1)\big|\vec x\big) =q_{k_1}$ does not depend on $\vec x$.
    
    \item \underline{Proof that \eqref{eq:def_causal_p} $\Rightarrow$ \eqref{eq:def_causal_p_recursive}:} Assume Eq.~\eqref{eq:def_causal_p}. The probability distribution $p(\pi,\vec a|\vec x)$ in the decomposition of $p(\vec a|\vec x)$ satisfies in particular, for $n=0$ and $n=1$, that for any $k_1$, $p\big((k_1)\big|\vec x\big) = p\big((k_1)\big)$ is independent of $\vec x$, and $p\big((k_1),a_{k_1}\big|\vec x\big) = p\big((k_1),a_{k_1}\big|x_{k_1}\big)$ only depends on $x_{k_1}$, and not on all the other inputs.
    Defining%
    \footnote{We assume for simplicity, in these definitions, that the denominators $q_{k_1}$ and $p\big((k_1),a_{k_1}\big|x_{k_1}\big)$ are nonzero. The terms for which these are zero do not contribute to the argument; some (irrelevant) values can always be given to $p_{k_1}(a_{k_1}|x_{k_1})$ and $p_{\mathcal{N}\backslash k_1,x_{k_1},a_{k_1}}(\vec a_{\mathcal{N}\backslash k_1}|\vec x_{\mathcal{N}\backslash k_1})$ in those cases, to make these valid and causal distributions. \label{ftn:nul_denom}}
    \begin{align}
        & q_{k_1}\coloneqq p\big((k_1)\big), \quad p_{k_1}(a_{k_1}|x_{k_1})\coloneqq p\big((k_1),a_{k_1}\big|x_{k_1}\big) \big/ q_{k_1}, \notag \\[1mm]
        & p_{\mathcal{N}\backslash k_1,x_{k_1},a_{k_1}}\big((k_2,\ldots,k_N),\vec a_{\mathcal{N}\backslash k_1}\big|\vec x_{\mathcal{N}\backslash k_1}\big)\coloneqq p\big((k_1,k_2,\ldots,k_N),\vec a\big|\vec x\big) \big/ p\big((k_1),a_{k_1}\big|x_{k_1}\big) \notag \\[1mm]
        \text{and} \quad & p_{\mathcal{N}\backslash k_1,x_{k_1},a_{k_1}}(\vec a_{\mathcal{N}\backslash k_1}|\vec x_{\mathcal{N}\backslash k_1})\coloneqq \sum_{(k_2,\ldots,k_N)} p_{\mathcal{N}\backslash k_1,x_{k_1},a_{k_1}}\big((k_2,\ldots,k_N),\vec a_{\mathcal{N}\backslash k_1}\big|\vec x_{\mathcal{N}\backslash k_1}\big), \label{eq:def_rec_proof1}
    \end{align}
    the decomposition of Eq.~\eqref{eq:def_causal_p} can then be written in exactly the same form as in Eq.~\eqref{eq:def_causal_p_recursive}.

    Clearly $q_{k_1}\ge 0$, $\sum_{k_1}q_{k_1}=1$, and $p_{k_1}(a_{k_1}|x_{k_1})$ is a valid (nonnegative and normalised) probability distribution. It remains to verify that $p_{\mathcal{N}\backslash k_1,x_{k_1},a_{k_1}}(\vec a_{\mathcal{N}\backslash k_1}|\vec x_{\mathcal{N}\backslash k_1})$ is a causal $(N-1)$-partite probability distribution (according to Abbott \emph{et al.}'s definition).

    To see this, notice that the last line in Eq.~\eqref{eq:def_rec_proof1} above gives precisely a decomposition of $p_{\mathcal{N}\backslash k_1,x_{k_1},a_{k_1}}(\vec a_{\mathcal{N}\backslash k_1}|\vec x_{\mathcal{N}\backslash k_1})$ of the form of Eq.~\eqref{eq:def_causal_p} for $(N-1)$-partite distributions, with a sum over all permutations $\pi'=(k_2,\ldots,k_n)$ of $\mathcal{N}'\coloneqq\mathcal{N}\backslash k_1$. Consider some $n\in \{1,\ldots,N-1\}$ and some subsequence $(k_2,\ldots,k_{n+1})$ of $\mathcal{N}'$. One then has
    \begin{align}
        p_{\mathcal{N}\backslash k_1,x_{k_1},a_{k_1}}\big((k_2,\ldots,k_n,k_{n+1}),\vec a_{k_2,\ldots,k_n}\big|\vec x_{\mathcal{N}'}\big) = p\big((k_1,k_2,\ldots,k_n,k_{n+1}),\vec a_{k_1,k_2,\ldots,k_n}\big|\vec x\big) \big/ p\big((k_1),a_{k_1}\big|x_{k_1}\big).
    \end{align}
    According to Eq.~\eqref{eq:def_causal_p}, which we assumed for $p(\vec a|\vec x)$, the numerator in the right-hand-side above does not depend on $\vec x_{\mathcal{N}\backslash\{k_1,k_2,\ldots,k_n\}}$ (and neither does the denominator). Hence, the left-hand-side does not depend on $\vec x_{\mathcal{N}\backslash\{k_1,k_2,\ldots,k_n\}} = \vec x_{\mathcal{N}'\backslash\{k_2,\ldots,k_n\}}$, so that $p_{\mathcal{N}\backslash k_1,x_{k_1},a_{k_1}}(\vec a_{\mathcal{N}\backslash k_1}|\vec x_{\mathcal{N}\backslash k_1})$ satisfies the constraint of Eq.~\eqref{eq:def_causal_p} for $(N-1)$-partite distributions. It is thus causal according to that definition, and therefore, by our induction hypothesis, it is causal according to Abbott \emph{et al.}'s definition, as required.
    
\end{itemize}

\subsection{Oreshkov and Giarmatzi's definition}

The first general definition for multipartite causal correlations that appeared in the literature was in fact proposed earlier by Oreshkov and Giarmatzi~\cite{Oreshkov16}. It says that causal correlations are those compatible with a causal structure described as a Strict Partial Order (SPO), rather than a full permutation as in our definition of Eq.~\eqref{eq:def_causal_p}.

\medskip

More specifically: an SPO between $N$ parties $A_i$ ($i\in \mathcal{N} \coloneqq \{1,\ldots,N\}$) is a binary relation $\prec$ which is irreflexive, transitive and antisymmetric. An SPO can in general be defined by the (consistent\footnote{Some consistency constraints are implied by the irreflexivity, transitivity and antisymmetry of the SPO. E.g., the pairwise relations $A_1 \prec A_2$, $A_2 \prec A_3$, $A_3 \prec A_1$ would be inconsistent.}) list of all pairwise relations for all $i \neq j$:%
\footnote{When writing such pairwise relations between $A_i$ and $A_j$, we will always assume $i \neq j$, without necessarily specifying it explicitly.}
either $A_i \prec A_j$ (understood as $A_i$ being in the causal past of $A_j$), $A_j \prec A_i$ ($A_j$ being in the causal past of $A_i$), or $A_i \nprec\nsucc A_j$ (neither party being in the causal past of the other one). We shall generically denote by $\kappa = \kappa(\mathcal{N})$ an SPO for the parties in $\mathcal{N}$. Note that any such SPO $\kappa(\mathcal{N})$ induces an SPO $\kappa(\mathcal{K})$ on any subset $\mathcal{K}$ of $\mathcal{N}$.

Rephrasing it slightly, Oreshkov and Giarmatzi's definition (Definition~2.3 in~\cite{Oreshkov16}) states that a correlation $p(\vec a|\vec x)$ is causal if and only if
\begin{align}
    \exists\, p(\kappa,\vec a|\vec x) \text{ s.t. } & p(\vec a|\vec x) = \sum_\kappa p(\kappa,\vec a|\vec x) \notag \\[-1mm]
    & \text{and} \ \forall\, \mathcal{K} \subsetneq \mathcal{N}, \ \forall\, \ell \in \mathcal{N}\backslash \mathcal{K}, \ \forall\, \bar{\kappa}=\bar{\kappa}(\mathcal{K}\cup\{\ell\}) \text{ for which } A_\ell\stackrel{\forall}{\nprec}\mathcal{K}, \notag \\[1mm]
    & \hspace{40mm} p\big(\bar{\kappa}(\mathcal{K}\cup\{\ell\}),\vec a_{\mathcal{K}}\big|\vec x\big) \text{ does not depend on } x_\ell,
    \label{eq:def_causal_p_OG}
\end{align}
where $p(\kappa,\vec a|\vec x)$ is a probability distribution (so that $p(\kappa,\vec a|\vec x)\ge 0$, $\sum_{\kappa,\vec a} p(\kappa,\vec a|\vec x)=1$) in which the random variable $\kappa = \kappa(\mathcal{N})$ takes values in all possible SPOs on $\mathcal{N}$, where the shorthand notation $A_\ell\stackrel{\forall}{\nprec}\mathcal{K}$ means $A_\ell \nprec A_k$ (i.e., $A_k \prec A_\ell$ or $A_k \nprec\nsucc A_\ell$) for all $k\in\mathcal{K}$, and where the last line involves the marginal distributions $p\big(\bar{\kappa}(\mathcal{K}\cup\{\ell\}),\vec a_{\mathcal{K}}\big|\vec x\big)$ obtained from $p(\kappa,\vec a_{\mathcal{K}}|\vec x)$ ($=\sum_{\vec a_{\mathcal{N}\backslash\mathcal{K}}}p(\kappa,\vec a_{\mathcal{K}},\vec a_{\mathcal{N}\backslash\mathcal{K}}|\vec x)$) by summing over all SPOs $\kappa=\kappa(\mathcal{N})$ which induce the SPO $\bar{\kappa}(\mathcal{K}\cup\{\ell\})$ on the subset $\mathcal{K}\cup\{\ell\}$: that is, $p\big(\bar{\kappa}(\mathcal{K}\cup\{\ell\}),\vec a_{\mathcal{K}}\big|\vec x\big) \coloneqq \sum_{\kappa(\mathcal{N})\text{ s.t. }\kappa(\mathcal{K}\cup\{\ell\}) = \bar{\kappa}(\mathcal{K}\cup\{\ell\})} p\big(\kappa(\mathcal{N}),\vec a_{\mathcal{K}}\big|\vec x\big)$.

\bigskip

That this definition is equivalent to that of Abbott \emph{et al.}, Eq.~\eqref{eq:def_causal_p_recursive} above, was in fact already proven by Oreshkov and Giarmatzi in~\cite{Oreshkov16}. 
Let us for convenience propose here a new version of this proof.

\begin{itemize}
    
    \item \underline{Proof that \eqref{eq:def_causal_p_recursive} $\Rightarrow$ \eqref{eq:def_causal_p_OG} [or \eqref{eq:def_causal_p} $\Rightarrow$ \eqref{eq:def_causal_p_OG}]:} Assume Eq.~\eqref{eq:def_causal_p_recursive}. We have just shown in the previous subsection that this is equivalent to assuming that $p(\vec a|\vec x)$ has a decomposition as in Eq.~\eqref{eq:def_causal_p}.
    Now, clearly a total order defined by some permutation $\pi$ is a particular case of an SPO $\kappa$, defined by the pairwise relations $A_i \prec A_j$ iff $i$ comes before $j$ in $\pi$ (for all $i\neq j$). Hence, the decomposition $p(\vec a|\vec x) = \sum_\pi p(\pi,\vec a|\vec x)$ is of the form $p(\vec a|\vec x) = \sum_\kappa p(\kappa,\vec a|\vec x)$, with $p(\kappa,\vec a|\vec x) = 0$ as soon as $\kappa$ contains a pairwise relation of the form $A_i \nprec\nsucc A_j$. It then remains to prove that the second and third lines of Eq.~\eqref{eq:def_causal_p_OG} are indeed satisfied.

    Consider $\mathcal{K} \subsetneq \mathcal{N}$, $\ell \in \mathcal{N}\backslash \mathcal{K}$, and some SPO $\bar{\kappa}=\bar{\kappa}(\mathcal{K}\cup\{\ell\})$ for which $A_\ell\stackrel{\forall}{\nprec}\mathcal{K}$. If $\bar{\kappa}$ contains any relation of the form $A_i \nprec\nsucc A_j$, then so does any SPO $\kappa(\mathcal{N})$ that induces $\bar{\kappa}(\mathcal{K}\cup\{\ell\})$ on the subset $\mathcal{K}\cup\{\ell\}$, and hence $p\big(\bar{\kappa}(\mathcal{K}\cup\{\ell\}),\vec a_{\mathcal{K}}\big|\vec x\big) = 0$, so that the last line of Eq.~\eqref{eq:def_causal_p_OG} is trivially satisfied. Otherwise, if $\bar{\kappa}$ contains no relation of the form $A_i \nprec\nsucc A_j$, then it can be equivalently described as a permutation $(k_1,\ldots,k_n,\ell)$ of $\mathcal{K}\cup\{\ell\}$ (where $n$ is the cardinality of $\mathcal{K}$, and noting that $A_\ell\stackrel{\forall}{\nprec}\mathcal{K}$ implies that $\ell$ must necessarily come last). The SPOs $\kappa=\kappa(\mathcal{N})$ that contribute to the decomposition $p(\vec a|\vec x) = \sum_\kappa p(\kappa,\vec a|\vec x)$ and that induce $\bar{\kappa}\equiv(k_1,\ldots,k_n,\ell)$ on the subset $\mathcal{K}\cup\{\ell\}$ are then the permutations $\pi$ of $\mathcal{N}$ that contain $(k_1,\ldots,k_n,\ell)$ as a subsequence---which we write $\pi \supset (k_1,\ldots,k_n,\ell)$---so that
    \begin{align}
        & p\big(\bar{\kappa}(\mathcal{K}\cup\{\ell\}),\vec a_{\mathcal{K}}\big|\vec x\big) = \sum_{\pi \supset (k_1,\ldots,k_n,\ell)} p(\pi,\vec a_{\mathcal{K}}|\vec x) \notag \\
        & \qquad = \hspace{-3mm}\sum_{\substack{(i_1,\ldots,i_m,\ell)\\ \quad\supset (k_1,\ldots,k_n,\ell)}} \hspace{-2mm} p\big((i_1,\ldots,i_m,\ell),\vec a_{\mathcal{K}}\big|\vec x\big) \ = \hspace{-2mm} \sum_{\substack{(i_1,\ldots,i_m,\ell)\\ \quad\supset (k_1,\ldots,k_n,\ell)}} \sum_{\substack{a_i:\\ \forall\,i\in\{i_1,\ldots,i_m\}\backslash\mathcal{K}}} p\big((i_1,\ldots,i_m,\ell),\vec a_{i_1,\ldots,i_m}\big|\vec x\big),
    \end{align}
    where the second sum above is over all supersequences $(i_1,\ldots,i_m,\ell)$ of $(k_1,\ldots,k_n,\ell)$ (with $n\le m<N$), and the last sum is over the outcomes $a_i$ of all parties with indices $i\in\{i_1,\ldots,i_m\}$ that were not already in $\mathcal{K}$.
    The assumption of Eq.~\eqref{eq:def_causal_p} implies that each probability in the last sums above do not depend on $x_\ell$, which implies that $p\big(\bar{\kappa}(\mathcal{K}\cup\{\ell\}),\vec a_{\mathcal{K}}\big|\vec x\big)$ itself does not depend on $x_\ell$---which indeed means that the last line of Eq.~\eqref{eq:def_causal_p_OG} is satisfied.

    \item \underline{Proof that \eqref{eq:def_causal_p_OG} $\Rightarrow$ \eqref{eq:def_causal_p_recursive}:}
    We prove this by induction on $N$. This clearly holds for $N=1$ (both conditions are trivial); let us then assume that Eq.~\eqref{eq:def_causal_p_OG} implies Eq.~\eqref{eq:def_causal_p_recursive} for $(N-1)$-partite distributions (with $N-1\geq 1$), and consider an $N$-partite distribution $p(\vec a|\vec x)$ that admits a decomposition $p(\vec a|\vec x) = \sum_\kappa p(\kappa,\vec a|\vec x)$ satisfying Eq.~\eqref{eq:def_causal_p_OG}.

    The first task, for each SPO $\kappa$ that appears in the decomposition, is to identify which party $A_{k_1}$ can be ``taken to come first'', so as to provide a decomposition as in Eq.~\eqref{eq:def_causal_p_recursive}. For this let us denote, for any given $\kappa = \kappa(\mathcal{N})$, by $\mathcal{S}_\kappa^\textsc{I}$ the (nonempty) set of parties that have no other party in their causal past---the ``first consecutive set'' according to the terminology of Ref.~\cite{Oreshkov16}.
    We will first show that the independence condition of Eq.~\eqref{eq:def_causal_p_OG} also holds when one further restricts the probabilities to SPOs with a given first consecutive set $\mathcal{S}$, i.e.:%
    \footnote{Here we adopt the notation that, for a given property $[\cdot]$ that a SPO may or may not satisfy, we write $p([\cdot],\vec a|\vec x) = \sum_{\kappa\text{ s.t.}\,[\cdot]} \,p(\kappa,\vec a|\vec x)$. E.g.\ in Eq.~\eqref{eq:lemma_proof_OG_equiv}, $p(\mathcal{S}_\kappa^\textsc{I}=\mathcal{S},\bar{\kappa}(\mathcal{K}\cup\{\ell\}),\vec a_{\mathcal{K}}|\vec x) = \sum_{\kappa\text{ s.t.}\,\mathcal{S}_\kappa^\textsc{I}=\mathcal{S}, \kappa(\mathcal{K}\cup\{\ell\})=\bar{\kappa}(\mathcal{K}\cup\{\ell\})} \,p(\kappa,\vec a_{\mathcal{K}}|\vec x)$. \\
    In the proof that follows, the notation $\mathcal{S}\stackrel{\exists}{\prec} A_j$ means that $\exists\,i\in\mathcal{S}$, $A_i\prec A_j$; the notation $\mathcal{S}\stackrel{\forall}{\nprec} A_j$ means that $\forall\,i\in\mathcal{S}$, $A_i\nprec A_j$.}
    \begin{align}
        & \forall\, \mathcal{S} \subseteq \mathcal{N}, \ \forall\, \mathcal{K} \subsetneq \mathcal{N}, \ \forall\, \ell \in \mathcal{N}\backslash \mathcal{K}, \ \forall\, \bar{\kappa}=\bar{\kappa}(\mathcal{K}\cup\{\ell\}) \text{ for which } A_\ell\stackrel{\forall}{\nprec}\mathcal{K}, \notag \\
        & \hspace{40mm} p\big(\mathcal{S}_\kappa^\textsc{I}=\mathcal{S},\bar{\kappa}(\mathcal{K}\cup\{\ell\}),\vec a_{\mathcal{K}}\big|\vec x\big) \text{ does not depend on } x_\ell. \label{eq:lemma_proof_OG_equiv}
    \end{align}

    \emph{Proof of Eq.~\eqref{eq:lemma_proof_OG_equiv}:}

    If $\ell\in\mathcal{S}$, then using the inclusion–exclusion principle one can write:%
    \footnote{More specifically, we use the fact that for some events $E$ and $\{F_j\}_{j\in\mathcal{I}}$ (for some finite set of indices $\mathcal{I}$), one has
    \begin{align}
        p\Big(E, \bigwedge_{j\in\mathcal{I}} F_j \Big) 
        = p\big(E\big) - p\Big(E, \bigvee_{j\in\mathcal{I}} (\neg F_j) \Big) 
        & = p\big(E\big) - p\Big(\bigvee_{j\in\mathcal{I}} (E\wedge \neg F_j) \Big) \notag \\
        & = p\big(E\big) - \sum_{\emptyset\neq \mathcal{J}\subseteq \mathcal{I}} (-1)^{|\mathcal{J}|-1}\, p\Big(\bigwedge_{j\in\mathcal{J}} (E\wedge \neg F_j) \Big) = \sum_{\mathcal{J}\subseteq \mathcal{I}} (-1)^{|\mathcal{J}|}\, p\Big(E,\bigwedge_{j\in\mathcal{J}} (\neg F_j) \Big), \notag
    \end{align}
    where the first equality on the second line is a direct application of the inclusion–exclusion principle.}
    \begin{align}
        p\big(\mathcal{S}_\kappa^\textsc{I}=\mathcal{S},\bar{\kappa}(\mathcal{K}\cup\{\ell\}),\vec a_{\mathcal{K}}\big|\vec x\big) & = p\Bigg(\bigwedge_{i,i'\in\mathcal{S}} \!\! \big(A_i\nprec\nsucc A_{i'}\big), \bigwedge_{j\in\mathcal{N}\backslash\mathcal{S}} \!\!\! \Big(\mathcal{S}\stackrel{\exists}{\prec} A_j\Big), \bar{\kappa}(\mathcal{K}\cup\{\ell\}),\vec a_{\mathcal{K}}\Bigg|\vec x\Bigg) \notag \\
        & = \sum_{\mathcal{J}\subseteq \mathcal{N}\backslash\mathcal{S}} (-1)^{|\mathcal{J}|}\ p\Bigg(\bigwedge_{i,i'\in\mathcal{S}} \!\! \big(A_i\nprec\nsucc A_{i'}\big), \bigwedge_{j\in\mathcal{J}} \Big(\mathcal{S}\stackrel{\forall}{\nprec} A_j\Big), \bar{\kappa}(\mathcal{K}\cup\{\ell\}),\vec a_{\mathcal{K}}\Bigg|\vec x\Bigg) \notag \\
        & = \sum_{\mathcal{J}\subseteq \mathcal{N}\backslash\mathcal{S}} (-1)^{|\mathcal{J}|} \sum_{\substack{\bar{\bar{\kappa}}(\mathcal{S}\cup\mathcal{J}\cup\mathcal{K})\text{ s.t. }\\[-1mm] \bigwedge_{i,i'\in\mathcal{S}} (A_i\nprec\nsucc A_{i'}), \bigwedge_{j\in\mathcal{J}} \big(\mathcal{S}\stackrel{\forall}{\nprec} A_j\big),\\ \bar{\bar{\kappa}}(\mathcal{K}\cup\{\ell\})=\bar{\kappa}(\mathcal{K}\cup\{\ell\})}} p\big(\bar{\bar{\kappa}}(\mathcal{S}\cup\mathcal{J}\cup\mathcal{K}),\vec a_{\mathcal{K}}\big|\vec x\big).
    \end{align}
    Now, recalling that $\ell\in\mathcal{S}$ here, one can see that $\bigwedge_{i,i'\in\mathcal{S}} (A_i\nprec\nsucc A_{i'})$ and $\bigwedge_{j\in\mathcal{J}} \big(\mathcal{S}\stackrel{\forall}{\nprec} A_j\big)$ imply that $A_\ell\stackrel{\forall}{\nprec}(\mathcal{S}\backslash\ell)$ and $A_\ell\stackrel{\forall}{\nprec}\mathcal{J}$, respectively. The constraint that $\bar{\bar{\kappa}}(\mathcal{K}\cup\{\ell\})=\bar{\kappa}(\mathcal{K}\cup\{\ell\})$ implies that the condition $A_\ell\stackrel{\forall}{\nprec}\mathcal{K}$, which was imposed on $\bar{\kappa}$, also holds for $\bar{\bar{\kappa}}$. Hence all the $\bar{\bar{\kappa}}$'s considered in the sum above satisfy $A_\ell\stackrel{\forall}{\nprec}(\mathcal{S}\cup\mathcal{J}\cup\mathcal{K}\backslash\ell)$.
    From Eq.~\eqref{eq:def_causal_p_OG} (with $\mathcal{K} \to \mathcal{S}\cup\mathcal{J}\cup\mathcal{K}\backslash\ell$, after marginalising over some of the outputs $a_k$) it then follows that the terms $p\big(\bar{\bar{\kappa}}(\mathcal{S}\cup\mathcal{J}\cup\mathcal{K}),\vec a_{\mathcal{K}}\big|\vec x\big)$ that appear in the sum, and therefore $p\big(\mathcal{S}_\kappa^\textsc{I}=\mathcal{S},\bar{\kappa}(\mathcal{K}\cup\{\ell\}),\vec a_{\mathcal{K}}\big|\vec x\big)$ itself, do not depend on $x_\ell$.
        
    \medskip
    
    Otherwise if $\ell\in\mathcal{N}\backslash\mathcal{S}$, then using again the inclusion–exclusion principle one can write:
    \begin{align}
        p\big(\mathcal{S}_\kappa^\textsc{I}=\mathcal{S},\bar{\kappa}(\mathcal{K}\cup\{\ell\}),\vec a_{\mathcal{K}}\big|\vec x\big) & = p\Bigg(\bigwedge_{i,i'\in\mathcal{S}} \!\! \big(A_i\nprec\nsucc A_{i'}\big), \Big(\mathcal{S}\stackrel{\exists}{\prec} A_\ell\Big), \bigwedge_{j\in\mathcal{N}\backslash\mathcal{S}\backslash\ell} \!\!\! \Big(\mathcal{S}\stackrel{\exists}{\prec} A_j\Big), \bar{\kappa}(\mathcal{K}\cup\{\ell\}),\vec a_{\mathcal{K}}\Bigg|\vec x\Bigg) \notag \\
        & = \sum_{\mathcal{J}\subseteq \mathcal{N}\backslash\mathcal{S}\backslash\ell} (-1)^{|\mathcal{J}|}\ p\Bigg(\bigwedge_{i,i'\in\mathcal{S}} \!\! \big(A_i\nprec\nsucc A_{i'}\big), \Big(\mathcal{S}\stackrel{\exists}{\prec} A_\ell\Big), \bigwedge_{j\in\mathcal{J}} \Big(\mathcal{S}\stackrel{\forall}{\nprec} A_j\Big), \bar{\kappa}(\mathcal{K}\cup\{\ell\}),\vec a_{\mathcal{K}}\Bigg|\vec x\Bigg) \notag \\
        & = \sum_{\mathcal{J}\subseteq \mathcal{N}\backslash\mathcal{S}\backslash\ell} (-1)^{|\mathcal{J}|} \hspace{-7mm} \sum_{\substack{\bar{\bar{\kappa}}(\mathcal{S}\cup\mathcal{J}\cup\mathcal{K}\cup\{\ell\})\text{ s.t. }\\ \bigwedge_{i,i'\in\mathcal{S}} (A_i\nprec\nsucc A_{i'}), \big(\mathcal{S}\stackrel{\exists}{\prec} A_\ell\big), \bigwedge_{j\in\mathcal{J}} \big(\mathcal{S}\stackrel{\forall}{\nprec} A_j\big),\\ \bar{\bar{\kappa}}(\mathcal{K}\cup\{\ell\})=\bar{\kappa}(\mathcal{K}\cup\{\ell\})}} \hspace{-7mm} p\big(\bar{\bar{\kappa}}(\mathcal{S}\cup\mathcal{J}\cup\mathcal{K}\cup\{\ell\}),\vec a_{\mathcal{K}}\big|\vec x\big).
    \end{align}
    Now, one can see that $\bigwedge_{i,i'\in\mathcal{S}} (A_i\nprec\nsucc A_{i'})$ and $\big(\mathcal{S}\stackrel{\exists}{\prec} A_\ell\big)$ together imply that $A_\ell\stackrel{\forall}{\nprec}\mathcal{S}$, and that $\big(\mathcal{S}\stackrel{\exists}{\prec} A_\ell\big)$ and $\bigwedge_{j\in\mathcal{J}} \big(\mathcal{S}\stackrel{\forall}{\nprec} A_j\big)$ together imply that $A_\ell\stackrel{\forall}{\nprec}\mathcal{J}$.
    The constraint that $\bar{\bar{\kappa}}(\mathcal{K}\cup\{\ell\})=\bar{\kappa}(\mathcal{K}\cup\{\ell\})$ again implies that the condition $A_\ell\stackrel{\forall}{\nprec}\mathcal{K}$ also holds for $\bar{\bar{\kappa}}$. Hence all the $\bar{\bar{\kappa}}$'s considered in the sum above satisfy $A_\ell\stackrel{\forall}{\nprec}(\mathcal{S}\cup\mathcal{J}\cup\mathcal{K})$.
    From Eq.~\eqref{eq:def_causal_p_OG} (with now $\mathcal{K} \to \mathcal{S}\cup\mathcal{J}\cup\mathcal{K}$, after marginalising over some of the outputs $a_k$) it then follows that the terms $p\big(\bar{\bar{\kappa}}(\mathcal{S}\cup\mathcal{J}\cup\mathcal{K}\cup\{\ell\}),\vec a_{\mathcal{K}}\big|\vec x\big)$ that appear in the sum, and therefore $p\big(\mathcal{S}_\kappa^\textsc{I}=\mathcal{S},\bar{\kappa}(\mathcal{K}\cup\{\ell\}),\vec a_{\mathcal{K}}\big|\vec x\big)$ itself, do not depend on $x_\ell$---which concludes the proof of Eq.~\eqref{eq:lemma_proof_OG_equiv}.

    \bigskip
    
    Now that we have this in place, let us introduce some arbitrary choice function $\underline{k}_\textsc{I}$ which, to each nonempty subset $\mathcal{S}\subseteq\mathcal{N}$, assigns one of its elements, $\underline{k}_\textsc{I}(\mathcal{S})\in\mathcal{S}$ (which will be ``taken to come first''); and for any SPO $\kappa$, let us define $\underline{k}_{\textsc{I},\kappa}\coloneqq \underline{k}_\textsc{I}(\mathcal{S}_\kappa^\textsc{I})$ (where $\mathcal{S}_\kappa^\textsc{I}$ is the ``first consecutive set'' of $\kappa$, as introduced above).
    
    Notice that for $\mathcal{K}=\emptyset$, Eq.~\eqref{eq:lemma_proof_OG_equiv} implies that $\forall\, \mathcal{S} \subseteq \mathcal{N}$, $p\big(\mathcal{S}_\kappa^\textsc{I}=\mathcal{S}\big|\vec x\big)$ does not depend on any $x_\ell$, i.e., it does not depend on $\vec x$ at all; and that for $\mathcal{K}=\{k_1\}\subseteq\mathcal{S}$ (so that $\mathcal{S}_\kappa^\textsc{I}=\mathcal{S}$ implies $A_\ell\nprec A_{k_1}$ for any $\ell\neq k_1$), Eq.~\eqref{eq:lemma_proof_OG_equiv} implies that $\forall\, \mathcal{S} \subseteq \mathcal{N}$, $\forall\, k_1\in\mathcal{S}$, $p\big(\mathcal{S}_\kappa^\textsc{I}=\mathcal{S},a_{k_1}\big|\vec x\big)$ does not depend on any $x_\ell$ with $\ell\neq k_1$, i.e., it only depends on $x_{k_1}$.
    It then follows that for any given $k_1\in\mathcal{N}$,
    \begin{alignat}{2}
        p\big(\underline{k}_{\textsc{I},\kappa}=k_1\big|\vec x\big) & = \sum_{\mathcal{S} \text{ s.t. } \underline{k}_\textsc{I}(\mathcal{S})= k_1} p\big(\mathcal{S}_\kappa^\textsc{I}=\mathcal{S}\big|\vec x\big) && \quad \text{ does not depend on } \vec x \notag \\
        \text{and} \quad p\big(\underline{k}_{\textsc{I},\kappa}=k_1,a_{k_1}\big|\vec x\big) & = \sum_{\mathcal{S} \text{ s.t. } \underline{k}_\textsc{I}(\mathcal{S})= k_1} p\big(\mathcal{S}_\kappa^\textsc{I}=\mathcal{S},a_{k_1}\big|\vec x\big) && \quad \text{ only depends on } x_{k_1}.
    \end{alignat}
    One can then define%
    \footnote{Again we assume for simplicity, in these definitions, that the denominators are nonzero. The terms for which these are zero do not contribute to the argument (see Footnote~\ref{ftn:nul_denom}).}
    \begin{align}
        & q_{k_1}\coloneqq p\big(\underline{k}_{\textsc{I},\kappa}=k_1\big), \quad p_{k_1}(a_{k_1}|x_{k_1})\coloneqq \frac{p\big(\underline{k}_{\textsc{I},\kappa}=k_1,a_{k_1}\big|x_{k_1}\big)}{q_{k_1}}, \notag \\[1mm]
        \text{for any SPO } \kappa'(\mathcal{N}\backslash k_1), \quad & p_{\mathcal{N}\backslash k_1,x_{k_1},a_{k_1}}(\kappa'(\mathcal{N}\backslash k_1),\vec a_{\mathcal{N}\backslash k_1}|\vec x_{\mathcal{N}\backslash k_1})\coloneqq \frac{p\big(\underline{k}_{\textsc{I},\kappa}=k_1,\kappa'(\mathcal{N}\backslash k_1),\vec a\big|\vec x\big)}{p\big(\underline{k}_{\textsc{I},\kappa}=k_1,a_{k_1}\big|x_{k_1}\big)}, \notag \\[1mm]
        \text{and} \quad & p_{\mathcal{N}\backslash k_1,x_{k_1},a_{k_1}}(\vec a_{\mathcal{N}\backslash k_1}|\vec x_{\mathcal{N}\backslash k_1})\coloneqq \sum_{\kappa'(\mathcal{N}\backslash k_1)} p_{\mathcal{N}\backslash k_1,x_{k_1},a_{k_1}}(\kappa'(\mathcal{N}\backslash k_1),\vec a_{\mathcal{N}\backslash k_1}|\vec x_{\mathcal{N}\backslash k_1}),
    \end{align}
    so that the decomposition $p(\vec a|\vec x) = \sum_\kappa p(\kappa,\vec a|\vec x)$ can be written as
    \begin{align}
        p(\vec a|\vec x) & = \sum_{k_1} p(\underline{k}_{\textsc{I},\kappa}=k_1,\vec a|\vec x) = \sum_{k_1} q_{k_1}\, p_{k_1}(a_{k_1}|x_{k_1}) \,p_{\mathcal{N}\backslash k_1,x_{k_1},a_{k_1}}(\vec a_{\mathcal{N}\backslash k_1}|\vec x_{\mathcal{N}\backslash k_1}),
    \end{align}
    i.e., in exactly the same form as in Eq.~\eqref{eq:def_causal_p_recursive}.
    Clearly $q_{k_1}\ge 0$, $\sum_{k_1}q_{k_1}=1$, and $p_{k_1}(a_{k_1}|x_{k_1})$ is a valid probability distribution. It remains to verify that each $(N-1)$-partite probability distribution $p_{\mathcal{N}\backslash k_1,x_{k_1},a_{k_1}}(\vec a_{\mathcal{N}\backslash k_1}|\vec x_{\mathcal{N}\backslash k_1})$ is causal (according to Abbott \emph{et al.}'s definition).
    
    To see this, consider $\mathcal{K}' \subsetneq \mathcal{N}'\coloneqq\mathcal{N}\backslash k_1$, $\ell \in \mathcal{N}'\backslash \mathcal{K}'$, and $\bar{\kappa}'=\bar{\kappa}'(\mathcal{K}'\cup\{\ell\})$ for which $A_\ell\stackrel{\forall}{\nprec}\mathcal{K}'$. Defining $\mathcal{K}\coloneqq\{k_1\}\cup\mathcal{K}'$, one has
    \begin{align}
        & p_{\mathcal{N}\backslash k_1,x_{k_1},a_{k_1}}\big(\bar{\kappa}'(\mathcal{K}'\cup\{\ell\}),\vec a_{\mathcal{K}'}\big|\vec x_{\mathcal{N}\backslash k_1}\big) \ = \ \frac{p\big(\underline{k}_{\textsc{I},\kappa}=k_1,\bar{\kappa}'(\mathcal{K}'\cup\{\ell\}),\vec a_{\mathcal{K}}\big|\vec x\big)}{p\big(\underline{k}_{\textsc{I},\kappa}=k_1,a_{k_1}\big|x_{k_1}\big)} \notag \\
        & \hspace{30mm} = \sum_{\substack{\mathcal{S} \text{ s.t. } \underline{k}_\textsc{I}(\mathcal{S})= k_1, \\ \bar{\kappa}(\mathcal{K}\cup\{\ell\}) \text{ s.t. } \bar{\kappa}(\mathcal{K}'\cup\{\ell\}) = \bar{\kappa}'(\mathcal{K}'\cup\{\ell\})}} \hspace{-5mm} p\big(\mathcal{S}_\kappa^\textsc{I}=\mathcal{S},\bar{\kappa}(\mathcal{K}\cup\{\ell\}),\vec a_{\mathcal{K}}\big|\vec x\big) \Bigg/ p\big(\underline{k}_{\textsc{I},\kappa}=k_1,a_{k_1}\big|x_{k_1}\big).
    \end{align}
    Now, $\bar{\kappa}(\mathcal{K}'\cup\{\ell\}) = \bar{\kappa}'(\mathcal{K}'\cup\{\ell\})$ implies that $\bar{\kappa}(\mathcal{K}\cup\{\ell\})$ is also such that $A_\ell\stackrel{\forall}{\nprec}\mathcal{K}'$. Furthermore, if $\mathcal{S}_\kappa^\textsc{I}=\mathcal{S}$, with $\underline{k}_\textsc{I}(\mathcal{S})= k_1$, then $k_1\in\mathcal{S}_\kappa^\textsc{I}$ and therefore $A_\ell\nprec A_{k_1}$ for all $\bar{\kappa}(\mathcal{K}\cup\{\ell\})$ that contribute to the sum. 
    All in all, these thus satisfy $A_\ell\stackrel{\forall}{\nprec}\{k_1\}\cup\mathcal{K}'=\mathcal{K}$.
    From Eq.~\eqref{eq:lemma_proof_OG_equiv} it follows that the terms $p\big(\mathcal{S}_\kappa^\textsc{I}=\mathcal{S},\bar{\kappa}(\mathcal{K}\cup\{\ell\}),\vec a_{\mathcal{K}}\big|\vec x\big)$ in the sum above do not depend on $x_\ell$; given that the same holds for the denominator $p\big(\underline{k}_{\textsc{I},\kappa}=k_1,a_{k_1}\big|x_{k_1}\big)$, we obtain that the term $p_{\mathcal{N}\backslash k_1,x_{k_1},a_{k_1}}\big(\bar{\kappa}'(\mathcal{K}'\cup\{\ell\}),\vec a_{\mathcal{K}'}\big|\vec x_{\mathcal{N}\backslash k_1}\big)$ itself does not depend on $x_\ell$, i.e., $p_{\mathcal{N}\backslash k_1,x_{k_1},a_{k_1}}(\vec a_{\mathcal{N}\backslash k_1}|\vec x_{\mathcal{N}\backslash k_1})$ satisfies Eq.~\eqref{eq:def_causal_p_OG}. By our induction hypothesis, it also satisfies Eq.~\eqref{eq:def_causal_p_recursive}, i.e., it is causal according to Abbott \emph{et al.}'s definition---which concludes the proof that Oreshkov and Giarmatzi's definition is equivalent to that of Abbott \emph{et al.}

\end{itemize}

\medskip

By unravelling our inductive proof that \eqref{eq:def_causal_p_OG} $\Rightarrow$ \eqref{eq:def_causal_p_recursive} above, one can see that each SPO $\kappa$ gets extended to a total order (a permutation) $\pi_\kappa$ of $\mathcal{N}$ that is ``compatible'' with $\kappa$, i.e., that preserves all pairwise relations of the form $A_i\prec A_j$.
Specifically, $\pi_\kappa=(k_1,k_2,\ldots,k_N)$ is obtained inductively as follows, using the choice function $\underline{k}_\textsc{I}$: $k_1=\underline{k}_\textsc{I}(\mathcal{S}_{\kappa(\mathcal{N})}^\textsc{I})$, $k_2=\underline{k}_\textsc{I}(\mathcal{S}_{\kappa(\mathcal{N}\backslash k_1)}^\textsc{I})$, $k_3=\underline{k}_\textsc{I}(\mathcal{S}_{\kappa(\mathcal{N}\backslash\{k_1,k2\})}^\textsc{I})$, etc. It can be seen that the decomposition $p(\vec a|\vec x) = \sum_\kappa p(\kappa,\vec a|\vec x)$ thus provides a new decomposition
\begin{align}
    p(\vec a|\vec x) = \sum_\kappa p(\kappa,\vec a|\vec x) = \sum_\pi \sum_{\kappa\text{ s.t. }\pi_\kappa=\pi} p(\kappa,\vec a|\vec x) = \sum_\pi p(\pi,\vec a|\vec x), \label{eq:decomp_extend_SPO}
\end{align}
with $p(\pi,\vec a|\vec x)\coloneqq \sum_{\kappa\text{ s.t. }\pi_\kappa=\pi} p(\kappa,\vec a|\vec x)$ satisfying Eq.~\eqref{eq:def_causal_p} (as it follows from the proof that \eqref{eq:def_causal_p_recursive} $\Rightarrow$ \eqref{eq:def_causal_p} in the previous subsection).

Note that compatible extensions of a given SPO to a total order are in general not unique. 
For the proof above to work, one could however not just take any arbitrary extension, independently for each SPO $\kappa$. Indeed it was important that the choice of the ``first party'' (when $\mathcal{S}_\kappa^\textsc{I}$ was not a singleton) only depended on the set $\mathcal{S}_\kappa^\textsc{I}$, rather than on $\kappa$ directly---i.e., for two SPOs $\kappa,\kappa'$ with the same first consecutive set $\mathcal{S}_\kappa^\textsc{I}=\mathcal{S}_{\kappa'}^\textsc{I}$, the (otherwise arbitrary) choice of the first party $\underline{k}_{\textsc{I},\kappa^{(\prime)}}$ must be the same. (That choice could also be probabilistic, but with probabilities that do not depend on the settings $\vec x$.) Otherwise one can construct counterexamples, for which extending each SPO independently to a total order leads to a decomposition $p(\vec a|\vec x) = \sum_\pi p(\pi,\vec a|\vec x)$ as in Eq.~\eqref{eq:decomp_extend_SPO} that does not satisfy Eq.~\eqref{eq:def_causal_p}.%
\footnote{For an explicit such counterexample: consider for instance 3 parties $A,B,C$ with the two SPOs $\kappa_0 = (A\nprec\nsucc B,A\prec C,B\nprec\nsucc C)$ and $\kappa_1 = (A\nprec\nsucc B,A\prec C,B\prec C)$, which both have the same first consecutive set $\mathcal{S}_{\kappa_0}^\textsc{I}=\mathcal{S}_{\kappa_1}^\textsc{I}=\{A,B\}$, and the distribution (with no output for any party and only an input $x=0,1$ for $A$) $p(\kappa|x)=\delta_{\kappa,\kappa_x}$, which satisfies Eq.~\eqref{eq:def_causal_p_OG}. 
If one were to extend $\kappa_0$ to $\pi_{\kappa_0} = (A,B,C)$ and $\kappa_1$ to $\pi_{\kappa_1} = (B,A,C)$, i.e.\ with different choices for the first party, then this would result in $p(\pi|x)=\left\{\begin{array}{ll}
    \delta_{\pi,(A,B,C)} & \text{if}\ x=0 \\
    \delta_{\pi,(B,A,C)} & \text{if}\ x=1
\end{array}\right.$, which does not satisfy Eq.~\eqref{eq:def_causal_p} (as for instance $p((k_1)=(A)|x) = \delta_{x,0}$ then depends on $x$).}

\subsection{Two more equivalent characterisations of causal correlations}
\label{app:more_causal_charact}

Here we provide two additional equivalent characterisations of multipartite causal correlations. The first one is that a correlation $p(\vec a|\vec x)$ is causal if and only if it can be decomposed in terms of (well-defined, normalised) probability distributions $\tilde{p}\big((k_n),a_{k_n}\big|(k_1,\ldots,k_{n-1}),\vec x_{k_1,\ldots,k_n},\vec a_{k_1,\ldots,k_{n-1}}\big)$ as
\begin{align}
    & p(\vec a|\vec x) = \sum_\pi p(\pi,\vec a|\vec x) \notag \\[-4mm]
    & \hspace{10mm} \text{with} \quad p\big(\pi=(k_1,\ldots,k_N),\vec a\big|\vec x\big) = \prod_{n=1}^N \tilde{p}\big((k_n),a_{k_n}\big|(k_1,\ldots,k_{n-1}),\vec x_{k_1,\ldots,k_n},\vec a_{k_1,\ldots,k_{n-1}}\big), \notag  \\
    & \hspace{20mm} \text{s.t. } \forall\,n, \ \sum_{a_{k_n}} \tilde{p}\big((k_n),a_{k_n}\big|(k_1,\ldots,k_{n-1}),\vec x_{k_1,\ldots,k_n},\vec a_{k_1,\ldots,k_{n-1}}\big) \text{ does not depend on } x_{k_n}. \label{eq:def_causal_p_v4}
\end{align}

\medskip

The last characterisation we provide is that a correlation $p(\vec a|\vec x)$ is causal if and only if it can be decomposed in terms of (well-defined, normalised) probability distributions $\hat{p}\big((k_{n+1}),a_{k_n}\big|(k_1,\ldots,k_n),\vec x_{k_1,\ldots,k_n},\vec a_{k_1,\ldots,k_{n-1}}\big)$---which reduce to just $\hat{p}\big((k_1)\big)$ for $n=0$ and to $\hat{p}\big(a_{k_N}\big|(k_1,\ldots,k_N),\vec x,\vec a_{\N\backslash k_N}\big)$ for $n=N$---as
\begin{align}
    & p(\vec a|\vec x) = \sum_\pi p(\pi,\vec a|\vec x) \notag \\[-4mm]
    & \hspace{10mm} \text{with} \quad p\big(\pi=(k_1,\ldots,k_N),\vec a\big|\vec x\big) = \prod_{n=0}^N \hat{p}\big((k_{n+1}),a_{k_n}\big|(k_1,\ldots,k_n),\vec x_{k_1,\ldots,k_n},\vec a_{k_1,\ldots,k_{n-1}}\big). \label{eq:def_causal_p_v5}
\end{align}

\bigskip

Each term $\tilde{p}\big((k_n),a_{k_n}\big|(k_1,\ldots,k_{n-1}),\vec x_{k_1,\ldots,k_n},\vec a_{k_1,\ldots,k_{n-1}}\big)$ in Eq.~\eqref{eq:def_causal_p_v4} corresponds to the probability for the ``next'' ($n^\text{th}$) party to be $A_{k_n}$ and to output $a_{k_n}$, given that the previous parties were (in that order) those labelled $(k_1,\ldots,k_{n-1})$, given all previous parties' as well as $A_{k_n}$'s inputs $\vec x_{k_1,\ldots,k_n}$, and given all previous parties' outputs $\vec a_{k_1,\ldots,k_{n-1}}$. Somewhat analogously, each term $\hat{p}\big((k_{n+1}),a_{k_n}\big|(k_1,\ldots,k_n),\vec x_{k_1,\ldots,k_n},\vec a_{k_1,\ldots,k_{n-1}}\big)$ in Eq.~\eqref{eq:def_causal_p_v5} corresponds to the probability that party $A_{k_n}$ outputs $a_{k_n}$ and that the next ($(n+1)^\text{th}$) party will be $A_{k_{n+1}}$, given that the order of the parties so far was $(k_1,\ldots,k_n)$, given all of those parties' inputs $\vec x_{k_1,\ldots,k_n}$, and given all previous parties' outputs $\vec a_{k_1,\ldots,k_{n-1}}$.
Both decompositions thus translate---in two slightly different manners---the idea that the causal order can be established step by step, with each of the parties' outputs and the choice of the ``next'' party depending only on previously available information (the order of the previous parties, all inputs so far, and all previous outputs).

As we will see, the characterisation of Eq.~\eqref{eq:def_causal_p_v4} will be useful in particular in our proofs that QC-QCs only generate causal correlations, and reciprocally that any causal correlation can be obtained from a QC-QC; see Appendices~\ref{app:corr_QC-QC} and~\ref{app:obtaining_any_causal_correl}.

\medskip

\begin{itemize}

    \item \underline{Proof that \eqref{eq:def_causal_p_v4} $\Rightarrow$ \eqref{eq:def_causal_p}:} For a distribution $p\big(\pi=(k_1,\ldots,k_N),\vec a\big|\vec x\big)$ that factorises as in Eq.~\eqref{eq:def_causal_p_v4}, one easily obtains, recursively (from $n=N-1$ down to $n=0$), that 
    \begin{align}
        & p\big((k_1,\ldots,k_n,k_{n+1}),\vec a_{k_1,\ldots,k_n}\big|\vec x\big) \notag \\[-1mm]
        & = \Big( \prod_{j=1}^n \tilde p\big((k_j),a_{k_j}\big|(k_1,\ldots,k_{j-1}),\vec x_{k_1,\ldots,k_j},\vec a_{k_1,\ldots,k_{j-1}}\big) \Big) \, \tilde p\big((k_{n+1})\big|(k_1,\ldots,k_n),\vec x_{k_1,\ldots,k_{n+1}},\vec a_{k_1,\ldots,k_n}\big) . \label{eq:proof_causal_p_v4}
    \end{align}
    Clearly the first product within brackets does not depend on $\vec x_{\mathcal{N}\backslash\{k_1,\ldots,k_n\}}$; the last term%
    \footnote{In order to lighten the notations we sometimes write $\tilde p\big((k_{n+1}),a_{k_{n+1}}\big|\cdot,\cdot,\cdot\big)$ for $\tilde p\big((k_{n+1}),a_{k_{n+1}}\big|(k_1,\ldots,k_n),\vec x_{k_1,\ldots,k_{n+1}},\vec a_{k_1,\ldots,k_n}\big)$.}
    $\tilde p\big((k_{n+1})\big|\cdot,\cdot,\cdot\big) \coloneqq \sum_{a_{k_{n+1}}} \tilde p\big((k_{n+1}),a_{k_{n+1}}\big|\cdot,\cdot,\cdot\big)$ also clearly does not depend on $\vec x_{\mathcal{N}\backslash\{k_1,\ldots,k_{n+1}\}}$, and is further required (from Eq.~\eqref{eq:def_causal_p_v4}) not to depend on $x_{k_{n+1}}$. All in all, $p\big((k_1,\ldots,k_n,k_{n+1}),\vec a_{k_1,\ldots,k_n}\big|\vec x\big)$ does not depend on $\vec x_{\mathcal{N}\backslash\{k_1,\ldots,k_n\}}$, as required in Eq.~\eqref{eq:def_causal_p}. 
    
    \item \underline{Proof that \eqref{eq:def_causal_p} $\Rightarrow$ \eqref{eq:def_causal_p_v4}:} From a decomposition as in Eq.~\eqref{eq:def_causal_p}, one can define $\tilde{p}\big((k_1),a_{k_1}\big|x_{k_1}\big) \coloneqq p\big((k_1),a_{k_1}\big|\vec x\big)$ (which indeed only depends on $x_{k_1}$) and (for $n=2,\ldots,N$)%
    \footnote{We once again assume for simplicity that the denominators are nonzero; the terms for which these are zero do not contribute to the decomposition of $p(\vec a|\vec x)$.}
    $\tilde{p}\big((k_n),a_{k_n}\big|(k_1,\ldots,k_{n-1}),\vec x_{k_1,\ldots,k_n},\vec a_{k_1,\ldots,k_{n-1}}\big) \coloneqq p\big((k_1,\ldots,k_n),\vec a_{k_1,\ldots,k_n}\big|\vec x\big)/p\big((k_1,\ldots,k_{n-1}),\vec a_{k_1,\ldots,k_{n-1}}\big|\vec x\big)$ (which indeed only depends on the inputs $\vec x_{k_1,\ldots,k_n}$). It is easy to see that these are normalised probability distributions whose product gives $p\big(\pi=(k_1,\ldots,k_N),\vec a\big|\vec x\big)$ as in the second line of Eq.~\eqref{eq:def_causal_p_v4}, and such that (for each $n$) $\sum_{a_{k_n}} \tilde{p}\big((k_n),a_{k_n}\big|\cdot,\cdot,\cdot\big)$ does not depend on $x_{k_n}$.

\end{itemize}

The proofs that \eqref{eq:def_causal_p_v5} $\Rightarrow$ \eqref{eq:def_causal_p} and \eqref{eq:def_causal_p} $\Rightarrow$ \eqref{eq:def_causal_p_v5} are quite similar: from Eq.~\eqref{eq:def_causal_p_v5} one easily obtains recursively that $p\big((k_1,\ldots,k_n,k_{n+1}),\vec a_{k_1,\ldots,k_n}\big|\vec x\big) = \prod_{j=0}^n \hat{p}\big((k_{j+1}),a_{k_j}\big|(k_1,\ldots,k_j),\vec x_{k_1,\ldots,k_j},\vec a_{k_1,\ldots,k_{j-1}}\big)$, which clearly does not depend on $\vec x_{\mathcal{N}\backslash\{k_1,\ldots,k_n\}}$; from Eq.~\eqref{eq:def_causal_p}, one can define the normalised probability distributions $\hat{p}\big((k_1)\big) \coloneqq p\big((k_1)\big|\vec x\big)$ (which indeed does not depend on $\vec  x$), $\hat{p}\big((k_{n+1}),a_{k_n}\big|(k_1,\ldots,k_n),\vec x_{k_1,\ldots,k_n},\vec a_{k_1,\ldots,k_{n-1}}\big) \coloneqq p\big((k_1,\ldots,k_n,k_{n+1}),\vec a_{k_1,\ldots,k_n}\big|\vec x\big) / p\big((k_1,\ldots,k_n),\vec a_{k_1,\ldots,k_{n-1}}\big|\vec x\big)$ (which indeed only depends on the inputs $\vec x_{k_1,\ldots,k_n}$), and $\hat{p}\big(a_{k_N}\big|(k_1,\ldots,k_N),\vec x,\vec a_{\N\backslash k_N}\big) \coloneqq p\big((k_1,\ldots,k_N),\vec a\big|\vec x\big) / p\big((k_1,\ldots,k_N),\vec a_{\N\backslash k_N}\big|\vec x\big)$, whose product reproduces Eq.~\eqref{eq:def_causal_p_v5}.

\subsection{Equivalent definitions for correlations with non-influenceable causal orders}
\label{app:subsec:equiv_def_NIO}

Recall that we defined correlations with non-influenceable causal order (Eq.~\eqref{eq:def_NIO}) by referring to our definition of causal correlations based on decompositions of the form $p(\vec a|\vec x) = \sum_\pi p(\pi,\vec a|\vec x)$.

One could however also refer to Oreshkov and Giarmatzi's definition, namely by defining correlations with NIO as those with a decomposition of the form $p(\vec a|\vec x) = \sum_\kappa p(\kappa,\vec a|\vec x)$, as in \eqref{eq:def_causal_p_OG}, which further satisfies
\begin{align}
    \forall\, \kappa, \ p(\kappa|\vec x)\ \text{does not depend on } \vec x. \label{eq:def_NIO_OG}
\end{align}

It is however easy to see, following the previous subsection, that the two definitions are equivalent. Indeed a total order $\pi$ is a particular case of an SPO $\kappa$, so that if a decomposition of the form $p(\vec a|\vec x) = \sum_\pi p(\pi,\vec a|\vec x)$ (satisfying Eq.~\eqref{eq:def_causal_p}) with $p(\pi|\vec x)$ being independent of $\vec x$ exists, then this readily provides a decomposition of the form $p(\vec a|\vec x) = \sum_\kappa p(\kappa,\vec a|\vec x)$ (satisfying Eq.~\eqref{eq:def_causal_p_OG}, as we have seen before) with $p(\kappa|\vec x)$ that does not depend on $\vec x$.
Conversely, if a decomposition of the form $p(\vec a|\vec x) = \sum_\kappa p(\kappa,\vec a|\vec x)$ (satisfying Eq.~\eqref{eq:def_causal_p_OG}) exists, then we have just seen above how to extend each SPO $\kappa$ to a total order $\pi$, and obtain a decomposition $p(\vec a|\vec x) = \sum_\pi p(\pi,\vec a|\vec x)$ as in Eq.~\eqref{eq:decomp_extend_SPO}, with $p(\pi,\vec a|\vec x)= \sum_{\kappa\text{ s.t. }\pi_\kappa=\pi} p(\kappa,\vec a|\vec x)$ satisfying Eq.~\eqref{eq:def_causal_p}; if $p(\kappa|\vec x)$ does not depend on $\vec x$, then neither does the distribution $p(\pi|\vec x)$ thus obtained.

\section{Tripartite causal correlations}

\subsection{The ``simplest'' tripartite scenario with dynamical causal order}
\label{app:simplest_N3}

Consider a tripartite scenario in which the three parties $A,B,C$ have binary inputs $x,y,z=0,1$, party $A$ has no output, while parties $B$ and $C$ have a binary output $b,c=0,1$ only when their input is 1 ($B$ and $C$ are in the so-called ``lazy scenario''). This scenario is simple enough that the polytopes of causal correlations ($\mathcal{P}_{\textup{causal}}$), of correlations with static causal order ($\mathcal{P}_{\textup{convFO}}$), as well as of all valid probability distributions ($\mathcal{P}_{\textup{all}}$) can be fully characterised:
\begin{itemize}
    \item The causal polytope $\mathcal{P}_{\textup{causal}}$ has 144 vertices: 112 of them are compatible with a fixed causal order---of the form $p(b,c|x,y,z) = \delta_{b,y\beta_x}\delta_{c,z\gamma_{x,y}}$ or $p(b,c|x,y,z) = \delta_{b,y\beta_{x,z}}\delta_{c,z\gamma_x}$, or both, for some $\beta_{x(,z)}, \gamma_{x(,y)}=0,1$---and 32 of them feature dynamical causal order---of the form $p(b,c|x,y,z) = \delta_{x,0}\delta_{b,y\beta_0}\delta_{c,z\gamma_{0,y}} + \delta_{x,1}\delta_{b,y\beta_{1,z}}\delta_{c,z\gamma_1}$ or $p(b,c|x,y,z) = \delta_{x,1}\delta_{b,y\beta_1}\delta_{c,z\gamma_{1,y}} + \delta_{x,0}\delta_{b,y\beta_{0,z}}\delta_{c,z\gamma_0}$, with $\beta_{x,z}$ and $\gamma_{x,y}$ having a nontrivial dependence on $z$ and $y$, resp.

    Solving the facet enumeration problem, we find that $\mathcal{P}_{\textup{causal}}$ has 8 facets of the ``Lazy Guess Your Neighbour's Input'' (LGYNI)~\cite{branciard16} form%
\footnote{Here we only write explicitly a single representative for each type of facet. The other facets ``of the same form'' are obtained by exchanging the parties $B$ and $C$, by flipping $A$'s input $x$, and/or by flipping $B$ and/or $C$'s outputs when their input is 1.}
    \begin{align}
        p(b=c=yz|x) \underset{\text{causal}}{\leq} \frac34
    \end{align}
    (for some fixed value $x$, with $p(b=c=yz|x) \coloneqq \frac14\sum_{y,z} p(b=c=yz|x,y,z)$), in addition to 8 trivial facets of the form $p(b|x,y=1,z) \ge 0$ and 8 trivial facets of the form $p(b,c|x,y=z=1) \ge 0$.
    
    \item The polytope $\mathcal{P}_{\textup{convFO}}$ only has the 112 vertices compatible with a fixed order given above.
    We find that it has the same facets as $\mathcal{P}_{\textup{causal}}$, plus 8 facets of the form
    \begin{align}
        I_3 \coloneqq p\big(x=0, c=yz\big) + p\big(x=1, b=yz\big) \underset{\text{convFO}}{\leq} \frac78,
    \end{align}
    as in Eqs.~\eqref{eq:def_I3} and~\eqref{eq:bnd_I3_FO} of the main text.

Let us present here a direct proof of the latter inequality.
Consider a fixed order in which $B$ comes before $C$, so that $p(b|x,y,z)$ does not depend on $z$: we shall write $p(b|x,y,z) = p(b|x,y,\cancel{z})$. We then have
\begin{align}
    I_3 & \coloneqq \underbrace{p\big(x=0, c=yz\big)}_{\leq p(x=0)=\frac12} + \underbrace{p\big(x=1, b=yz\big)}_{=\frac18\sum_{yz}p(b=yz|x=1,y,z)} \notag \\
    & \leq \frac12 + \frac18 \sum_{z=0,1} \big[ \underbrace{p\big(b=0|x=1,y=0,\cancel{z}\big)}_{\leq 1} + p\big(b=z|x=1,y=1,\cancel{z}\big) \big] \leq \frac12 + \frac14 + \frac18 = \frac78.
\end{align}
The same bound holds, by symmetry, for any fixed order in which $C$ comes before $B$, and by linearity, for any convex combination of fixed orders: $I_3 \leq_{\text{convFO}} \frac{7}{8}$.

    \item The polytope $\mathcal{P}_{\textup{all}}$ of all valid probability distributions is only delimited by the 16 trivial facets of the form $p(b|x,y=1,z) \ge 0$ and $p(b,c|x,y=z=1) \ge 0$.
    Solving the vertex enumeration problem, we find that it has the same 144 vertices as $\mathcal{P}_{\textup{causal}}$, plus 112 non-causal vertices that feature 2-way signalling, of the form $p(b,c|x,y,z) = \delta_{b,y\beta_{x,z}}\delta_{c,z\gamma_{x,y}}$, with both $\beta_{0,z}$ and $\gamma_{0,y}$, and/or both $\beta_{1,z}$ and $\gamma_{1,y}$, having a nontrivial dependence on $z$ and $y$, resp.
    
\end{itemize}

\bigskip

Notice that the scenario considered here is the ``smallest'' one---in the sense of minimising the number of parties and the number of inputs and outputs for each party \emph{individually}%
\footnote{Note however that one could also consider even ``smaller'' scenarios where certain \emph{combinations} of inputs, for the different parties, are further restricted. In fact, the expression $I_3$ does not involve (in any nontrivial way) terms with $(x,z)=(0,0)$, nor with $(x,y)=(1,0)$. Hence one can already distinguish dynamical orders in scenarios where these combinations of inputs never appear.\label{ftn:even_smaller}}---that can feature dynamical causal order. With two parties causal correlations indeed reduce to convex mixtures of the two possible orders as in Eq.~\eqref{eq:causal_N_2}; if one of the three parties had no input then it could always be considered to come last,%
\footnote{More specifically: if one party, say $A_N$, has no input, then starting from a probability distribution $p\big(\pi=(k_1,\ldots,k_N),\vec a\big|\vec x\big)$ one can define $p'\big(\pi=(k_1,\ldots,k_{N-1},N),\vec a\big|\vec x\big)\coloneqq \sum_{i=1}^N p\big(\pi=(k_1,\ldots,k_{i-1},N,k_i,\ldots,k_{N-1}),\vec a\big|\vec x\big)$, such that party $A_N$ always comes last in $\pi=(k_1,\ldots,k_{N-1},N)$.
\\
Similarly, if one party, say $A_1$, has no output, then starting from $p\big(\pi=(k_1,\ldots,k_N),\vec a\big|\vec x\big)$ one can define $p'\big(\pi=(1,k_2,\ldots,k_N),\vec a\big|\vec x\big)\coloneqq \sum_{i=1}^N p\big(\pi=(k_2,\ldots,k_i,1,k_{i+1},\ldots,k_N),\vec a\big|\vec x\big)$, such that party $A_1$ always comes first in $\pi=(1,k_2,\ldots,k_N)$.
\\
One can verify that in both cases, if the original distribution $p(\pi,\vec a|\vec x)$ satisfies the causal constraints of Eq.~\eqref{eq:def_causal_p}, then so does $p'(\pi,\vec a|\vec x)$; if $p(\pi,\vec a|\vec x)$ satisfies the NIO constraints of Eq.~\eqref{eq:def_NIO}, then so does $p'(\pi,\vec a|\vec x)$; and if $p(\pi,\vec a|\vec x)$ satisfies the NIO$'$ constraints of Eq.~\eqref{eq:def_NIO'}, then so does $p'(\pi,\vec a|\vec x)$. \label{ftn:no_IO}}
and causal correlations would reduce to convex mixtures of the two possible orders for the other two parties; and if a second party had no output then all probability distributions (for just the single remaining output) could be obtained causally, with the two parties with no output coming first.
It is in fact rather remarkable that the dynamical property appears already in a tripartite scenario where one of the parties has no output.

\subsection{$\mathcal{P}_{\textup{NIO}}$ and $\mathcal{P}_{\textup{NIO}'}$ reduce to $\mathcal{P}_{\textup{convFO}}$ for $N = 3$}
\label{app:NIO_convFO_N3}

While, as just seen, a tripartite scenario can exhibit dynamical causal order, we show here that it cannot exhibit ``dynamical but non-influenceable causal order''---that is, correlations with non-influenceable causal order (in $\mathcal{P}_{\textup{NIO}}$) reduce to convex mixtures of correlations with fixed orders (i.e., they are in $\mathcal{P}_{\textup{convFO}}$).

To see this, notice that for $N=3$, the causality conditions of Eq.~\eqref{eq:def_causal_p} imply in particular that for all $\pi=(k_1,k_2,k_3)$,%
\footnote{Notice that the following argument does not work in general for $N\ge 4$---unless one party has no input or one party has no output, using the observation made in the previous footnote---because $p\big((k_1,k_2),a_{k_1}\big|\vec x\big)$ cannot be readily identified with $p(\pi,a_{k_1}|\vec x)$ as we do in Eq.~\eqref{eq:causal_p_N3}, and the terms $p(\vec a|\vec x,\pi)$ in Eq.~\eqref{eq:convex_decomp_pi_N3} are in general not compatible with the causal order $\pi$. \\
As an explicit counter-example, one can take for instance the NIO correlation in Eq.~\eqref{eq:saturate_NIO} of the next appendix, with its causal decomposition as proposed just below Eq.~\eqref{eq:saturate_NIO}. For these one gets $p(a,b,c,d|x,y,z,t,\pi{=}(A,B,C,D)) = \delta_{a=b=xy}\,\delta_{c,0}\,\delta_{d,zt}$ and 
$p(a,b,c,d|x,y,z,t,\pi{=}(A,B,D,C)) = \delta_{a,x\bar{y}}\,\delta_{b,\bar{x}y}\,\delta_{d,0}\,\delta_{c,zt}$ (with $\bar{x}\coloneqq 1-x$, $\bar{y}\coloneqq 1-y$), which are not compatible with any fixed causal order.\label{ftn:NIO_N3}}
\begin{align}
    & p\big((k_1,k_2),a_{k_1}\big|\vec x\big) = p(\pi,a_{k_1}|\vec x) \ \text{does not depend on } \vec x_{k_2,k_3} \notag \\
    \text{and}\ & p\big((k_1,k_2,k_3),a_{k_1},a_{k_2}\big|\vec x\big) = p(\pi,a_{k_1},a_{k_2}|\vec x) \ \text{does not depend on } x_{k_3}.
 \label{eq:causal_p_N3}
\end{align}
Assuming the non-influenceability condition of Eq.~\eqref{eq:def_NIO} and writing
\begin{align}
    p(\vec a|\vec x) = \sum_\pi p(\pi,\vec a|\vec x) = \sum_\pi p(\pi) \, p(\vec a|\vec x,\pi), \label{eq:convex_decomp_pi_N3}
\end{align}
we then have that, for $\pi=(k_1,k_2,k_3)$, 

\begin{align}
    & p(a_{k_1}|\vec x,\pi) = \frac{p(\pi,a_{k_1}|\vec x)}{p(\pi)} \ \text{does not depend on } \vec x_{k_2,k_3}, \notag \\
    & p(a_{k_1},a_{k_2}|\vec x,\pi) = \frac{p(\pi,a_{k_1},a_{k_2}|\vec x)}{p(\pi)} \ \text{does not depend on } x_{k_3}, 
\end{align}
so that $p(\vec a|\vec x,\pi)$ is compatible with the fixed causal order defined by $\pi$, and Eq.~\eqref{eq:convex_decomp_pi_N3} indeed provides a convex decomposition of $p(\vec a|\vec x)$ onto correlations compatible with fixed causal orders.

\medskip

We further note that for $N=3$, the ``non-influenceable coarse-grained order'' condition of Eq.~\eqref{eq:def_NIO'} reduces to the mere ``non-influenceable order'' condition of Eq.~\eqref{eq:def_NIO} (since Eq.~\eqref{eq:def_NIO'} implies that $p\big((k_1,k_2,k_3)\big|\vec x\big) = p\big((k_1,k_2)\big|\vec x\big) = p\big((\{k_1\},k_2)\big|\vec x\big)$ does not depend on $\vec x$). Hence, $\mathcal{P}_{\textup{NIO}'}$ also reduces in the tripartite case to $\mathcal{P}_{\textup{convFO}}$.

\section{Proofs of the different causal bounds on $I_4 \coloneqq p\big(a=b=xy , d=zt\big) + p\big(\neg(a=b=xy) , c=zt\big)$}
\label{app:proofs_bounds}

We prove in this appendix the various causal bounds on the expression $I_4$ that we introduced in the main text to distinguish between correlations in $\mathcal{P}_{\textup{convFO}}$, $\mathcal{P}_{\textup{NIO}}$, $\mathcal{P}_{\textup{NIO}'}$ and $\mathcal{P}_{\textup{causal}}$.

Recall that we consider here a 4-partite scenario with parties denoted $A,B,C,D$, with binary inputs $x,y,z,t=0,1$, each party having a trivial output $a,b,c,d=0$ if their input is 0, or a binary output $a,b,c,d=0,1$ if their input is 1 (the so-called ``lazy scenario'').
This is the ``smallest'' scenario (with the minimal number of parties, and of inputs and outputs for each party individually\footnote{Note, as in Footnote~\ref{ftn:even_smaller}, that one can in fact still further restrict the number of allowed \emph{combinations} of inputs for the different parties. E.g., the expression $I_4$ does not involve (in any nontrivial way) terms with $(z,t)=(0,0)$, nor with $(x,y,z,t)=(0,0,1,0)$. Hence one can already distinguish correlations with dynamical but non-influenceable causal order in scenarios where these combinations of inputs never appear.
Restricting to such smaller spaces of correlations (and perhaps also restricting to the probabilities of certain combinations of outputs only, e.g., only those that appear nontrivially in $I_4$), may help to get a better characterisation of the different polytopes of correlations under consideration.}) in which one can find correlations with dynamical but non-influenceable causal order: if one of the four parties has no output or no input at all, then using the observation from Footnote~\ref{ftn:no_IO}, a similar argument to that for $N=3$ parties (just above) shows that correlations in $\mathcal{P}_{\textup{NIO}}$ and $\mathcal{P}_{\textup{NIO}'}$ are necessarily convex combinations of correlations with fixed order (cf.\ also Footnote~\ref{ftn:NIO_N3}).

\subsection{For (convex mixtures of) fixed orders: $I_4 \leq_{\textnormal{convFO}} \frac{15}{16}$}
\label{app:proof_convFO_bound}

By linearity and convexity, to obtain the bound on $I_4$ for correlations in $\mathcal{P}_{\textup{convFO}}$, it is sufficient to calculate the bound for each possible fixed order independently.
Consider therefore a correlation $p(a,b,c,d|x,y,z,t)$ compatible with a given fixed order $\pi$ (described as a permutation of the parties).
We will first distinguish whether $C$ comes before or after $D$ in $\pi$:

\begin{itemize}

    \item if $C$ comes before $D$ in $\pi$, then let us write
    \begin{align}
        I_4 & = p\big(a=b=xy , d=zt\big) + p\big(\neg(a=b=xy) , c=zt\big) \notag \\
        & \leq p\big(a=b=xy\big) + p\big(\neg(a=b=xy) , c=zt\big) \notag \\
        & \qquad = \frac{1}{16}\sum_{x,y,z,t=0,1} \Big[ p\big(a=b=xy\big|x,y,z,t\big) + p\big(\neg(a=b=xy) , c=zt\big|x,y,z,t\big) \Big] \notag \\
        & \qquad = \frac{1}{16}\sum_{x,y,t} \Big[ p\big(a=b=xy\big|x,y,z=0,t\big) + p\big(\neg(a=b=xy) , \cancel{c=0}\big|x,y,z=0,t\big) \notag \\[-3mm]
        & \hspace{25mm} + p\big(a=b=xy\big|x,y,z=1,t\big) + p\big(\neg(a=b=xy) , c=t\big|x,y,z=1,t\big) \Big] \notag \\
        & \qquad = \frac{1}{16}\sum_{x,y,t} \Big[ 1 + p\big(a=b=xy\big|x,y,z=1,t\big) + p\big(\neg(a=b=xy) , c=t\big|x,y,z=1,t\big) \Big], \label{eq:proof_FO_C_before_D}
    \end{align}
    where we used the fact that for an input $0$ (here, $z=0$), the output is necessarily $0$ ($c=0$).
    
    Let us further distinguish here the different cases according to which party comes last in $\pi$:
    \begin{itemize}
        \item If $D$ comes last in $\pi$, then%
        \footnote{In the proofs we shall indicate that a probability does not depend on a given input, according to some causality condition, by striking out the corresponding input; on the subsequent lines we will simply not write that input anymore.}
        \begin{align}
            I_4 & \leq \frac{1}{16}\sum_{x,y,t} \Big[ 1 + p\big(a=b=xy\big|x,y,z=1,\cancel{t}\big) + p\big(\neg(a=b=xy) , c=t\big|x,y,z=1,\cancel{t}\big) \Big] \notag \\
            & \qquad = \frac{1}{16}\sum_{x,y} \Big[ 2 + 2\,p\big(a=b=xy\big|x,y,z=1\big) + p\big(\neg(a=b=xy)\big|x,y,z=1\big) \Big] \notag \\
            & \qquad = \frac{1}{16}\sum_{x,y} \Big[ 3+p\big(a=b=xy\big|x,y,z=1\big) \Big] = \frac14 \Big(3 + \frac14\sum_{x,y} p\big(a=b=xy\big|x,y,z=1\big) \Big) \leq \frac14 \Big(3 + \frac34 \Big) = \frac{15}{16}, \label{eq:proof_FO_C_before_D_Dlast}
        \end{align}
        where the last inequality just follows from the LGYNI inequality~\cite{branciard16} for $A$ and $B$ (for which conditioning on the input $z=1$ of another party does not change the causal bound).
        \item If $B$ comes last in $\pi$, then
        \begin{align}
            I_4 & \leq \frac{1}{16}\sum_t \Big[ 4 + \underbrace{p\big(a=b=0\big|x=y=0,z=1,t\big)}_{1} + \underbrace{p\big(\neg(a=b=0) , c=t\big|x=y=0,z=1,t\big)}_{0} \notag \\
        & \hspace{15mm} + p\big(a=b=0\big|x=0,y=1,z=1,t\big) + \underbrace{p\big(\neg(a=b=0) , c=t\big|x=0,y=1,z=1,t\big)}_{\leq \,p(\neg(a=b=0)|x=0,y=1,z=1,t)} \notag \\
        & \hspace{15mm} + \underbrace{p\big(a=b=0\big|x=1,y=0,z=1,t\big)}_{p(a=0|x=1,\cancel{y=0},z=1,t)} + \underbrace{p\big(\neg(a=b=0) , c=t\big|x=1,y=0,z=1,t\big)}_{p(a=1 , c=t|x=1,\cancel{y=0},z=1,t)} \notag \\
        & \hspace{15mm} + \underbrace{p\big(a=b=1\big|x=y=1,z=1,t\big) + p\big(\neg(a=b=1) , c=t\big|x=y=1,z=1,t\big)}_{p(a=b=1, c=\bar{t}|x=y=1,z=1,t) + p(c=t|x=\cancel{y}=1,z=1,\cancel{t})} \Big] \notag \\
        & \leq \frac{1}{16}\sum_t \Big[ 6 + p\big(a=0\big|x=z=1,t\big) + p\big(a=1,c=t\big|x=z=1,t\big) \notag \\[-3mm]
        & \hspace{20mm} + \underbrace{p\big(a=b=1,c=\bar{t}\big|x=y=z=1,t\big)}_{\leq \,p(a=1,c=\bar{t}|x=\cancel{y}=z=1,t)} + \,p\big(c=t\big|x=z=1\big) \Big] \notag \\
        & \leq \frac{1}{16}\sum_t \Big[ 7 + p\big(c=t\big|x=z=1\big) \Big]  = \frac{15}{16} \label{eq:proof_FO_C_before_D_Blast}
        \end{align}
        (where we used the notation $\bar{t}\coloneqq 1-t$).
        \item If $A$ comes last in $\pi$, then by symmetry (with respect to $A \leftrightarrow B$), we also get the bound $I_4 \leq \frac{15}{16}$.
    \end{itemize}
    
    \item if $D$ comes before $C$ in $\pi$, then let us write (considering the case where $x=y=0$ separately, and noting that $a=b=xy$ always holds in this case):
    \begin{align}
        I_4 & = p\big(a=b=xy , d=zt\big) + p\big(\neg(a=b=xy) , c=zt\big) \notag \\
        & = \frac{1}{16} \Bigg[ \sum_{z,t=0,1} p\big(d=zt\big|x=y=0,\cancel{z},t\big) \notag \\[-2mm]
        & \hspace{10mm} + \sum_{\substack{x,y,z,t=0,1: \\ (x,y)\neq(0,0)}} \Big[ \underbrace{p\big(a=b=xy , d=zt\big|x,y,z,t\big) + p\big(\neg(a=b=xy) , c=zt\big|x,y,z,t\big)}_{\leq \, 1} \Big] \Bigg] \notag \\
        & \leq \frac{1}{16} \Big[ \underbrace{{\textstyle \sum_z} \,p\big(d=0\big|x=y=0,t=0\big)}_{2} + \underbrace{{\textstyle \sum_z} \,p\big(d=z\big|x=y=0,t=1\big)}_{1} + 12 \Big] = \frac{15}{16}. \label{eq:proof_FO_D_before_C}
    \end{align}
    
\end{itemize}

All in all, for any fixed causal order, and by linearity for any convex combination of fixed causal orders, we obtained the bound
\begin{align}
    I_4 \underset{\text{convFO}}{\leq} \frac{15}{16},
\end{align}
as in Eq.~\eqref{eq:bnd_I4_FO}.

This bound can be reached for instance by the correlation (compatible with any fixed order where $C$ comes before $D$) $p(a,b,c,d|x,y,z,t) = \delta_{a,0}\,\delta_{b,0}\,\delta_{c,0}\,\delta_{d,zt}$.

\subsection{For correlations with non-influenceable causal orders: $I_4 \leq_{\textnormal{NIO}} \frac{31}{32}$}
\label{app:proof_NIO_bound}

Consider a causal correlation $p(\vec a|\vec x)$, for which there exists $p(\pi,\vec a|\vec x)$ that satisfies Eq.~\eqref{eq:def_causal_p}. One can then write
\begin{align}
    I_4 = \sum_\pi I_\pi \quad \text{with} \quad I_\pi \coloneqq p\big(\pi,a=b=xy , d=zt\big) + p\big(\pi,\neg(a=b=xy) , c=zt\big) . \label{eq:def_I_pi}
\end{align}
Following similar calculations as in the different cases with fixed order above, we see that:
\begin{itemize}
    \item If $C$ comes before $D$ in $\pi$, then (similarly to Eq.~\eqref{eq:proof_FO_C_before_D})
    \begin{align}
        I_\pi & \leq \frac{1}{16}\sum_{x,y,t} \Big[ p\big(\pi\big|x,y,z=0,t\big) + p\big(\pi,a=b=xy\big|x,y,z=1,t\big) + p\big(\pi,\neg(a=b=xy) , c=t\big|x,y,z=1,t\big) \Big].
    \end{align}
    \begin{itemize}
        \item If $D$ comes last in $\pi$, then it follows from the causality constraints of Eq.~\eqref{eq:def_causal_p} that $p(\pi,a,b,c|x,y,z,t)$, and hence $p(\pi,a,b|x,y,z,t)$ and $p(\pi|x,y,z,t)$, do not depend on $t$. We then have (similarly to Eq.~\eqref{eq:proof_FO_C_before_D_Dlast})
        \begin{align}
            I_\pi & \leq \frac{1}{16}\sum_{x,y,t} \Big[ p\big(\pi\big|x,y,z=0,\cancel{t}\big) + p\big(\pi,a=b=xy\big|x,y,z=1,\cancel{t}\big) + p\big(\pi,\neg(a=b=xy) , c=t\big|x,y,z=1,\cancel{t}\big) \Big] \notag \\
                & \qquad = \frac{1}{16}\sum_{x,y} \Big[ 2\,p\big(\pi\big|x,y,z=0\big) + p\big(\pi\big|x,y,z=1\big) + p\big(\pi,a=b=xy\big|x,y,z=1\big) \Big] \notag \\
                & \qquad = \frac{1}{8}\sum_{x,y,z} p\big(\pi\big|x,y,z\big) - \frac{1}{16}\sum_{x,y} p\big(\pi,\neg(a=b=xy)\big|x,y,z=1\big).
        \end{align}
        If, in addition, $A$ comes before $B$ in $\pi$, then we have that
        \begin{align}
            & \sum_{x,y} p\big(\pi,\neg(a=b=xy)\big|x,y,z=1\big) \notag \\
            & \quad = p\big(\pi,a=1\big|x=1,y=0,z=1\big) + \underbrace{p\big(\pi,b=1\big|x=0,y=1,z=1\big)}_{\ge 0} + \underbrace{p\big(\pi,\neg(a=b=1)\big|x=1,y=1,z=1\big)}_{\ge p(\pi,a=0|x=1,y=1,z=1)} \notag \\[-2mm]
            & \quad \ge \sum_y p\big(\pi,a=\bar{y}\big|x=1,y,z=1\big),
        \end{align}
        so that
        \begin{align}
            I_\pi & \leq \frac{1}{8}\sum_{x,y,z} p\big(\pi\big|x,y,z\big) - \frac{1}{16}\sum_y p\big(\pi,a=\bar{y}\big|x=1,y,z=1\big).
        \end{align}
        Similarly, if $B$ comes before $A$ in $\pi$ instead (still in addition to $D$ coming last), then
        \begin{align}
            I_\pi & \leq \frac{1}{8}\sum_{x,y,z} p\big(\pi\big|x,y,z\big) - \frac{1}{16}\sum_x p\big(\pi,b=\bar{x}\big|x,y=1,z=1\big).
        \end{align}
    
        \item If $B$ comes last in $\pi$, then $p(\pi,a,c|x,y,z,t)$ does not depend on $y$ and we have (similarly to Eq.~\eqref{eq:proof_FO_C_before_D_Blast})
        \begin{align}
            I_\pi & \leq \frac{1}{16}\sum_t \Big[ p\big(\pi\big|x=0,\cancel{y=0},z=0,t\big) + p\big(\pi\big|x=0,\cancel{y=0},z=1,t\big) \notag \\[-3mm]
            & \hspace{15mm} + p\big(\pi\big|x=0,\cancel{y=1},z=0,t\big) + p\big(\pi\big|x=0,\cancel{y=1},z=1,t\big) \notag \\
            & \hspace{15mm} + p\big(\pi\big|x=1,\cancel{y=0},z=0,t\big) + p\big(\pi,a=0\big|x=1,\cancel{y=0},z=1,t\big) + p\big(\pi,a=1,c=t\big|x=1,\cancel{y=0},z=1,t\big) \notag \\
            & \hspace{15mm} + p\big(\pi\big|x=1,\cancel{y=1},z=0,t\big) + p\big(\pi,a=1,c=\bar{t}\big|x=1,\cancel{y=1},z=1,t\big) + p\big(\pi,c=t\big|x=1,\cancel{y=1},z=1,t\big) \Big] \notag \\
            & \qquad = \frac{1}{16}\sum_t \Big[ 2\,p\big(\pi\big|x=z=0,t\big) + 2\,p\big(\pi\big|x=0,z=1,t\big) + 2\,p\big(\pi\big|x=1,z=0,t\big) \notag \\[-3mm]
            & \hspace{80mm} + p\big(\pi\big|x=z=1,t\big) + p\big(\pi,c=t\big|x=z=1,t\big) \Big] \notag \\
            & \qquad = \frac{1}{8}\sum_{x,z,t} p\big(\pi\big|x,z,t\big) - \frac{1}{16}\sum_t p\big(\pi,c=\bar{t}\big|x=z=1,t\big).
        \end{align}
        
        \item If $A$ comes last in $\pi$, then by symmetry the same bound holds (just replacing $x$ by $y$ in the above expression):
        \begin{align}
            I_\pi & \leq \frac{1}{8}\sum_{y,z,t} p\big(\pi\big|y,z,t\big) - \frac{1}{16}\sum_t p\big(\pi,c=\bar{t}\big|y=z=1,t\big).
        \end{align}
    \end{itemize}

    \item If $D$ comes before $C$ in $\pi$, then (similarly to Eq.~\eqref{eq:proof_FO_D_before_C})
    \begin{align}
        I_\pi & \leq \frac{1}{16} \Big[ \sum_{z=0,1} \big[ p\big(\pi\big|x=y=0,z,t=0\big) + p\big(\pi,d=z\big|x=y=0,z,t=1\big) \big] + \sum_{\substack{x,y,z,t=0,1: \\ (x,y)\neq(0,0)}} p\big(\pi\big|x,y,z,t\big) \Big] \notag \\
        & \qquad = \frac{1}{16} \sum_{x,y,z,t} p\big(\pi\big|x,y,z,t\big) - \frac{1}{16} \sum_{z} p\big(\pi,d=\bar{z}\big|x=y=0,z,t=1\big).
    \end{align}
    Now, it follows from the causality constraints of Eq.~\eqref{eq:def_causal_p} that $p(\pi{:}D{\prec}C,d|x,y,z,t) \coloneqq \sum_{\pi{:}D{\prec}C} p(\pi,d|x,y,z,t)$ and $p(\pi{:}D{\prec}C|x,y,z,t) \coloneqq \sum_{\pi{:}D{\prec}C} p(\pi|x,y,z,t)$ do not depend on $z$.%
    \footnote{As indicated by the conditions `$\pi{:}D{\prec}C$', the sums $\sum_{\pi{:}D{\prec}C}$ are over all permutations in which $D$ comes before $C$. To see that $p(\pi{:}D{\prec}C,d|x,y,z,t)$ does not depend on $z$, one can write $p(\pi{:}D{\prec}C,d|x,y,z,t) = p\big((k_1,k_2,k_3)=(A,B,D),d\big|x,y,\cancel{z},t\big) + p\big((k_1,k_2,k_3)=(B,A,D),d\big|x,y,\cancel{z},t\big) + p\big((k_1,k_2)=(A,D),d\big|x,\cancel{y},\cancel{z},t\big) + p\big((k_1,k_2)=(B,D),d\big|\cancel{x},y,\cancel{z},t\big) + p\big((k_1)=(D),d\big|\cancel{x},\cancel{y},\cancel{z},t\big)$, where all independencies (struck-out inputs) follow from Eq.~\eqref{eq:def_causal_p}.}
    We then have (again, similarly to Eq.~\eqref{eq:proof_FO_D_before_C})
    \begin{align}
        \sum_{\pi{:}D{\prec}C} I_\pi & \leq \frac{1}{16} \sum_{x,y,z,t} p\big(\pi{:}D{\prec}C\big|x,y,\cancel{z},t\big) - \frac{1}{16} \sum_{z} p\big(\pi{:}D{\prec}C,d=\bar{z}\big|x=y=0,\cancel{z},t=1\big) \notag \\
        & \qquad = \frac{1}{8} \sum_{x,y,t} p\big(\pi{:}D{\prec}C\big|x,y,t\big) - \frac{1}{16} p\big(\pi{:}D{\prec}C\big|x=y=0,t=1\big).
    \end{align}
    
\end{itemize}

All in all, we obtain
\begin{align}
I_4 = & \sum_\pi I_\pi = \sum_{\substack{\pi:D\text{ last}, \\ A\prec B}} I_\pi + \sum_{\substack{\pi:D\text{ last}, \\ B\prec A}} I_\pi + \sum_{\substack{\pi{:}C{\prec}D, \\ A\text{ last}}} I_\pi + \sum_{\substack{\pi{:}C{\prec}D, \\ B\text{ last}}} I_\pi + \sum_{\pi{:}D{\prec}C} I_\pi \notag \\
\leq & \!\scalebox{0.85}{${\displaystyle \sum_{\substack{\pi:D\text{ last}, \\ A\prec B}}}$} \!\Big[ {\textstyle \frac{1}{8}}\scalebox{0.85}{${\displaystyle \sum_{x,y,z}}$}\, p(\pi|x,y,z) - {\textstyle \frac{1}{16}}\scalebox{0.85}{${\displaystyle \sum_y}$}\, p(\pi,a=\bar{y}|x=1,y,z=1) \Big] \!+ \!\scalebox{0.85}{${\displaystyle \sum_{\substack{\pi:D\text{ last}, \\ B\prec A}}}$} \!\Big[ {\textstyle \frac{1}{8}}\scalebox{0.85}{${\displaystyle \sum_{x,y,z}}$}\, p(\pi|x,y,z) - {\textstyle \frac{1}{16}}\scalebox{0.85}{${\displaystyle \sum_x}$}\, p(\pi,b=\bar{x}|x,y=z=1) \Big] \notag \\
& + \!\scalebox{0.85}{${\displaystyle \sum_{\substack{\pi{:}C{\prec}D, \\ B\text{ last}}}}$} \!\Big[ {\textstyle \frac{1}{8}}\scalebox{0.85}{${\displaystyle \sum_{x,z,t}}$}\, p(\pi|x,z,t) - {\textstyle \frac{1}{16}}\scalebox{0.85}{${\displaystyle \sum_t}$}\, p(\pi,c=\bar{t}|x=z=1,t) \Big] + \!\scalebox{0.85}{${\displaystyle \sum_{\substack{\pi{:}C{\prec}D, \\ A\text{ last}}}}$} \!\Big[ {\textstyle \frac{1}{8}}\scalebox{0.85}{${\displaystyle \sum_{y,z,t}}$}\, 
p(\pi|y,z,t) - {\textstyle \frac{1}{16}}\scalebox{0.85}{${\displaystyle \sum_t}$}\, p(\pi,c=\bar{t}|y=z=1,t) \Big] \notag \\
& + {\textstyle \frac{1}{8}}\scalebox{0.85}{${\displaystyle \sum_{x,y,t}}$}\, p(\pi{:}D{\prec}C|x,y,t) - {\textstyle \frac{1}{16}} p(\pi{:}D{\prec}C|x=y=0,t=1) \notag \\
& \quad =  1 - \frac{1}{16} \Big( J_{C\prec D} + p\big(\pi{:}D{\prec}C\big|x=y=0,t=1\big) \Big), \label{eq:bnd_I4_JCD_pDC}
\end{align}
with
\begin{align}
J_{C\prec D} \coloneqq & \sum_{\substack{\pi:D\text{ last}, \\ A\prec B}} \sum_y p\big(\pi,a=\bar{y}\big|x=1,y,z=1\big) + \sum_{\substack{\pi:D\text{ last}, \\ B\prec A}} \sum_x p\big(\pi,b=\bar{x}\big|x,y=z=1\big) \notag \\
& \hspace{8mm} + \sum_{\substack{\pi{:}C{\prec}D, \\ B\text{ last}}} \sum_t p\big(\pi,c=\bar{t}\big|x=z=1,t\big) + \sum_{\substack{\pi{:}C{\prec}D, \\ A\text{ last}}} \sum_t p\big(\pi,c=\bar{t}\big|y=z=1,t\big) \notag \\[3mm]
= & \sum_y \underbrace{p\big(\pi=(A,B,C,D),a=\bar{y}\big|x=1,y,\cancel{z=1}\big)}_{\ge\,p(\pi=(A,B,C,D)|x=1,y)-p((k_1,k_2)=(A,B),a=y|x=1,\cancel{y})} + \sum_{\substack{\pi=(A,C,B,D), \\ (C,A,B,D)}} \underbrace{{\textstyle \sum_y}\, p\big(\pi,a=\bar{y}\big|x=1,\cancel{y},z=1\big)}_{p(\pi|x=z=1)} \notag \\[1mm]
& + \sum_x \underbrace{p\big(\pi=(B,A,C,D),b=\bar{x}\big|x,y=\cancel{z}=1\big)}_{\ge\,p(\pi=(B,A,C,D)|x,y=1)-p((k_1,k_2)=(B,A),b=x|\cancel{x},y=1)} + \sum_{\substack{\pi=(B,C,A,D), \\ (C,B,A,D)}} \underbrace{{\textstyle \sum_x}\, p\big(\pi,b=\bar{x}\big|\cancel{x},y=z=1\big)}_{p(\pi|y=z=1)} \notag \\[1mm]
& + \sum_t p\big(\pi=(C,D,A,B),c=\bar{t}\big|\cancel{x}=z=1,t\big) + \sum_{\substack{\pi=(C,A,D,B), \\ (A,C,D,B)}} \underbrace{{\textstyle \sum_t}\, p\big(\pi,c=\bar{t}\big|x=z=1,\cancel{t}\big)}_{p(\pi|x=z=1)} \notag \\[1mm]
& + \sum_t p\big(\pi=(C,D,B,A),c=\bar{t}\big|\cancel{y}=z=1,t\big) + \sum_{\substack{\pi=(C,B,D,A), \\ (B,C,D,A)}} \underbrace{{\textstyle \sum_t}\, p\big(\pi,c=\bar{t}\big|y=z=1,\cancel{t}\big)}_{p(\pi|y=z=1)} \notag \\[3mm]
\ge & \sum_y p\big(\pi=(A,B,C,D)\big|x=1,y\big) + \sum_x p\big(\pi=(B,A,C,D)\big|x,y=1\big) \notag \\
& \quad - p\big((k_1,k_2)=(A,B)\big|x=1\big) - p\big((k_1,k_2)=(B,A)\big|y=1\big) \notag \\[1mm]
& \quad + p\big((k_1,k_2)=(A,C)\big|x=\cancel{z}=1\big) + p\big((k_1,k_2)=(C,A)\big|\cancel{x}=z=1\big) \notag \\[1mm]
& \quad + p\big((k_1,k_2)=(B,C)\big|y=\cancel{z}=1\big) + p\big((k_1,k_2)=(C,B)\big|\cancel{y}=z=1\big) \notag \\[1mm]
& \quad + \underbrace{{\textstyle \sum_t}\, p\big((k_1,k_2)=(C,D),c=\bar{t}\big|z=1,\cancel{t}\big)}_{p((k_1,k_2)=(C,D)|z=1)} \notag \\[3mm]
& = \sum_y p\big(\pi=(A,B,C,D)\big|x=1,y\big) + \sum_x p\big(\pi=(B,A,C,D)\big|x,y=1\big) \notag \\
& \quad - p\big((k_1,k_2)=(A,B)\big|x=1\big) - p\big((k_1,k_2)=(B,A)\big|y=1\big) \notag \\[1mm]
& \quad + p\big((k_1,k_2)=(A,C)\big|x=1\big) + p\big((k_1,k_2)=(B,C)\big|y=1\big) + p\big((k_1)=(C)\big|\cancel{z=1}\big), \label{app:eq_intermediate_NIO_bound}
\end{align}
where we repeatedly used the causality constraints of Eq.~\eqref{eq:def_causal_p} to get rid of some dependencies and simplify the expressions.

\bigskip

Note that we have not used the non-influenceable order (NIO) assumption of Eq.~\eqref{eq:def_NIO} yet: the derivations above hold for all causal correlations.
Assuming it now---that is, assuming $p(\pi|x,y,z,t) = p(\pi)$---we then have:
\begin{align}
    & J_{C\prec D} + p\big(\pi{:}D{\prec}C\big|x=y=0,t=1\big) \ge \frac12 J_{C\prec D} + p\big(\pi{:}D{\prec}C\big) \notag \\
    & \quad \ge p\big(\pi=(A,B,C,D)\big) + p\big(\pi=(B,A,C,D)\big) - \frac12 p\big((k_1,k_2)=(A,B)\big) - \frac12 p\big((k_1,k_2)=(B,A)\big) \notag \\
    & \qquad + \frac12 p\big((k_1,k_2)=(A,C)\big) + \frac12 p\big((k_1,k_2)=(B,C)\big) + \frac12 p\big((k_1)=(C)\big) + p\big(\pi{:}D{\prec}C\big) \notag \\[2mm]
    & \qquad = \underbrace{p\big(\pi=(A,B,C,D)\lor(B,A,C,D)\big) + p\big((k_1,k_2)=(A,C)\lor(B,C)\big) + p\big((k_1)=(C)\big) + p\big(\pi{:}D{\prec}C\big)}_{=\, 1} \notag \\
    & \qquad \quad - \frac12 \Big[ \underbrace{p\big((k_1,k_2)=(A,B)\lor(B,A)\lor(A,C)\lor(B,C)\big) + p\big((k_1)=(C)\big)}_{\le 1} \Big] \ge \frac12, \label{app:eq_intermediate_NIO_bound2}
\end{align}
where we used the shorthand notations $p\big(\pi=(A,B,C,D)\lor(B,A,C,D)\big) = p\big(\pi=(A,B,C,D)\big) + p\big(\pi=(B,A,C,D)\big)$, $p\big((k_1,k_2)=(A,C)\lor(B,C)\big) = p\big((k_1,k_2)=(A,C)\big) + p\big((k_1,k_2)=(B,C)\big)$ etc. Recalling Eq.~\eqref{eq:bnd_I4_JCD_pDC}, we then obtain, for correlations with NIO:
\begin{align}
    I_4 \leq 1 - \frac{1}{16} \Big( J_{C\prec D} + p\big(\pi{:}D{\prec}C\big|x=y=0,t=1\big) \Big) \underset{\text{NIO}}{\leq} \frac{31}{32},
\end{align}
as in Eq.~\eqref{eq:bnd_I4_NIO}.

\medskip

This bound can be reached for instance by the correlation
\begin{align}
    & p(a,b,c,d|x,y,z,t) \notag \\[-1mm]
    & \qquad = \underbrace{\left\{
    \begin{array}{ll}
        \delta_{a,0} & \text{if}\ x=0 \\
        \frac12 & \text{if}\ x=1
    \end{array}\right.}_{p(a|x)}
    \underbrace{\left\{
    \begin{array}{ll}
        \delta_{b,0} & \text{if}\ y=0 \\
        \frac12 & \text{if}\ y=1, x=0 \\
        \delta_{b,a} & \text{if}\ y=1, x=1
    \end{array}\right.}_{p(b|x,y,a)} 
    \underbrace{\left\{
    \begin{array}{ll}
        \frac12\,\delta_{c,0}\,\delta_{d,zt} + \frac12\,\delta_{d,0}\,\delta_{c,zt} & \text{if}\ x=y=0 \\[1mm]
        \left\{
        \begin{array}{ll}
            \delta_{c,0}\,\delta_{d,zt} & \text{if}\ a=b=xy \\
            \delta_{d,0}\,\delta_{c,zt} & \text{if}\ \neg(a=b=xy)
        \end{array}\right. & \text{if}\ \neg(x=y=0)
    \end{array}\right.}_{p(c,d|x,y,z,t,a,b)}, \label{eq:saturate_NIO}
\end{align}
which can easily be seen to be causal. In fact, an explicit causal decomposition as in Eq.~\eqref{eq:def_causal_p} can be obtained by taking $p(\pi,a,b,c,d|x,y,z,t)$ to be of the same form as $p(a,b,c,d|x,y,z,t)$ above, after just multiplying the terms $\delta_{c,0}\,\delta_{d,zt}$ by $\delta_{\pi,(A,B,C,D)}$ and the terms $\delta_{d,0}\,\delta_{c,zt}$ by $\delta_{\pi,(A,B,D,C)}$.%
\footnote{
One may notice that the marginal correlation
$p(a,b|x,y) = \left\{
\begin{array}{llcll}
    \delta_{a,0}\,\delta_{b,0} & \text{if}\ x=y=0 & \ ; \ & \frac12\,\delta_{a,0} & \text{if}\ x=0,y=1 \\
    \frac12\,\delta_{b,0} & \text{if}\ x=1,y=0 & \ ; \ & \frac12\,\delta_{a,b} & \text{if}\ x=y=1
\end{array}\right.$
(which does not depend on $z,t$) is non-signalling---and in fact, even Bell-local~\cite{bell64} (as is any non-signalling correlation in the ``lazy'' scenario, where each party has nontrivial outputs for at most one input): $p(a,b|x,y) = \sum_{\lambda=0,1}\frac12\delta_{a,\lambda x}\delta_{b,\lambda y}$---so one could also decompose this correlation onto different orders for $A$ and $B$. \label{ftn:decomp_saturate_I4NIO}}
For such a decomposition, it is then easy to see that $p\big(\pi=(A,B,C,D)\big|x,y,z,t\big) = p\big(\pi=(A,B,D,C)\big|x,y,z,t\big) = \frac12$ does not depend on the inputs $x,y,z,t$, which indeed proves that the correlation is in $\mathcal{P}_{\textup{NIO}}$---and not in $\mathcal{P}_{\textup{convFO}}$, since it violates the bound $I_4 \leq_{\text{convFO}} \frac{15}{16}$.

\subsection{For correlations with non-influenceable coarse-grained causal orders: $I_4 \leq_{\textnormal{NIO}'} \frac{47}{48}$}
\label{app:proof_NIO'_bound}

Let us now relax the NIO assumption, and just consider the ``non-influenceable coarse-grained causal orders'' (NIO$'$) assumption of Eq.~\eqref{eq:def_NIO'}, that $p\big((\mathcal{K}_{n-1},k_n)\big|\vec x\big) = p\big((\mathcal{K}_{n-1},k_n)\big)$ does not depend on $\vec x$.
Note that it implies in particular, here, that
\begin{align}
    & \hspace{-3mm} p\big(\pi{:}D{\prec}C\big|x=y=0,t=1\big) \notag \\
    & = p\big((\mathcal{K}_2,k_3)=(\{A,B\},D)\big|\cancel{x=y=0,t=1}\big) + p\big((\mathcal{K}_2,k_3)=(\{A,D\},B)\big|\cancel{x=y=0,t=1}\big) \notag \\
    & \quad + p\big((\mathcal{K}_2,k_3)=(\{B,D\},A)\big|\cancel{x=y=0,t=1}\big) + p\big((\mathcal{K}_2,k_3)=(\{A,D\},C)\big|\cancel{x=y=0,t=1}\big) \notag \\
    & \quad + p\big((\mathcal{K}_2,k_3)=(\{B,D\},C)\big|\cancel{x=y=0,t=1}\big) + p\big((\mathcal{K}_1,k_2)=(\{D\},C)\big|\cancel{x=y=0,t=1}\big) \ = \ p\big(\pi{:}D{\prec}C\big)
\end{align}
does not depend on the inputs $x,y,t$.

Coming back to Eq.~\eqref{app:eq_intermediate_NIO_bound}---which, as we emphasised, holds for all causal correlations---we can now write (instead of Eq.~\eqref{app:eq_intermediate_NIO_bound2})
\begin{align}
    & J_{C\prec D} + p\big(\pi{:}D{\prec}C\big|x=y=0,t=1\big) \ge \frac13 J_{C\prec D} + p\big(\pi{:}D{\prec}C\big) \notag \\
    & \quad \ge \frac13 \Big[ p\big(\pi=(A,B,C,D)\big|x=1,y=0\big) + p\big(\pi=(B,A,C,D)\big|x=0,y=1\big) + p\big((\mathcal{K}_2,k_3)=(\{A,B\},C)\big|\cancel{x=y=1}\big) \notag \\[-1mm]
    & \qquad \qquad - p\big((k_1,k_2)=(A,B)\big|\cancel{x=1}\big) - p\big((k_1,k_2)=(B,A)\big|\cancel{y=1}\big) \notag \\
    & \qquad \qquad + p\big((k_1,k_2)=(A,C)\big|\cancel{x=1}\big) + p\big((k_1,k_2)=(B,C)\big|\cancel{y=1}\big) + p\big((k_1)=(C)\big|\cancel{z=1}\big) + \hspace{-10mm} \underbrace{3\,p\big(\pi{:}D{\prec}C\big)}_{\ge\,p(\pi{:}D{\prec}C)+2\,p((\mathcal{K}_2,k_3)=(\{A,B\},D))\qquad} \hspace{-10mm} \Big] \notag \\
    & \quad \ge \frac13 \Bigg[ \underbrace{p\big((\mathcal{K}_2,k_3)=(\{A,B\},C)\big) + p\big((k_1,k_2)=(A,C)\lor(B,C)\big) + p\big((k_1)=(C)\big) + p\big(\pi{:}D{\prec}C\big)}_{=\, 1} \notag \\
    & \qquad \qquad + p\big((\mathcal{K}_2,k_3)=(\{A,B\},D)\big) - \Big( \underbrace{p\big((k_1,k_2)=(A,B)\big) - p\big(\pi=(A,B,C,D)\big|x=1,y=0\big)}_{p(\pi=(A,B,D,C)|x=1,y=0)} \Big) \notag \\
    & \qquad \qquad + p\big((\mathcal{K}_2,k_3)=(\{A,B\},D)\big) - \Big( \underbrace{p\big((k_1,k_2)=(B,A)\big) - p\big(\pi=(B,A,C,D)\big|x=0,y=1\big)}_{p(\pi=(B,A,D,C)|x=0,y=1)} \Big) \Bigg] \notag \\[1mm]
    & \qquad = \frac13 \Big[ 1 + p\big(\pi=(B,A,D,C)\big|x=1,y=0\big) + p\big(\pi=(A,B,D,C)\big|x=0,y=1\big) \Big] \ge \frac13 .
\end{align}
Recalling again Eq.~\eqref{eq:bnd_I4_JCD_pDC}, we then obtain, for correlations with non-influenceable coarse-grained causal orders:
\begin{align}
    I_4 \leq 1 - \frac{1}{16} \Big( J_{C\prec D} + p\big(\pi{:}D{\prec}C\big|x=y=0,t=1\big) \Big) \underset{\text{NIO}'}{\leq} \frac{47}{48},
\end{align}
as in Eq.~\eqref{eq:bnd_I4_NIO'}.

\medskip

This bound can be reached for instance by the correlation 
\begin{align}
    p\big(a,b,c,d|x,y,z,t\big) = & \,\frac{2}{3}
    \left\{ \begin{array}{ll}
        \delta_{a,0}\,\delta_{b,0}\,\delta_{c,0}\,\delta_{d,zt} & \text{if}\ x=0 \\[1mm]
        \frac{1}{2}\,\delta_{b,y} \left\{ \begin{array}{ll}
        \delta_{c,0}\,\delta_{d,zt} & \text{if}\ y=a \\
        \delta_{d,0}\,\delta_{c,zt} & \text{if}\ y=\bar{a}
        \end{array} \right. & \text{if}\ x=1
    \end{array} \right. \,+\, \frac{1}{3}\,\delta_{b,y}\,\delta_{a,xy}
    \left\{ \begin{array}{ll}
        \delta_{d,0}\,\delta_{c,zt} & \text{if}\ x=0 \\[1mm]
        \delta_{c,0}\,\delta_{d,zt} & \text{if}\ x=1
    \end{array} \right. , \label{eq:saturate_NIO'}
\end{align}
 which can easily be seen to be causal. To see that it is in ${\cal P}_{\textup{NIO}'}$, notice that one can decompose it as
\begin{align}
    p\big((k_1,k_2,k_3,k_4),a,b,c,d|x,y,z,t\big) = & \frac{2}{3}\,\delta_{(k_1,k_2),(A,B)}
    \left\{ \begin{array}{ll}
        \delta_{a,0}\,\delta_{b,0}\,\delta_{(k_3,k_4),(C,D)}\,\delta_{c,0}\,\delta_{d,zt} & \text{if}\ x=0 \\[1mm]
        \frac{1}{2}\,\delta_{b,y} \left\{ \begin{array}{ll}
        \delta_{(k_3,k_4),(C,D)}\,\delta_{c,0}\,\delta_{d,zt} & \text{if}\ y=a \\
        \delta_{(k_3,k_4),(D,C)}\,\delta_{d,0}\,\delta_{c,zt} & \text{if}\ y=\bar{a}
        \end{array} \right. & \text{if}\ x=1
    \end{array} \right. \notag \\[1mm]
    & + \frac{1}{3}\,\delta_{(k_1,k_2),(B,A)}\,\delta_{b,y}\,\delta_{a,xy}
    \left\{ \begin{array}{ll}
        \delta_{(k_3,k_4),(D,C)}\,\delta_{d,0}\,\delta_{c,zt} & \text{if}\ x=0 \\[1mm]
        \delta_{(k_3,k_4),(C,D)}\,\delta_{c,0}\,\delta_{d,zt} & \text{if}\ x=1
    \end{array} \right. ,
\end{align}
such that
\begin{align}
     p\big((k_1,k_2,k_3,k_4)|x,y,z,t\big) = & \left\{ \begin{array}{ll}
    \frac{2}{3}\,\delta_{(k_1,k_2,k_3,k_4),(A,B,C,D)} + \frac{1}{3}\,\delta_{(k_1,k_2,k_3,k_4),(B,A,D,C)} & \text{if}\ x=0 \\[1mm]
    \frac{1}{3}\,\delta_{(k_1,k_2,k_3,k_4),(A,B,C,D)} + \frac{1}{3}\,\delta_{(k_1,k_2,k_3,k_4),(A,B,D,C)} + \frac{1}{3}\,\delta_{(k_1,k_2,k_3,k_4),(B,A,C,D)} & \text{if}\ x=1
\end{array} \right. .
\end{align}
It is then easy to verify that $p((k_1)=(A)|x,y,z,t)=p((k_1,k_2)=(A,B)|x,y,z,t)=\frac23$, $p((k_1)=(B)|x,y,z,t)=p((k_1,k_2)=(B,A)|x,y,z,t)=\frac13$, $p((\mathcal{K}_2,k_3)=(\{A,B\},C)|x,y,z,t)=p((\mathcal{K}_3,k_4)=(\{A,B,C\},D)|x,y,z,t)=\frac23$ and $p((\mathcal{K}_2,k_3)=(\{A,B\},D)|x,y,z,t)=p((\mathcal{K}_3,k_4)=(\{A,B,D\},C)|x,y,z,t)=\frac13$ do not depend on $x,y,z,t$ (with all other probabilities $p\big((\mathcal{K}_{n-1},k_n)\big|\vec x\big)$ being null), which indeed proves that the correlation is in $\mathcal{P}_{\textup{NIO}'}$---and not in $\mathcal{P}_{\textup{NIO}}$, since it violates the bound $I_4 \leq_{\text{NIO}} \frac{31}{32}$.

\subsection{For general causal correlations: $I_4 \leq_{\textnormal{causal}} 1$}
\label{app:proof_causal_bound}

We conclude this appendix with the bound for general causal correlations, which is simply the trivial algebraic bound:
\begin{align}
    I_4 & = p\big(a=b=xy , d=zt\big) + p\big(\neg(a=b=xy) , c=zt\big) \leq p\big(a=b=xy\big) + p\big(\neg(a=b=xy)\big) = 1.
\end{align}

An example of a causal correlation that indeed reaches this algebraic maximum is
\begin{align}
    p(a,b,c,d|x,y,z,t) = \delta_{a,0}\,\delta_{b,0} \left\{ \begin{array}{ll}
    \delta_{c,0}\,\delta_{d,zt} & \text{if}\ xy=0 \\[1mm]
    \delta_{d,0}\,\delta_{c,zt} & \text{if}\ xy=1
\end{array} \right.,
\end{align}
which can be decomposed as
\begin{equation}
p((k_1,k_2,k_3,k_4),a,b,c,d|x,y,z,t) = \delta_{(k_1,k_2),(A,B)}\,\delta_{a,0}\,\delta_{b,0} \left\{ \begin{array}{ll}
    \!\!\delta_{(k_3,k_4),(C,D)}\,\delta_{c,0}\,\delta_{d,zt} & \text{if}\ xy=0 \\[1mm]
    \!\!\delta_{(k_3,k_4),(D,C)}\,\delta_{d,0}\,\delta_{c,zt} & \text{if}\ xy=1
\end{array} \right. ,
\end{equation}
in which the causal order between $C$ and $D$ depends on the product of inputs $xy$ (indeed the causal order has to be ``influenceable'' so as to violate the previous bounds).

\section{Description and characterisation of our different classes of quantum circuits}
\label{app:sec_classes}

In this appendix we provide more details on the description and characterisation of the different classes of quantum circuits with classical and quantum control of causal order considered in the paper. These are described within the wider framework of process matrices~\cite{oreshkov12}. We start by recalling, for completeness, the validity constraints for general process matrices that the characterisations of some of our classes refer to.

\subsection{Validity constraints for general process matrices}
\label{app:subsec_validity_cstr}

The process matrix framework was first introduced to characterise the set of all possible correlations that can be established (in a consistent manner) by a number of parties implementing operations that are locally described by the quantum formalism (as quantum instruments~\cite{davies70}), but without assuming \emph{a priori} that these are applied in any specific, well-defined causal order~\cite{oreshkov12}. Under some reasonable assumptions (in particular: dependency of the probabilities on the realised operations only rather than on the full quantum instruments; linearity of the probabilities to preserve the interpretation of probabilistic mixtures of operations and coarse-graining of outcomes), it was shown that all such correlations could be written in the form of the generalised Born rule of Eq.~\eqref{eq:Born_rule}, for some ``process matrix'' $W \in \mathcal{L}(\mathcal{H}^{A_{\mathcal{N}}^{IO}})$ (where $\mathcal{N}=\{1,\ldots,N\}$ refers to the set of parties, each having input/output spaces $\mathcal{H}^{A_k^I}/\mathcal{H}^{A_k^O}$, and with $\mathcal{H}^{A_{\mathcal{N}}^{IO}} = \bigotimes_{k\in\N}\mathcal{H}^{A_k^{IO}} = \bigotimes_{k\in\N}\mathcal{H}^{A_k^I}\otimes\mathcal{H}^{A_k^O}$).

Requiring that the probabilities thus obtained are nonnegative, even when the parties share other auxiliary systems and their instruments can also act on these, imposes that the process matrix must be positive semidefinite (PSD): $W\ge 0$. Recalling that the elements of a quantum instrument must sum to a trace-preserving (TP) map, and that the TP condition for a map $\M_k: \L(\HS^{A_k^I}) \to \L(\HS^{A_k^O})$ translates in terms of its Choi matrix into $\Tr_{A_k^O} M_k = \id^{A_k^I}$, then the requirement that the probabilities must be normalised, on the other hand, can be phrased as follows:
\begin{align}
    \forall\, M_1\in\L(\HS^{A_1^{IO}}),\ldots,M_N\in\L(\HS^{A_N^{IO}}) \ \text{ such that } \ \Tr_{A_1^O} M_1 = \id^{A_1^I},\ldots,\Tr_{A_N^O} M_N = \id^{A_N^I}, \ \ \Tr [\Big(\bigotimes_k M_k\Big)^{\!T} \, W] = 1. \label{eq:cstr_W}
\end{align}
This was shown to be equivalent to imposing some linear and affine constraints on $W$, which can be written as~\cite{Araujo15,wechs19}%
\footnote{We note that another equivalent formulation of these constraints was also given in terms of ``allowed'' and ``forbidden'' terms in a Hilbert-Schmidt basis decomposition of $W$~\cite{oreshkov12,Oreshkov16}.}
\begin{align}
    \forall \,\emptyset \subsetneq \K \subseteq \N, \, {}_{\prod_{k\in\K}[1-A_k^O]}\Tr_{A_{\N\backslash\K}^{IO}}W = 0 \quad \text{and} \quad \Tr W = d_\N^O, \label{eq:validityW}
\end{align}
where we used the notation ${}_{[1-A_k^O]}W\coloneqq W-(\Tr_{A_k^O} W)\otimes\frac{\id^{A_k^O}}{d_k^O}$ (with $d_k^O$ denoting the dimension of $\HS^{A_k^O}$), such that ${}_{\prod_{k\in\K}[1-A_k^O]}W = {}_{[1-A_{k_1}^O]}({}_{[1-A_{k_2}^O]}(\cdots({}_{[1-A_{k_{|\K|}}^O]}W)))$ for $\K=\{k_1,k_2,\ldots,k_{|\K|}\}$, and where $d_\N^O \coloneqq \prod_{k\in\N}d_k^O$ is the product of the dimensions of all parties' output spaces.

As recalled in the main text, another version of the process matrix framework was introduced in Ref.~\cite{araujo17}, with some ``open'' past and future spaces $\HS^P, \HS^F$. Here the process matrix acts as a ``supermap''~\cite{Chiribella08supermaps} that transforms the parties' local maps to a new quantum map from $\HS^P$ to $\HS^F$, given by the further generalised Born rule of Eq.~\eqref{eq:Born_rule_PF}. Requiring that completely positive (CP) maps thereby get transformed to a CP map (even when involving auxiliary systems: the supermap must be ``completely CP-preserving'') imposes again that $W$ must be PSD. The requirement that TP maps (corresponding e.g.\ to the sum of the parties' instrument elements) get transformed to a TP map (the supermap must be ``TP-preserving''), on the other hand, can be written here as in Eq.~\eqref{eq:cstr_W}, just replacing the last equality by $\Tr_F [(\bigotimes_k M_k)^T \, W] = \id^P$. It can be shown to be equivalent to the following linear and affine constraints:%
\footnote{These can be verified to be equivalent to the set of constraints obtained (after certain simplifications) from Eq.~\eqref{eq:cstr_W}, when one attributes $\HS^P$ and $\HS^F$ to two additional parties with trivial input and output spaces, respectively.}
\begin{align}
    \forall \,\emptyset \subsetneq \K \subseteq \N, \, {}_{\prod_{k\in\K}[1-A_k^O]}\Tr_{A_{\N\backslash\K}^{IO}F}W = 0 \quad \text{and} \quad \Tr_{A_{\N}^{IO}F}W = d_\N^O \, \id^P. \label{eq:validityW_PF}
\end{align}

\medskip

Let us emphasise again that the validity constraints of Eqs.~\eqref{eq:validityW} and~\eqref{eq:validityW_PF} correspond to the requirement that the generalised Born rule of Eq.~\eqref{eq:Born_rule} gives the constant value 1, or that the alternative version of Eq.~\eqref{eq:Born_rule_PF} returns a trace-preserving global map $M_{\vec a|\vec x}^{PF}$, whenever one plugs in TP external operations (obtained e.g.\ as the sum of the elements of the parties' instruments).
When dealing with process matrices, one is often interested in the linear parts of the validity constraints, ignoring the affine part, i.e.\ ignoring the constraint $\Tr W = d_\N^O$ in Eq.~\eqref{eq:validityW}, or only imposing $\Tr_{A_{\N}^{IO}F}W \propto \id^P$ in Eq.~\eqref{eq:validityW_PF}. One then refers to matrices that are valid process matrices \emph{up to normalisation} only%
\footnote{Notice that this includes in particular the null matrix, $W=0$.}%
---as in the characterisations of our classes \textup{\textsf{QC-NICC}} and \textup{\textsf{QC-NIQC}}, see Eqs.~\eqref{eq:cstr_NICC} and~\eqref{eq:cstr_NIQC}.
Requiring validity up to normalisation only, for a given $W$, therefore has the following interpretation: when plugging in TP external operations, the generalised Born rules of Eq.~\eqref{eq:Born_rule} or~\eqref{eq:Born_rule_PF} must return a constant value, or a global induced map whose output state has a constant trace, respectively, whatever the choice of the TP external operations.

\subsection{Quantum circuits with classical control of causal order (QC-CCs)}
\label{app:subsec_QCCC_classes}

Let us now come to a more in-depth presentation of QC-CCs, and of the subclasses of QC-convFOs and QC-NICCs that we considered in this work. Secs.~\ref{app:QC-CC} (general QC-CCs) and~\ref{app:QC-convFO} (QC-convFOs) below are essentially just reviews of previous work, to set up the stage for the presentation and characterisation of our new class of QC-NICCs in Sec.~\ref{app:def_QC-NICC}.

\medskip

QC-CCs were introduced and characterised in Ref.~\cite{wechs21}, which we refer to for more details. One technical ingredient that we will use below is the so-called \emph{link product}~\cite{Chiribella08,Chiribella09}. It is defined, for two matrices $M_1^{XY}\in\L(\HS^{XY})$ and $M_2^{XY}\in\L(\HS^{YZ})$ that share a common space $\HS^Y$, as
\begin{align}
    M_1^{XY} * M_2^{YZ} \coloneqq & \big( \id^{XZ} \otimes \bbra{\id}^{YY} \big) (M_1^{XY} \otimes M_2^{YZ}) \big( \id^{XZ} \otimes \kket{\id}^{YY} \big) \notag \\
    =& \Tr_{Y} \big[ \big( (M_1^{XY})^{T_Y} \otimes \id^Z \big) \big( \id^X \otimes M_2^{YZ} \big) \big] \quad \in \L\big(\HS^{XZ}\big), \label{eq:def_mixed_link_product}
\end{align}
where $\kket{\id}^{YY}\coloneqq \sum_j \ket{i}^Y\otimes\ket{i}^Y$ denotes the unnormalised maximally entangled state (defined for two copies of the space $\HS^Y$ with computational basis $\{\ket{i}^Y\}_i$), and where $\Tr_{Y}$ and ${}^{T_Y}$ denote the partial trace and partial transpose over $\HS^Y$, respectively. The link product corresponds in the Choi picture to the composition of maps: if $M_1^{XY}$ and $M_2^{YZ}$ are the Choi matrices of two linear maps $\M_1: \L(\HS^X)\to\L(\HS^Y)$ and $\M_2: \L(\HS^Y)\to\L(\HS^Z)$, then $M_1^{XY} * M_2^{YZ}$ is the Choi matrix of their composition $\M_2 \circ \M_1: \L(\HS^X)\to\L(\HS^Z)$. (The same interpretation in terms of composition also applies if the output and input spaces of $\M_1$ and $\M_2$ do not match~\cite{Chiribella08,Chiribella09,wechs21}.)

The link product is commutative (up to a re-ordering of the tensor products) and associative (provided each Hilbert space is involved at most twice in an $n$-fold link product---as will always be the case here). It reduces to a standard tensor product $M_1^X * M_2^Z = M_1^X \otimes M_2^Z$ if the space $\HS^Y$ is trivial, and to an inner product $M_1^Y * M_2^Y = \Tr [(M_1^Y)^T M_2^Y]$ if the spaces $\HS^X$ and $\HS^Z$ are trivial.
Noticing that the Choi matrix of the preparation of a state $\rho$ coincides with the density matrix $\rho$ itself, one can write the output state of a map $\M$ (with Choi matrix $M$) applied to $\rho$ as the composition $\M(\rho) = \rho*M$.
Notice also that the generalised Born rule of Eq.~\eqref{eq:Born_rule_PF} can be written in a simple form as a link product, namely: $M_{\vec a|\vec x}^{PF} = \big(\bigotimes_k M^{A_k}_{a_k|x_k}\big) * W$. \label{page:born_rul_PF_link_prod}

\subsubsection{General QC-CCs}
\label{app:QC-CC}

\paragraph*{Construction of a QC-CC.}

As briefly introduced in the main text, a QC-CC consists in the application of a number of ``internal operations'', each of which determines which ``external operation'' (implemented by one of the $N$ parties $A_k$) is to be applied next---with each of the latter being applied once and only once by the end of the protocol.

One way to understand this is to consider that each internal operation corresponds to a quantum instrument $\{\M_{(k_1,\ldots,k_n)}^{\to k_{n+1}}\}_{k_{n+1}\notin\{k_1,\ldots,k_n\}}$, whose classical output value $k_{n+1}$ indicates the label of the next external operation to be applied, conditioned on the previous operations being those labelled $(k_1,\ldots,k_n)$, in that order. 
The instrument corresponding to the first internal operation consists of CP maps $\M_{\emptyset}^{\to k_1}: \L(\HS^P) \to \L(\HS^{A_{k_1}^I\alpha_1})$ that transform the input state in the ``global past'' space $\HS^P$ into a state in the input space $\HS^{A_{k_1}^I}$ of the first operation to be applied, as well as in some auxiliary ``memory'' space $\HS^{\alpha_1}$.%
\footnote{The auxiliary ``memory'' systems are not touched by the external operations. Notice that these may just be trivial.}
The subsequent internal operations, for $n=1,\ldots,N-1$, consist of CP maps $\M_{(k_1,\ldots,k_n)}^{\to k_{n+1}}: \L(\HS^{A_{k_n}^O\alpha_n}) \to \L(\HS^{A_{k_{n+1}}^I\alpha_{n+1}})$ that transform (conditioned on the previous external operations being $(k_1,\ldots,k_n)$, in that order) the output state of the latest external operation, together with the memory system, into a state in the input space $\HS^{A_{k_{n+1}}^I}$ of the next operation to be applied, as well as in some new memory space $\HS^{\alpha_{n+1}}$. Finally, the last internal operation consists of just a single CPTP map $\M_{(k_1,\ldots,k_N)}^{\to F}: \L(\HS^{A_{k_N}^O\alpha_N}) \to \L(\HS^F)$ (since there is no choice of next operation to be made) that transforms (conditioned on all external operations having been applied in the order $(k_1,\ldots,k_N)$) the output state of the last external operation, together with the memory system, into a state in the ``global future'' space $\HS^F$.

\medskip

\begin{figure*}[t]
\centering
\includegraphics[width=\columnwidth]{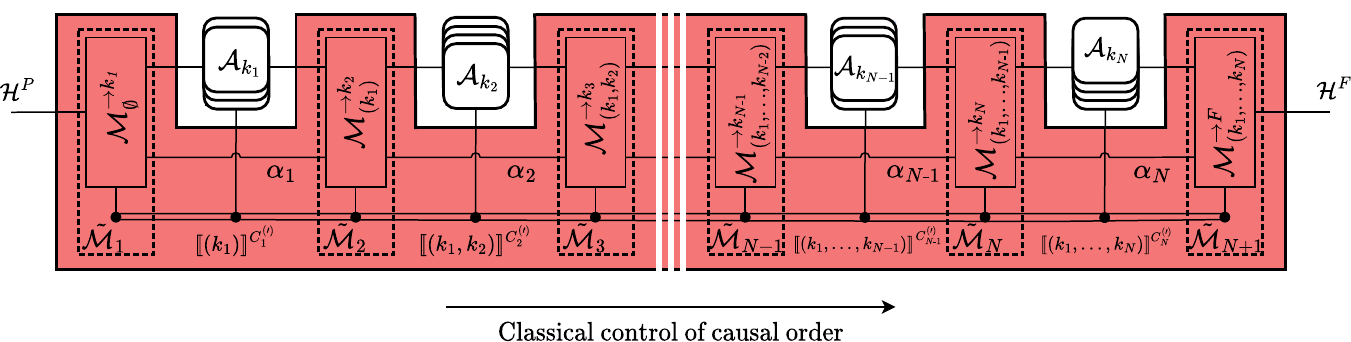}
\caption{Graphical representation of a QC-CC: a more detailed version than Fig.~\ref{fig:QCCC_graph} (clarifying here the internal structure of the circuit operations $\tilde{\M}_n$), reproduced from Fig.~9 of Ref.~\cite{wechs21}.}
\label{fig:QCCC_graph_full}
\end{figure*}

Another equivalent way to describe such a protocol---as depicted in Fig.~\ref{fig:QCCC_graph} of the main text, or with more details in Fig.~\ref{fig:QCCC_graph_full} here---is to encode the output of these instruments into a physical (effectively classical) system and consider the internal operations to also act on this system, which is used to control the subsequent operations. This can be done by introducing, for each time-slot $t_n$ ($n=1,\ldots,N$) at which an external operation is to be applied, a Hilbert space $\HS^{C_n}$ carrying orthogonal control states of the form
\begin{align}
    [\![(k_1,\ldots,k_n)]\!]^{C_n} \coloneqq \ketbra{(k_1,\ldots,k_n)}{(k_1,\ldots,k_n)}^{C_n}, \label{eq:def_classical_Cn}
\end{align}
which keeps track of all previous external operations, as well as of the current one to be applied.

In order to define a controlled version of the external operations, it is assumed for simplicity that all input Hilbert spaces $\HS^{A_k^I}$ (all output Hilbert spaces $\HS^{A_k^O}$, respectively) of all external operations are isomorphic. One can then introduce, for each time-slot $t_n$, a ``generic'' input Hilbert space $\HS^{\tilde{A}_n^I}$ (a ``generic'' output Hilbert space $\HS^{\tilde{A}_n^O}$, resp.), isomorphic to all $\HS^{A_k^I}$'s (isomorphic to all $\HS^{A_k^O}$'s, resp.). In this way, when an external operation%
\footnote{Here for simplicity we allow for some double-use of notation, using the same $A_k$ for the Choi matrix of an external operation and the party that implements it, as considered in the main text. The operations $A_k$ (in the Choi picture) considered here correspond to the individual CP maps $M_{a_k|x_k}^{A_k}$ considered in the main text to be part of each party's instrument.}
$\A_{k_n}:\L(\HS^{A_{k_n}^I})\to\L(\HS^{A_{k_n}^O})$ (with Choi matrix $A_{k_n}\in\L(\HS^{A_{k_n}^{IO}})$) is applied at time-slot $t_n$, one can identify it formally with an operation of the type $\tilde{\A}_{k_n}:\L(\HS^{\tilde{A}_n^I})\to\L(\HS^{\tilde{A}_n^O})$ (with Choi matrix $\tilde{A}_{k_n}:\L(\HS^{\tilde{A}_n^{IO}})$). This allows one to write all the external operations that may be applied at time-slot $t_n$ as acting in the same generic input and output spaces, and to embed these into some ``larger'' controlled operation $\tilde{\A}_n$, which applies the desired operation $\tilde{\A}_{k_n}$ depending on the state of the control (more specifically, on its last label $k_n$):
\begin{align}
\tilde \A_n \coloneqq & \sum_{(k_1,\ldots,k_n)} \tilde\A_{k_n} \otimes \bigpi_{(k_1,\ldots,k_n)}^{C_n \to C_n'}\ : \ \L(\HS^{\tilde A_n^I C_n}) \to \L(\HS^{\tilde A_n^O C_n'}), \label{eq:tilde_An_QCCC}
\end{align}
where for later convenience we introduced a copy $\HS^{C_n'}$ of the Hilbert space $\HS^{C_n}$, and where $\bigpi_{(k_1,\ldots,k_n)}^{C_n \to C_n'}$ is the (classical) map that projects the control system onto the state $[\![(k_1,\ldots,k_n)]\!]^{C_n}$, while re-labelling the control system $C_n$ to $C_n'$.

Similarly, one can formally write the internal operations $\M_{(k_1,\ldots,k_n)}^{\to k_{n+1}}: \L(\HS^{A_{k_n}^O\alpha_n}) \to \L(\HS^{A_{k_{n+1}}^I\alpha_{n+1}})$ (for each $n=1,\ldots,N-1$) as maps of the type $\tilde{\M}_{(k_1,\ldots,k_n)}^{\to k_{n+1}}: \L(\HS^{\tilde{A}_n^O\alpha_n}) \to \L(\HS^{\tilde{A}_{n+1}^I\alpha_{n+1}})$ acting in the generic Hilbert space introduced above. These can then be embedded into some ``larger'' operation $\tilde{\M}_{n+1}$ that also involves the control system: upon receiving the latter in the state $[\![(k_1,\ldots,k_n)]\!]^{C_n'}$ (following the operation $\tilde \A_n$), $\tilde{\M}_{n+1}$ applies the operations $\tilde{\M}_{(k_1,\ldots,k_n)}^{\to k_{n+1}}$ and updates the control system into the corresponding state $[\![(k_1,\ldots,k_n,k_{n+1})]\!]^{C_n}$. Formally:
\begin{align}
\tilde \M_{n+1} \coloneqq & \sum_{(k_1,\ldots,k_n,k_{n+1})} \tilde\M_{(k_1,\ldots,k_n)}^{\to k_{n+1}} \otimes \bigpi_{(k_1,\ldots,k_n),k_{n+1}}^{C_n' \to C_{n+1}} \ : \ \L(\HS^{\tilde A_n^O \alpha_nC_n'}) \to \L(\HS^{\tilde A_{n+1}^I \alpha_{n+1}C_{n+1}}), \label{eq:tilde_Mn1_QCCC}
\end{align}
where $\bigpi_{(k_1,\ldots,k_n),k_{n+1}}^{C_n' \to C_{n+1}}: \L(\HS^{C_n'}) \to \L(\HS^{C_{n+1}})$ is the (classical) map that projects the control system onto $[\![(k_1,\ldots,k_n)]\!]^{C_n'}$ and updates it to $[\![(k_1,\ldots,k_n,k_{n+1})]\!]^{C_{n+1}}$.
For the first and last internal operations, one similarly defines
\begin{align}
\tilde \M_1 \coloneqq  & \sum_{k_1} \tilde\M_\emptyset^{\to k_1} \otimes \bigpi_{\emptyset,k_1}^{\emptyset \to C_1} \ : \ \L(\HS^P) \to \L(\HS^{\tilde A_1^I \alpha_1C_1}) \label{eq:tilde_M1_QCCC} \\
\text{and} \quad \tilde \M_{N+1} \coloneqq & \sum_{(k_1,\ldots,k_N)} \tilde\M_{(k_1,\ldots,k_N)}^{\to F} \otimes \bigpi_{(k_1,\ldots,k_N),F}^{C_N' \to \emptyset} \ : \ \L(\HS^{\tilde A_N^O \alpha_NC_N'}) \to \L(\HS^F), \label{eq:tilde_MN1_QCCC}
\end{align}
where $\bigpi_{\emptyset,k_1}^{\emptyset \to C_1}$ and $\bigpi_{(k_1,\ldots,k_N),F}^{C_N' \to \emptyset}$ are the maps that create the initial control states $[\![(k_1)]\!]^{C_1}$ and that project onto the final control state $[\![(k_1,\ldots,k_N)]\!]^{C_N'}$, respectively.

In the Choi representation, all these embedding operations are written
\begin{align}
\tilde A_n = & \sum_{(k_1,\ldots,k_n)} \tilde A_{k_n} \otimes [\![(k_1,\ldots,k_n)]\!]^{C_n} \otimes [\![(k_1,\ldots,k_n)]\!]^{C_n'} \ \in \L(\HS^{\tilde A_n^I C_n\tilde A_n^O C_n'}), \label{eq:Choi_An_QCCC} \\[2mm]
\tilde M_1 = & \sum_{k_1} \tilde M_\emptyset^{\to k_1} \otimes [\![(k_1)]\!]^{C_1} \ \in \L(\HS^{P\tilde A_1^I \alpha_1C_1}), \label{eq:def_tilde_M1_QCCC}\\[2mm]
\tilde M_{n+1} = & \sum_{(k_1,\ldots,k_n,k_{n+1})} \tilde M_{(k_1,\ldots,k_n)}^{\to k_{n+1}} \otimes [\![(k_1,\ldots,k_n)]\!]^{C_n'} \otimes [\![(k_1,\ldots,k_n,k_{n+1})]\!]^{C_{n+1}} \ \in \L(\HS^{\tilde A_n^O \alpha_nC_n'\tilde A_{n+1}^I \alpha_{n+1}C_{n+1}}) \notag \\[-3mm]
& \hspace{110mm} (\text{for }n=1,\ldots,N-1), \label{eq:Choi_Mn1_QCCC}\\[2mm]
\tilde M_{N+1} =& \sum_{(k_1,\ldots,k_N)} \tilde M_{(k_1,\ldots,k_N)}^{\to F} \otimes [\![(k_1,\ldots,k_N)]\!]^{C_N'} \ \in \L(\HS^{\tilde A_N^O \alpha_NC_N'F}). \label{eq:def_tilde_MN1_QCCC}
\end{align}
These allow one, in particular, to conveniently evaluate the state of all systems at any point in the circuit. Consider, for that, inputting some state $\rho\in\L(\HS^P)$ in the global past of the circuit. The states $\varrho_{(n)}'$ before and $\varrho_{(n+1)}$ after each operation $\tilde{\M}_{n+1}$ are found (recursively) to be
\begin{align}
\varrho_{(0)}' & = \rho \quad \in \L(\HS^P), \label{eq:link_prod_rho_tilde_0} \\[3mm]
\varrho_{(n)}' & = \rho * \tilde M_1 * \tilde A_1 * \tilde M_2 * \cdots * \tilde M_n * \tilde A_n \notag \\
& = \sum_{(k_1,\ldots,k_n)} \big( \rho \otimes A_{k_1} \otimes \cdots \otimes A_{k_n} \big) * \big(M_\emptyset^{\to k_1} * M_{(k_1)}^{\to k_2} * \cdots * M_{(k_1,\ldots,k_{n-1})}^{\to k_n} * \kketbra{\id}{\id}^{A^O_{k_n}\tilde A_n^O}\big) \otimes [\![(k_1,\ldots,k_n)]\!]^{C_n'} \notag \\[-2mm]
& \hspace{100mm} \in \L(\HS^{\tilde A_n^O\alpha_n C_n'}) \quad \text{ for } n\ge 1 \label{eq:link_prod_rho_tilde_M1_tilde_An}
\end{align}
and
\begin{align}
\varrho_{(n+1)} & = \rho * \tilde M_1 * \tilde A_1 * \tilde M_2 * \cdots * \tilde M_n * \tilde A_n * \tilde M_{n+1} \notag \\
& = \sum_{(k_1,\ldots,k_n,k_{n+1})} \big( \rho \otimes A_{k_1} \otimes \cdots \otimes A_{k_n} \big) * \big( M_\emptyset^{\to k_1} * M_{(k_1)}^{\to k_2} * \cdots * M_{(k_1,\ldots,k_{n-1})}^{\to k_n} * M_{(k_1,\ldots,k_n)}^{\to k_{n+1}} * \kketbra{\id}{\id}^{A^I_{k_{n+1}}\tilde A_{n+1}^I} \big) \notag \\[-2mm]
& \hspace{50mm} \otimes [\![(k_1,\ldots,k_n,k_{n+1})]\!]^{C_{n+1}} \ \in \L(\HS^{\tilde A_{n+1}^I\alpha_{n+1} C_{n+1}}) \quad \text{ for } n\le N-1, \label{eq:link_prod_rho_tilde_M1_tilde_Mn1} \\[3mm]
\varrho_{(N+1)} & = \rho * \tilde M_1 * \tilde A_1 * \tilde M_2 * \cdots * \tilde M_N * \tilde A_N * \tilde M_{N+1} \notag \\
& = \sum_{(k_1,\ldots,k_N)} \big( \rho \otimes A_{k_1} \otimes \cdots \otimes A_{k_N} \big) * \big( M_\emptyset^{\to k_1} * M_{(k_1)}^{\to k_2} * \cdots * M_{(k_1,\ldots,k_{N-1})}^{\to k_N} * M_{(k_1,\ldots,k_N)}^{\to F} \big) \notag \\[-1mm]
& = \big( \rho \otimes A_1 \otimes \cdots \otimes A_N \big) * \Big( \sum_{(k_1,\ldots,k_N)} M_\emptyset^{\to k_1} * M_{(k_1)}^{\to k_2} * \cdots * M_{(k_1,\ldots,k_{N-1})}^{\to k_N} * M_{(k_1,\ldots,k_N)}^{\to F} \Big) \quad \in \L(\HS^F), \label{eq:link_prod_rho_tilde_M1_tilde_MN1}
\end{align}
where $\kketbra{\id}{\id}^{A^O_{k_n}\tilde A_n^O}$ and $\kketbra{\id}{\id}^{A^I_{k_{n+1}}\tilde A_{n+1}^I}$ in Eqs.~\eqref{eq:link_prod_rho_tilde_M1_tilde_An} and~\eqref{eq:link_prod_rho_tilde_M1_tilde_Mn1} simply represent some trivial mappings from the external operations' output and input spaces $\HS^{A^O_{k_n}}$ and $\HS^{A^I_{k_{n+1}}}$ to the corresponding generic output and input spaces $\HS^{\tilde A_n^O}$ and $\HS^{A^I_{k_{n+1}}\tilde A_{n+1}^I}$ (notice how the tildes were removed between the first and second lines in each equation).

\bigskip

\paragraph*{Trace-preserving conditions.}

In order to define a TP-preserving quantum circuit---such that when one plugs in TP external operations, the induced global map remains TP---one requires all internal operations $\tilde{\M}_{n+1}$ to preserve the trace of their input state (they must be TP on their \emph{effective} input space, in which their input state $\varrho_{(n)}'$ may effectively be found). In fact, a stronger condition is imposed here: the control system must evolve in a consistent way, so that for any $n\le N-1$ and any given sequence $(k_1,\ldots,k_n)$, the probabilities of finding the control in the state $[\![(k_1,\ldots,k_n,k_{n+1})]\!]^{C_{n+1}}$ after the operation $\tilde{\M}_{n+1}$ (and of the operations $A_{k_1},\ldots,A_{k_n}$ to be applied, when these are part of quantum instruments) sum up, for all possible $k_{n+1}\notin\{k_1,\ldots,k_n\}$, to the probability of finding the control in the state $[\![(k_1,\ldots,k_n)]\!]^{C_n'}$ before $\tilde{\M}_{n+1}$ (and of the operations $A_{k_1},\ldots,A_{k_n}$ to be applied).
From Eqs.~\eqref{eq:link_prod_rho_tilde_M1_tilde_An} and~\eqref{eq:link_prod_rho_tilde_M1_tilde_Mn1}, these probabilities are
\begin{align}
& \sum_{k_{n+1}} \Tr[(\id^{\tilde A_{n+1}^I \alpha_{n+1}} \otimes [\![ (k_1,\ldots,k_n,k_{n+1}) ]\!]^{C_{n+1}})\varrho_{(n+1)}] \notag \\[-1mm]
& \hspace{15mm} = \sum_{k_{n+1}} \Tr\big[ \big( \rho \otimes A_{k_1} \otimes \cdots \otimes A_{k_n} \big) * \big( M_\emptyset^{\to k_1} * M_{(k_1)}^{\to k_2} * \cdots * M_{(k_1,\ldots,k_{n-1})}^{\to k_n} * M_{(k_1,\ldots,k_n)}^{\to k_{n+1}} \big) \big] \notag \\
& \hspace{15mm} = \Tr\big[\big(\rho \otimes A_{k_1} \otimes \cdots \otimes A_{k_n} \big)^T \sum_{k_{n+1}} \Tr_{A_{k_{n+1}}^I\alpha_{n+1}} ( M_\emptyset^{\to k_1} * M_{(k_1)}^{\to k_2} * \cdots * M_{(k_1,\ldots,k_{n-1})}^{\to k_n} * M_{(k_1,\ldots,k_n)}^{\to k_{n+1}} ) \big]. \label{eq:tr_varrho_n_1_QCCC}
\end{align}
and
\begin{align}
\Tr\big[(\id^{\tilde A^O_n\alpha_n} \otimes [\![ (k_1,\ldots,k_n) ]\!]^{C_n'})\varrho_{(n)}'\big] = \Tr\big[ \big( \rho \otimes A_{k_1} \otimes \cdots \otimes A_{k_n} \big) * \big(M_\emptyset^{\to k_1} * M_{(k_1)}^{\to k_2} * \cdots * M_{(k_1,\ldots,k_{n-1})}^{\to k_n} \big) \big] \notag \\
 = \Tr\big[\big(\rho \otimes A_{k_1} \otimes \cdots \otimes A_{k_n} \big)^T \big( \Tr_{\alpha_n} ( M_\emptyset^{\to k_1} * M_{(k_1)}^{\to k_2} * \cdots * M_{(k_1,\ldots,k_{n-1})}^{\to k_n} \big) \big]. \label{eq:tr_varrho_n_QCCC}
\end{align}
As these two expressions must be equal for all possible $\rho \otimes A_{k_1} \otimes \cdots \otimes A_{k_n}$, which span the whole space $\L(\HS^{PA_{\{k_1,\ldots,k_n\}}^{IO}})$, then the second terms inside both final traces must themselves be equal:
\begin{align}
\sum_{k_{n+1}} \Tr_{A_{k_{n+1}}^I\alpha_{n+1}} ( M_\emptyset^{\to k_1} * M_{(k_1)}^{\to k_2} * \cdots * M_{(k_1,\ldots,k_{n-1})}^{\to k_n} * M_{(k_1,\ldots,k_n)}^{\to k_{n+1}} ) = \Tr_{\alpha_n} ( M_\emptyset^{\to k_1} * M_{(k_1)}^{\to k_2} * \cdots * M_{(k_1,\ldots,k_{n-1})}^{\to k_n} ) \otimes \id^{A_{k_n}^O}. \label{eq:QCCC_TP_n}
\end{align}
For the limiting cases $n=0$ and $n=N$, we correspondingly obtain the constraints:
\begin{align}
\sum_{k_1} \Tr_{A_{k_1}^I\alpha_1} ( M_\emptyset^{\to k_1} ) = \id^P \label{eq:QCCC_TP_0}
\end{align}
and
\begin{align}
\Tr_F ( M_\emptyset^{\to k_1} * M_{(k_1)}^{\to k_2} * \cdots * M_{(k_1,\ldots,k_{N-1})}^{\to k_N} * M_{(k_1,\ldots,k_N)}^{\to F} ) = \Tr_{\alpha_N} ( M_\emptyset^{\to k_1} * M_{(k_1)}^{\to k_2} * \cdots * M_{(k_1,\ldots,k_{N-1})}^{\to k_N} ) \otimes \id^{A_{k_N}^O}. \label{eq:QCCC_TP_N}
\end{align}

\bigskip

\paragraph*{Process matrix description and characterisation.}

The expression for $\varrho_{N+1}$ in Eq.~\eqref{eq:link_prod_rho_tilde_M1_tilde_MN1} allows one to readily obtain the process matrix description of the QC-CC under consideration. Indeed, it follows from the generalised Born rule of Eq.~\eqref{eq:Born_rule_PF} (or its form given on p.~\pageref{page:born_rul_PF_link_prod} after the introduction of the link product) that the final state in the global future should be $\varrho_{N+1} = ( \rho \otimes A_1 \otimes \cdots \otimes A_N ) * W$. 
Since the $\rho \otimes A_1 \otimes \cdots \otimes A_N$'s span the whole space $\L(\HS^{PA_\N^{IO}})$, there is only one matrix $W$ that gives the correct output state through this expression. From Eq.~\eqref{eq:link_prod_rho_tilde_M1_tilde_MN1} we can thus directly identify the process matrix of the QC-CC under consideration to be 
\begin{align}
    W = \sum_{(k_1,\ldots,k_N)} M_\emptyset^{\to k_1} * M_{(k_1)}^{\to k_2} * \cdots * M_{(k_1,\ldots,k_{N-1})}^{\to k_N} * M_{(k_1,\ldots,k_N)}^{\to F} \quad \in \L(\HS^{PA_\N^{IO}F}). \label{eq:W_QCCC}
\end{align}

\medskip

Defining, for all $n=1,\ldots,N$ and all $(k_1,\ldots,k_n)$,
\begin{align}
    W_{(k_1,\ldots,k_n)} & \coloneqq \Tr_{\alpha_n} \big( M_\emptyset^{\to k_1} * M_{(k_1)}^{\to k_2} * \cdots * M_{(k_1,\ldots,k_{n-1})}^{\to k_n} \big) \quad \in \L(\HS^{PA_{\{k_1,\ldots,k_{n-1}\}}^{IO}A_{k_n}^I}), \label{eq:def_Wk1_kn_QCCC} \\
    W_{(k_1,\ldots,k_N,F)} & \coloneqq M_\emptyset^{\to k_1} * M_{(k_1)}^{\to k_2} * \cdots * M_{(k_1,\ldots,k_{N-1})}^{\to k_N} * M_{(k_1,\ldots,k_N)}^{\to F} \quad \in \L(\HS^{PA_\N^{IO}F}), \label{eq:def_Wk1_kN_F_QCCC}
\end{align}
then it follows directly from Eq.~\eqref{eq:W_QCCC} and the TP conditions of Eqs.~\eqref{eq:QCCC_TP_n}--\eqref{eq:QCCC_TP_N} that the process matrix $W \, (\ge 0)$ of a QC-CC has a characterisation as in Eq.~\eqref{eq:charact_W_QCCC_decomp}.

Conversely, from any decomposition of a matrix $W$ in terms of PSD matrices $W_{(k_1,\ldots,k_N,F)}$ and $W_{(k_1,\ldots,k_n)}$ as in Eq.~\eqref{eq:charact_W_QCCC_decomp}, one can explicitly reconstruct (in a non-unique manner) the internal operations $M_{(k_1,\ldots,k_n)}^{\to k_{n+1}}$ and $M_{(k_1,\ldots,k_N)}^{\to F}$ of a QC-QC (satisfying the required TP conditions), whose composition reproduces $W_{(k_1,\ldots,k_n)}$, $W_{(k_1,\ldots,k_N,F)}$ and $W$ as in Eqs.~\eqref{eq:W_QCCC}--\eqref{eq:def_Wk1_kN_F_QCCC} above. This shows that Eq.~\eqref{eq:charact_W_QCCC_decomp} is also a sufficient condition for $W$ to be a QC-CC.
An explicit procedure for such a reconstruction was detailed in Appendix~B\,2\,c of Ref.~\cite{wechs21}.

\subsubsection{Convex mixtures of quantum circuits with fixed causal order (QC-convFOs)}
\label{app:QC-convFO}

\paragraph*{Quantum circuits with fixed causal order (QC-FOs).}

QC-FOs have been described and characterised previously, under the alternative names of channels with memory~\cite{Kretschmann05}, quantum strategies~\cite{Gutoski06}, or quantum combs~\cite{Chiribella08,Chiribella09}. One may describe them in a similar and in fact much simpler way to QC-CCs above, without the need to introduce the control system. For a given fixed order $(k_1,\ldots,k_N)$, the internal operations would simply be of the form $\tilde{\M}_{n+1} = \tilde{\M}_{(k_1,\ldots,k_n)}^{\to k_{n+1}}$. The TP conditions they satisfy are as in Eqs.~\eqref{eq:QCCC_TP_n}--\eqref{eq:QCCC_TP_N}, just without the sums $\sum_{k_{n+1}}, \sum_{k_1}$. One can identify their process matrix representation in the same way as we did above, and one simply gets
\begin{align}
    W = W_{(k_1,\ldots,k_N,F)} = M_\emptyset^{\to k_1} * M_{(k_1)}^{\to k_2} * \cdots * M_{(k_1,\ldots,k_{N-1})}^{\to k_N} * M_{(k_1,\ldots,k_N)}^{\to F} \quad \in \L(\HS^{PA_\N^{IO}F}). \label{eq:W_QCFO}
\end{align}
From the TP conditions, and defining%
\footnote{Notice that in contrast to the general QC-CC case, the matrices $W_{(k_1,\ldots,k_n)}$ defined here are all valid process matrices (indeed, one can readily verify explicitly that the constraints of Eq.~\eqref{eq:validityW_PF} are satisfied). This implies in particular that the matrices $W_{(k_1,\ldots,k_{n})}^{(k_{n+1},\ldots,k_N)}$ that appear in the QC-convFO decomposition of Eq.~\eqref{eq:charact_W_QC_convFO_decomp}, defined below as $W_{(k_1,\ldots,k_{n})}^{(k_{n+1},\ldots,k_N)}\coloneqq q_{\pi=(k_1,\ldots,k_N)}\tilde{W}_{(k_1,\ldots,k_n)}$, are also valid process matrices up to normalisation. \label{ftn:valid_Wk1_kn_FO}}
$W_{(k_1,\ldots,k_n)}$ as in Eq.~\eqref{eq:def_Wk1_kn_QCCC} above, one can readily see that the process matrix $W = W_{(k_1,\ldots,k_N,F)} \, (\ge 0)$ of a quantum circuit with fixed causal order $(k_1,\ldots,k_N)$ is necessarily such that for all $n = 1, \ldots, N$, there exist PSD matrices $W_{(k_1,\ldots,k_n)} \in {\cal L}(\HS^{PA_{\{k_1,\ldots,k_{n-1}\}}^{IO} A_{k_n}^I})$ such that
\begin{align}
    \left\{
    \begin{array}{l}
        \Tr_F W = \Tr_F W_{(k_1,\ldots,k_N,F)} = W_{(k_1,\ldots,k_N)} \otimes \id^{A_{k_N}^O}, \\[3mm]
        \forall \, n=1,\ldots,N-1, \ \ \Tr_{A_{k_{n+1}}^I} W_{(k_1,\ldots,k_{n+1})} = W_{(k_1,\ldots,k_{n})} \otimes \id^{A_{k_n}^O}, \\[3mm]
        \Tr_{A_{k_1}^I} W_{(k_1)} = \id^P.
    \end{array}
    \right. \label{eq:charact_W_QC_FO_decomp}
\end{align}
Conversely, for any PSD matrix $W \in \L(\HS^{PA_\N^{IO}F})$ satisfying the above constraint, one can construct (in a similar way to the case of general QC-CCs) internal circuit operations $\tilde{\M}_{n+1} = \tilde{\M}_{(k_1,\ldots,k_n)}^{\to k_{n+1}}$, whose composition as in Eq.~\eqref{eq:W_QCFO} gives precisely $W$~\cite{Gutoski06,Chiribella09,wechs21}. This shows that Eq.~\eqref{eq:charact_W_QC_FO_decomp} is also a sufficient characterisation of the class of QC-FOs.

\bigskip

\paragraph*{Convex mixtures of QC-FOs (QC-convFOs).}

The process matrix representation of a convex mixture of QC-FOs is then of the form $W = \sum_{\pi=(k_1,\ldots,k_N)} q_\pi\, \tilde{W}_{(k_1,\ldots,k_N,F)}$, with $q_\pi\ge 0$, $\sum_\pi q_\pi = 1$, and each $\tilde{W}_{(k_1,\ldots,k_N,F)}$ satisfying the constraints of Eq.~\eqref{eq:charact_W_QC_FO_decomp}.
By defining the subnormalised process matrices $W_{(k_1,\ldots,k_N,F)}\coloneqq q_{\pi=(k_1,\ldots,k_N)}\tilde{W}_{(k_1,\ldots,k_N,F)}$ and $W_{(k_1,\ldots,k_{n})}^{(k_{n+1},\ldots,k_N)}\coloneqq q_{\pi=(k_1,\ldots,k_N)}\tilde{W}_{(k_1,\ldots,k_n)}$ (with $\tilde{W}_{(k_1,\ldots,k_n)}$ obtained from the QC-FO decomposition of $\tilde{W}_{(k_1,\ldots,k_N,F)}$), one readily gets the characterisation of Eq.~\eqref{eq:charact_W_QC_convFO_decomp} for the class of QC-convFOs.
The fact that Eq.~\eqref{eq:charact_W_QC_convFO_decomp} is a sufficient condition for $W$ to be a QC-convFO also follows straightforwardly from the sufficiency of Eq.~\eqref{eq:charact_W_QC_FO_decomp} above.

\bigskip

\paragraph*{An alternative description of QC-convFOs.}

We note that one could present an alternative description of QC-convFOs, by encoding each order in the mixture into a classical system that is then used to control all operations and their order---as in a general QC-CC, except that the complete order $(k_1,\ldots,k_N)$ would be decided, and encoded in the control system, from the very beginning (and independently of the global input state of the circuit).

This approach is the one we will follow to define the class of quantum circuits with superposition of fixed causal orders (QC-supFO), after turning the classical control system into a quantum control,  whose initial state will encode the complete order; see Sec.~\ref{app:def_QC-supFO} below.

\subsubsection{Quantum circuits with non-influenceable classical control of causal order (QC-NICCs)}
\label{app:def_QC-NICC}

Let us now come to our new class of QC-NICCs, defined in Sec.~\ref{sec:QC-NICC} of the main text as follows:
\begin{align*}
  \parbox{0.85\linewidth}{%
    \emph{QC-NICCs are the QC-CCs for which there exists an implementation such that, at any intermediate time step, the state of the control system is independent of the choice of previously applied external operations and of the potential initial state preparation in the global past,}%
  }
\end{align*}
with here the external operations referring to the parties' whole instruments (or simply a quantum channel they apply) rather than the individual CP maps that comprise them, and the state of the control system being considered after summing over all these individual CP maps.

The states $\varrho_{(n)}$, $\varrho_{(n)}'$ of all systems at any intermediate time step $t_n$ (for $n=1,\ldots,N$, just before or just after the operation $\tilde\A_n$) were given in Eqs.~\eqref{eq:link_prod_rho_tilde_M1_tilde_An} and~\eqref{eq:link_prod_rho_tilde_M1_tilde_Mn1}. Considering the case where all external operations $A_k$ are TP (equivalently, summing over the CP maps that constitute the parties' instruments), the reduced state of the control system is obtained as%
\footnote{In Eq.~\eqref{eq:reduced_state_Cn_QCCC} we used the TP property $\Tr_{A_{k_n}^O}A_{k_n}=\id^{A_{k_n}^I}$ to remove $A_{k_n}$ from the expression of $\Tr_{\tilde A_n^O\alpha_n} \varrho_{(n)}'$. Since we formally get the same reduced state of the control before and after the operation $\tilde\A_n$, we ignore here the distinction between systems $C_n$ and $C_n'$. \label{ftn:same_state_Cn}}
\begin{align}
\Tr_{\tilde A_n^I\alpha_n} \varrho_{(n)} = \Tr_{\tilde A_n^O\alpha_n} \varrho_{(n)}' & = \!\sum_{(k_1,\ldots,k_n)}\!\! \Tr[\big( \rho \!\otimes\! A_{k_1} \!\otimes\! \cdots \!\otimes\! A_{k_{n-1}} \big) * \big( M_\emptyset^{\to k_1} \!*\! M_{(k_1)}^{\to k_2} \!*\! \cdots \!*\! M_{(k_1,\ldots,k_{n-1})}^{\to k_n} \big)] [\![(k_1,\ldots,k_n)]\!]^{C_n^{(\prime)}} \notag \\ 
& = \sum_{(k_1,\ldots,k_n)} \Tr[\big( \rho \otimes A_{k_1} \otimes \cdots \otimes A_{k_{n-1}} \big) * W_{(k_1,\ldots,k_n)} ] \, [\![(k_1,\ldots,k_n)]\!]^{C_n^{(\prime)}}, \label{eq:reduced_state_Cn_QCCC}
\end{align}
with $W_{(k_1,\ldots,k_n)}$ as defined in Eq.~\eqref{eq:def_Wk1_kn_QCCC}, and contributing to the decomposition of the process matrix $W$ of the QC-CC under consideration, as in Eq.~\eqref{eq:charact_W_QCCC_decomp}.

The condition that the state of the control system should be independent of $\rho$ and of the previously applied external operations translates into the constraint that for all $(k_1,\ldots,k_n)$,
\begin{align}
\Tr[\big( \rho \otimes A_{k_1} \otimes \cdots \otimes A_{k_{n-1}} \big) * W_{(k_1,\ldots,k_n)} ] 
& \text{ does not depend on } \rho, A_{k_1}, \ldots, A_{k_{n-1}} \notag \\
& \quad \text{ whenever } A_{k_1}, \ldots, A_{k_{n-1}} \text{ are TP (and $\Tr\rho=1$)}.
\end{align}
Recalling the discussion at the end of Sec.~\ref{app:subsec_validity_cstr}, this is equivalent to
\begin{align}
    \text{all} \ W_{(k_1,\ldots,k_n)} \text{'s are valid process matrices (up to normalisation)},
\end{align}
as in Eq.~\eqref{eq:cstr_NICC}.

Conversely, as we recalled above, from the decomposition of a matrix $W$ in terms of PSD matrices $W_{(k_1,\ldots,k_N,F)}$ and $W_{(k_1,\ldots,k_n)}$ as in Eq.~\eqref{eq:charact_W_QCCC_decomp}, one can reconstruct some internal operations $M_{(k_1,\ldots,k_n)}^{\to k_{n+1}}$ (and $M_{(k_1,\ldots,k_N)}^{\to F}$) of a QC-CC such that $\Tr_{\alpha_n} \big( M_\emptyset^{\to k_1} * M_{(k_1)}^{\to k_2} * \cdots * M_{(k_1,\ldots,k_{n-1})}^{\to k_n} \big) = W_{(k_1,\ldots,k_n)}$. 
If each $W_{(k_1,\ldots,k_n)}$ is proportional to a valid process matrix, then the reduced state of the control system at time step $t_n$, as given in Eq.~\eqref{eq:reduced_state_Cn_QCCC} above, does not depend on $\rho$, nor on the operations $A_k$ when these are TP. Hence, the existence of a decomposition as in Eq.~\eqref{eq:charact_W_QCCC_decomp} in terms of matrices $W_{(k_1,\ldots,k_n)}$ that are proportional to valid process matrices is also a sufficient condition for $W$ to be the process matrix of a QC-NICC.

\subsubsection{Inclusion relations}
\label{app:subsubsec_QCCC_inclusions}

It is quite clear, from the definition of the \textup{\textsf{QC-NICC}} class just above, that $\textup{\textsf{QC-NICC}} \subset \textup{\textsf{QC-CC}}$.
To see that $\textup{\textsf{QC-convFO}} \subset \textup{\textsf{QC-NICC}}$, consider a process matrix $W\in\textup{\textsf{QC-convFO}}$, with a decomposition in terms of PSD matrices (and even valid process matrices, up to normalisation; see Footnote~\ref{ftn:valid_Wk1_kn_FO}) $W_{(k_1,\ldots,k_N,F)}$ and $W_{(k_1,\ldots,k_n)}^{(k_{n+1},\ldots,k_N)}$ as in Eq.~\eqref{eq:charact_W_QC_convFO_decomp}; defining, for each $(k_1,\ldots,k_n)$, $W_{(k_1,\ldots,k_n)}\coloneqq \sum_{(k_{n+1},\ldots,k_N)} W_{(k_1,\ldots,k_n)}^{(k_{n+1},\ldots,k_N)}$, one can easily see that these satisfy the QC-CC decomposition of Eq.~\eqref{eq:charact_W_QCCC_decomp}, and that these are themselves valid process matrices (up to normalisation), which proves that $W \in \textup{\textsf{QC-NICC}}$.
Hence, we obtain the general inclusion relations of Eq.~\eqref{eq:inclusions_QC-CCs}:
\begin{align}
    \text{\textup{\textsf{QC-convFO}}} \ \subset \ \text{\textup{\textsf{QC-NICC}}} \ \subset \ \text{\textup{\textsf{QC-CC}}}.
\end{align}

\medskip

It is straightforward to see that all these classes reduce to the same one (which also matches the whole class of valid process matrices, see Sec.~\ref{app:subsec_validity_cstr}) for $N=1$.

For $N=2$ already, with a nontrivial global past space $\HS^P$, one finds a strict separation between \text{\textup{\textsf{QC-NICC}}} and \text{\textup{\textsf{QC-CC}}}: an example of a process matrix in $\text{\textup{\textsf{QC-CC}}}\backslash\text{\textup{\textsf{QC-NICC}}}$ was given by the ``classical switch'' $W_{\textup{CS}}$, Eq.~\eqref{eq:CS-example} of the main text. If $\HS^P$ is trivial on the other hand, \text{\textup{\textsf{QC-NICC}}} and \text{\textup{\textsf{QC-CC}}} coincide for $N=2$ (see the proof below), and one needs to go to $N\ge 3$ to see a strict separation.

As for \text{\textup{\textsf{QC-convFO}}} vs \text{\textup{\textsf{QC-NICC}}}, these can be seen to coincide for $N\le 3$ (proof below). A strict separation can be found for $N\ge4$: an example of a process matrix in $\text{\textup{\textsf{QC-NICC}}}\backslash\text{\textup{\textsf{QC-convFO}}}$, for $N=4$ with trivial global past and future spaces $\HS^P, \HS^F$, was given by $W_{\textup{NICC}}$ from Eq.~\eqref{eq:QC-NICC-example}.

\begin{itemize}

    \item \underline{Proof that $\text{\textup{\textsf{QC-NICC}}}=\text{\textup{\textsf{QC-CC}}}$ for $N=2$ and a trivial $\HS^P$:} Given the general inclusion $\text{\textup{\textsf{QC-NICC}}}\subset\text{\textup{\textsf{QC-CC}}}$, it remains to prove the inclusion in the reverse direction. \label{proof:NICC_CC_N2_noP}
    
    Consider for that the process matrix $W$ of a QC-CC in that scenario. It decomposes in terms of PSD matrices $W_{(k_1,k_2,F)}$, $W_{(k_1,k_2)}$ and $W_{(k_1)}$ as in Eq.~\eqref{eq:charact_W_QCCC_decomp}---satisfying in particular $\Tr_{A_{k_2}^I} W_{(k_1,k_2)} = W_{(k_1)}\otimes\id^{A_{k_1}^O}$. Now, according to Eq.~\eqref{eq:validityW_PF}, the validity constraints, up to normalisation, for $W_{(k_1,k_2)} \in \L(\HS^{A_{k_1}^{IO}A_{k_2}^I})$ (with a trivial $\HS^P$) reduce to ${}_{[1-A_{k_1}^O]}\Tr_{A_{k_2}^I} W_{(k_1,k_2)} = 0$, which is indeed implied by $\Tr_{A_{k_2}^I} W_{(k_1,k_2)} = W_{(k_1)}\otimes\id^{A_{k_1}^O}$. Hence, the $W_{(k_1,k_2)}$'s are valid process matrices up to normalisation, which implies that $W$ is a QC-NICC, and therefore that $\text{\textup{\textsf{QC-CC}}}\subset\text{\textup{\textsf{QC-NICC}}}$ in the scenario under consideration. (Recall that it is enough here to check the validity for the $W_{(k_1,k_2)}$'s; in any case, the validity constraints for $W_{(k_1)} \in \L(\HS^{A_{k_1}^I})$, with no $\HS^P$, are trivially satisfied.)

    \item \underline{Proof that $\text{\textup{\textsf{QC-convFO}}}=\text{\textup{\textsf{QC-NICC}}}$ for $N\le 3$:} Given the general inclusion $\text{\textup{\textsf{QC-convFO}}}\subset\text{\textup{\textsf{QC-NICC}}}$, it remains to prove the inclusion in the reverse direction; furthermore, it is enough to consider $N=3$, since any scenario with fewer parties can be considered as a 3-partite scenario where some party(-ies) have trivial input and output spaces. \label{proof:convFO_NICC_N3}

    Consider, therefore, the process matrix $W$ of a tripartite ($N=3$) QC-NICC. It decomposes in terms of PSD matrices $W_{(k_1,k_2,k_3,F)}$, $W_{(k_1,k_2,k_3)}$, $W_{(k_1,k_2)}$ and $W_{(k_1)}$ as in Eq.~\eqref{eq:charact_W_QCCC_decomp}, with all of these being valid process matrices (up to normalisation). Defining $W_{(k_1,k_2,k_3)}^\emptyset\coloneqq W_{(k_1,k_2,k_3)}\ge 0$ and $W_{(k_1,k_2)}^{(k_3)}\coloneqq W_{(k_1,k_2)}\ge 0$ (for $k_3\neq k_1,k_2$), one can see that these (together with $W_{(k_1,k_2,k_3,F)}$) satisfy the first two lines of Eq.~\eqref{eq:charact_W_QC_convFO_decomp}, as well as the third constraint thereof, for $n=2$. Furthermore, the validity of $W_{(k_1,k_2)}\in\L(\HS^{PA_{k_1}^{IO}A_{k_2}^I})$ implies (according to Eq.~\eqref{eq:validityW_PF}) that ${}_{[1-A_{k_1}^O]}\Tr_{A_{k_2}^I} W_{(k_1,k_2)} = 0$, i.e., $\Tr_{A_{k_2}^I} W_{(k_1,k_2)}^{(k_3)} = \frac{1}{d_{k_1}^O}\Tr_{A_{k_1}^OA_{k_2}^I} W_{(k_1,k_2)}^{(k_3)}\otimes\id^{A_{k_1}^O}$, so that by defining $W_{(k_1)}^{(k_2,k_3)}\coloneqq \frac{1}{d_{k_1}^O}\Tr_{A_{k_1}^OA_{k_2}^I} W_{(k_1,k_2)}^{(k_3)}\ge 0$, the third constraint of Eq.~\eqref{eq:charact_W_QC_convFO_decomp} gets also satisfied for $n=1$. The validity of $W_{(k_1,k_2)}$ also implies that ${}_{[1-P]}\Tr_{A_{k_1}^{IO}A_{k_2}^I} W_{(k_1,k_2)} = 0$, i.e., $\Tr_{A_{k_1}^I} W_{(k_1)}^{(k_2,k_3)} = (\frac{1}{d^P}\Tr W_{(k_1)}^{(k_2,k_3)})\id^P$, so that final constraint of Eq.~\eqref{eq:charact_W_QC_convFO_decomp} is also satisfied, with $q_{\pi=(k_1,k_2,k_3)} = \frac{1}{d^P}\Tr W_{(k_1)}^{(k_2,k_3)} \ge 0$, such that $\sum_\pi q_\pi = \sum_{(k_1,k_2,k_3)} \frac{1}{d^P}\Tr W_{(k_1)}^{(k_2,k_3)} 
    = \sum_{k_1} \frac{1}{d^Pd_{k_1}^O} \sum_{k_2}\Tr W_{(k_1,k_2)} = \frac{1}{d^P} \sum_{k_1} \Tr W_{(k_1)} = 1$ as required (where we used $\sum_{k_2}\Tr W_{(k_1,k_2)}=\Tr\,(W_{(k_1)}\otimes\id^{A_{k_1}^O})$ and $\sum_{k_1} \Tr W_{(k_1)} = d^P$, which follow from Eq.~\eqref{eq:charact_W_QCCC_decomp}). 

    We thus obtained a decomposition of $W$ that satisfies all constraints of Eq.~\eqref{eq:charact_W_QC_convFO_decomp}, which shows that $W$ is a QC-convFO, and therefore that $\text{\textup{\textsf{QC-NICC}}}\subset\text{\textup{\textsf{QC-convFO}}}$, for the (at most) tripartite case considered here.
    
\end{itemize}

\subsubsection{Explicit decompositions of our specific examples of QC-CCs}
\label{app:decomp_ex_QCCCs}

To close this section on QC-CCs, we now provide explicit decompositions of our examples of QC-CCs from Sec.~\ref{subsubsec:examples_QCCCs}, in the appropriate form so as to match the characterisation of the class that they belong to.

\bigskip

Our first example, $W_{\textup{convFO}} = \frac12\ketbra{\psi}{\psi}^{A^I} \otimes \kketbra{\id}{\id}^{A^OB^I}\otimes \id^{B^O} + \frac12\ketbra{\psi}{\psi}^{B^I} \otimes \kketbra{\id}{\id}^{B^OA^I}\otimes \id^{A^O}$ from Eq.~\eqref{eq:QC-convFO-example}, is clearly of the QC-convFO form of Eq.~\eqref{eq:charact_W_QC_convFO_decomp}, with
\begin{align}
\left\{
    \begin{array}{rclcrcl}
W_{(A,B)}^\emptyset &=& {\textstyle\frac12}\ketbra{\psi}{\psi}^{A^I} \otimes \kketbra{\id}{\id}^{A^OB^I}, 
&& W_{(B,A)}^\emptyset &=& {\textstyle\frac12}\ketbra{\psi}{\psi}^{B^I} \otimes \kketbra{\id}{\id}^{B^OA^I}, \\[2mm]
W_{(A)}^{(B)} &=& {\textstyle\frac12}\ketbra{\psi}{\psi}^{A^I}, 
&& W_{(B)}^{(A)} &=& {\textstyle\frac12}\ketbra{\psi}{\psi}^{B^I}, \\[2mm]
\multicolumn{3}{r}{q_{(A,B)}} &=& \multicolumn{3}{l}{q_{(B,A)} \,=\, {\textstyle\frac12}}
    \end{array}
\right. \label{eq:decomp_WconvFO}
\end{align}
(and with $W_{(k_1,k_2,F)} = W_{(k_1,k_2)}^\emptyset \otimes \id^{A_{k_2}^O}$, since $\HS^F$ is trivial).

\bigskip

Our second example, $W_{\textup{NICC}} = \frac12 W_{+\alpha} \otimes \ketbra{\psi}{\psi}^{C^I} \otimes \kketbra{\id}{\id}^{C^O D^I} \otimes \id^{D^{O}} + \frac12 W_{-\alpha}\otimes \ketbra{\psi}{\psi}^{D^I} \otimes \kketbra{\id}{\id}^{D^O C^I} \otimes \id^{C^{O}}$ from Eq.~\eqref{eq:QC-NICC-example}, admits a QC-NICC decomposition of the form of Eq.~\eqref{eq:charact_W_QCCC_decomp}, with for instance%
\footnote{It is clear here that other decompositions are also possible, e.g., involving terms with $B\prec A$ rather than with $A\prec B$ (notice that $A$ and $B$ come in parallel in the implementation of Fig.~\ref{fig:QC-NICC-example}, which is compatible with both orders), or with a mixture of the orders $A\prec B$ and $B\prec A$.}
the valid process matrices (up to normalisation, as required by the QC-NICC condition of Eq.~\eqref{eq:cstr_NICC})
\begin{align}
\left\{
    \begin{array}{rclcrcl}
W_{(A,B,C,D)} &=& {\textstyle\frac12} W_{+\alpha} \otimes \ketbra{\psi}{\psi}^{C^I} \otimes \kketbra{\id}{\id}^{C^O D^I}, 
&& W_{(A,B,D,C)} &=& {\textstyle\frac12} W_{-\alpha}\otimes \ketbra{\psi}{\psi}^{D^I} \otimes \kketbra{\id}{\id}^{D^O C^I}, \\[2mm]
W_{(A,B,C)} &=& {\textstyle\frac12} W_{+\alpha} \otimes \ketbra{\psi}{\psi}^{C^I}, 
&& W_{(A,B,D)} &=& {\textstyle\frac12} W_{-\alpha}\otimes \ketbra{\psi}{\psi}^{D^I}, \\[2mm]
\multicolumn{3}{r}{W_{(A,B)}} &=& \multicolumn{3}{l}{{\textstyle\frac14} \id^{A^{IO}B^I},} \\[2mm]
\multicolumn{3}{r}{W_{(A)}} &=& \multicolumn{3}{l}{{\textstyle\frac12} \id^{A^I},}
    \end{array}
\right.
 \label{app:dec_WNICC}
\end{align}
and all other $W_{(k_1,\ldots,k_n)}=0$ (and again with $W_{(k_1,\ldots,k_4,F)} = W_{(k_1,\ldots,k_4)} \otimes \id^{A_{k_4}^O}$, since $\HS^F$ is trivial).
Notice that while the $W_{\pm\alpha}$'s are valid process matrices, the fact that they are causally nonseparable prevents one from obtaining a QC-convFO decomposition of $W_{\textup{NICC}}$ (and indeed no such decomposition exists, as can be checked via SDP).

\bigskip

Our third example, $W_{\textup{CS}} = \ketbra{0}{0}^P \otimes \ketbra{\psi}{\psi}^{A^I} \otimes \kketbra{\id}{\id}^{A^OB^I} \otimes \id^{B^O} + \ketbra{1}{1}^P\otimes \ketbra{\psi}{\psi}^{B^I} \otimes \kketbra{\id}{\id}^{B^OA^I} \otimes \id^{A^O}$ from Eq.~\eqref{eq:CS-example}, can be decomposed in the general QC-CC form of Eq.~\eqref{eq:charact_W_QCCC_decomp} as follows:
\begin{align}
\left\{
    \begin{array}{rclcrcl}
W_{(A,B)} &=& \ketbra{0}{0}^P \otimes \ketbra{\psi}{\psi}^{A^I} \otimes \kketbra{\id}{\id}^{A^OB^I}, && W_{(B,A)} &=& \ketbra{1}{1}^P\otimes \ketbra{\psi}{\psi}^{B^I} \otimes \kketbra{\id}{\id}^{B^OA^I}, \\[2mm]
W_{(A)} &=& \ketbra{0}{0}^P \otimes \ketbra{\psi}{\psi}^{A^I}, && W_{(B)} &=& \ketbra{1}{1}^P\otimes \ketbra{\psi}{\psi}^{B^I}
    \end{array}
\right.
\label{app:dec_WCS}
\end{align}
(and once again with $W_{(k_1,k_2,F)} = W_{(k_1,k_2)} \otimes \id^{A_{k_2}^O}$, since $\HS^F$ is trivial).
Notice that none of the matrices in the above decomposition are valid process matrices; indeed it can be shown via SDP that no decomposition of the form of Eq.~\eqref{eq:charact_W_QCCC_decomp} exists for $W_{\textup{CS}}$ with only valid $W_{(k_1(,k_2))}$'s, which implies that $W_{\textup{CS}}$ is not a QC-NICC.

\subsection{Quantum circuits with quantum control of causal order (QC-QCs)}
\label{app:subsec_QCQC_classes}

We now move to the more in-depth presentation of QC-QCs and of our new subclasses of QC-supFOs and QC-NIQC.

\medskip

QC-QCs were introduced and characterised in Ref.~\cite{wechs21}. The basic idea was to turn the classical control of QC-CCs into a quantum control, so as to allow for coherently controlled superpositions of different causal orders.

To describe such coherent superpositions, it is more convenient to work in the ``pure'' picture of quantum theory: we will deal here with pure quantum states described by ket vectors of the form $\ket{\psi}\in\HS$ (rather than with density matrices $\rho\in\L(\HS)$), that evolve through pure linear operators of the form $V:\HS^X\to\HS^Y$ (rather than through CP maps $\M:\L(\HS^X)\to\L(\HS^Y)$). Correspondingly, we will use the ``pure'' version of the Choi–Jamiołkowski isomorphism and of the link product. Namely, for any given operator $V:\HS^X\to\HS^Y$, we define its Choi vector as the ``double-ket'' vector
\begin{align}
    \kket{V}^{XY} \coloneqq (\id^X\otimes V) \kket{\id}^{XX} = \sum_i \ket{i}^X\otimes V\ket{i}^X \ \in \HS^{XY} \label{eq:pure_Choi_def}
\end{align}
(with $\{\ket{i}^X\}_i$ denoting the computational basis of $\HS^X$, $\kket{\id}^{XX} \coloneqq \sum_i \ket{i}^X\otimes\ket{i}^X$); and for two vectors $\kket{V_1}\in\HS^{XY}$ and $\kket{V_2}\in\HS^{YZ}$ that share a common space $\HS^Y$, we define their pure link product as~\cite{wechs21}
\begin{align}
    \kket{V_1}^{XY} * \kket{V_2}^{YZ} \coloneqq & \big( \id^{XZ} \otimes \bbra{\id}^{YY} \big) \big(\kket{V_1}^{XY} \otimes \kket{V_2}^{YZ}\big) = \big( (\kket{V_1}^{XY})^{T_Y} \otimes \id^Z \big) \kket{V_2}^{YZ} \quad \in \HS^{XZ}. \label{eq:def_pure_link_product}
\end{align}
The pure link product has the same commutativity and associativity properties as in the mixed state, and similarly reduces to a tensor product if $\HS^Y$ is trivial, or to an inner product if $\HS^X$ and $\HS^Z$ are trivial. 
As in the mixed case, when $\kket{V_1}^{XY}$ and $\kket{V_2}^{YZ}$ are the Choi vectors of two linear operators $V_1:\HS^X\to\HS^Y$ and $V_2:\HS^Y\to\HS^Z$, then their link product is the Choi vector of their composition (or simply their product, here). 

\subsubsection{General QC-QCs}
\label{app:introQCQC}

\paragraph*{Construction of a QC-QC.}

\begin{figure*}[t]
\centering
\includegraphics[width=1\columnwidth]{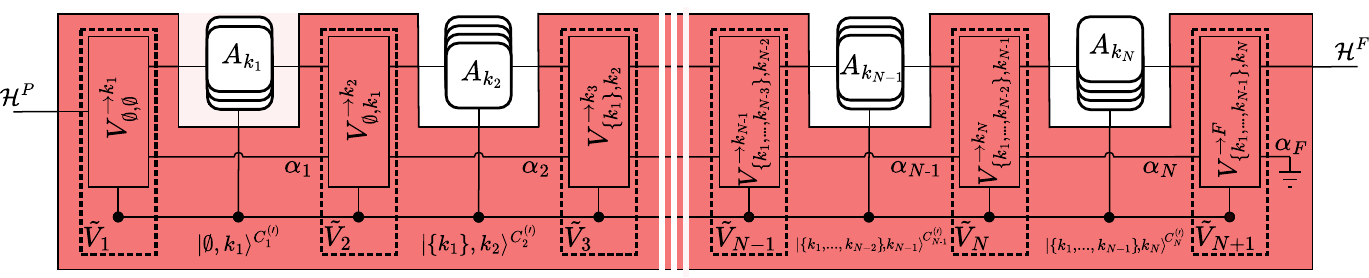}
\caption{Graphical representation of a QC-QC: a more detailed version than Fig.~\ref{fig:QCQC_graph} (clarifying here the internal structure of the circuit operations $\tilde{V}_n$), reproduced from Fig.~10 of Ref.~\cite{wechs21}.}
\label{fig:QCQC_graph_full}
\end{figure*}

The general construction of a QC-QC, depicted in Fig.~\ref{fig:QCQC_graph} of the main text or with more details in Fig.~\ref{fig:QCQC_graph_full} here, is quite similar to that of a QC-CC; we will simply highlight here the essential differences, referring again to Ref.~\cite{wechs21} for more details.

As just mentioned, here we work in the ``pure'' picture, so we now describe both internal and external operations as linear operators $\tilde V_1 : \HS^P \to \HS^{\tilde A_1^I \alpha_1C_1}$, $\tilde V_{n+1} : \HS^{\tilde A_n^O \alpha_nC_n'} \to \HS^{\tilde A_{n+1}^I \alpha_{n+1}C_{n+1}}$ (for $n=1,\ldots,N-1$), $\tilde V_{N+1} : \HS^{\tilde A_N^O \alpha_NC_N'} \to \HS^{F\alpha_F}$ (instead of $\tilde \M_{n+1} : \L(\HS^{\tilde A_n^O \alpha_nC_n'}) \to \L(\HS^{\tilde A_{n+1}^I \alpha_{n+1}C_{n+1}})$ etc.), and $A_k: \HS^{A_k^I}\to\HS^{A_k^O}$ (instead of $\A_k:\L(\HS^{A_k^I})\to\L(\HS^{A_k^O})$).%
\footnote{We again introduce ``generic'' input (respectively output) spaces $\HS^{\tilde A_n^I}$ (resp. $\HS^{\tilde A_n^O}$) for the external operations to be applied at time-slot $t_n$, assuming that all their input (resp. output) spaces are isomorphic. \\
Notice also that we introduce an additional auxiliary system $\alpha_F$ in the global future. This system will ultimately be traced out (see Fig.~\ref{fig:QCQC_graph}) so as to still allow for some mixtures in the general definition of QC-QCs. \label{ftn:abuse_notation} \\
And we again allow for a slight conflict of notation with $A_k$. Here in the pure picture, $A_k: \HS^{A_k^I}\to\HS^{A_k^O}$ (with Choi vector $\kket{A_k}\in \HS^{A_k^{IO}}$) can be understood as a Kraus operator of the operation applied by party $A_k$ (rather than the Choi matrix of their CP map, as we wrote in the context of QC-CCs).}

The structure of the control system is somewhat different here. While in the QC-CC case the (effectively classical) control states were keeping track of the whole order of the previous operations, for QC-QCs the control system at each time-slot $t_n$ only encodes the unordered set $\K_{n-1}$ of the $n-1$ previous operations, as well as the label $k_n (\notin\K_{n-1})$ of the current operation. Thus, the (orthogonal) basis states of the control system take the form
\begin{align}
    \ket{\K_{n-1},k_n}^{C_n^{(\prime)}} \label{eq:def_quantum_Cn}
\end{align}
(instead of $[\![(k_1,\ldots,k_n)]\!]^{C_n^{(\prime)}} \coloneqq \ketbra{(k_1,\ldots,k_n)}{(k_1,\ldots,k_n)}^{C_n^{(\prime)}}$ as in Eq.~\eqref{eq:def_classical_Cn}).
Such states control which external operation $A_{k_n}$ is to be applied at time-slot $t_n$ by embedding these into the larger controlled operation
\begin{align}
\tilde A_n \coloneqq & \sum_{\K_{n-1},k_n} \tilde A_{k_n} \otimes \ket{\K_{n-1},k_n}^{C_n'}\bra{\K_{n-1},k_n}^{C_n}: \ \HS^{\tilde A_n^I C_n} \to \HS^{\tilde A_n^O C_n'} \label{eq:tilde_An_QCQC}
\end{align}
(instead of Eq.~\eqref{eq:tilde_An_QCCC}).
They similarly control the internal operations through%
\footnote{We use the short-hand notations $\K_{n-1}\cup k_n=\K_{n-1}\cup \{k_n\}$ and (as used before already) ${\cal K}_n \backslash k_n={\cal K}_n \backslash\{k_n\}$.}
\begin{align}
\tilde V_1 & \coloneqq \sum_{k_1} \tilde V_{\emptyset,\emptyset}^{\to k_1} \otimes \ket{\emptyset,k_1}^{C_1} : && \HS^P \to \HS^{\tilde A_1^I \alpha_1C_1}, \label{eq:def_V1_QCQC} \\
\tilde V_{n+1} & \coloneqq \sum_{\substack{\K_{n-1}, \\ k_n,k_{n+1}}} \tilde V_{\K_{n-1},k_n}^{\to k_{n+1}} \otimes \ket{\K_{n-1}\cup k_n,k_{n+1}}^{C_{n+1}}\bra{\K_{n-1},k_n}^{C_n'}: \hspace{-5mm} && \HS^{\tilde A_n^O \alpha_nC_n'} \to \HS^{\tilde A_{n+1}^I \alpha_{n+1}C_{n+1}}, \label{eq:tilde_Vn1_QCQC} \\
\tilde V_{N+1} & \coloneqq \sum_{k_N} \tilde V_{\N\backslash k_N,k_N}^{\to F} \otimes \bra{\N\backslash k_N,k_N}^{C_N'}: && \HS^{\tilde A_N^O \alpha_NC_N'} \to \HS^{F\alpha_F} \label{eq:def_VN1_QCQC}
\end{align}
(instead of Eqs.~\eqref{eq:tilde_Mn1_QCCC}--\eqref{eq:tilde_MN1_QCCC}), where the operators $\tilde V_{\emptyset,\emptyset}^{\to k_1}$, $\tilde V_{\K_{n-1},k_n}^{\to k_{n+1}}$ and $\tilde V_{\N\backslash k_N,k_N}^{\to F}$ act on the target and auxiliary memory systems in an analogous (now coherently) controlled manner to $\tilde \M_{\emptyset}^{\to k_1}$, $\tilde \M_{(k_1,\ldots,k_n)}^{\to k_{n+1}}$ and $\tilde \M_{(k_1,\ldots,k_N)}^{\to F}$ in the QC-CC case.
Notice in particular how the specific position of $k_n$ gets erased in Eq.~\eqref{eq:tilde_Vn1_QCQC}, by merging (in the control state) $\K_{n-1}$ and $k_n$ into the unordered set $\K_{n-1}\cup k_n$; forgetting about the specific order of the past operations will allow different orders to interfere.

The state of all systems at any point in the circuit is simply obtained by composing the above internal and external operations---which is conveniently done in terms of their Choi vectors, using the pure link product. Inputting some initial state $\ket{\psi}\in\HS^P$ in the global past of the circuit, and defining (for any subset $\K_n$ of $\N$) the vectors $\ket{\psi,A_{\K_n}}\coloneqq\ket{\psi}\bigotimes_{k\in\K_n}\kket{A_k}\in\HS^{PA_{\K_n}^{IO}}$, the states $\ket{\varphi'_{(n)}}$ before and $\ket{\varphi_{(n+1)}}$ after each operation $\tilde{V}_{n+1}$ are found to be
\begin{align}
\ket{\varphi_{(0)}'} & = \ket{\psi} \quad \in \HS^P, \label{eq:varphi0} \\[2mm]
\ket{\varphi_{(n)}'} & = \ket{\psi} * \kket{\tilde V_1} * \kket{\tilde A_1} * \kket{\tilde V_2} * \cdots * \kket{\tilde V_n} * \kket{\tilde A_n} \notag \\[1mm]
& = \sum_{(k_1,\ldots,k_n)} \big(\ket{\psi} \otimes \kket{A_{k_1}} \otimes \cdots \otimes \kket{A_{k_n}} \big) * \big( \kket{V_{\emptyset,\emptyset}^{\to k_1}} * \kket{V_{\emptyset,k_1}^{\to k_2}} * \cdots * \kket{V_{\{k_1,\ldots,k_{n-2}\},k_{n-1}}^{\to k_n}} * \kket{\id}^{A^O_{k_n}\tilde A^O_n} \big) \notag \\[-3mm]
& \hspace{125mm} \otimes \ket{\{k_1,\ldots,k_{n-1}\},k_n}^{C_n'} \notag \\[1mm]
& = \sum_{\K_n, \, k_n \in \K_n} \big( \ket{\psi,A_{\K_n}} * \ket{w_{(\K_n\backslash k_n,k_n)}} * \kket{\id}^{A^O_{k_n}\tilde A^O_n} \big) \otimes \ket{\K_n\backslash k_n,k_n}^{C_n'} \quad \in \HS^{\tilde A_n^O\alpha_n C_n'} \quad \text{ for } n\ge 1 \label{eq:link_prod_psi_tilde_V1_tilde_An}
\end{align}
and
\begin{align}
\ket{\varphi_{(n+1)}} & = \ket{\psi} * \kket{\tilde V_1} * \kket{\tilde A_1} * \cdots * \kket{\tilde V_n} * \kket{\tilde A_n} * \kket{\tilde V_{n+1}} \notag \\[1mm]
& = \sum_{(k_1,\ldots,k_n,k_{n+1})} \big(\ket{\psi} \otimes \kket{A_{k_1}} \otimes \cdots \otimes \kket{A_{k_n}} \big) * \big( \kket{V_{\emptyset,\emptyset}^{\to k_1}} * \cdots \notag \\[-3mm]
& \hspace{30mm} \cdots * \kket{V_{\{k_1,\ldots,k_{n-2}\},k_{n-1}}^{\to k_n}} * \kket{V_{\{k_1,\ldots,k_{n-1}\},k_n}^{\to k_{n+1}}} * \kket{\id}^{A^I_{k_{n+1}}\tilde A^I_{n+1}} \big) \otimes \ket{\{k_1,\ldots,k_n\},k_{n+1}}^{C_{n+1}} \notag \\[2mm]
& = \sum_{\K_n, \,k_{n+1} \notin \K_n} \big( \ket{\psi,A_{\K_n}} * \ket{w_{(\K_n,k_{n+1})}} * \kket{\id}^{A^I_{k_{n+1}}\tilde A^I_{n+1}} \big) \otimes \ket{\K_n,k_{n+1}}^{C_{n+1}} \notag \\[-3mm]
& \hspace{90mm} \in \HS^{\tilde A_{n+1}^I\alpha_{n+1} C_{n+1}} \quad \text{ for } n\le N-1, \label{eq:link_prod_psi_tilde_V1_tilde_Vn1} \\[3mm]
\ket{\varphi_{(N+1)}} & = \ket{\psi,A_\N} * \ket{w_{(\N,F)}} \quad \in \HS^{F\alpha_F} \label{eq:link_prod_psi_tilde_V1_tilde_VN1}
\end{align}
(instead of Eqs.~\eqref{eq:link_prod_rho_tilde_0}--\eqref{eq:link_prod_rho_tilde_M1_tilde_MN1}), with 
\begin{align}
\ket{w_{(\K_{n-1},k_n)}} & \coloneqq \sum_{(k_1,\ldots,k_{n-1})\in\K_{n-1}} \kket{V_{\emptyset,\emptyset}^{\to k_1}} * \kket{V_{\emptyset,k_1}^{\to k_2}} * \cdots * \kket{V_{\{k_1,\ldots,k_{n-2}\},k_{n-1}}^{\to k_n}} \quad \in \HS^{PA_{\K_{n-1}}^{IO} A_{k_n}^I\alpha_n}, \label{eq:def_w_Knm1_kn} \\[2mm]
\ket{w_{(\N,F)}} & \coloneqq \sum_{(k_1,\ldots,k_N)} \kket{V_{\emptyset,\emptyset}^{\to k_1}} * \kket{V_{\emptyset,k_1}^{\to k_2}} * \cdots * \kket{V_{\{k_1,\ldots,k_{N-2}\},k_{N-1}}^{\to k_N}} * \kket{V_{\{k_1,\ldots,k_{N-1}\},k_N}^{\to F}} \quad \in \HS^{PA_{\N}^{IO} F\alpha_F}. \label{eq:def_w_KN_F}
\end{align}

\bigskip

\paragraph*{Trace-preserving conditions.}

The TP conditions correspond here to the requirement that all internal operations $\tilde V_{n+1}$ preserve the norm of their input state $\ket{\varphi_{(n)}'}$: they must act as isometries on their effective input space.
The squared norms of $\ket{\varphi_{(n)}'}$ and $\ket{\varphi_{(n+1)}}$ (for $1\leq n\leq N-1$) can be obtained from Eqs.~\eqref{eq:link_prod_psi_tilde_V1_tilde_An}--\eqref{eq:link_prod_psi_tilde_V1_tilde_Vn1} above as
\begin{align}
\braket*{\varphi_{(n)}'}{\varphi_{(n)}'} & = \sum_{\K_n, \, k_n \in \K_n} \Tr\Big[\ketbra{\psi,A_{\K_n}}{\psi,A_{\K_n}} * \ketbra{w_{(\K_n\backslash k_n,k_n)}}{w_{(\K_n\backslash k_n,k_n)}}\Big] \notag \\
& = \sum_{\K_n} \Tr\Big[\big(\ketbra{\psi,A_{\K_n}}{\psi,A_{\K_n}}\big)^T \Big(\sum_{k_n\in\K_n}\Tr_{\alpha_n}\ketbra{w_{(\K_n\backslash k_n,k_n)}}{w_{(\K_n\backslash k_n,k_n)}} \otimes\id^{A_{k_n}^O}\Big)\Big], \\[2mm]
\braket{\varphi_{(n+1)}}{\varphi_{(n+1)}} & = \sum_{\K_n, \, k_{n+1} \notin \K_n} \Tr\Big[\ketbra{\psi,A_{\K_n}}{\psi,A_{\K_n}} * \ketbra{w_{(\K_n,k_{n+1})}}{w_{(\K_n,k_{n+1})}}\Big] \notag \\
& = \sum_{\K_n} \Tr\Big[\big(\ketbra{\psi,A_{\K_n}}{\psi,A_{\K_n}}\big)^T \Big(\sum_{k_{n+1}\notin\K_n}\Tr_{A_{k_{n+1}}^I\alpha_{n+1}}\ketbra{w_{(\K_n,k_{n+1})}}{w_{(\K_n,k_{n+1})}} \Big)\Big].
\end{align}
Equality between these two expressions, for any possible choice of $\ket{\psi}$ and of the $A_k$'s, implies that the last terms in round brackets must themselves be equal, for any $\K_n$:
\begin{align}
\sum_{k_{n+1}\notin\K_n}\Tr_{A_{k_{n+1}}^I\alpha_{n+1}}\ketbra{w_{(\K_n,k_{n+1})}}{w_{(\K_n,k_{n+1})}} = \sum_{k_n\in\K_n}\Tr_{\alpha_n}\ketbra{w_{(\K_n\backslash k_n,k_n)}}{w_{(\K_n\backslash k_n,k_n)}} \otimes\id^{A_{k_n}^O}. \label{eq:QCQC_TP_n}
\end{align}
For the extremal cases $n=0$ and $n=N$, the corresponding constraints are
\begin{align}
& \sum_{k_1\in\N}\Tr_{A_{k_1}^I\alpha_1}\ketbra{w_{(\emptyset,k_1)}}{w_{(\emptyset,k_1)}} = \id^P \label{eq:QCQC_TP_0} \\
\text{and} \qquad & \Tr_{F\alpha_F}\ketbra{w_{(\N,F)}}{w_{(\N,F)}} = \sum_{k_N\in\N}\Tr_{\alpha_N}\ketbra{w_{(\N\backslash k_N,k_N)}}{w_{(\N\backslash k_N,k_N)}} \otimes\id^{A_{k_N}^O}. \label{eq:QCQC_TP_N}
\end{align}

\bigskip

\paragraph*{Process matrix description and characterisation.}

From the expression for $\ket{\varphi_{(N+1)}}$ in Eq.~\eqref{eq:link_prod_psi_tilde_V1_tilde_VN1}, one gets the (in general, mixed) global output state of the whole QC-QC, after tracing out $\alpha_F$:
\begin{align}
    \varrho_{N+1} = \Tr_{\alpha_F} \ketbra{\varphi_{(N+1)}}{\varphi_{(N+1)}} = \ketbra{\psi,A_\N}{\psi,A_\N} * \big(\Tr_{\alpha_F} \ketbra{w_{(\N,F)}}{w_{(\N,F)}}\big) \quad \in \L(\HS^F).
\end{align}
With this, one can readily identify the process matrix that describes the QC-QC under consideration (as we did in the QC-CC case), namely
\begin{align}
    W = \Tr_{\alpha_F} \ketbra{w_{(\N,F)}}{w_{(\N,F)}} \quad \in \L(\HS^{PA_\N^{IO}F}), \label{eq:W_QCQC}
\end{align}
with $\ket{w_{(\N,F)}}$ defined in Eq.~\eqref{eq:def_w_KN_F} above, as a coherent superposition of terms corresponding to different causal orders.

\medskip

Defining, for all $n=1,\ldots,N$, all strict subsets $\K_{n-1}$ of $\N$ and all $k_n\in\N\backslash\K_{n-1}$,
\begin{align}
    W_{({\cal K}_{n-1},k_n)} & \coloneqq \Tr_{\alpha_n}\ketbra{w_{(\K_{n-1},k_n)}}{w_{(\K_{n-1},k_n)}} \quad \in {\cal L}(\HS^{PA_{{\cal K}_{n-1}}^{IO} A_{k_n}^I}), \label{eq:def_W_Kn1_kn}
\end{align}
it then follows directly from Eq.~\eqref{eq:W_QCQC} and the TP conditions of Eqs.~\eqref{eq:QCQC_TP_n}--\eqref{eq:QCQC_TP_N} that the process matrix $W \, (\ge 0)$ of a QC-QC has a characterisation as in Eq.~\eqref{eq:charact_W_QCQC_decomp}.

Conversely, from any decomposition of a process matrix $W$ in terms of PSD matrices $W_{({\cal K}_{n-1},k_n)}$ satisfying Eq.~\eqref{eq:charact_W_QCQC_decomp}, one can explicitly reconstruct (in a non-unique manner) the internal operations $V_{\K_{n-1},k_n}^{\to k_{n+1}}$ and $V_{\N\setminus k_N, k_N}^{\to F}$ of a QC-QC (satisfying the required TP conditions), whose composition reproduces $\ket{w_{(\N,F)}}$ and, subsequently, $W$ as in Eqs.~\eqref{eq:def_w_KN_F} and~\eqref{eq:W_QCQC} above.
This shows that Eq.~\eqref{eq:charact_W_QCQC_decomp} is also a sufficient condition for $W$ to be a QC-QC. An explicit procedure for such a reconstruction was detailed in Appendix~B\,3\,c of~\cite{wechs21}.

\subsubsection{Quantum circuits with superposition of fixed causal orders (QC-supFO)}
\label{app:def_QC-supFO}

As succinctly introduced in the main text, we define the class of quantum circuits with superposition of fixed causal orders as a modification of the general QC-QC construction above, in which the control system at any time-slot $t_n$ now encodes (in addition to the past and current operations) the full causal order $(k_{n+1},\ldots,k_N)$ of the upcoming external operations that are still to be applied. The weight of each total order $\pi=(k_1,\ldots,k_N)$ is fixed from the very beginning to some value $q_\pi$ (such that $q_\pi\ge 0$, $\sum_\pi q_\pi=1$). When updating the control system, the internal operations may again erase the order of previous operations, but they should not modify the weight of each remaining ``future'' order $(k_{n+2},\ldots,k_N)$.

The construction, the derivation of the TP conditions, the process matrix description and the characterisation of QC-supFOs align quite closely to the case of general QC-QCs recalled above. We will just point out the differences.

\bigskip

\paragraph*{Construction of a QC-supFO.}

As just said, the control system should now also encode the order of future operations; we thus take its (orthogonal) basis states to be of the form
\begin{align}
    \ket{\K_{n-1},k_n;(k_{n+1},\ldots,k_N)}^{C_n^{(\prime)}}, \label{eq:def_quantum_Cn_supFO}
\end{align}
with $k_n,k_{n+1},\ldots,k_N\not\in\K_{n-1}$, instead of just $\ket{\K_{n-1},k_n}^{C_n^{(\prime)}}$ as in Eq.~\eqref{eq:def_quantum_Cn}.

These states control the external and internal operations through
\begin{align}
\tilde A_n \coloneqq & \sum_{\substack{\K_{n-1},\\ (k_n,k_{n+1},\ldots,k_N)}} \tilde A_{k_n} \otimes \ket{\K_{n-1},k_n;(k_{n+1},\ldots,k_N)}^{C_n'}\bra{\K_{n-1},k_n;(k_{n+1},\ldots,k_N)}^{C_n} \label{eq:tilde_An_QC-supFO}
\end{align}
and
\begin{align}
\tilde V_1 & \coloneqq \sum_{(k_1,k_2,\ldots,k_N)} \tilde V_{\emptyset,\emptyset}^{\to k_1;(k_2,\ldots,k_N)} \otimes \ket{\emptyset,k_1;(k_2,\ldots,k_N)}^{C_1}, \label{eq:def_V1_QC-supFO} \\[1mm]
\tilde V_{n+1} & \coloneqq \!\!\!\sum_{\substack{\K_{n-1}, \\ (k_n,k_{n+1},k_{n+2},\ldots,k_N)}} \!\!\!\!\!\!\!\!\! \tilde V_{\K_{n-1},k_n}^{\to k_{n+1};(k_{n+2},\ldots,k_N)} \otimes \ket{\K_{n-1}\cup k_n,k_{n+1};(k_{n+2},\ldots,k_N)}^{C_{n+1}}\bra{\K_{n-1},k_n;(k_{n+1},k_{n+2},\ldots,k_N)}^{C_n'}, \label{eq:tilde_Vn1_QC-supFO} \\
\tilde V_{N+1} & \coloneqq \ \sum_{k_N} \ \tilde V_{\N\backslash k_N,k_N}^{\to F} \otimes \bra{\N\backslash k_N,k_N;\emptyset}^{C_N'}, \label{eq:def_VN1_QC-supFO}
\end{align}
instead of Eqs.~\eqref{eq:tilde_An_QCQC}--\eqref{eq:def_VN1_QCQC} for the general QC-QC case. Notice that the operators $\tilde V_{\emptyset,\emptyset}^{\to k_1;(k_2,\ldots,k_N)}$ and $\tilde V_{\K_{n-1},k_n}^{\to k_{n+1};(k_{n+2},\ldots,k_N)}$ that act on the target and auxiliary memory systems can depend in general on the order of future operations, as encoded in the system $C_{n(+1)}^{(\prime)}$ that controls them.

Composing the above internal and external operations, we obtain the state of all systems before and after each operation $\tilde{V}_{n+1}$:
\begin{align}
\ket{\varphi_{(0)}'} & = \ket{\psi}, \label{eq:varphi0_supFO} \\[2mm]
\ket{\varphi_{(n)}'} & = \ket{\psi} * \kket{\tilde V_1} * \kket{\tilde A_1} * \kket{\tilde V_2} * \cdots * \kket{\tilde V_n} * \kket{\tilde A_n} \notag \\[1mm]
& = \sum_{(k_1,\ldots,k_N)} \big(\ket{\psi} \otimes \kket{A_{k_1}} \otimes \cdots \otimes \kket{A_{k_n}} \big) \notag \\[-4mm]
& \hspace{30mm} * \big( \kket{V_{\emptyset,\emptyset}^{\to k_1;(k_2,\ldots,k_N)}} * \kket{V_{\emptyset,k_1}^{\to k_2;(k_3,\ldots,k_N)}} * \cdots * \kket{V_{\{k_1,\ldots,k_{n-2}\},k_{n-1}}^{\to k_n;(k_{n+1},\ldots,k_N)}} * \kket{\id}^{A^O_{k_n}\tilde A^O_n} \big) \notag \\
& \hspace{100mm} \otimes \ket{\{k_1,\ldots,k_{n-1}\},k_n;(k_{n+1},\ldots,k_N)}^{C_n'} \notag \\[1mm]
& = \sum_{\substack{\K_n, \, k_n \in \K_n,\\ (k_{n+1},\ldots,k_N)\not\in\K_n}} \big( \ket{\psi,A_{\K_n}} * \ket{w_{(\K_n\backslash k_n,k_n)}^{(k_{n+1},\ldots,k_N)}} * \kket{\id}^{A^O_{k_n}\tilde A^O_n} \big) \otimes \ket{\K_n\backslash k_n,k_n;(k_{n+1},\ldots,k_N)}^{C_n'} \label{eq:link_prod_psi_tilde_V1_tilde_An_supFO}
\end{align}
and
\begin{align}
\ket{\varphi_{(n+1)}} & = \ket{\psi} * \kket{\tilde V_1} * \kket{\tilde A_1} * \cdots * \kket{\tilde V_n} * \kket{\tilde A_n} * \kket{\tilde V_{n+1}} \notag \\[1mm]
& = \sum_{(k_1,\ldots,k_N)} \big(\ket{\psi} \otimes \kket{A_{k_1}} \otimes \cdots \otimes \kket{A_{k_n}} \big) \notag \\[-4mm]
& \hspace{30mm}  * \big( \kket{V_{\emptyset,\emptyset}^{\to k_1;(k_2,\ldots,k_N)}} * \cdots * \kket{V_{\{k_1,\ldots,k_{n-2}\},k_{n-1}}^{\to k_n;(k_{n+1},\ldots,k_N)}} * \kket{V_{\{k_1,\ldots,k_{n-1}\},k_n}^{\to k_{n+1};(k_{n+2},\ldots,k_N)}} * \kket{\id}^{A^I_{k_{n+1}}\tilde A^I_{n+1}} \big) \notag \\
& \hspace{95mm} \otimes \ket{\{k_1,\ldots,k_n\},k_{n+1};(k_{n+2},\ldots,k_N)}^{C_{n+1}} \notag \\[2mm]
& = \sum_{\substack{\K_n,\\ (k_{n+1},k_{n+2},\ldots,k_N)\not\in\K_n}} \big( \ket{\psi,A_{\K_n}} * \ket{w_{(\K_n,k_{n+1})}^{(k_{n+2},\ldots,k_N)}} * \kket{\id}^{A^I_{k_{n+1}}\tilde A^I_{n+1}} \big) \otimes \ket{\K_n,k_{n+1};(k_{n+2},\ldots,k_N)}^{C_{n+1}}, \label{eq:link_prod_psi_tilde_V1_tilde_Vn1_supFO} \\[3mm]
\ket{\varphi_{(N+1)}} & = \ket{\psi,A_\N} * \ket{w_{(\N,F)}}
\label{eq:link_prod_psi_tilde_V1_tilde_VN1_supFO}
\end{align}
with
\begin{align}
\ket{w_{(\K_{n-1},k_n)}^{(k_{n+1},\ldots,k_N)}} & \coloneqq \sum_{(k_1,\ldots,k_{n-1})\in\K_{n-1}} \kket{V_{\emptyset,\emptyset}^{\to k_1;(k_2,\ldots,k_N)}} * \kket{V_{\emptyset,k_1}^{\to k_2;(k_3,\ldots,k_N)}} * \cdots * \kket{V_{\{k_1,\ldots,k_{n-2}\},k_{n-1}}^{\to k_n;(k_{n+1},\ldots,k_N)}}, \label{eq:def_w_Knm1_kn_supFO} \\[2mm]
\ket{w_{(\N,F)}} & \coloneqq \sum_{(k_1,\ldots,k_N)} \kket{V_{\emptyset,\emptyset}^{\to k_1;(k_2,\ldots,k_N)}} * \kket{V_{\emptyset,k_1}^{\to k_2;(k_3,\ldots,k_N)}} * \cdots * \kket{V_{\{k_1,\ldots,k_{N-2}\},k_{N-1}}^{\to k_N;\emptyset}} * \kket{V_{\{k_1,\ldots,k_{N-1}\},k_N}^{\to F}} \label{eq:def_w_KN_F_supFO}
\end{align}
instead of Eqs.~\eqref{eq:varphi0}--\eqref{eq:def_w_KN_F}.

\bigskip

\paragraph*{Trace-preserving conditions.} 

As specified above, the weight of each total order $\pi=(k_1,\ldots,k_N)$ of all parties, which is encoded in the state $\ket{\emptyset,k_1;(k_2,\ldots,k_N)}^{C_1}$ of the control system right after $\tilde V_1$, is fixed to some predefined value $q_\pi$. This weight---which can be interpreted as the probability that the control system is found in the state corresponding to that order (if it were measured in the appropriate basis)---can be obtained from the expression of $\varphi_{(1)}$ in Eq.~\eqref{eq:link_prod_psi_tilde_V1_tilde_Vn1_supFO}:
\begin{align}
    & \Big|\big(\id^{\tilde A_1^I\alpha_1}\otimes\bra{\emptyset,k_1;(k_2,\ldots,k_N)}^{C_1}\big)\ket{\varphi_{(1)}}\Big|^2 = \Big|\ket{\psi} * \ket{w_{(\emptyset,k_1)}^{(k_2,\ldots,k_N)}}\Big|^2 \notag \\
    & \hspace{30mm} = \Tr\Big[\big(\ketbra{\psi}{\psi}\big)^T \Big(\Tr_{A_{k_1}^I\alpha_1}\ketbra{w_{(\emptyset,k_1)}^{(k_2,\ldots,k_N)}}{w_{(\emptyset,k_1)}^{(k_2,\ldots,k_N)}}\Big)\Big] = q_{\pi=(k_1,\ldots,k_N)}.
\end{align}
Imposing that this equality holds for all possible (normalised) input states $\ket{\psi}\in\HS^P$ requires that
\begin{align}
& \Tr_{A_{k_1}^I\alpha_1}\ketbra{w_{(\emptyset,k_1)}^{(k_2,\ldots,k_N)}}{w_{(\emptyset,k_1)}^{(k_2,\ldots,k_N)}} = q_{\pi=(k_1,\ldots,k_N)} \,\id^P, \label{eq:QCQC_TP_0_supFO}
\end{align}
instead of the TP constraint of Eq.~\eqref{eq:QCQC_TP_0} in the general QC-QC case.

Additionally, we impose that the internal operations $\tilde V_{n+1}$ preserve the weight of each future order $(k_{n+1},\ldots,k_N)$.
The total weight attached to that order before $\tilde V_{n+1}$ is obtained as the sum over $k_n$ of the weights corresponding to $(k_n,k_{n+1},\ldots,k_N)$ in the state $\ket{\varphi_{(n)}'}$. Namely, for $\K_n=\N\backslash\{k_{n+1},\ldots,k_N\}$ (using Eq.~\eqref{eq:link_prod_psi_tilde_V1_tilde_An_supFO}):
\begin{align}
    & \sum_{k_n\in\K_n} \Big|\big(\id^{\tilde A_n^O\alpha_n}\otimes\bra{\K_n\backslash k_n,k_n;(k_{n+1},\ldots,k_N)}^{C_n'}\big)\ket{\varphi_{(n)}'}\Big|^2 = \sum_{k_n\in\K_n} \Big|\ket{\psi,A_{\K_n}} * \ket{w_{(\K_n\backslash k_n,k_n)}^{(k_{n+1},\ldots,k_N)}}\Big|^2 \notag \\
    & \hspace{40mm} = \Tr\Big[\big(\ketbra{\psi,A_{\K_n}}{\psi,A_{\K_n}}\big)^T \Big(\sum_{k_n\in\K_n}\Tr_{\alpha_n}\ketbra{w_{(\K_n\backslash k_n,k_n)}^{(k_{n+1},\ldots,k_N)}}{w_{(\K_n\backslash k_n,k_n)}^{(k_{n+1},\ldots,k_N)}} \otimes\id^{A_{k_n}^O}\Big)\Big].
\end{align}
The weight of $(k_{n+1},\ldots,k_N)$ after $\tilde V_{n+1}$ on the other hand is simply (using Eq.~\eqref{eq:link_prod_psi_tilde_V1_tilde_Vn1_supFO})
\begin{align}
    & \Big|\big(\id^{\tilde A_{n+1}^I\alpha_{n+1}}\otimes\bra{\K_n,k_{n+1};(k_{n+2},\ldots,k_N)}^{C_{n+1}}\big)\ket{\varphi_{(n+1)}}\Big|^2 = \Big|\ket{\psi,A_{\K_n}} * \ket{w_{(\K_n,k_{n+1})}^{(k_{n+2},\ldots,k_N)}}\Big|^2 \notag \\
    & \hspace{40mm} = \Tr\Big[\big(\ketbra{\psi,A_{\K_n}}{\psi,A_{\K_n}}\big)^T \Big(\Tr_{A_{k_{n+1}}^I\alpha_{n+1}}\ketbra{w_{(\K_n,k_{n+1})}^{(k_{n+2},\ldots,k_N)}}{w_{(\K_n,k_{n+1})}^{(k_{n+2},\ldots,k_N)}}\Big)\Big].
\end{align}
Imposing that these two expressions are equal, for any $\ket{\psi}$ and any $A_{\K_n}$, implies that one must have (for any $\K_n$, any $k_{n+1},\ldots,k_N\notin\K_n$)
\begin{align}
    \Tr_{A_{k_{n+1}}^I\alpha_{n+1}}\ketbra{w_{(\K_n,k_{n+1})}^{(k_{n+2},\ldots,k_N)}}{w_{(\K_n,k_{n+1})}^{(k_{n+2},\ldots,k_N)}} = \sum_{k_n\in\K_n}\Tr_{\alpha_n}\ketbra{w_{(\K_n\backslash k_n,k_n)}^{(k_{n+1},\ldots,k_N)}}{w_{(\K_n\backslash k_n,k_n)}^{(k_{n+1},\ldots,k_N)}} \otimes\id^{A_{k_n}^O}, \label{eq:QCQC_TP_n_supFO}
\end{align}
instead of Eq.~\eqref{eq:QCQC_TP_n}.
For the final case of $n=N$, the corresponding constraint is
\begin{align}
\Tr_{F\alpha_F}\ketbra{w_{(\N,F)}}{w_{(\N,F)}} = \sum_{k_N\in\N}\Tr_{\alpha_N}\ketbra{w_{(\N\backslash k_N,k_N)}^\emptyset}{w_{(\N\backslash k_N,k_N)}^\emptyset} \otimes\id^{A_{k_N}^O}. \label{eq:QCQC_TP_N_supFO}
\end{align}
instead of Eq.~\eqref{eq:QCQC_TP_N}.

\bigskip

\paragraph*{Process matrix description and characterisation.}

As in the general QC-QC case, from the expression for $\ket{\varphi_{(N+1)}}$ in Eq.~\eqref{eq:link_prod_psi_tilde_V1_tilde_VN1_supFO}, one gets the global output state of the whole QC-supFO, after tracing out $\alpha_F$:
\begin{align}
    \varrho_{N+1} = \Tr_{\alpha_F} \ketbra{\varphi_{(N+1)}}{\varphi_{(N+1)}} = \ketbra{\psi,A_\N}{\psi,A_\N} * \big(\Tr_{\alpha_F} \ketbra{w_{(\N,F)}}{w_{(\N,F)}}\big) \quad \in \L(\HS^F).
\end{align}
One can then identify the process matrix that describes the QC-supFO under consideration as
\begin{align}
    W = \Tr_{\alpha_F} \ketbra{w_{(\N,F)}}{w_{(\N,F)}} \quad \in \L(\HS^{PA_\N^{IO}F}), \label{eq:W_QCQC_supFO}
\end{align}
with $\ket{w_{(\N,F)}}$ defined in Eq.~\eqref{eq:def_w_KN_F_supFO} above.

\medskip

Defining, for all $n=1,\ldots,N$, all $\K_{n-1}\subsetneq\N$ and all $(k_n,k_{n+1},\ldots,k_N)\notin\K_{n-1}$,
\begin{align}
    W_{({\cal K}_{n-1},k_n)}^{(k_{n+1},\ldots,k_N)} & \coloneqq \Tr_{\alpha_n}\ketbra{w_{(\K_{n-1},k_n)}^{(k_{n+1},\ldots,k_N)}}{w_{(\K_{n-1},k_n)}^{(k_{n+1},\ldots,k_N)}} \quad \in {\cal L}(\HS^{PA_{{\cal K}_{n-1}}^{IO} A_{k_n}^I}),
    \label{eq:def_submatrices_QCSupFO}
\end{align}
then it follows directly from Eq.~\eqref{eq:W_QCQC_supFO} and the TP conditions of Eqs.~\eqref{eq:QCQC_TP_0_supFO}, \eqref{eq:QCQC_TP_n_supFO} and~\eqref{eq:QCQC_TP_N_supFO} that the process matrix $W \, (\ge 0)$ of a QC-supFO has a characterisation as in Eq.~\eqref{eq:charact_W_QC_SupFO_decomp}.
Notice that this characterisation implies in particular that the $W_{({\cal K}_{n-1},k_n)}^{(k_{n+1},\ldots,k_N)}$'s are valid process matrices, up to normalisation---as was the case for the $W_{(k_1,\ldots,k_n)}^{(k_{n+1},\ldots,k_N)}$'s in the decomposition of a QC-convFO (cf.\ Footnote~\ref{ftn:valid_Wk1_kn_FO}).

Conversely, from any decomposition of a process matrix $W$ in terms of PSD matrices $W_{({\cal K}_{n-1},k_n)}^{(k_{n+1},\ldots,k_N)}$ satisfying Eq.~\eqref{eq:charact_W_QC_SupFO_decomp}, one can explicitly reconstruct (in a non-unique manner) the internal operations $V_{\K_{n-1},k_n}^{\to k_{n+1};(k_{n+2},\ldots,k_N)}$ and $V_{\N\setminus k_N, k_N}^{\to F}$ of a QC-supFO, whose composition (through Eqs.~\eqref{eq:def_w_KN_F} and~\eqref{eq:W_QCQC} above) reproduces the process matrix $W$.
For this, one can indeed follow a very similar procedure to that for QC-QCs, detailed in Ref.~\cite{wechs21}---we will not repeat it explicitly here: it suffices to add the superscripts ${}^{((k_{n+1},)k_{n+2},\ldots,k_N)}$ to all objects introduced in the construction ($\ket{w_{(\K_{n-1},k_n)}^i}$, $\ket{w_{(\K_{n-1},k_n)}}$, $\ket{\omega_{\K_n}}$, $\Omega_{\K_n}$, $\ket{\omega_{\K_n}^+}$, $\ket{V_{\K_n}^{k_{n+1}}}$ and finally $\kket{V_{\K_{n-1},k_n}^{\to k_{n+1}}}$: see Appendix~B\,3\,c of~\cite{wechs21}). The fact that the matrices $W_{({\cal K}_{n-1},k_n)}^{(k_{n+1},\ldots,k_N)}$ satisfy Eq.~\eqref{eq:charact_W_QC_SupFO_decomp} ensures that the internal operations thus constructed satisfy the TP conditions of Eqs.~\eqref{eq:QCQC_TP_0_supFO}, \eqref{eq:QCQC_TP_n_supFO} and~\eqref{eq:QCQC_TP_N_supFO}, as required.
This shows that Eq.~\eqref{eq:charact_W_QC_SupFO_decomp} is also a sufficient condition for $W$ to be the process matrix of a QC-supFO.

\bigskip

\paragraph*{Another perspective: \textup{\textsf{QC-supFO}} as a recursive generalisation of the \textup{\textsf{Sup}} class from Ref.~\cite{Liu23}.}
\label{app:discuss_Sup}

Another subclass of QC-QCs was introduced in Ref.~\cite{Liu23}, which could also have been seen as a potential candidate to define quantum circuits with superpositions of fixed orders---it was indeed simply called \textup{\textsf{Sup}} by the authors of~\cite{Liu23}. We will argue however that its definition has an undesirable feature that should rule it out as such a meaningful definition of a relevant class of circuits, and that by generalising the \textup{\textsf{Sup}} class so as to avoid this unwanted feature, we naturally recover our \textup{\textsf{QC-supFO}} class.

\medskip

Formally, the \textup{\textsf{Sup}} class was defined as follows: $W \in \mathcal{L}(\mathcal{H}^{PA_\mathcal{N}^{IO}F})$ is in \textup{\textsf{Sup}} iff $W \ge 0$, and
\begin{align}
    \left\{
    \begin{array}{l}
        \Tr_F W = \sum_\pi q_\pi \, \tilde{W}_\pi, \\[2mm]
        \forall \, \pi, \tilde{W}_\pi \in \text{\textup{\textsf{QC-FO}}}_\pi, \ q_\pi\ge 0, \, \sum_\pi q_\pi = 1.
    \end{array}
    \right. \label{eq:def_Sup}
\end{align}
where $\text{\textup{\textsf{QC-FO}}}_\pi$ is the class of quantum circuits with fixed causal order $\pi$, with a trivial $\HS^F$ here---i.e., for $\pi=(k_1,\ldots,k_N)$, of PSD matrices $W$ that admit a decomposition as in Eq.~\eqref{eq:charact_W_QC_FO_decomp}, with the first line replaced by just $W = W_{(k_1,\ldots,k_N)} \otimes \id^{A_{k_N}^O}$.
This class was meant to generalise the ``static'' quantum switch, in which the control system is initially prepared in some predefined quantum state corresponding to a superposition of fixed orders and ultimately sent to $\HS^F$ in the global future, so that tracing $\HS^F$ out leaves the rest of the parties in an incoherent mixture of processes with fixed causal order. It can easily be seen from the definition above that $\textup{\textsf{QC-convFO}} \subset \textup{\textsf{Sup}} \subset \textup{\textsf{QC-supFO}}$.

\bigskip

The unwanted feature we identify in the definition of \textup{\textsf{Sup}} above is that it crucially depends on the labelling of $\HS^F$. More specifically, recall first that considering the future space $\HS^F$ as ``open'' or attributing it to an additional party, with no output space, leads to formally equivalent frameworks~\cite{araujo17,wechs21}.
Now, if one were to rename $\HS^F$ and consider it as the input space of a new party, treated on equal footing with the other ones, then the definition of the 
\textup{\textsf{Sup}} class above would change.

Concretely, it can be seen for example that the process matrix of the ``static'' quantum switch, $W_{\textup{supFO}}$ from Eq.~\eqref{eq:QC-supFO-example}, is in \textup{\textsf{Sup}}. However, if one were to attribute $\HS^F$---or even just $\HS^{F_\text{c}}$---to another party (call it $C$) and write $W_{\textup{supFO}}=\ketbra{w_\textup{supFO}}{w_\textup{supFO}}$ with $\ket{w_\textup{supFO}} = {\textstyle\frac{1}{\sqrt{2}}} \ket{\psi}^{A^I} \otimes \kket{\id}^{A^OB^I} \otimes 
\kket{\id}^{B^OF_\text{t}} \otimes \ket{0}^{C^I} + {\textstyle\frac{1}{\sqrt{2}}}\ket{\psi}^{B^I} \otimes \kket{\id}^{B^OA^I} \otimes \kket{\id}^{A^OF_\text{t}} \otimes \ket{1}^{C^I}$ instead of Eq.~\eqref{eq:QC-supFO-example}, then one would conclude that $W_{\textup{supFO}}\notin\text{\textup{\textsf{Sup}}}$---even though the situation is physically exactly the same as the one in which one concludes that $W_{\textup{supFO}}\in\text{\textup{\textsf{Sup}}}$.

This observation leads us to argue that any physically meaningful definition of a class of quantum circuits should not depend on the labellings of the different parties, including (crucially) parties in the global past or future.%
\footnote{Unless the class one wishes to define refers explicitly to a certain labelling of the parties. E.g.\ the class $\text{\textup{\textsf{QC-FO}}}_{\pi=(A,B)}$ of quantum circuits with the fixed causal order $(A,B)$ is of course different from the class $\text{\textup{\textsf{QC-FO}}}_{\pi=(B,A)}$ obtained by swapping the two parties' labels.}
In particular, if two parties have no output space, the class should not depend on whose input space if called $\HS^F$; similarly if two parties have no input space, the class should not depend on whose output space if called $\HS^P$.
One can verify that all of our 6 classes introduced in the main text satisfy this requirement (see Appendix~\ref{app:subsec:P/F-label indep} for a proof sketch). We will now see how one can modify the \textup{\textsf{Sup}} class so as to also satisfy it.

\bigskip

For this, recall that the defining property of \textup{\textsf{Sup}} is that if one traces out $\HS^F$, one should be left with an incoherent mixture of fixed causal orders. If one attributes $\HS^F$ to another party, we would hence like that there is different party in the future, such that if one traces them out, the rest should be in an incoherent mixture of fixed orders.

The key insight we can take from the example of $W_{\textup{supFO}}$ just discussed is that in order to conclude that $W_{\textup{supFO}}\in\text{\textup{\textsf{Sup}}}$, it may not be enough to just trace one future party out; after tracing out one party, one may need to check whether one is left in a situation where another party can be considered to be in the future of the remaining ones, such that tracing it out leaves the other ones in an incoherent mixture of orders---or more generally, whether one is left in an (incoherent) mixture of such situations, with fixed weights. This suggests considering a recursive generalisation of the \textup{\textsf{Sup}} class that would formalise this.

As it turns out, following this idea provides an alternative characterisation of our class \textup{\textsf{QC-supFO}}. More formally: let us add subscripts and write $\textup{\textsf{QC-supFO}}_{P,\N,F}$ to specify the global past Hilbert space $\HS^P$, the set of parties $\N$, and the global future Hilbert space $\HS^F$ considered in the definition of \textup{\textsf{QC-supFO}}. We show below that $\textup{\textsf{QC-supFO}} = \textup{\textsf{QC-supFO}}_{P,\N,F}$ has the following recursive characterisation, which formalises precisely our suggestion above for generalising the \textup{\textsf{Sup}} class: 
\begin{align}
         \textup{\textsf{QC-supFO}}_{P,\N,F} & = 
         \left\{ W\in\L(\HS^{PA_\N^{IO} F})\, \middle|
    \begin{array}{c}
          W\ge 0, \, \Tr_F W = \sum_{k_N\in\N} \, q^{[k_N]}\, \tilde W^{[k_N]}\otimes\id^{A_{k_N}^O}, \\[2mm]
         q^{[k_N]}\ge 0, \ \sum_{k_N} q^{[k_N]} = 1, \\[2mm]
         \tilde W^{[k_N]} \in \textup{\textsf{QC-supFO}}_{P,\N\backslash k_N,A_{k_N}^I}
    \end{array}
    \right\}
    \quad \text{for } |\N|\ge 1, \notag \\[3mm]
    \textup{\textsf{QC-supFO}}_{P,\emptyset,F} & = \Big\{W\in\L(\HS^{PF})\, \Big| \, W\ge 0, \, \Tr_F W = \id^P \Big\}. \label{eq:rec_charact_SupFO}
\end{align}
This characterisation could also be taken as a definition of our class \textup{\textsf{QC-supFO}}, which can therefore be seen, from this perspective, as a natural generalisation of the \textup{\textsf{Sup}} class of Ref.~\cite{Liu23}.
It can indeed be checked that if two parties $F_1$, $F_2$ have no output Hilbert spaces, then $\textup{\textsf{QC-supFO}}_{P,\N\cup F_1,F_2} = \textup{\textsf{QC-supFO}}_{P,\N\cup F_2,F_1}$:%
\footnote{One can show recursively that if $W\in \textup{\textsf{QC-supFO}}_{P,\N\cup F_1,F_2}$, then $\Tr_{F_1}W\in \textup{\textsf{QC-supFO}}_{P,\N,F_2}$. This implies that $\Tr_{F_1}W$ has a decomposition as in the very top line of Eq.~\eqref{eq:rec_charact_SupFO}, with only one term $\tilde{W}^{[F_2]} = \Tr_{F_1}W \in \textup{\textsf{QC-supFO}}_{P,\N,F_2}$, which means that $W\in \textup{\textsf{QC-supFO}}_{P,\N\cup F_2,F_1}$. See also the sketch for a direct, non-recursive proof in Appendix~\ref{app:subsec:P/F-label indep}.}
the above characterisation does not depend on which of $F_1$, $F_2$ is labelled $F$, thus indeed resolving the issue with the \textup{\textsf{Sup}} class that we pinpointed above.

\bigskip

\noindent\underline{Proof of Eq.~\eqref{eq:rec_charact_SupFO}:}
Let us denote by $\textup{\textsf{QC-supFO}}_{P,\N,F}'$ the RHS of the first equation in~\eqref{eq:rec_charact_SupFO} (adding the prime recursively, to $\textup{\textsf{QC-supFO}}_{P,\N\backslash k_N,A_{k_N}^I}$ inside the curly bracket and to $\textup{\textsf{QC-supFO}}_{P,\emptyset,F}$ on the second line of~\eqref{eq:rec_charact_SupFO} as well)---so that our task here is to prove that $\textup{\textsf{QC-supFO}}_{P,\N,F}' = \textup{\textsf{QC-supFO}}_{P,\N,F}$, with $\textup{\textsf{QC-supFO}}_{P,\N,F}$ defined by Eq.~\eqref{eq:charact_W_QC_SupFO_decomp}. 

\begin{itemize}

    \item For $|\N|=N=1$ ($\N=\{A\}$), it is easy to see that the set $\textup{\textsf{QC-supFO}}_{P,\{A\},F}' = \{ W\in\L(\HS^{PA^{IO} F}) | W\ge 0, \Tr_F W = \tilde W^{[A]}\otimes\id^{A^O}, \Tr_{A^I}\tilde W^{[A]} = \id^P \}$ coincides with $\textup{\textsf{QC-supFO}}_{P,\{A\},F}$ as characterised by Eq.~\eqref{eq:charact_W_QC_SupFO_decomp}. (These in fact also coincide with the set of all valid process matrices, as characterised through Eq.~\eqref{eq:validityW_PF}.)

    \item Assume that $\textup{\textsf{QC-supFO}}_{P,\N',F}' = \textup{\textsf{QC-supFO}}_{P,\N',F}$ for all $\N'$ such that $|\N'|=N-1\ge 1$.
    
    Consider then $W\in\textup{\textsf{QC-supFO}}_{P,\N,F}'$, with $|\N|=N$; by definition $\Tr_F W$ can be decomposed as a convex mixture
    \begin{align}
        \Tr_F W = \sum_{k_N\in\N} q^{[k_N]}\, \tilde W^{[k_N]}\otimes\id^{A_{k_N}^O} \label{eq:proof_rec_SupFO_1}
    \end{align}
    with each $\tilde W^{[k_N]} \in \textup{\textsf{QC-supFO}}_{P,\N\backslash k_N,A_{k_N}^I}' = \textup{\textsf{QC-supFO}}_{P,\N\backslash k_N,A_{k_N}^I}$ (by the induction hypothesis), which therefore has a decomposition in terms of PSD matrices $\tilde W_{({\cal K}_{n-1},k_n)}^{(k_{n+1},\ldots,k_{N-1});[k_N]} \in {\cal L}(\HS^{PA_{{\cal K}_{n-1}}^{IO} A_{k_n}^I})$ as in Eq.~\eqref{eq:charact_W_QC_SupFO_decomp}:
\begin{align}
    \left\{
    \begin{array}{l}
        \Tr_{A_{k_N}^I} \tilde W^{[k_N]} = \sum_{k_{N-1} \in \N\backslash k_N} \tilde W_{({\cal N} \backslash \{k_{N-1},k_N\},k_{N-1})}^{\emptyset;[k_N]}\otimes \id^{A_{k_{N-1}}^O}, \\[3mm]
        \forall \, \emptyset \subsetneq {\cal K}_n \subsetneq \N\backslash k_N, \, \forall \, (k_{n+1},\ldots,k_{N-1}) \in \N\backslash k_N\backslash{\cal K}_n, \\[1mm]
        \qquad \Tr_{A_{k_{n+1}}^I} \tilde W_{({\cal K}_n,k_{n+1})}^{(k_{n+2},\ldots,k_{N-1});[k_N]} = \sum_{k_n \in {\cal K}_n} \tilde W_{({\cal K}_n \backslash k_n,k_n)}^{(k_{n+1},\ldots,k_{N-1});[k_N]}\otimes \id^{A_{k_n}^O}, \\[3mm]
        \forall \, \pi'=(k_1,\ldots,k_{N-1}) \in {\cal N}\backslash k_N, \ \Tr_{A_{k_1}^I} \tilde W_{(\emptyset,k_1)}^{(k_2,\ldots,k_{N-1});[k_N]} = q_{\pi'}^{[k_N]} \,\id^P, \ q_{\pi'}^{[k_N]}\ge 0, \, \sum_{\pi'} q_{\pi'}^{[k_N]} = 1.
    \end{array}
    \right. \label{eq:proof_rec_SupFO_2}
\end{align}
Defining $W_{({\cal N} \backslash k_N,k_N)}^{\emptyset}\coloneqq q^{[k_N]}\, \tilde W^{[k_N]}$, $W_{({\cal K}_{n-1},k_n)}^{(k_{n+1},\ldots,k_{N-1},k_N)}\coloneqq q^{[k_N]}\, \tilde W_{({\cal K}_{n-1},k_n)}^{(k_{n+1},\ldots,k_{N-1});[k_N]}$ (for $n=1,\ldots,N-1$) and $q_{\pi=(k_1,\ldots,k_{N-1},k_N)}\coloneqq q^{[k_N]}\, q_{\pi'=(k_1,\ldots,k_{N-1})}^{[k_N]}$ (such that $q_\pi\ge0, \sum_\pi q_\pi=1$), one can see that Eqs.~\eqref{eq:proof_rec_SupFO_1} and~\eqref{eq:proof_rec_SupFO_2} together provide a decomposition of $W$ as in Eq.~\eqref{eq:charact_W_QC_SupFO_decomp}, which proves that $W\in\textup{\textsf{QC-supFO}}_{P,\N,F}$.

Conversely, consider $W\in\textup{\textsf{QC-supFO}}_{P,\N,F}$, with a decomposition as in Eq.~\eqref{eq:charact_W_QC_SupFO_decomp}.
Defining, for each $k_N$, $q^{[k_N]}\coloneqq \sum_{(k_1,\ldots,k_{N-1})\in\N\backslash k_N} q_{\pi=( k_1,\ldots,k_{N-1},k_N)}$ and $\tilde W^{[k_N]}\coloneqq W_{({\cal N} \backslash k_N,k_N)}^{\emptyset}/q^{[k_N]}$ (when $q^{[k_N]}>0$; otherwise, $\tilde W^{[k_N]}\coloneqq 0$), then one obtains a decomposition of $\Tr_F W$ as in Eq.~\eqref{eq:proof_rec_SupFO_1} above, with $q^{[k_N]}\ge 0$, $ \sum_{k_N} q^{[k_N]} = 1$, and one can further see that each $\tilde W^{[k_N]}$ admits itself a decomposition of the form of Eq.~\eqref{eq:charact_W_QC_SupFO_decomp}, in terms of PSD matrices $\tilde W_{({\cal K}_{n-1},k_n)}^{(k_{n+1},\ldots,k_{N-1});[k_N]} \coloneqq W_{({\cal K}_{n-1},k_n)}^{(k_{n+1},\ldots,k_{N-1},k_N)}/q^{[k_N]}$ and with weights $q_{\pi'=(k_1,\ldots,k_{N-1})}^{[k_N]}\coloneqq q_{\pi=(k_1,\ldots,k_{N-1},k_N)}/q^{[k_N]}\, $ in the final constraint. This implies that each $\tilde W^{[k_N]}\in\textup{\textsf{QC-supFO}}_{P,\N\backslash k_N,F}=\textup{\textsf{QC-supFO}}_{P,\N\backslash k_N,F}'$, and hence that $W\in\textup{\textsf{QC-supFO}}_{P,\N,F}'$.

We thus have that $\textup{\textsf{QC-supFO}}_{P,\N,F}'=\textup{\textsf{QC-supFO}}_{P,\N,F}$. This, by recursion, concludes the proof of Eq.~\eqref{eq:rec_charact_SupFO}.

\end{itemize}

\subsubsection{Quantum circuits with non-influenceable quantum control of causal order (QC-NIQCs)}
\label{app:def_QC-NIQC}

The last class we introduced is that of QC-NIQCs, defined in Sec.~\ref{sec:QC-NIQC} of the main text as follows:
\begin{align*}
  \parbox{0.85\linewidth}{%
    \emph{QC-NIQCs are the QC-QCs for which there exists an implementation such that, at any intermediate time step, the classical weights of each term in the basis $\{\ket{\K_{n-1},k_n}^{C_n}\}_{\K_{n-1},k_n}$ that defines the controllisation---i.e.\ the diagonal elements of the state of the control system in that basis---are independent of the choice of previously applied external operations and of the potential initial state preparation in the global past.}%
  }
\end{align*}
(with, as for QC-NICCs, the external operations referring to the parties' whole instruments rather than the individual CP maps that compose them, and the state of the control system being considered after summing over all these individual CP maps---and over all their Kraus operators when one starts from a description in the ``pure'' picture).

The state $\ket{\varphi_{(n)}}$ of all systems at any intermediate time step $t_n$ (for $n=1,\ldots,N$, just before%
\footnote{Considering the case where the external operations are TP, the reduced state of the control system after the operator $\tilde A_n$ (and after summing over the corresponding Kraus operators) will be the same, as in Eq.~\eqref{eq:reduced_state_Cn_QCCC} (see also Footnote~\ref{ftn:same_state_Cn}).}
applying the operator $\tilde A_n$) is given by Eq.~\eqref{eq:link_prod_psi_tilde_V1_tilde_Vn1}. From this, the weight of each component $\ket{\K_{n-1},k_n}^{C_n}$ of the control system is obtained as
\begin{align}
    \Big\|\,\ket{\psi,A_{\K_{n-1}}} * \ket{w_{(\K_{n-1},k_n)}}\,\Big\|^2 = \Tr\Big[\Big(\ketbra{\psi}{\psi}\bigotimes_{k\in\K_{n-1}}\kketbra{A_k}{A_k}\Big) * W_{({\cal K}_{n-1},k_n)}\Big],
    \label{eq:proba_control_QCQC}
\end{align}
with $W_{({\cal K}_{n-1},k_n)} = \Tr_{\alpha_n}\ketbra{w_{(\K_{n-1},k_n)}}{w_{(\K_{n-1},k_n)}}$ as defined in Eq.~\eqref{eq:def_W_Kn1_kn}.
Summing over the Kraus operators $A_k$ so as to obtain a CPTP map for each party---whose Choi matrix, abusing the notation slightly, we still denote by $A_k$ (see Footnote~\ref{ftn:abuse_notation})---the requirement expressed above translates into the constraint that for all $\K_{n-1},k_n$,
\begin{align}
\Tr\Big[\Big( \ketbra{\psi}{\psi} \bigotimes_{k\in\K_{n-1}} A_k \Big) * W_{(\K_{n-1},k_n)} \Big] 
& \text{ does not depend on $\ket{\psi}$ nor on the $A_k$'s,} \notag \\[-4mm]
& \quad \text{ whenever all } A_k\text{'s are TP (and $\ket{\psi}$ is normalised)}. \label{eq:NIQC_indep_weight}
\end{align}
Recalling the discussion at the end of Sec.~\ref{app:subsec_validity_cstr}, this is equivalent to
\begin{align}
    \text{all} \ W_{(\K_{n-1},k_n)} \text{'s are valid process matrices (up to normalisation)},
\end{align}
as in Eq.~\eqref{eq:cstr_NIQC}.

Conversely, as we recalled above, from the decomposition of a matrix $W$ in terms of PSD matrices $W_{({\cal K}_{n-1},k_n)}$ as in Eq.~\eqref{eq:charact_W_QCQC_decomp}, one can reconstruct some internal operations $V_{\K_{n-1},k_n}^{\to k_{n+1}}$ and $V_{\N\setminus k_N, k_N}^{\to F}$ of a QC-QC, whose appropriate compositions implement $W$ and recover (through Eqs.~\eqref{eq:def_w_Knm1_kn} and~\eqref{eq:def_W_Kn1_kn}) the $W_{({\cal K}_{n-1},k_n)}$'s.
If each $W_{({\cal K}_{n-1},k_n)}$ is proportional to a valid process matrix, then the weight of each component $\ket{\K_{n-1},k_n}^{C_n}$ of the control system in that implementation, as given by the first expression in Eq.~\eqref{eq:NIQC_indep_weight} above, does not depend on $\ket{\psi}$, nor on the operations $A_k$ when these are TP. Hence, the existence of a decomposition as in Eq.~\eqref{eq:charact_W_QCQC_decomp} in terms of matrices $W_{({\cal K}_{n-1},k_n)}$ that are proportional to valid process matrices is also a sufficient condition for $W$ to be the process matrix of a QC-NIQC.

\subsubsection{Inclusion relations}
\label{app:subsubsec_QCQC_inclusions}

Let us start here by noting that each of the three classes of quantum circuits with quantum control considered above contains the corresponding class with classical control as a subclass: $\text{\textup{\textsf{QC-convFO}}} \subset \text{\textup{\textsf{QC-supFO}}}$, $\text{\textup{\textsf{QC-NICC}}} \subset \text{\textup{\textsf{QC-NIQC}}}$ and $\text{\textup{\textsf{QC-CC}}} \subset \text{\textup{\textsf{QC-QC}}}$.
In the latter case for instance, this can be seen by starting from a QC-CC decomposition in terms of PSD matrices $W_{(k_1,\ldots,k_N,F)}$ and $W_{(k_1,\ldots,k_n)}$ as in Eq.~\eqref{eq:charact_W_QCCC_decomp}, and defining $W_{({\cal K}_{n-1},k_n)}\coloneqq\sum_{(k_1,\ldots,k_{n-1})\in\K_{n-1}}W_{(k_1,\ldots,k_n)}$, which indeed readily provide a QC-QC decomposition as in Eq.~\eqref{eq:charact_W_QCQC_decomp}. For $\text{\textup{\textsf{QC-convFO}}} \subset \text{\textup{\textsf{QC-supFO}}}$, a similar construction works, after just adding the appropriate superscripts $(k_{n+1},\ldots,k_N)$. For $\text{\textup{\textsf{QC-NICC}}} \subset \text{\textup{\textsf{QC-NIQC}}}$, we further note that if $W_{(k_1,\ldots,k_n)}$'s are valid process matrices (up to normalisation), then so are the $W_{({\cal K}_{n-1},k_n)}$'s just defined.

Quite trivially, for $N=1$ all six classes reduce to the same one---which is also the whole class of all valid process matrices. For $N=2$, with a nontrivial global future space $\HS^F$, one already obtains a strict separation between the three QC-QC classes and their QC-CC counterparts: indeed the ``static quantum switch'', $W_{\textup{supFO}}$ from Eq.~\eqref{eq:QC-supFO-example}, provides in that scenario an example in \textup{\textsf{QC-supFO}} (hence also in \textup{\textsf{QC-NIQC}} and \textup{\textsf{QC-QC}}, see below) but not in \textup{\textsf{QC-CC}} (nor in any of its subclasses).
If $\HS^F$ is trivial on the other hand, one can easily see that for $N=2$ each of the three QC-QC classes reduces to its QC-CC counterpart; one then needs to go to $N\ge 3$ to see a strict separation.

\bigskip

Let us now compare the three (sub)classes of QC-QCs.

It is again quite clear, from the definition of the \textup{\textsf{QC-NIQC}} class just above, that $\textup{\textsf{QC-NIQC}} \subset \textup{\textsf{QC-QC}}$. To see that $\textup{\textsf{QC-supFO}} \subset \textup{\textsf{QC-NIQC}}$, consider a process matrix $W\in\textup{\textsf{QC-supFO}}$, with a decomposition in terms of PSD matrices (and even valid process matrices, up to normalisation) $W_{({\cal K}_{n-1},k_n)}^{(k_{n+1},\ldots,k_N)}$ as in Eq.~\eqref{eq:charact_W_QC_SupFO_decomp}; defining, for each $(\K_{n-1},k_n)$, $W_{(\K_{n-1},k_n)}\coloneqq \sum_{(k_{n+1},\ldots,k_N)} W_{(\K_{n-1},k_n)}^{(k_{n+1},\ldots,k_N)}$, one can easily see that these satisfy the QC-QC decomposition of Eq.~\eqref{eq:charact_W_QCQC_decomp}, and that these are themselves valid process matrices (up to normalisation), which proves that $W \in \textup{\textsf{QC-NIQC}}$. 
Hence, we obtain the general inclusion relations of Eq.~\eqref{eq:inclusions_QC-QCs}:
\begin{align}
    \text{\textup{\textsf{QC-supFO}}} \ \subset \ \text{\textup{\textsf{QC-NIQC}}} \ \subset \ \text{\textup{\textsf{QC-QC}}}. 
\end{align}

\medskip

For $N=2$ already, with a nontrivial global past space $\HS^P$, one finds a strict separation between \text{\textup{\textsf{QC-NIQC}}} and \text{\textup{\textsf{QC-QC}}}. An example of a process matrix in $\text{\textup{\textsf{QC-QC}}}\backslash\text{\textup{\textsf{QC-NIQC}}}$ was given by the quantum switch $W_{\textup{QS}}$, Eq.~\eqref{eq:QC-QC-example} of the main text (for $N=2$ indeed, with a nontrivial space $\HS^P$ but also a nontrivial space $\HS^F$ to avoid falling back into the \textup{\textsf{QC-CC}} class; see above). If $\HS^P$ is trivial on the other hand, \text{\textup{\textsf{QC-NIQC}}} and \text{\textup{\textsf{QC-QC}}} coincide for $N=2$ (the proof being quite similar to that for $\text{\textup{\textsf{QC-NICC}}}=\text{\textup{\textsf{QC-CC}}}$ in the same scenario, see p.~\pageref{proof:NICC_CC_N2_noP}), and one needs to go to $N\ge 3$ to see a strict separation.

As for \text{\textup{\textsf{QC-supFO}}} vs \text{\textup{\textsf{QC-NIQC}}}, these can be seen to coincide for $N\le 3$ (again, the proof is quite similar to that for $\text{\textup{\textsf{QC-convFO}}}=\text{\textup{\textsf{QC-NICC}}}$ in the same $N\leq 3$ case, on p.~\pageref{proof:convFO_NICC_N3}). A strict separation can be found for $N\ge4$: an example of a process matrix in $\text{\textup{\textsf{QC-NIQC}}}\backslash\text{\textup{\textsf{QC-supFO}}}$ (which was also not in \textup{\textsf{QC-CC}}), for $N=4$ with trivial global past and future spaces $\HS^P, \HS^F$, was given by $W_{\textup{NIQC}}$ from Eq.~\eqref{eq:QC-NICC-example}. \label{proof:convFO_NIQC_N3}

\subsubsection{Explicit decompositions of our specific examples of QC-QCs}
\label{app:decomp_ex_QCQCs}

We finish this section on QC-QCs by providing explicit decompositions of our examples of QC-QCs from Sec.~\ref{subsubsec:examples_QCQCs}, in the appropriate form so as to match the characterisation of the class that they belong to.

\bigskip

The process matrix $W_{\textup{supFO}}=\ketbra{w_\textup{supFO}}{w_\textup{supFO}}$, with $\ket{w_\textup{supFO}} = \frac{1}{\sqrt{2}} \ket{\psi}^{A^I} \otimes \kket{\id}^{A^OB^I} \otimes  \kket{\id}^{B^OF_\text{t}} \otimes \ket{0}^{F_\text{c}} + \frac{1}{\sqrt{2}}\ket{\psi}^{B^I} \otimes \kket{\id}^{B^OA^I} \otimes \kket{\id}^{A^OF_\text{t}} \otimes \ket{1}^{F_\text{c}}$ of the ``static'', or ``non-dynamical'' quantum switch, Eq.~\eqref{eq:QC-supFO-example}, can be decomposed in the QC-supFO form of Eq.~\eqref{eq:charact_W_QC_SupFO_decomp}, with
\begin{align}
\left\{
    \begin{array}{rclcrcl}
W_{(\{A\},B)}^\emptyset &=& {\textstyle\frac12}\ketbra{\psi}{\psi}^{A^I} \otimes \kketbra{\id}{\id}^{A^OB^I}, 
&& W_{(\{B\},A)}^\emptyset &=& {\textstyle\frac12}\ketbra{\psi}{\psi}^{B^I} \otimes \kketbra{\id}{\id}^{B^OA^I}, \\[2mm]
W_{(\emptyset,A)}^{(B)} &=& {\textstyle\frac12}\ketbra{\psi}{\psi}^{A^I}, 
&& W_{(\emptyset,B)}^{(A)} &=& {\textstyle\frac12}\ketbra{\psi}{\psi}^{B^I}, \\[2mm]
\multicolumn{3}{r}{q_{(A,B)}} &=& \multicolumn{3}{l}{q_{(B,A)} \,=\, {\textstyle\frac12}.}
    \end{array}
\right.
\end{align}
Notice that this is (up to very slightly different notations) the same decomposition as that given for $W_{\textup{convFO}}$ in Eq.~\eqref{eq:decomp_WconvFO}. This is because $\Tr_F W_{\textup{supFO}}=W_{\textup{convFO}}$, and because the QC-supFO constraints for $\Tr_F W_{\textup{supFO}}$ reduce (in the $N=2$ case, with no more $\HS^F$) to the QC-convFO constraints. Besides, it is also well-known---and easy to see---that $W_{\textup{supFO}}$ is not a QC-CC (noting that $W_{\textup{supFO}}=\ketbra{w_\textup{supFO}}{w_\textup{supFO}}$ is rank-1, so that any decomposition of $W_{\textup{supFO}}$ directly into PSD matrices can only contain terms that are proportional to itself).

\bigskip

Our example of a QC-NIQC, $W_{\textup{NIQC}} = \frac12 W_0\otimes \ketbra{\psi}{\psi}^{C^I} \otimes \kketbra{\id}{\id}^{C^OD^I} \otimes \id^{D^O} + \frac12 W_1\otimes \ketbra{\psi}{\psi}^{D^I} \otimes \kketbra{\id}{\id}^{D^OC^I} \otimes \id^{C^O}$ from Eq.~\eqref{eq:QC-NIQC-example},
with $W_0$ and $W_1$ two (valid) process matrices defined in Eq.~\eqref{eq:W0/W1} and satisfying ${\textstyle\frac12} W_0 + {\textstyle\frac12} W_1 = \frac{1}{8}(2 \,\id^{\otimes 4}+ 2 \,Z\id Z\id + XZX\id - YZY\id + X\id XZ - Y\id YZ)^{A^{IO}B^{IO}}$,
admits a QC-NIQC decomposition of the form of Eq.~\eqref{eq:charact_W_QCQC_decomp}, with the valid process matrices (up to normalisation, as required by the QC-NIQC condition of Eq.~\eqref{eq:cstr_NIQC})
\begin{align}
\left\{
    \begin{array}{rclrcl}
W_{(\{A,B,C\},D)} &=& {\textstyle\frac12} W_0 \otimes \ketbra{\psi}{\psi}^{C^I} \otimes \kketbra{\id}{\id}^{C^O D^I}, & W_{(\{A,B,D\},C)} &=& {\textstyle\frac12} W_1\otimes \ketbra{\psi}{\psi}^{D^I} \otimes \kketbra{\id}{\id}^{D^O C^I}, \\[3mm]
W_{(\{A,B\},C)} &=& {\textstyle\frac12} W_0 \otimes \ketbra{\psi}{\psi}^{C^I}, & W_{(\{A,B\},D)} &=& {\textstyle\frac12} W_1\otimes \ketbra{\psi}{\psi}^{D^I}, \\[3mm]
W_{(\{A\},B)} &=& \frac{1}{8}(\id^{\otimes 3}+ Z\id Z + XZX - YZY)^{A^{IO}B^I}, & W_{(\{B\},A)} &=& \frac{1}{8}(\id^{\otimes 3} + ZZ\id + XXZ - YYZ)^{A^IB^{IO}}, \\[3mm]
W_{(\emptyset,A)} &=& {\textstyle\frac14} \id^{A^I}, & W_{(\emptyset,B)} &=& {\textstyle\frac14} \id^{B^I},
    \end{array}
\right.
 \label{app:dec_WNIQC}
\end{align}
and all other $W_{(\K_{n-1},k_n)}=0$. Notice that while the $W_{0/1}$'s are valid process matrices, the fact that they are causally nonseparable prevents one from obtaining a QC-supFO decomposition of $W_{\textup{NIQC}}$ (and indeed no such decomposition exists, as can be checked via SDP).

\bigskip

Finally, the process matrix $W_{\textup{QS}}=\ketbra{w_\textup{QS}}{w_\textup{QS}}$, with $\ket{w_\textup{QS}} = \ket{0}^P \otimes \ket{\psi}^{A^I} \otimes \kket{\id}^{A^OB^I} \otimes \kket{\id}^{B^OF_\text{t}} \otimes  \ket{0}^{F_\text{c}} + \ket{1}^P \otimes \ket{\psi}^{B^I} \otimes \kket{\id}^{B^OA^I} \otimes \kket{\id}^{A^OF_\text{t}} \otimes \ket{1}^{F_\text{c}}$ of the ``dynamical'' quantum switch, Eq.~\eqref{eq:QC-QC-example}, can be decomposed in the general QC-QC form of Eq.~\eqref{eq:charact_W_QCQC_decomp} as follows:
\begin{align}
\left\{
    \begin{array}{rclcrcl}
W_{(\{A\},B)} &=& \ketbra{0}{0}^P \otimes \ketbra{\psi}{\psi}^{A^I} \otimes \kketbra{\id}{\id}^{A^OB^I}, && W_{(\{B\},A)} &=& \ketbra{1}{1}^P\otimes \ketbra{\psi}{\psi}^{B^I} \otimes \kketbra{\id}{\id}^{B^OA^I}, \\[2mm]
W_{(\emptyset,A)} &=& \ketbra{0}{0}^P \otimes \ketbra{\psi}{\psi}^{A^I}, && W_{(\emptyset,B)} &=& \ketbra{1}{1}^P\otimes \ketbra{\psi}{\psi}^{B^I}.
    \end{array}
\right.
\label{app:dec_WQS}
\end{align}
Notice that this is again essentially the same decomposition as that given for the process matrix of the classical switch $W_{\textup{CS}}$ in Eq.~\eqref{app:dec_WCS}. This is because $\Tr_F W_{\textup{QS}}=W_{\textup{CS}}$, and because the QC-QC constraints for $\Tr_F W_{\textup{QS}}$ reduce (in the $N=2$ case, with no more $\HS^F$) to the QC-CC constraints. 
As in the case of $W_{\textup{CS}}$, none of the matrices in the above decomposition are valid process matrices; indeed it can be shown via SDP that no decomposition of the form of Eq.~\eqref{eq:charact_W_QCQC_decomp} exists for $W_{\textup{QS}}$ with only valid $W_{(\{k_1\},k_2)}$'s and $W_{(\emptyset,k_1)}$'s, which implies that $W_{\textup{QS}}$ is not a QC-NIQC.
Besides, it is also well-known---and again easy to see---that $W_{\textup{QS}}$ is not a QC-CC (e.g., by contradiction: otherwise, feeding the state $\ket{+}=\frac{\ket{0}+\ket{1}}{\sqrt{2}}$ into $\HS^P$ would otherwise imply that $W_{\textup{supFO}}$ is a QC-CC).

\subsection{Sketch of the proof that our classes are ``$P/F$-label independent''}
\label{app:subsec:P/F-label indep}

We claimed above that if two or more parties have no output space, or two or more parties have no input space, then the definitions of our classes of processes (\textup{\textsf{QC-CC}}, \textup{\textsf{QC-convFO}}, \textup{\textsf{QC-NICC}}, \textup{\textsf{QC-QC}}, \textup{\textsf{QC-supFO}} and \textup{\textsf{QC-NIQC}}) does not depend on which party is labelled $F$, or which party is labelled $P$. Let us just briefly sketch the arguments that allow one to verify this. We will use here the notation $(k_1,\ldots,k_n)\underset{j}{\cup}\ell$ for an ordered list $(k_1,\ldots,k_n)$, into which an element $\ell$ is inserted at the $j^\text{th}$ position: $(k_1,\ldots,k_n)\underset{j}{\cup}\ell\coloneqq(k_1,\ldots,k_{j-1},\ell,k_j,\ldots,k_n)$.

\medskip

Consider first the case where one party in $\N$, say $A_f$, has no output space.

Suppose that a QC-CC admits a decomposition in terms of PSD matrices $W_{(k_1,\ldots,k_N,F)}$ and $W_{(k_1,\ldots,k_n)}$ as in Eq.~\eqref{eq:charact_W_QCCC_decomp}. Then it can be verified that it also has a similar decomposition to Eq.~\eqref{eq:charact_W_QCCC_decomp}, in which the roles of the spaces $F$ and $A_f^I$ are exchanged, in terms of the PSD matrices $\tilde W_{(k_1,\ldots,k_{N-1},F,f)} \coloneqq \sum_{j=1}^N W_{(k_1,\ldots,k_{N-1},F)\underset{j}{\cup}f}$, $\tilde W_{(k_1,\ldots,k_{N-1},F)} \coloneqq \sum_{j=1}^N \Tr_{A_f^I} W_{(k_1,\ldots,k_{N-1},F)\underset{j}{\cup}f}$ and $\tilde W_{(k_1,\ldots,k_n)} \coloneqq W_{(k_1,\ldots,k_n)} + \sum_{j=1}^n \Tr_{A_f^I} W_{(k_1,\ldots,k_n)\underset{j}{\cup}f}$ (for $(k_1,\ldots,k_n)\subseteq\N\backslash f$, and with all other nonspecified matrices that should \emph{a priori} appear in the decomposition being null).

Similarly, suppose that a QC-convFO admits a decomposition in terms of PSD matrices $W_{(k_1,\ldots,k_N,F)}$ and $W_{(k_1,\ldots,k_n)}^{(k_{n+1},\ldots,k_N)}$, and with the weights $q_\pi$ as in Eq.~\eqref{eq:charact_W_QC_convFO_decomp}. Then it can be seen that it also has a similar decomposition to Eq.~\eqref{eq:charact_W_QC_convFO_decomp}, in which the roles of the spaces $F$ and $A_f^I$ are exchanged, in terms of the PSD matrices $\tilde W_{(k_1,\ldots,k_{N-1},F,f)} \coloneqq \sum_{j=1}^N W_{(k_1,\ldots,k_{N-1},F)\underset{j}{\cup}f}$, $\tilde W_{(k_1,\ldots,k_{N-1},F)}^\emptyset \coloneqq \sum_{j=1}^N \Tr_{A_f^I} W_{(k_1,\ldots,k_{N-1},F)\underset{j}{\cup}f}$ and $\tilde W_{(k_1,\ldots,k_n)}^{(k_{n+1},\ldots,k_{N-1},F)} \coloneqq \sum_{j=1}^{N-n} W_{(k_1,\ldots,k_n)}^{(k_{n+1},\ldots,k_{N-1})\underset{j}{\cup}f} + \sum_{j=1}^n \Tr_{A_f^I} W_{(k_1,\ldots,k_n)\underset{j}{\cup}f}^{(k_{n+1},\ldots,k_{N-1})}$, and with the weights $\tilde q_{\pi=(k_1,\ldots,k_{N-1},F)} \coloneqq \sum_{j=1}^N q_{\pi=(k_1,\ldots,k_{N-1})\underset{j}{\cup}f}$.

For a QC-QC whose process matrix $W$ has a decomposition in terms of PSD matrices $W_{(\K_{n-1},k_n)}$ as in Eq.~\eqref{eq:charact_W_QCQC_decomp}, it can be verified that it also has a similar decomposition to Eq.~\eqref{eq:charact_W_QCQC_decomp}, in which the roles of the spaces $F$ and $A_f^I$ are exchanged, in terms of the PSD matrices $\tilde W_{(\N\backslash f,F)} \coloneqq \Tr_{A_f^I} W$ and $\tilde W_{(\K_{n-1},k_n)} \coloneqq W_{(\K_{n-1},k_n)} + \Tr_{A_f^I} W_{(\K_{n-1}\cup f,k_n)}$ (for subsets $\K_{n-1}\subseteq\N\backslash f$ and $k_n\ne f$).

As for a QC-supFO whose process matrix $W$ has a decomposition in terms of PSD matrices $W_{(\K_{n-1},k_n)}^{(k_{n+1},\ldots,k_N)}$ and with the weights $q_\pi$ as in Eq.~\eqref{eq:charact_W_QC_SupFO_decomp}, it can be seen to also admit a similar decomposition to Eq.~\eqref{eq:charact_W_QC_SupFO_decomp}, in which the roles of the spaces $F$ and $A_f^I$ are exchanged, in terms of the PSD matrices $\tilde W_{(\N\backslash f,F)}^\emptyset \coloneqq \Tr_{A_f^I} W$ and $\tilde W_{(\K_{n-1},k_n)}^{(k_{n+1},\ldots,k_{N-1},F)} \coloneqq \sum_{j=1}^{N-n} W_{(\K_{n-1},k_n)}^{(k_{n+1},\ldots,k_{N-1})\underset{j}{\cup}f} + \Tr_{A_f^I} W_{(\K_{n-1}\cup f,k_n)}^{(k_{n+1},\ldots,k_{N-1})}$ and with the weights $\tilde q_{\pi=(k_1,\ldots,k_{N-1},F)} \coloneqq \sum_{j=1}^N q_{\pi=(k_1,\ldots,k_{N-1})\underset{j}{\cup}f}$.

\medskip

Consider now the case where one party in $\N$, say $A_p$, has no input space.

Suppose that a QC-CC admits a decomposition in terms of PSD matrices $W_{(k_1,\ldots,k_N,F)}$ and $W_{(k_1,\ldots,k_n)}$ as in Eq.~\eqref{eq:charact_W_QCCC_decomp}. Then it can be verified that it also has a similar decomposition to Eq.~\eqref{eq:charact_W_QCCC_decomp}, in which the roles of the spaces $P$ and $A_p^O$ are exchanged, in terms of the PSD matrices $\tilde W_{(P,k_2,\ldots,k_N,F)} \coloneqq \sum_{j=1}^N W_{(k_2,\ldots,k_N,F)\underset{j}{\cup}p}$, $\tilde W_{(P,k_2,\ldots,k_n)} \coloneqq \sum_{j=1}^{n-1} W_{(k_2,\ldots,k_n)\underset{j}{\cup}p} + \id^{A_p^O} \otimes W_{(k_2,\ldots,k_n)}$ (for $(k_1,\ldots,k_n)\subseteq\N\backslash p$) and $\tilde W_{(P)} \coloneqq \id^{A_p^O}$.

Similarly, suppose that a QC-convFO admits a decomposition in terms of PSD matrices $W_{(k_1,\ldots,k_N,F)}$ and $W_{(k_1,\ldots,k_n)}^{(k_{n+1},\ldots,k_N)}$, and with the weights $q_\pi$ as in Eq.~\eqref{eq:charact_W_QC_convFO_decomp}. Then it can be seen that it also has a similar decomposition to Eq.~\eqref{eq:charact_W_QC_convFO_decomp}, in which the roles of the spaces $P$ and $A_p^O$ are exchanged, in terms of the PSD matrices $\tilde W_{(P,k_2,\ldots,k_N,F)} \coloneqq \sum_{j=1}^N W_{(k_2,\ldots,k_N,F)\underset{j}{\cup}p}$, $\tilde W_{(P,k_2,\ldots,k_n)}^{(k_{n+1},\ldots,k_N)} \coloneqq \sum_{j=1}^{n-1} W_{(k_2,\ldots,k_n)\underset{j}{\cup}p}^{(k_{n+1},\ldots,k_N)} + \id^{A_p^O} \otimes \sum_{j=1}^{N-n+1} W_{(k_2,\ldots,k_n)}^{(k_{n+1},\ldots,k_N)\underset{j}{\cup}p}$ and $\tilde W_{(P)}^{(k_2,\ldots,k_N)} \coloneqq \tilde q_{\pi=(P,k_2,\ldots,k_N)} \id^{A_p^O}$, with the weights $\tilde q_{\pi=(P,k_2,\ldots,k_N)} \coloneqq \sum_{j=1}^N q_{\pi=(k_2,\ldots,k_N)\underset{j}{\cup}p}$.

For a QC-QC whose process matrix $W$ has a decomposition in terms of PSD matrices $W_{(\K_{n-1},k_n)}$ as in Eq.~\eqref{eq:charact_W_QCQC_decomp}, it can be verified that it also has a similar decomposition to Eq.~\eqref{eq:charact_W_QCQC_decomp}, in which the roles of the spaces $P$ and $A_p^O$ are exchanged, in terms of the PSD matrices $\tilde W_{(P\cup\K_{n-2},k_n)} \coloneqq W_{(p\cup\K_{n-2},k_n)} + \id^{A_p^O} \otimes W_{(\K_{n-2},k_n)}$ (for subsets $\K_{n-2}\subseteq\N\backslash p$, $k_n\ne p$) and $\tilde W_{(\emptyset,P)} \coloneqq \id^{A_p^O}$.

As for a QC-supFO whose process matrix $W$ has a decomposition in terms of PSD matrices $W_{(\K_{n-1},k_n)}^{(k_{n+1},\ldots,k_N)}$ and with the weights $q_\pi$ as in Eq.~\eqref{eq:charact_W_QC_SupFO_decomp}, it can be seen to also admit a similar decomposition to Eq.~\eqref{eq:charact_W_QC_SupFO_decomp}, in which the roles of the spaces $P$ and $A_p^O$ are exchanged, in terms of the PSD matrices $\tilde W_{(P\cup\K_{n-2},k_n)}^{(k_{n+1},\ldots,k_N)} \coloneqq W_{(p\cup\K_{n-2},k_n)}^{(k_{n+1},\ldots,k_N)} + \id^{A_p^O} \otimes \sum_{j=1}^{N-n+1} W_{(\K_{n-2},k_n)}^{(k_{n+1},\ldots,k_N)\underset{j}{\cup}p}$ and $\tilde W_{(\emptyset,P)}^{(k_2,\ldots,k_N)} \coloneqq \tilde q_{\pi=(P,k_2,\ldots,k_N)} \id^{A_p^O}$, with the weights $\tilde q_{\pi=(P,k_2,\ldots,k_N)} \coloneqq \sum_{j=1}^N q_{\pi=(k_2,\ldots,k_N)\underset{j}{\cup}p}$.

\medskip

For QC-NICCs and QC-NIQCs, it then remains to note that in both the QC-CC and QC-QC cases, if one assumes that the $W_{\dots}$ matrices are all valid process matrices (up to normalisation), then so are all the $\tilde{W}_{\dots}$ matrices defined above.

\subsection{A graphical representation of the SDP constraints for our various classes of circuits}
\label{app:subsec:graphical_rep}

Looking at the SDP constraints that characterise each of the classes we considered in this work (see Eqs.~\eqref{eq:charact_W_QCCC_decomp}, \eqref{eq:charact_W_QC_convFO_decomp}, \eqref{eq:charact_W_QCQC_decomp}, and~\eqref{eq:charact_W_QC_SupFO_decomp}), one may notice some similar structure in the different decompositions.
Here we propose a graphical representation of these constraints that exploits their common features and helps visualise the structure of the different classes.

Each of the decompositions under consideration involves PSD matrices in $\L(\HS^{PA_{\K_{n-1}}^{IO}A_{k_n}^I})$ (for some subset $\K_{n-1}\subsetneq\N$ and some $k_n\notin\K_{n-1}$) and in $\L(\HS^{PA_{\N}F})$, as well as the identity matrix $\id^P\in\L(\HS^P)$. The idea of the representation we propose is to organise these PSD matrices in a graph, defined as follows.
\begin{itemize}
    \item The nodes are the various PSD matrices that appear in the decomposition. These are organised in $N+2$ layers: matrices in some $\L(\HS^{PA_{\K_{n-1}}^{IO}A_{k_n}^I})$ appear on the $n^\text{th}$ layer, for each $n=1,\ldots,N$ (from the bottom up), matrices in $\L(\HS^{PA_{\N}F})$ appear on the $(n=N+1)^\text{th}$ (top) layer, while the $\id$'s (i.e.\ the $\id^P$'s; we will drop the superscript~${}^P$) appear on the $(n=0)^\text{th}$ (bottom) layer.
    \item For any SDP constraint of the general form $\big(\sum_{k_{n+1}}\big) \Tr_{A_{k_{n+1}}^I} W_{(\cdots,k_{n+1})} = \big(\sum_{k_n}\big) W_{(\cdots,k_n)}\otimes \id^{A_{k_n}^O}$, with $1\le n\le N-1$, all matrices $W_{(\cdots,k_{n+1})}$ appearing on the LHS of the equality (and on the $(n+1)^\text{th}$ layer in the graph) are connected to all matrices $W_{(\cdots,k_n)}$ appearing on the RHS of the equality (and on the $n^\text{th}$ layer in the graph).
    \item For any SDP constraint of the form $\Tr_F W_{(\cdots)} = \big(\sum_{k_N}\big) W_{(\cdots,k_N)}\otimes \id^{A_{k_N}^O}$, for some $W_{(\cdots)}\in\L(\HS^{PA_{\N}F})$, the matrix $W_{(\cdots)}$ (that appears on the top layer in the graph) is connected to all matrices $W_{(\cdots,k_N)}$ appearing on the RHS of the equality (and on the $N^\text{th}$ layer in the graph).
    \item For any SDP constraint of the form $\big(\sum_{k_1}\big) \Tr_{A_{k_1}^I} W_{(\cdots,k_1)} = q_{(\cdots)}\id^P$ (possibly with $q_{(\cdots)}=1$), all matrices $W_{(\cdots,k_1)}$ appearing on the LHS of the equality (and on the $(n=1)^\text{th}$ layer in the graph) are connected to $q_{(\cdots)}\id$ (that appears on the bottom layer in the graph).
\end{itemize}

\medskip

\begin{figure}[hbtp]
    \includegraphics[width=.5\columnwidth]{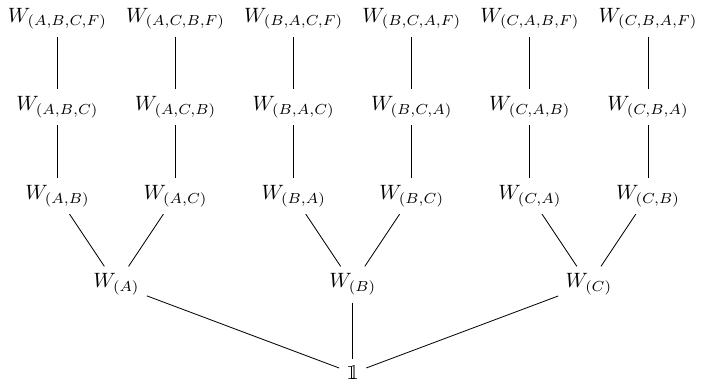}
    \centering
    \caption{Graphical representation of the SDP constraints of Eq.~\eqref{eq:charact_W_QCCC_decomp}, for the \textup{\textsf{QC-CC}} class with $N=3$. For the \textup{\textsf{QC-NICC}} class, all matrices that appear in the graph must be valid process matrices (up to normalisation), see Eq.~\eqref{eq:cstr_NICC}.}
    \label{fig:graph_QCCC_N3}
\end{figure}

For concreteness, let us start by illustrating this for the \textup{\textsf{QC-CC}} class, characterised in Eq.~\eqref{eq:charact_W_QCCC_decomp}. From the bottom to the top: the matrix $\id$ ($=\id^P$, on the bottom layer) is connected to all $W_{(k_1)}$'s (on the $(n=1)^\text{th}$ layer); then each $W_{(k_1,\ldots,k_n)}$ (for $n=1,\ldots,N-1$, on the $n^\text{th}$ layer) is connected to all $W_{(k_1,\ldots,k_n,k_{n+1})}$'s (with the same first labels $(k_1,\ldots,k_n)$, on the $(n+1)^\text{th}$ layer); finally, each $W_{(k_1,\ldots,k_N)}$ (on the $N^\text{th}$ layer) is connected to $W_{(k_1,\ldots,k_N,F)}$ (on the top layer). We thus obtain a branch-graph structure, as illustrated in Fig.~\ref{fig:graph_QCCC_N3} for the $N=3$ case.

\begin{figure}[hbtp]
    \includegraphics[width=.5\columnwidth]{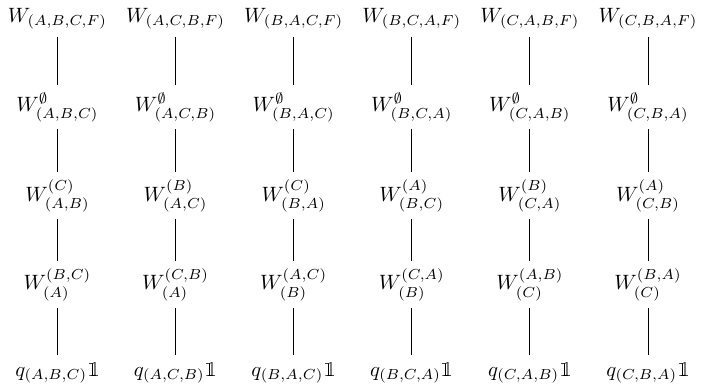}
    \centering
    \caption{Graphical representation of the SDP constraints of Eq.~\eqref{eq:charact_W_QC_convFO_decomp}, for the \textup{\textsf{QC-convFO}} class with $N=3$.}
    \label{fig:graph_QCconvFO_N3}
\end{figure}

\begin{figure}[hbtp]
    \centering
    \includegraphics[width=.5\columnwidth]{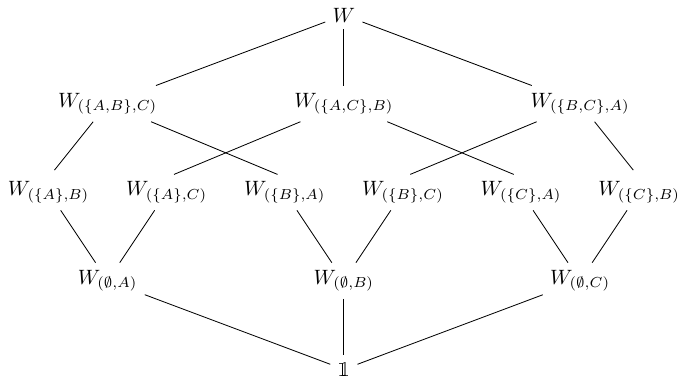}\\
    \vspace{5mm}
    \includegraphics[width=\textwidth]{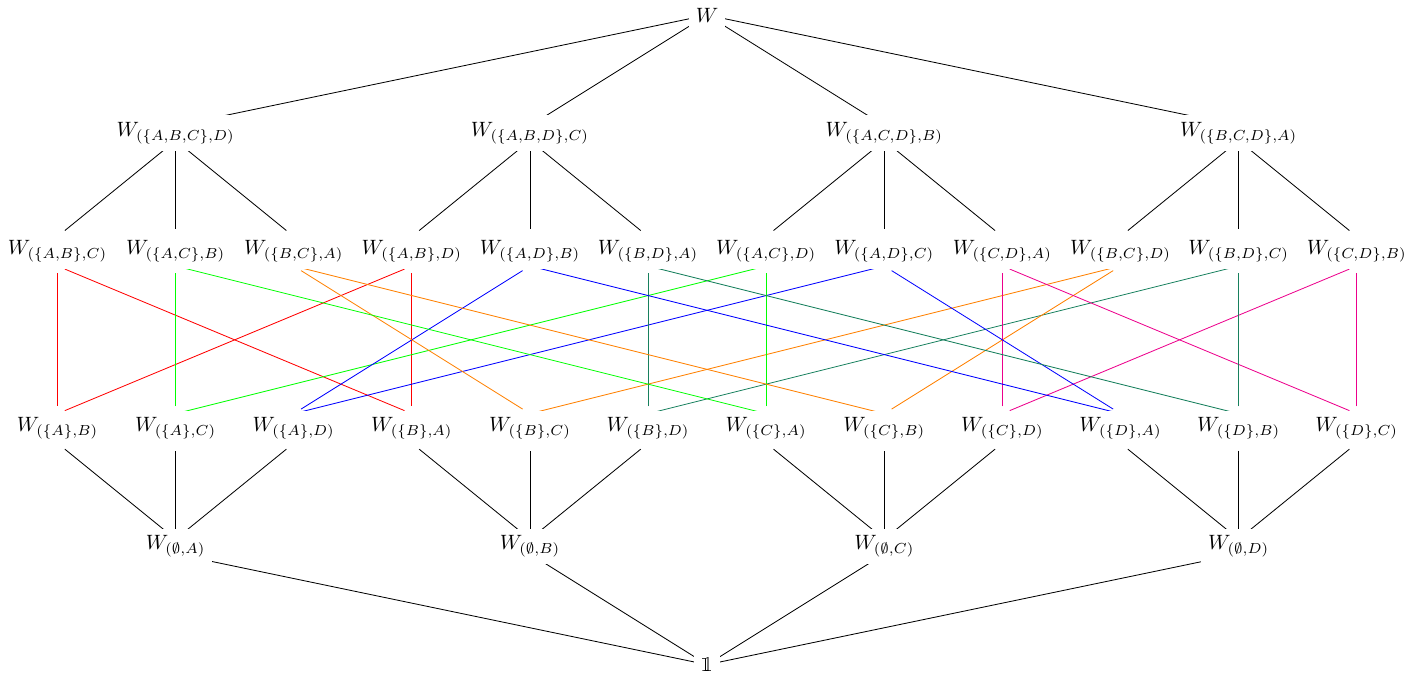}
    \caption{Graphical representation of the SDP constraints of Eq.~\eqref{eq:charact_W_QCQC_decomp}, for the \textup{\textsf{QC-QC}} class with $N=3$ (top) and $N=4$ (bottom, where we drew the edges between the middle two layers with colours to help visualise which matrices are connected together, in the different ``branches''). For the \textup{\textsf{QC-NIQC}} class, all matrices that appear in the graph must be valid process matrices (up to normalisation), see Eq.~\eqref{eq:cstr_NIQC}.
    \\
    Notice the upside-down symmetry of the graph: flipping the graph upside-down and relabelling the nodes according to $W_{(\K_{n-1},k_n)} \leftrightarrow W_{(\N\backslash\K_{n-1}\backslash k_n,k_n)}$, $W\leftrightarrow\id$ gives the same graph.
    Interestingly, one may also notice a similar structure to that of the ``branch graph''~\cite{vanrietvelde22} that one can build from a possible generic routed quantum circuit representation of QC-QCs~\cite{grothus25} (compare e.g.\ the top figure here to the black subgraph of Figure~4.5 (right) from~\cite{grothus25}). Specifically: by taking the branch graph of ${\cal G}_\text{QC-QC}(N)$ from~\cite{grothus25}, keeping only the solid black arrows (``strong parent relations''), bypassing all nodes corresponding to the branches $\textup{\textbf{V}}_{n+1}^{\K_n}$ for $n=1,\ldots,N-1$ (i.e.\ replacing the chains of arrows of the form $\textup{\textbf{A}}_{k_n}^{\K_n\backslash k_n}\to\textup{\textbf{V}}_{n+1}^{\K_n}\to\textup{\textbf{A}}_{k_{n+1}}^{\K_n}$ by just $\textup{\textbf{A}}_{k_n}^{\K_n\backslash k_n}\to\textup{\textbf{A}}_{k_{n+1}}^{\K_n}$), and renaming $\textup{\textbf{A}}_{k_n}^{\K_{n-1}}\to W_{(\K_{n-1},k_n)}$, $\textup{\textbf{V}}_1^\emptyset\to\id$, $\textup{\textbf{V}}_{N+1}^\N\to W$, one gets the same graph as the one defined here for QC-QCs. Clarifying the connection between these two graphical representations is left for future investigations.}
    \label{fig:graph_QCQC}
\end{figure}

\begin{figure}[hbtp]
    \centering
    \includegraphics[width=.5\columnwidth]{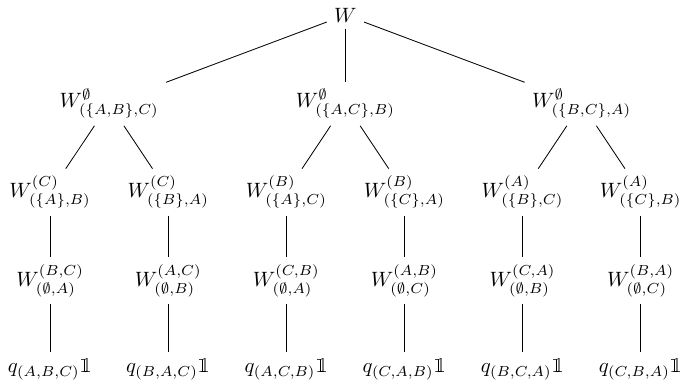}
    \caption{Graphical representation of the SDP constraints of Eq.~\eqref{eq:charact_W_QC_SupFO_decomp}, for the \textup{\textsf{QC-supFO}} class with $N=3$. \\
    Notice that this graph is the upside-down version of that of the \textup{\textsf{QC-CC}} class in Fig.~\ref{fig:graph_QCCC_N3} (that one obtains by flipping the graph here and relabelling the nodes according to $W\to\id$, $W_{({\cal K}_{n-1},k_n)}^{(k_{n+1},\ldots,k_N)} \to W_{(k_N,\ldots,k_n)}$, $q_{(k_1,\ldots,k_N)}\id\to W_{(k_N,\ldots,k_1,F)}$).}
    \label{fig:graph_QCsupFO_N3}
\end{figure}

Following the above construction, we can similarly build the graphs that represent the SDP constraints for our other classes: these are shown in Fig.~\ref{fig:graph_QCconvFO_N3} for the \textup{\textsf{QC-convFO}} class (with $N=3$), in Fig.~\ref{fig:graph_QCQC} for the \textup{\textsf{QC-QC}} class (with $N=3$ and $4$), and in Fig.~\ref{fig:graph_QCsupFO_N3} for the \textup{\textsf{QC-supFO}} class (with $N=3$). These representations help visualise certain features of each class: e.g., that the \textup{\textsf{QC-convFO}} class involves independent orders, in parallel; that the graph of the \textup{\textsf{QC-QC}} class has an ``upside-down symmetry''; that the graph of the \textup{\textsf{QC-supFO}} class is the ``upside-down version'' of that of the \textup{\textsf{QC-CC}} class (reflecting the comment made a couple of paragraphs below Eq.~\eqref{eq:charact_W_QC_SupFO_decomp}).

\bigskip

Conversely, rather than building the graph corresponding to a given set of SDP constraints, one can take a given graph to define a set of SDP constraints, and thereby a class of quantum circuits. 

An ``SDP-constraints-defining graph'' must be of a certain form:
\begin{itemize}
    \item It must be a layered graph, with $N+2$ layers for a class of $N$-slot circuits, each being referred to by an integer $n=0,\ldots,N+1$.
    \item To each node on a layer $n=1,\ldots,N$, one attaches an $(n-1)-$partite set of parties $\K_{n-1}\subsetneq\N$, a party labelled by $k_n\notin\K_{n-1}$ (these could be the same for two different nodes on the same layer), and a PSD matrix in $\L(\HS^{PA_{\K_{n-1}}^{IO}A_{k_n}^I})$. As a general convention these matrices are denoted $W_{\cdots}^{\cdots}$, where the subscript contains $n$ parties: the first $n-1$ parties altogether form $\K_{n-1}$, while the last party in the subscript is $k_n$ (so that the notations themselves define $\K_{n-1},k_n$ unambiguously).
    \item To each node on layer $n=N+1$ one attaches a PSD matrix in $\L(\HS^{PA_\N^{IO}F})$.
    \item To each node on layer $n=0$ one attaches a PSD matrix in $\L(\HS^{P})$---typically, $\id$ or $q_{\cdots}\id$, for some $q_{\cdots}\ge 0$.
    \item Edges only connect nodes in subsequent layers. The edges between any two subsequent layers $n,n+1$ (for $n=0,\ldots,N$) define a so-called ``branched relation''~\cite{vanrietvelde22}: the sets ${\cal S}_n, {\cal S}_{n+1}$ of matrices in the lower and upper layers are partitioned into ${\cal S}_n = \sqcup_j {\cal S}_{n,j}^\uparrow$ and ${\cal S}_{n+1} = \sqcup_j {\cal S}_{n+1,j}^\downarrow$, with the ${\cal S}_{n,j}^\uparrow$'s and the ${\cal S}_{n+1,j}^\downarrow$'s being in one-to-one correspondence, such that for each $j$, all matrices in ${\cal S}_{n,j}^\uparrow$ are connected to all matrices in ${\cal S}_{n+1,j}^\downarrow$.%
    \footnote{Each layer (each set of matrices ${\cal S}_n$), for $n=1,\ldots,N$, is thus partitioned in two different ways, ${\cal S}_n = \sqcup_j {\cal S}_{n,j}^\uparrow = \sqcup_k {\cal S}_{n,k}^\downarrow$. The number of subsets in each partition can be different.}
    Each pair ${\cal B}_{n,j}=({\cal S}_{n,j}^\uparrow,{\cal S}_{n+1,j}^\downarrow)$ defines a so-called ``branch''.\\
    Furthermore, for $n=1,\ldots,N-1$, it is required that in any branch ${\cal B}_{n,j}=({\cal S}_{n,j}^\uparrow,{\cal S}_{n+1,j}^\downarrow)$, the sets $\K_{n-1}\cup k_n$ attached to all matrices in ${\cal S}_{n,j}^\uparrow$ are the same, and are the same as all sets $\K_n$ attached to all matrices in ${\cal S}_{n+1,j}^\downarrow$.%
    \footnote{Recall from the second bullet point above that a subset $\K_{n-1}$ and a party $k_n$ are attached to each node in ${\cal S}_{n,j}^\uparrow$; similarly, a subset $\K_n$ and a party $k_{n+1}$ are attached to each node in ${\cal S}_{n+1,j}^\downarrow$.}
\end{itemize}
Any graph of the above form then defines a set of linear constraints (hence, SDP constraints when one further requires that all matrices are PSD) as follows: 
\begin{itemize}
    \item for all matrices $W_i$ of the top ($n=N+1$) layer: $W = \sum_i W_i$;
    \item for $n=1,\ldots,N-1$: for any branch ${\cal B}_{n,j}=({\cal S}_{n,j}^\uparrow,{\cal S}_{n+1,j}^\downarrow)$, $\sum_{W_{(\cdots,k_{n+1})}^{\cdots}\in{\cal S}_{n+1,j}^\downarrow} \Tr_{A_{k_{n+1}}^I}W_{(\cdots,k_{n+1})}^{\cdots} = \sum_{W_{(\cdots,k_n)}^{\cdots}\in{\cal S}_{n,j}^\uparrow} W_{(\cdots,k_n)}^{\cdots}\otimes\id^{A_{k_n}^O}$;%
    \footnote{The last requirement on the general form of an ``SDP-constraints-defining graph'', as stated above, ensures that all terms (on both sides of the equality) live in the same space.}
    \item for all matrices $W_i$ of the bottom ($n=0$) layer: $\sum_i W_i = \id^P$.
\end{itemize}
The set of all $W$'s that satisfy these constraints forms a certain class of process matrices, thus directly defined from the graph under consideration.

\medskip

One can verify that the graphs in Figs.~\ref{fig:graph_QCCC_N3} to~\ref{fig:graph_QCsupFO_N3} are indeed of the required form specified above, and that starting from these graphs, one obtains the constraints that characterise the \textup{\textsf{QC-CC}}, \textup{\textsf{QC-convFO}}, \textup{\textsf{QC-QC}} and \textup{\textsf{QC-supFO}} classes, as given in Eqs.~\eqref{eq:charact_W_QCCC_decomp}, \eqref{eq:charact_W_QC_convFO_decomp}, \eqref{eq:charact_W_QCQC_decomp}, and~\eqref{eq:charact_W_QC_SupFO_decomp}, respectively (for the specified values of $N$). An example of a class only defined through its graph representation will be given in Appendix~\ref{app:new_class} (see Fig.~\ref{fig:graph_QCNIO_N4_diff_nstar}); this will illustrate how convenient the graph representation is, compared to the explicit specification of all constraints separately, and how it allows one to nicely visualise their structure.

Other insights can be gained from the graphical representation proposed here. As an example, the fact that the \textup{\textsf{QC-supFO}} class can be given a recursive characterisation (recall Eq.~\eqref{eq:rec_charact_SupFO}) can be seen from its graph representation (indeed, removing the top node of the graph for an $N$-partite QC-supFO leaves one with subgraphs of the same form, for $(N-1)$-partite QC-supFOs).
It would be interesting to further explore which properties of the classes can be conveniently visualised from the graphs directly, or how comparing graphs may help comparing classes. Exploring the full potential of this representation is left as an interesting avenue for future research.

\section{Correlations induced by quantum circuits with classical or quantum control of causal order}

In this appendix, we prove the claims that each of the classes of quantum circuits with classical or quantum control of causal order considered in the paper generate correlations in a given set.

Recall that these are given by the generalised Born rule of Eq.~\eqref{eq:Born_rule} (considering here trivial global past and future spaces $\HS^P, \HS^F$): $p(\vec a|\vec x) = \Tr\big[\big(M_{\vec a|\vec x})^{T} \, W\big]$ with $\vec x = (x_1,\ldots,x_N)$, $\vec a = (a_1,\ldots,a_N)$, where for convenience we introduced here the notation $M_{\vec a|\vec x} \coloneqq \bigotimes_{k=1}^N M_{a_k|x_k}$ (not to be confused with the resulting global map $M_{\vec a|\vec x}^{PF}$ of Eq.~\eqref{eq:Born_rule_PF}, in the presence of $\HS^P, \HS^F$), and where each $\{M_{a_k|x_k}\}_{a_k}$, for each $x_k$, denotes a quantum instrument (in the Choi picture).

\subsection{Correlations induced by QC-CCs}

Let us start with the correlations induced by QC-CCs. As described in the previous appendix, in these circuits one can identify some control system that encodes in a classical manner the causal order (which may be established dynamically). It is then rather straightforward to see the connection between the (sub)class of QC-CCs under consideration and the type of correlations that can be generated, and to construct the causal decompositions $p(\vec a|\vec x) = \sum_\pi p(\pi,\vec a|\vec x)$ explicitly.

\subsubsection{Correlations induced by QC-CCs are in ${\cal P}_\textup{causal}$}
\label{app:corr_QC-CC}

A process matrix $W\in\textup{\textsf{QC-CC}}$ has a decomposition as in Eq.~\eqref{eq:charact_W_QCCC_decomp}; recalling that we consider here some trivial global past ($\HS^P$) and future ($\HS^F$) spaces, we have in particular that $W = \sum_{(k_1,\ldots,k_N)} W_{(k_1,\ldots,k_N)} \otimes \id^{A_{k_N}^O}$.

Let us then define
\begin{align}
    p\big((k_1,\ldots,k_N),\vec a\big|\vec x\big) \coloneqq \Tr[(M_{\vec a|\vec x})^T(W_{(k_1,\ldots,k_N)}\otimes\id^{A_{k_N}^O})], \ \text{such that} \ p(\vec a|\vec x) = \sum_{(k_1,\ldots,k_N)} p\big((k_1,\ldots,k_N),\vec a\big|\vec x\big). \label{eq:proof_QCCC_causal_decomp1}
\end{align}
From Eq.~\eqref{eq:charact_W_QCCC_decomp} and using the trace-preserving property that $\sum_{a_k} \Tr_{A_k^O} M_{a_k|x_k} = \id^{A_k^I}$ for each instrument, it is easy to see, recursively, that (for all $n = 0,\ldots,N-1$ and all $(k_1,\ldots,k_n,k_{n+1})$)
\begin{align}
    p\big((k_1,\ldots,k_n,k_{n+1}),\vec a_{k_1,\ldots,k_n}\big|\vec x\big) & = \sum_{\substack{(k_{n+2},\ldots,k_N)\\ \vec a_{k_{n+1},k_{n+2},\ldots,k_N}}} p\big((k_1,\ldots,k_n,k_{n+1},k_{n+2},\ldots,k_N),\vec a\big|\vec x\big) \notag \\
    & = \Tr[\big(M_{\vec a_{k_1,\ldots,k_n}|\vec x_{k_1,\ldots,k_n}}\big)^T\big(\Tr_{A_{k_{n+1}}^I}W_{(k_1,\ldots,k_{n+1})}\big)]. \label{eq:proof_QCCC_causal_decomp2}
\end{align}

Indeed, for $n = N-1$,
\begin{align}
    p\big((k_1,\ldots,k_N),\vec a_{k_1,\ldots,k_{N-1}}\big|\vec x\big) & = \sum_{a_{k_N}} \Tr[(M_{\vec a|\vec x})^T(W_{(k_1,\ldots,k_N)}\otimes\id^{A_{k_N}^O})] \notag \\
    & = \Tr\,\! \Big[\Big(M_{\vec a_{k_1,\ldots,k_{N-1}}|\vec x_{k_1,\ldots,k_{N-1}}} \otimes \underbrace{{\textstyle \sum_{a_{k_N}}} \Tr_{A_{k_N}^O} M_{a_{k_N}|x_{k_N}}}_{\id^{A_{k_N}^I}}\Big)^T \, W_{(k_1,\ldots,k_N)}\Big] \notag \\
    & = \Tr [\big(M_{\vec a_{k_1,\ldots,k_{N-1}}|\vec x_{k_1,\ldots,k_{N-1}}}\big)^T \big(\Tr_{A_{k_N}^I} W_{(k_1,\ldots,k_N)}\big)]. \label{eq:proof_QCCC_causal_decomp3}
\end{align}
Assuming then that Eq.~\eqref{eq:proof_QCCC_causal_decomp2} holds for some value $n = 1,\ldots,N-1$, we have, for $n' = n-1$,
\begin{align}
    p\big((k_1,\ldots,k_{n'},k_{n'+1}),\vec a_{k_1,\ldots,k_{n'}}\big|\vec x\big) & = \sum_{a_{k_{n'+1}}, k_{n'+2}} p\big((k_1,\ldots,k_{n'},k_{n'+1},k_{n'+2}),\vec a_{k_1,\ldots,k_{n'},k_{n'+1}}\big|\vec x\big) \notag \\
    & = \sum_{a_{k_{n'+1}}, k_{n'+2}} \Tr[\big(M_{\vec a_{k_1,\ldots,k_{n'+1}}|\vec x_{k_1,\ldots,k_{n'+1}}}\big)^T\big(\Tr_{A_{k_{n'+2}}^I}W_{(k_1,\ldots,k_{n'+1},k_{n'+2})}\big)] \notag \\
    & = \Tr\,\! \Big[\big(M_{\vec a_{k_1,\ldots,k_{n'}}|\vec x_{k_1,\ldots,k_{n'}}}\!\!\otimes\!\!\!\sum_{a_{k_{n'+1}}}\!\!\!M_{a_{k_{n'+1}}|x_{k_{n'+1}}} \big)^T \Big(\underbrace{\!\sum_{k_{n'+2}} \!\Tr_{A_{k_{n'+2}}^I}\!\!W_{(k_1,\ldots,k_{n'+1},k_{n'+2})}\!\!}_{W_{(k_1,\ldots,k_{n'+1})}\otimes\id^{A_{k_{n'+1}}^O}}\Big)\Big] \notag \\
    & = \Tr\,\! \Big[\big(M_{\vec a_{k_1,\ldots,k_{n'}}|\vec x_{k_1,\ldots,k_{n'}}}\otimes\underbrace{{\textstyle \sum_{a_{k_{n'+1}}}} \Tr_{A_{k_{n'+1}}^O} M_{a_{k_{n'+1}}|x_{k_{n'+1}}}}_{\id^{A_{k_{n'+1}}^I}} \big)^T \,W_{(k_1,\ldots,k_{n'+1})} \Big] \notag \\
    & = \Tr[\big(M_{\vec a_{k_1,\ldots,k_{n'}}|\vec x_{k_1,\ldots,k_{n'}}}\big)^T\big(\Tr_{A_{k_{n'+1}}^I}W_{(k_1,\ldots,k_{n'+1})}\big)], \label{eq:proof_QCCC_causal_decomp4}
\end{align}
which shows that Eq.~\eqref{eq:proof_QCCC_causal_decomp2} holds for the value $n' = n-1$, and indeed recursively proves that it holds for all $n = 0,\ldots,N-1$.

\medskip

Eq.~\eqref{eq:proof_QCCC_causal_decomp2} implies that $p\big((k_1,\ldots,k_n,k_{n+1}),\vec a_{k_1,\ldots,k_n}\big|\vec x\big)$ does not depend on $\vec x_{\mathcal{N}\backslash\{k_1,\ldots,k_n\}}$, so that the decomposition $p(\vec a|\vec x) = \sum_{(k_1,\ldots,k_N)} p\big((k_1,\ldots,k_N),\vec a\big|\vec x\big)$ from Eq.~\eqref{eq:proof_QCCC_causal_decomp1} satisfies Eq.~\eqref{eq:def_causal_p}---which formally proves that the correlations $p(\vec a|\vec x)$ obtained from any QC-CC are causal.

\subsubsection{Correlations induced by QC-convFOs are in ${\cal P}_\textup{convFO}$}
\label{app:corr_QC-convFO}

By definition a quantum circuit with fixed causal order can only generate correlations that are compatible with the same fixed causal order.

It then straightforwardly follows, by linearity, that any quantum circuit in \textup{\textsf{QC-convFO}} can only generate correlations in ${\cal P}_\textup{convFO}$.

\subsubsection{Correlations induced by QC-NICCs are in ${\cal P}_\textup{NIO}$}
\label{app:corr_QC-NICC}

Recall that according to Eq.~\eqref{eq:cstr_NICC}, process matrices in \textup{\textsf{QC-NICC}} are QC-CCs with a decomposition in which all $W_{(k_1,\ldots,k_N)}$'s (and hence all $W_{(k_1,\ldots,k_N)}\otimes\id^{A_{k_N}^O}$'s) are proportional to valid process matrices; let us write $W_{(k_1,\ldots,k_N)}\otimes\id^{A_{k_N}^O} = q_{(k_1,\ldots,k_N)}\,\tilde W_{(k_1,\ldots,k_N)}$, for some fixed weights $q_{(k_1,\ldots,k_N)} (\ge 0)$ and some valid (now properly normalised) process matrices $\tilde W_{(k_1,\ldots,k_N)}$.

Recall also that valid process matrices $\tilde W$ satisfy $\sum_{\vec a} \Tr[\big(M_{\vec a|\vec x} \big)^T \, \tilde W] = 1$, independently of $\vec x$.
Hence, for a process matrix in \textup{\textsf{QC-NICC}} and its appropriate decomposition, $p\big((k_1,\ldots,k_N),\vec a\big|\vec x\big) \coloneqq \Tr[(M_{\vec a|\vec x})^T(W_{(k_1,\ldots,k_N)}\otimes\id^{A_{k_N}^O})]$ as defined in Eq.~\eqref{eq:proof_QCCC_causal_decomp1} above (and which, as we showed already, provides a causal decomposition for $p(\vec a|\vec x)$ satisfying Eq.~\eqref{eq:def_causal_p}) is then such that $p\big((k_1,\ldots,k_N)\big|\vec x\big) = \sum_{\vec a} p\big((k_1,\ldots,k_N),\vec a\big|\vec x\big) = \sum_{\vec a} \Tr[(M_{\vec a|\vec x})^T(W_{(k_1,\ldots,k_N)}\otimes\id^{A_{k_N}^O})] = q_{(k_1,\ldots,k_N)}$, independently of $\vec x$ (notice that this implies that $\sum_{(k_1,\ldots,k_N)} q_{(k_1,\ldots,k_N)} = 1$). This shows (recalling Eq.~\eqref{eq:def_NIO}) that the correlation $p(\vec a|\vec x)$ is indeed in ${\cal P}_\textup{NIO}$.

\subsection{Correlations induced by QC-QCs}

Although QC-QCs can be causally nonseparable~\cite{oreshkov12}, and therefore feature some form of indefinite causal orders, it was shown in Ref.~\cite{wechs21} that they can only generate causal correlations.
This result is less trivial than for QC-CCs. We will present below a new way to write its proof, which provides a causal decomposition of the form $p(\vec a|\vec x) = \sum_\pi p(\pi,\vec a|\vec x)$, as in Eq.~\eqref{eq:def_causal_p}, and thereby makes more explicit (and may perhaps provide a better intuition of) how the causal order arises in the induced correlations.

\subsubsection{Correlations induced by QC-QCs are in ${\cal P}_\textup{causal}$}
\label{app:corr_QC-QC}

Consider the process matrix $W$ of a QC-QC, which has a decomposition as in Eq.~\eqref{eq:charact_W_QCQC_decomp} (once again, considering here trivial global past and future spaces $\HS^P, \HS^F$), and let us define, $\forall \, n=1,\ldots,N, \ \forall \, \mathcal{K}_n \subseteq \mathcal{N}, \ \forall \, k_n \in \mathcal{K}_n$,
\begin{align}
    s_{({\cal K}_n\backslash k_n,k_n)}(\vec a_{{\cal K}_n}|\vec x_{{\cal K}_n}) & \coloneqq \Tr[(M_{\vec a_{{\cal K}_n}|\vec x_{{\cal K}_n}})^T\big(W_{({\cal K}_n\backslash k_n,k_n)}\otimes\id^{A_{k_n}^O}\big)] \ge 0 \notag \\
    \text{and} \quad f_{{\cal K}_n}(\vec a_{{\cal K}_n}|\vec x_{{\cal K}_n}) & \coloneqq \!\!\sum_{k_n\in{\cal K}_n}\!\! s_{({\cal K}_n\backslash k_n,k_n)}(\vec a_{{\cal K}_n}|\vec x_{{\cal K}_n}) = \Tr\Big[(M_{\vec a_{{\cal K}_n}|\vec x_{{\cal K}_n}})^T\Big(\sum_{k_n\in{\cal K}_n}\!\! W_{({\cal K}_n\backslash k_n,k_n)}\otimes\id^{A_{k_n}^O}\Big)\Big] \ge 0. \label{eq:def_s_f}
\end{align}

With these,%
\footnote{The watchful reader may notice that we introduce here the same definitions for $s_{({\cal K}_n\backslash k_n,k_n)}(\vec a_{{\cal K}_n}|\vec x_{{\cal K}_n})$ and $f_{{\cal K}_n}(\vec a_{{\cal K}_n}|\vec x_{{\cal K}_n})$ as in the proof of Ref.~\cite{wechs21}. Since it is always clear what the arguments $(\vec a_{{\cal K}_n}|\vec x_{{\cal K}_n})$ of $s_{({\cal K}_n\backslash k_n,k_n)}$ and $f_{{\cal K}_n}$ must be, to lighten the notations we shall sometimes just write $s_{({\cal K}_n\backslash k_n,k_n)}(\cdot|\cdot)$ and $f_{{\cal K}_n}(\cdot|\cdot)$, as e.g.\ in Eq.~\eqref{eq:proof_QCQC_causal_correl}.\\
For simplicity we assume in the upcoming equations that all $f_{\{k_1,\ldots,k_{n-1}\}}(\cdot|\cdot)>0$; otherwise the corresponding summands in Eq.~\eqref{eq:proof_QCQC_causal_correl}, i.e.\ $p((k_1,\ldots,k_N),\vec a|\vec x)$, are simply 0. \label{ftn:nonzero_denom_1}}
we easily obtain, iteratively, that
\begin{align}
    p(\vec a|\vec x) & = \sum_{k_N} s_{({\cal N}\backslash k_N,k_N)}(\vec a|\vec x) = \sum_{k_N} f_{{\cal N}\backslash k_N}(\cdot|\cdot) \frac{s_{({\cal N}\backslash k_N,k_N)}(\cdot|\cdot)}{f_{{\cal N}\backslash k_N}(\cdot|\cdot)} \notag \\
    & = \!\!\sum_{(k_{N-1},k_N)}\!\!\!\!\! s_{({\cal N}\backslash \{k_{N-1},k_N\},k_{N-1})}(\cdot|\cdot) \frac{s_{({\cal N}\backslash k_N,k_N)}(\cdot|\cdot)}{f_{{\cal N}\backslash k_N}(\cdot|\cdot)} = \!\!\sum_{k_{N-1},k_N}\!\!\!\! f_{{\cal N}\backslash \{k_{N-1},k_N\}}(\cdot|\cdot) \frac{s_{({\cal N}\backslash \{k_{N-1},k_N\},k_{N-1})}(\cdot|\cdot)}{f_{{\cal N}\backslash \{k_{N-1},k_N\}}(\cdot|\cdot)} \frac{s_{({\cal N}\backslash k_N,k_N)}(\cdot|\cdot)}{f_{{\cal N}\backslash k_N}(\cdot|\cdot)} \notag \\
    & = \!\!\sum_{(k_{N-2},k_{N-1},k_N)}\!\!\!\! s_{({\cal N}\backslash \{k_{N-2},k_{N-1},k_N\},k_{N-2})}(\cdot|\cdot) \, \frac{s_{({\cal N}\backslash \{k_{N-1},k_N\},k_{N-1})}(\cdot|\cdot)}{f_{{\cal N}\backslash \{k_{N-1},k_N\}}(\cdot|\cdot)} \, \frac{s_{({\cal N}\backslash k_N,k_N)}(\cdot|\cdot)}{f_{{\cal N}\backslash k_N}(\cdot|\cdot)} \notag \\[1mm]
    & = \ \cdots \notag \\
    & = \sum_{(k_1,\ldots,k_N)} \underbrace{s_{(\emptyset,k_1)}(\cdot|\cdot)}_{\tilde p((k_1),a_{k_1}|x_{k_1})} \ \underbrace{\frac{s_{(\{k_1\},k_2)}(\cdot|\cdot)}{f_{\{k_1\}}(\cdot|\cdot)}}_{\tilde p((k_2),a_{k_2}|(k_1),\vec x_{k_1,k_2},a_{k_1})} 
    \hspace{-10mm} \underbrace{\frac{s_{(\{k_1,k_2\},k_3)}(\cdot|\cdot)}{f_{\{k_1,k_2\}}(\cdot|\cdot)}}_{\substack{\tilde p((k_3),a_{k_3}|(k_1,k_2),\\ \hspace{25mm}\vec x_{k_1,k_2,k_3},\vec a_{k_1,k_2})}} 
    \hspace{-3mm} \cdots \hspace{-5mm}
    \underbrace{\frac{s_{(\{k_1,\ldots,k_{N-1}\},k_N)}(\cdot|\cdot)}{f_{\{k_1,\ldots,k_{N-1}\}}(\cdot|\cdot)}}_{\substack{\tilde p((k_N),a_{k_N}|(k_1,\ldots,k_{N-1}),\\ \hspace{30mm}\vec x_{k_1,\ldots,k_N},\vec a_{k_1,\ldots,k_{N-1}})}} \notag \\
    & = \sum_{(k_1,\ldots,k_N)} \prod_{n=1}^N \tilde p\big((k_n),a_{k_n}\big|(k_1,\ldots,k_{n-1}),\vec x_{k_1,\ldots,k_n},\vec a_{k_1,\ldots,k_{n-1}}\big) = \sum_{(k_1,\ldots,k_N)} p\big((k_1,\ldots,k_N),\vec a\big|\vec x\big), \label{eq:proof_QCQC_causal_correl}
\end{align}
where we defined, for $n=1,\ldots,N$ (and with $f_\emptyset(\vec a_\emptyset|\vec x_\emptyset)\coloneqq 1$),
\begin{align}
    & \tilde p\big((k_n),a_{k_n}\big|(k_1,\ldots,k_{n-1}),\vec x_{k_1,\ldots,k_n},\vec a_{k_1,\ldots,k_{n-1}}\big) \coloneqq \frac{s_{(\{k_1,\ldots,k_{n-1}\},k_n)}(\vec a_{k_1,\ldots,k_n}|\vec x_{k_1,\ldots,k_n})}{f_{\{k_1,\ldots,k_{n-1}\}}(\vec a_{k_1,\ldots,k_{n-1}}|\vec x_{k_1,\ldots,k_{n-1}})} \ge 0 \notag \\
    \text{and}\ \ & p\big((k_1,\ldots,k_N),\vec a\big|\vec x\big) \coloneqq \prod_{n=1}^N \tilde p\big((k_n),a_{k_n}\big|(k_1,\ldots,k_{n-1}),\vec x_{k_1,\ldots,k_n},\vec a_{k_1,\ldots,k_{n-1}}\big) \ge 0. \label{eq:def_tilde_p}
\end{align}

Using the trace-preserving property ($\sum_{a_k} \Tr_{A_k^O} M_{a_k|x_k} = \id^{A_k^I}$), we have that
\begin{align}
    & \tilde p\big((k_n)\big|(k_1,\ldots,k_{n-1}),\vec x_{k_1,\ldots,k_n},\vec a_{k_1,\ldots,k_{n-1}}\big) \coloneqq \sum_{a_{k_n}} \tilde p\big((k_n),a_{k_n}\big|(k_1,\ldots,k_{n-1}),\vec x_{k_1,\ldots,k_n},\vec a_{k_1,\ldots,k_{n-1}}\big) \notag \\[-2mm]
    & \hspace{25mm} = \frac{\sum_{a_{k_n}} \Tr[(M_{\vec a_{k_1,\ldots,k_n}|\vec x_{k_1,\ldots,k_n}})^T(W_{(\{k_1,\ldots,k_{n-1}\},k_n)}\otimes\id^{A_{k_n}^O})]}{f_{\{k_1,\ldots,k_{n-1}\}}(\vec a_{k_1,\ldots,k_{n-1}}|\vec x_{k_1,\ldots,k_{n-1}})} \notag \\
    & \hspace{25mm} = \frac{\Tr\Big[(M_{\vec a_{k_1,\ldots,k_{n-1}}|\vec x_{k_1,\ldots,k_{n-1}}}\otimes \overbrace{{\textstyle \sum_{a_{k_n}}}\Tr_{A_{k_n}^O}M_{a_{k_n}|x_{k_n}}}^{\id^{A_{k_n}^I}}\,)^T\,W_{(\{k_1,\ldots,k_{n-1}\},k_n)}\Big]}{f_{\{k_1,\ldots,k_{n-1}\}}(\vec a_{k_1,\ldots,k_{n-1}}|\vec x_{k_1,\ldots,k_{n-1}})} \notag \\
    & \hspace{25mm} = \frac{\Tr\Big[(M_{\vec a_{k_1,\ldots,k_{n-1}}|\vec x_{k_1,\ldots,k_{n-1}}})^T\,\big(\Tr_{A_{k_n}^I} W_{(\{k_1,\ldots,k_{n-1}\},k_n)}\big)\Big]}{f_{\{k_1,\ldots,k_{n-1}\}}(\vec a_{k_1,\ldots,k_{n-1}}|\vec x_{k_1,\ldots,k_{n-1}})} \label{eq:marginal_tilde_p}
\end{align}
does not depend on $x_{k_n}$.
Using the constraints of Eq.~\eqref{eq:charact_W_QCQC_decomp}, Eq.~\eqref{eq:marginal_tilde_p} further leads, for $n\ge 2$ and with $\mathcal{K}_{n-1}\coloneqq \{k_1,\ldots,k_{n-1}\}$, to
\begin{align}
    & \sum_{k_n, a_{k_n}} \tilde p\big((k_n),a_{k_n}\big|(k_1,\ldots,k_{n-1}),\vec x_{k_1,\ldots,k_n},\vec a_{k_1,\ldots,k_{n-1}}\big) \notag \\[-3mm]
    & \hspace{25mm} = \frac{\Tr\Big[(M_{\vec a_{\mathcal{K}_{n-1}}|\vec x_{\mathcal{K}_{n-1}}})^{\,T}\big(\sum_{k_n\notin \mathcal{K}_{n-1}}\Tr_{A_{k_n}^I}\!W_{(\mathcal{K}_{n-1},k_n)}\big) \Big]}{\Tr\Big[(M_{\vec a_{\mathcal{K}_{n-1}}|\vec x_{\mathcal{K}_{n-1}}})^{\,T}\big(\sum_{k_{n-1}'\in\mathcal{K}_{n-1}} W_{(\mathcal{K}_{n-1}\backslash k_{n-1}',k_{n-1}')}\otimes\id^{A_{k_{n-1}'}^O}\big)\Big]} = 1, \label{eq:norm_tilde_p}
\end{align}
and similarly for $n=1$, $\sum_{k_1, a_{k_1}} \tilde p\big((k_1),a_{k_1}\big|x_{k_1}\big) = \sum_{k_1} \Tr[W_{(\emptyset,k_1)}] = 1$.

Hence Eq.~\eqref{eq:proof_QCQC_causal_correl} provides a decomposition of $p(\vec a|\vec x)$ as in Eq.~\eqref{eq:def_causal_p_v4} of Appendix~\ref{app:more_causal_charact}, with each $\tilde p\big((k_n),a_{k_n}\big|\cdot,\cdot,\cdot\big)$ satisfying the necessary constraints, which indeed proves that the correlations induced by a QC-QC are causal.

\subsubsection{Correlations induced by QC-supFOs are in ${\cal P}_\textup{convFO}$}
\label{app:corr_QC-supFO}

Consider now a process matrix $W \in \textup{\textsf{QC-supFO}}$, which has a decomposition as in Eq.~\eqref{eq:charact_W_QC_SupFO_decomp} (still with some trivial $\HS^P, \HS^F$), and let us now define, $\forall \, n=1,\ldots,N, \ \forall \, \mathcal{K}_n \subseteq \mathcal{N}, \ \forall \, k_n \in \mathcal{K}_n, \ \forall \, (k_{n+1},\ldots,k_N) \in \mathcal{N}\setminus \mathcal{K}_n$,
\begin{align}
    s_{({\cal K}_n\backslash k_n,k_n)}^{(k_{n+1},\ldots,k_N)}(\vec a_{{\cal K}_n}|\vec x_{{\cal K}_n}) & \coloneqq \Tr[(M_{\vec a_{{\cal K}_n}|\vec x_{{\cal K}_n}})^T\big(W_{({\cal K}_n\backslash k_n,k_n)}^{(k_{n+1},\ldots,k_N)}\otimes\id^{A_{k_n}^O}\big)] \ge 0 \notag \\
    \text{and} \quad f_{{\cal K}_n}^{(k_{n+1},\ldots,k_N)}(\vec a_{{\cal K}_n}|\vec x_{{\cal K}_n}) & \coloneqq \!\!\sum_{k_n\in{\cal K}_n}\!\! s_{({\cal K}_n\backslash k_n,k_n)}^{(k_{n+1},\ldots,k_N)}(\vec a_{{\cal K}_n}|\vec x_{{\cal K}_n}) = \Tr\Big[(M_{\vec a_{{\cal K}_n}|\vec x_{{\cal K}_n}})^T\Big(\sum_{k_n\in{\cal K}_n}\!\! W_{({\cal K}_n\backslash k_n,k_n)}^{(k_{n+1},\ldots,k_N)}\otimes\id^{A_{k_n}^O}\Big)\Big] \ge 0. \label{eq:def_s_f_2}
\end{align}

Following similar calculations as in the first few lines of Eq.~\eqref{eq:proof_QCQC_causal_correl} (just adding the superscripts ${}^{(k_{n+1},\ldots,k_N)}$ wherever we need them), we can write here%
\footnote{In Eq.~\eqref{eq:proof_QCsupFO_convFO_correl} we again assume, for simplicity, that all $f_{\{k_1,\ldots,k_{n-1}\}}^{(k_n,\ldots,k_N)}(\cdot|\cdot)>0$; otherwise, the corresponding summands are simply 0. In the first line of Eq.~\eqref{eq:def_tilde_p_supFO} we assume that $q_\pi>0$; otherwise $\tilde p_{\pi,k_1}(a_{k_1}|x_{k_1})$ can be taken to be any valid conditional probability distribution. \label{ftn:nonzero_denom_2}}
\begin{align}
    p(\vec a|\vec x) & = \sum_{\pi=(k_1,\ldots,k_N)} 
    \underbrace{s_{(\emptyset,k_1)}^{(k_2,\ldots,k_N)}(\cdot|\cdot)}_{q_\pi \, \tilde p_{\pi,k_1}(a_{k_1}|x_{k_1})} \ \underbrace{\frac{s_{(\{k_1\},k_2)}^{(k_3,\ldots,k_N)}(\cdot|\cdot)}{f_{\{k_1\}}^{(k_2,\ldots,k_N)}(\cdot|\cdot)}}_{\tilde p_{\pi,k_2}(a_{k_2}|\vec x_{k_1,k_2},a_{k_1})} 
    \ \underbrace{\frac{s_{(\{k_1,k_2\},k_3)}^{(k_4,\ldots,k_N)}(\cdot|\cdot)}{f_{\{k_1,k_2\}}^{(k_3,\ldots,k_N)}(\cdot|\cdot)}}_{\substack{\tilde p_{\pi,k_3}(a_{k_3}|\vec x_{k_1,k_2,k_3}, \\ \hspace{18mm}\vec a_{k_1,k_2})}} 
    \ \cdots 
    \underbrace{\frac{s_{(\{k_1,\ldots,k_{N-1}\},k_N)}^\emptyset(\cdot|\cdot)}{f_{\{k_1,\ldots,k_{N-1}\}}^{(k_N)}(\cdot|\cdot)}}_{\substack{\tilde p_{\pi,k_N}(a_{k_N}|\vec x_{k_1,\ldots,k_N}, \\ \hspace{20mm}\vec a_{k_1,\ldots,k_{N-1}})}}
    \notag \\
    & = \sum_{\pi=(k_1,\ldots,k_N)} q_\pi \prod_{n=1}^N \tilde p_{\pi,k_n}(a_{k_n}|\vec x_{k_1,\ldots,k_n},\vec a_{k_1,\ldots,k_{n-1}}) = \sum_{\pi=(k_1,\ldots,k_N)} q_\pi \, p_\pi(\vec a|\vec x), \label{eq:proof_QCsupFO_convFO_correl}
\end{align}
with, for $\pi=(k_1,\ldots,k_N)$,
\begin{align}
    & \tilde p_{\pi,k_1}(a_{k_1}|x_{k_1}) \coloneqq \frac{s_{(\emptyset,k_1)}^{(k_2,\ldots,k_N)}(a_{k_1}|x_{k_1})}{q_\pi} \ge 0, \notag \\
    & \tilde p_{\pi,k_n}(a_{k_n}|\vec x_{k_1,\ldots,k_n}, \vec a_{k_1,\ldots,k_{n-1}}) \coloneqq \frac{s_{(\{k_1,\ldots,k_{n-1}\},k_n)}^{(k_{n+1},\ldots,k_N)}(\vec a_{k_1,\ldots,k_n}|\vec x_{k_1,\ldots,k_n})}{f_{\{k_1,\ldots,k_{n-1}\}}^{(k_n,\ldots,k_N)}(\vec a_{k_1,\ldots,k_{n-1}}|\vec x_{k_1,\ldots,k_{n-1}})} \ge 0 \ \ \ \text{for} \ n= 2,\ldots,N, \notag \\
    \text{and}\ \ & p_\pi(\vec a|\vec x) \coloneqq \prod_{n=1}^N \tilde p_{\pi,k_n}(a_{k_n}|\vec x_{k_1,\ldots,k_n},\vec a_{k_1,\ldots,k_{n-1}}) \ge 0. \label{eq:def_tilde_p_supFO}
\end{align}

Now, using again the trace-preserving property together with the constraints of Eq.~\eqref{eq:charact_W_QC_SupFO_decomp}, we have that
\begin{align}
    & \sum_{a_{k_1}} \tilde p_{\pi,k_1}(a_{k_1}|x_{k_1}) = \frac{\Tr\Big[(\overbrace{{\textstyle \sum_{a_{k_1}}} \Tr_{A_{k_1}^O} M_{a_{k_1}|x_{k_1}}}^{\id^{A_{k_1}^I}})^T\,W_{(\emptyset,k_1)}^{(k_2,\ldots,k_N)}\Big]}{q_\pi} = \frac{\Tr\big[W_{(\emptyset,k_1)}^{(k_2,\ldots,k_N)}\big]}{q_\pi} = 1
\end{align}
and for $n\ge 2$, with $\mathcal{K}_{n-1}\coloneqq \{k_1,\ldots,k_{n-1}\}$ (similarly to Eqs.~\eqref{eq:marginal_tilde_p} and~\eqref{eq:norm_tilde_p} before)
\begin{align}
    \sum_{a_{k_n}} \tilde p_{\pi,k_n}(a_{k_n}|\vec x_{k_1,\ldots,k_n}, \vec a_{k_1,\ldots,k_{n-1}}) & = \frac{\Tr\Big[(M_{\vec a_{k_1,\ldots,k_{n-1}}|\vec x_{k_1,\ldots,k_{n-1}}}\!\otimes\! \overbrace{{\textstyle \sum_{a_{k_n}}} \!\Tr_{A_{k_n}^O} \!M_{a_{k_n}|x_{k_n}}}^{\id^{A_{k_n}^I}})^T\,W_{(\{k_1,\ldots,k_{n-1}\},k_n)}^{(k_{n+1},\ldots,k_N)}\Big]}{f_{\{k_1,\ldots,k_{n-1}\}}^{(k_n,\ldots,k_N)}(\vec a_{k_1,\ldots,k_{n-1}}|\vec x_{k_1,\ldots,k_{n-1}})} \notag \\
    & = \frac{\Tr\Big[(M_{\vec a_{{\cal K}_{n-1}}|\vec x_{{\cal K}_{n-1}}})^{\,T}\big(\Tr_{A_{k_n}^I}W_{({\cal K}_{n-1},k_n)}^{(k_{n+1},\ldots,k_N)}\big)\Big]}{\Tr[(M_{\vec a_{{\cal K}_{n-1}}|\vec x_{{\cal K}_{n-1}}})^{\,T}\big(\sum_{k_{n-1}'\in{\cal K}_{n-1}} \!W_{\!({\cal K}_{n-1}\backslash k_{n-1}',k_{n-1}')}^{(k_n,\ldots,k_N)}\!\otimes\!\id^{A_{k_{n-1}'}^O}\big)]} = 1. \label{eq:sum_akn_tilde_p_supFO}
\end{align}

Hence for each $n=1,\ldots,N$, $\tilde p_{\pi,k_n}(a_{k_n}|\vec x_{k_1,\ldots,k_n},\vec a_{k_1,\ldots,k_{n-1}})$ defines a valid (normalised) probability distribution for party $A_{k_n}$'s output, which only depends on the inputs and outputs of the parties in the past of $A_{k_n}$, according to the order $\pi=(k_1,\ldots,k_N)$, and of $A_{k_n}$'s own input.
The definition of $p_\pi(\vec a|\vec x)$ in Eq.~\eqref{eq:def_tilde_p_supFO} then implies that $p_\pi(\vec a_{k_1,\ldots,k_n}|\vec x) = \prod_{j=1}^n \tilde p_{\pi,k_j}(a_{k_j}|\vec x_{k_1,\ldots,k_j},\vec a_{k_1,\ldots,k_{j-1}})$ does not depend on $\vec x_{k_{n+1},\ldots,k_N}$, i.e., that $p_\pi(\vec a|\vec x)$ is compatible with the fixed causal order $\pi$.

From the decomposition of Eq.~\eqref{eq:proof_QCsupFO_convFO_correl}, we then conclude that process matrices in \textup{\textsf{QC-supFO}} only give correlations that are convex mixtures of correlations with fixed orders, i.e., correlations in ${\cal P}_\textup{convFO}$.

\subsubsection{Correlations induced by QC-NIQCs are in ${\cal P}_{\textup{NIO}'}$}
\label{app:corr_QC-NIQC}

Consider a process matrix $W$ in \textup{\textsf{QC-NIQC}} (still with some trivial $\HS^P, \HS^F$), i.e., according to Eq.~\eqref{eq:cstr_NIQC}, a QC-QC with a decomposition as in Eq.~\eqref{eq:charact_W_QCQC_decomp}, in which all $W_{({\cal K}_{n-1},k_n)}$'s are proportional to valid process matrices.

As we have seen in Appendix~\ref{app:corr_QC-QC} above, the correlations induced by such a process matrix are causal, and an explicit causal decomposition for these can be constructed as in Eq.~\eqref{eq:proof_QCQC_causal_correl}.
For this decomposition, let us calculate (for any $n=1,\ldots,N$, any $\mathcal{K}_{n-1} \subsetneq \mathcal{N}$ and any $k_n \in \mathcal{N}\setminus \mathcal{K}_{n-1}$, using in particular Eq.~\eqref{eq:proof_causal_p_v4}):
\begin{align}
    & p\big((\mathcal{K}_{n-1},k_n),\vec a_{\mathcal{K}_{n-1},k_n}\big|\vec x\big) \coloneqq \!\sum_{(k_1,\ldots,k_{n-1}) \in \mathcal{K}_{n-1}} \!\!\!\!\!\!p\big((k_1,\ldots,k_{n-1},k_n),\vec a_{k_1,\ldots,k_{n-1},k_n}\big|\vec x\big) = \!\sum_{(k_1,\ldots,k_{n-1}) \in \mathcal{K}_{n-1}} \prod_{j=1}^n \tilde p\big((k_j),a_{k_j}\big|\cdot,\cdot,\cdot\big) \notag \\
    & \hspace{5mm} = \sum_{(k_1,\ldots,k_{n-1}) \in \mathcal{K}_{n-1}} \cancel{s_{(\emptyset,k_1)}(\cdot|\cdot)} \ \frac{s_{(\{k_1\},k_2)}(\cdot|\cdot)}{\cancel{f_{\{k_1\}}(\cdot|\cdot)}} \ \frac{s_{(\{k_1,k_2\},k_3)}(\cdot|\cdot)}{f_{\{k_1,k_2\}}(\cdot|\cdot)} \ \cdots \ \frac{s_{(\{k_1,\ldots,k_{n-2}\},k_{n-1})}(\cdot|\cdot)}{f_{\{k_1,\ldots,k_{n-2}\}}(\cdot|\cdot)} \ \frac{s_{(\{k_1,\ldots,k_{n-1}\},k_n)}(\cdot|\cdot)}{f_{\{k_1,\ldots,k_{n-1}\}}(\cdot|\cdot)} \notag \\
    & \hspace{5mm} = \hspace{-3mm} \sum_{\substack{\mathcal{K}_2 \subseteq \mathcal{K}_{n-1},\\(k_3,\ldots,k_{n-1}) \in \mathcal{K}_{n-1}\backslash\mathcal{K}_2}} \hspace{-5mm} \Big( \cancel{{\textstyle \sum_{k_2\in\mathcal{K}_2}} \, s_{(\mathcal{K}_2\backslash k_2,k_2)}(\cdot|\cdot)} \Big) \ \frac{s_{(\mathcal{K}_2,k_3)}(\cdot|\cdot)}{\cancel{f_{\mathcal{K}_2}(\cdot|\cdot)}} \ \frac{s_{(\mathcal{K}_2\cup\{k_3\},k_4)}(\cdot|\cdot)}{f_{\mathcal{K}_2\cup\{k_3\}}(\cdot|\cdot)} \ \cdots \notag \\[-5mm]
    & \hspace{95mm} \cdots \ \frac{s_{(\mathcal{K}_2\cup\{k_3,\ldots,k_{n-2}\},k_{n-1})}(\cdot|\cdot)}{f_{\mathcal{K}_2\cup\{k_3,\ldots,k_{n-2}\}}(\cdot|\cdot)} \ \frac{s_{(\mathcal{K}_2\cup\{k_3,\ldots,k_{n-1}\},k_n)}(\cdot|\cdot)}{f_{\mathcal{K}_2\cup\{k_3,\ldots,k_{n-1}\}}(\cdot|\cdot)} \notag \\[1mm]
    & \hspace{5mm} = \sum_{\substack{\mathcal{K}_3 \subseteq \mathcal{K}_{n-1},\\(k_4,\ldots,k_{n-1}) \in \mathcal{K}_{n-1}\backslash\mathcal{K}_3}} \Big( \cancel{{\textstyle \sum_{k_3\in\mathcal{K}_3}} \, s_{(\mathcal{K}_3\backslash k_3,k_3)}(\cdot|\cdot)} \Big) \, \frac{s_{(\mathcal{K}_3,k_4)}(\cdot|\cdot)}{\cancel{f_{\mathcal{K}_3}(\cdot|\cdot)}} \cdots \frac{s_{(\mathcal{K}_3\cup\{k_4,\ldots,k_{n-2}\},k_{n-1})}(\cdot|\cdot)}{f_{\mathcal{K}_3\cup\{k_4,\ldots,k_{n-2}\}}(\cdot|\cdot)} \, \frac{s_{(\mathcal{K}_3\cup\{k_4,\ldots,k_{n-1}\},k_n)}(\cdot|\cdot)}{f_{\mathcal{K}_3\cup\{k_4,\ldots,k_{n-1}\}}(\cdot|\cdot)} \notag \\
    & \hspace{5mm} = \ \ \cdots \ \ = \sum_{\substack{\mathcal{K}_{n-2} \subseteq \mathcal{K}_{n-1},\\ k_{n-1} \in \mathcal{K}_{n-1}\backslash\mathcal{K}_{n-2}}} \Big( \cancel{{\textstyle \sum_{k_{n-2}\in\mathcal{K}_{n-2}}} \, s_{(\mathcal{K}_{n-2}\backslash k_{n-2},k_{n-2})}(\cdot|\cdot)} \Big) \ \frac{s_{(\mathcal{K}_{n-2},k_{n-1})}(\cdot|\cdot)}{\cancel{f_{\mathcal{K}_{n-2}}(\cdot|\cdot)}} \ \frac{s_{(\mathcal{K}_{n-2}\cup\{k_{n-1}\},k_n)}(\cdot|\cdot)}{f_{\mathcal{K}_{n-2}\cup\{k_{n-1}\}}(\cdot|\cdot)} \notag \\
    & \hspace{5mm} = \Big( \cancel{\sum_{k_{n-1}\in\mathcal{K}_{n-1}} \, s_{(\mathcal{K}_{n-1}\backslash k_{n-1},k_{n-1})}(\cdot|\cdot)} \Big) \ \frac{s_{(\mathcal{K}_{n-1},k_n)}(\cdot|\cdot)}{\cancel{f_{\mathcal{K}_{n-1}}(\cdot|\cdot)}} \ = \ s_{(\mathcal{K}_{n-1},k_n)}(\cdot|\cdot) \notag \\
    & \hspace{5mm} = \Tr[(M_{\vec a_{{\cal K}_{n-1}\cup k_n}|\vec x_{{\cal K}_{n-1}\cup k_n}})^T\big(W_{({\cal K}_{n-1},k_n)}\otimes\id^{A_{k_n}^O}\big)].
\end{align}

By assumption all $W_{({\cal K}_{n-1},k_n)}$'s are valid up to normalisation, so that we can write $W_{({\cal K}_{n-1},k_n)}\otimes\id^{A_{k_n}^O} = q_{({\cal K}_{n-1},k_n)} \tilde W_{({\cal K}_{n-1},k_n)}$, for some weights $q_{({\cal K}_{n-1},k_n)} (\ge 0)$ and some (properly normalised) valid process matrices $\tilde W_{({\cal K}_{n-1},k_n)}$, which by definition satisfy $\sum_{\vec a_{{\cal K}_{n-1}\cup k_n}} \Tr[\big(M_{\vec a_{{\cal K}_{n-1}\cup k_n}|\vec x_{{\cal K}_{n-1}\cup k_n}} \big)^T \, \tilde W_{({\cal K}_{n-1},k_n)}] = 1$, independently of $\vec x_{{\cal K}_{n-1}\cup k_n}$. Hence we obtain that
\begin{align}
    p\big((\mathcal{K}_{n-1},k_n)\big|\vec x\big) & = \sum_{\vec a_{{\cal K}_{n-1}\cup k_n}} p\big((\mathcal{K}_{n-1},k_n),\vec a_{\mathcal{K}_{n-1},k_n}\big|\vec x\big) \notag \\
    & = q_{({\cal K}_{n-1},k_n)} \sum_{\vec a_{{\cal K}_{n-1}\cup k_n}} \Tr[(M_{\vec a_{{\cal K}_{n-1}\cup k_n}|\vec x_{{\cal K}_{n-1}\cup k_n}})^T\,\tilde W_{({\cal K}_{n-1},k_n)}] = q_{({\cal K}_{n-1},k_n)}
\end{align}
does not depend on $\vec x$ (notice that $\sum_{({\cal K}_{n-1},k_n)} q_{({\cal K}_{n-1},k_n)} = 1$). This proves that the correlations generated by the process matrix $W$ in \textup{\textsf{QC-NIQC}} are indeed in ${\cal P}_{\textup{NIO}'}$.

\subsubsection{Defining a new class whose correlations are in ${\cal P}_\textup{NIO}$}
\label{app:new_class}

While we showed above that correlations induced by QC-NICCs are in ${\cal P}_{\textup{NIO}}$ (as one may have expected), we could only show that correlations induced by QC-NIQCs are in ${\cal P}_{\textup{NIO}'}$, an (in general) strict superset of ${\cal P}_{\textup{NIO}}$. This is due to the structure of the quantum control system, which does not encode the full order of past operations, but only their unordered list. It remains an open question whether one can find correlations from QC-NIQCs that are in ${\cal P}_{\textup{NIO}'}\backslash{\cal P}_{\textup{NIO}}$.

One may also wonder whether one could define any other subclass of QC-NIQCs, larger than that of QC-NICCs, for which we could prove that the correlations are in ${\cal P}_{\textup{NIO}}$. We propose here such a subclass, inspired directly from the structure of the proofs above, by introducing precisely what was missing to prove that the correlations are in ${\cal P}_{\textup{NIO}}$ rather than just in ${\cal P}_{\textup{NIO}'}$. The basic intuition is that in order to prove that the correlations are in ${\cal P}_{\textup{NIO}}$, we need each full order $(k_1,\ldots,k_N)$ to appear ``somewhere in the decomposition'', and to have a valid process matrix (up to normalisation) attached to it (so as to generate a probability for that order that does not depend on the instruments).

\bigskip

Specifically, let us first define, for some fixed value of $n^* = 1,\ldots,N$, the class $\textup{\textsf{QC-NIO}}_{n^*}$ to be the set of $W \in \mathcal{L}(\mathcal{H}^{PA_\mathcal{N}^{IO}F})$, $W \ge 0$ such that there exist PSD matrices $W_{(k_1,\ldots,k_n)} \in {\cal L}(\HS^{PA_{\{k_1,\ldots,k_{n-1}\}}^{IO} A_{k_n}^I})$ (for all $n = 1, \ldots, n^*-1$ and all $(k_1,\ldots,k_n)$), $W_{(k_1,\ldots,k_{n^*})}^{(k_{n^*+1},\ldots,k_N)} \in {\cal L}(\HS^{PA_{\{k_1,\ldots,k_{n^*-1}\}}^{IO} A_{k_{n^*}}^I})$ (for all $(k_1,\ldots,k_{n^*},\ldots,k_N)$), $W_{({\cal K}_{n-1},k_n)}^{(k_{n+1},\ldots,k_N)} \in {\cal L}(\HS^{PA_{{\cal K}_{n-1}}^{IO} A_{k_n}^I})$ (for all strict subsets ${\cal K}_{n-1}$ of ${\cal N}$ with $n> n^*$ and all $(k_n, k_{n+1},\ldots,k_N) \in {\cal N}\backslash{\cal K}_{n-1}$) such that, defining $W_{(k_1,\ldots,k_{n^*})} \coloneqq \sum_{(k_{n^*+1},\ldots,k_N)} W_{(k_1,\ldots,k_{n^*})}^{(k_{n^*+1},\ldots,k_N)}$ and $W_{({\cal K}_{n^*} \backslash k_{n^*},k_{n^*})}^{(k_{n^*+1},\ldots,k_N)} \coloneqq \sum_{(k_1,\ldots,k_{n^*-1})\in{\cal K}_{n^*} \backslash k_{n^*}} W_{(k_1,\ldots,k_{n^*-1},k_{n^*})}^{(k_{n^*+1},\ldots,k_N)}$, we have
\begin{align}
    \left\{
    \begin{array}{l}
        \Tr_F W = \sum_{k_N \in {\cal N}} W_{({\cal N} \backslash k_N,k_N)}^{\emptyset}\otimes \id^{A_{k_N}^O}, \\[3mm]
        \forall \, n= n^*, \ldots,N-1, \ \forall \, {\cal K}_n \subsetneq {\cal N}, \, \forall \, (k_{n+1},\ldots,k_N) \in {\cal N}\backslash{\cal K}_n, \\[1mm]
        \hspace{15mm} \Tr_{A_{k_{n+1}}^I} W_{({\cal K}_n,k_{n+1})}^{(k_{n+2},\ldots,k_N)} = \sum_{k_n \in {\cal K}_n} W_{({\cal K}_n \backslash k_n,k_n)}^{(k_{n+1},\ldots,k_N)}\otimes \id^{A_{k_n}^O}, \\[3mm]
        \forall \, n = 1, \ldots, n^*-1, \ \forall \, (k_1, \ldots, k_n), \\[1mm]
        \hspace{15mm} \sum_{k_{n+1}} \Tr_{A_{k_{n+1}}^I} W_{(k_1,\ldots,k_n,k_{n+1})} = W_{(k_1,\ldots,k_n)} \otimes \id^{A_{k_n}^O}, \\[3mm]
        \sum_{k_1 \in {\cal N}} \Tr_{A_{k_1}^I} W_{(k_1)} = \id^P,
    \end{array}
    \right. \label{eq:W_NIOn_decomp}
\end{align}
and such that
\begin{equation}
    \text{all} \ W_{(k_1,\ldots,k_{n^*})}^{(k_{n^*+1},\ldots,k_N)} \text{'s are valid process matrices (up to normalisation).}\footnotemark
\end{equation}
\addtocounter{footnote}{-1}
\footnotetext{Notice that this, together with the constraints of Eq.~\eqref{eq:W_NIOn_decomp}, implies that all $W_{(k_1,\ldots,k_n)}$'s, for $n = 1, \ldots, n^*$, are also valid process matrices, up to normalisation.}%
As an illustration, the above set of constraints is depicted in Fig.~\ref{fig:graph_QCNIO_N4_n2} for the case of $N=4,n^*=2$, using the graphical representation introduced in Appendix~\ref{app:subsec:graphical_rep}. 

\begin{sidewaysfigure}
\centering
\includegraphics[width=\textwidth]{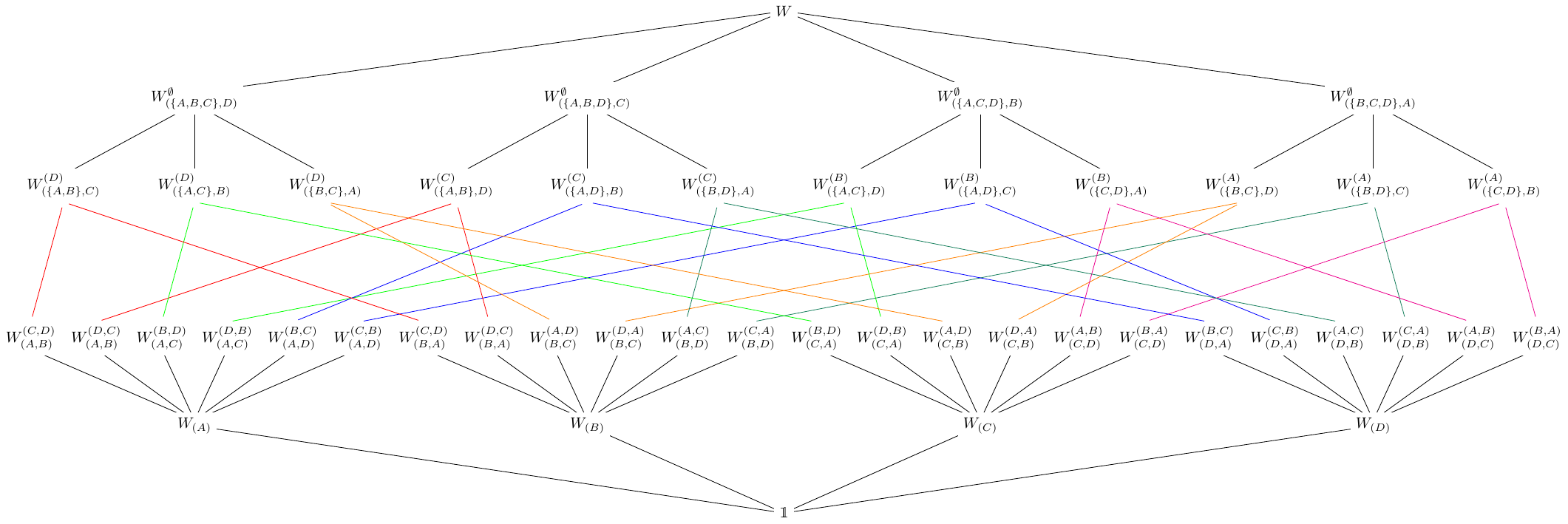}%
\vspace{1cm}
\caption{Graphical representation (as introduced in Appendix~\ref{app:subsec:graphical_rep}) of the SDP constraints of Eq.~\eqref{eq:W_NIOn_decomp}, for the $\textup{\textsf{QC-NIO}}_{n^*}$ class with $N=4$ and $n^*=2$. All matrices of the form $W_{(k_1,k_2)}^{(k_3,k_4)}$, in the layer corresponding to $n^*=2$, are assumed to be valid process matrices (up to normalisation)---which, here, implies the same for the matrices $W_{(k_1)}$ in the lower layer. Compared to the graph for the general \textup{\textsf{QC-QC}} class (for $N=4$ as well, shown in Fig.~\ref{fig:graph_QCQC}, bottom), one can see that the matrices on the $n^*=2$ layer split in two matrices each (corresponding to the two possible orders of the future parties)---to help visualise the changes, we kept the same colours as in Fig.~\ref{fig:graph_QCQC}~(bottom).
    \\
    For $N=4, n^*=3$ the structure of the graph would look symmetric to the present one, just upside-down: it is the matrices on the $n^*=3$ layer that would split in two matrices each (corresponding to the two possible orders of the past parties).}
\label{fig:graph_QCNIO_N4_n2}
\end{sidewaysfigure}

One can see that these classes $\textup{\textsf{QC-NIO}}_{n^*}$ interpolate between \textup{\textsf{QC-NICC}} (obtained%
\footnote{More precisely: with a trivial $\HS^F$ (as we consider when only looking at correlations) we get exactly \textup{\textsf{QC-NICC}} for $n^*=N$. With a nontrivial $\HS^F$, for $n^*=N$ we get in fact a larger class than \textup{\textsf{QC-NICC}}, for which the first two lines of Eq.~\eqref{eq:charact_W_QCCC_decomp} are replaced by $\Tr_F W = \sum_{(k_1,\ldots,k_N)} W_{(k_1,\ldots,k_N)} \otimes \id^{A_{k_N}^O}$. (This larger class includes in particular the class \textup{\textsf{Sup}} from Ref.~\cite{Liu23}, characterised as in Eq.~\eqref{eq:def_Sup}; it thus contains QC-QCs that are not QC-CCs.)}
for $n^*=N$) and \textup{\textsf{QC-supFO}} (obtained for $n^*=1$).%
\footnote{Notice that for $N=3$, the intermediate value $n^*=2$ also recovers the 
\textup{\textsf{QC-supFO}} class. Indeed, from Eq.~\eqref{eq:W_NIOn_decomp} one gets \emph{a priori} the same constraints as in Eq.~\eqref{eq:charact_W_QCQC_decomp}, for the \textup{\textsf{QC-QC}} class. The validity conditions of the $W_{(k_1,k_2)}^{(k_3)}$'s also imply here that the $W_{(k_1)}$'s are valid (up to normalisation). We thus obtain the characterisation of the \textup{\textsf{QC-NIQC}} class, which, in the $N=3$ case, reduces to \textup{\textsf{QC-supFO}} (cf p.~\pageref{proof:convFO_NIQC_N3}).} 
From these we define the class $\textup{\textsf{QC-NIO}}$ to be their convex hull:
\begin{align}
    \textup{\textsf{QC-NIO}} = \textup{conv}\Big(\bigcup_{n^*=1}^N \textup{\textsf{QC-NIO}}_{n^*}\Big),
\end{align}
which thus includes both \textup{\textsf{QC-NICC}} and \textup{\textsf{QC-supFO}}.
We will now show that process matrices in \textup{\textsf{QC-NIO}} only generate correlations in ${\cal P}_\textup{NIO}$.

\bigskip

By convexity, it suffices to prove that the same claim holds true for process matrices in $\textup{\textsf{QC-NIO}}_{n^*}$, for any fixed value of $n^*$.%
\footnote{Since the claim is already known to be true for $n^*=N$ (the \textup{\textsf{QC-NICC}} class) and for $n^*=1$ (the \textup{\textsf{QC-supFO}} class), we can restrict in the proof to $1<n^*<N$, to avoid bothering with these limit cases.}

Consider therefore a process matrix $W\in\textup{\textsf{QC-NIO}}_{n^*}$ with a decomposition as in Eq.~\eqref{eq:W_NIOn_decomp} with the $W_{(k_1,\ldots,k_{n^*})}^{(k_{n^*+1},\ldots,k_N)}$'s being proportional to valid process matrices, and with now some trivial $\HS^P$ and $\HS^F$.
Similarly to what we did before, let us start by defining here, for all subsets ${\cal K}_n$ of ${\cal N}$ with $n\ge n^*$, all $k_n\in\K_n$ and all $(k_{n+1},\ldots,k_N) \in {\cal N}\backslash{\cal K}_n$,
\begin{align}
    s_{({\cal K}_n\backslash k_n,k_n)}^{(k_{n+1},\ldots,k_N)}(\vec a_{{\cal K}_n}|\vec x_{{\cal K}_n}) & \coloneqq \Tr[(M_{\vec a_{{\cal K}_n}|\vec x_{{\cal K}_n}})^T\big(W_{({\cal K}_n\backslash k_n,k_n)}^{(k_{n+1},\ldots,k_N)}\otimes\id^{A_{k_n}^O}\big)] \ge 0 \notag \\
    \text{and} \quad f_{{\cal K}_n}^{(k_{n+1},\ldots,k_N)}(\vec a_{{\cal K}_n}|\vec x_{{\cal K}_n}) & \coloneqq \!\!\sum_{k_n\in{\cal K}_n}\!\! s_{({\cal K}_n\backslash k_n,k_n)}^{(k_{n+1},\ldots,k_N)}(\vec a_{{\cal K}_n}|\vec x_{{\cal K}_n}) = \Tr\Big[(M_{\vec a_{{\cal K}_n}|\vec x_{{\cal K}_n}})^T\Big(\sum_{k_n\in{\cal K}_n}\!\! W_{({\cal K}_n\backslash k_n,k_n)}^{(k_{n+1},\ldots,k_N)}\otimes\id^{A_{k_n}^O}\Big)\Big] \ge 0.
\end{align}
Following the first steps of Eq.~\eqref{eq:proof_QCQC_causal_correl}, while adding the superscripts ${}^{(k_{n+1},\ldots,k_N)}$ as in Eq.~\eqref{eq:proof_QCsupFO_convFO_correl}, we can write%
\footnote{Still assuming, for simplicity, that all denominators are nonzero. Cf.\ Footnotes~\ref{ftn:nonzero_denom_1} and~\ref{ftn:nonzero_denom_2}.}
\begin{align}
    p(\vec a|\vec x) & = \sum_{(k_{n^*},\ldots,k_N)} 
    s_{(\N\backslash\{k_{n^*},\ldots,k_N\},k_{n^*})}^{(k_{n^*+1},\ldots,k_N)}(\cdot|\cdot)
    \ \frac{s_{(\N\backslash\{k_{n^*+1},\ldots,k_N\},k_{n^*+1})}^{(k_{n^*+2},\ldots,k_N)}(\cdot|\cdot)}{f_{\N\backslash\{k_{n^*+1},\ldots,k_N\}}^{(k_{n^*+1},\ldots,k_N)}(\cdot|\cdot)}
    \ \cdots 
    \ \frac{s_{(\N\backslash k_N,k_N)}^\emptyset(\cdot|\cdot)}{f_{\N\backslash k_N}^{(k_N)}(\cdot|\cdot)}. \label{eq:proof_QC-NIO_1}
\end{align}

Let us then further define, for all $\pi=(k_1,\ldots,k_{n^*},\ldots,k_N)$,
\begin{align}
    \tilde{p}(\pi,\vec a_{k_1,\ldots,k_{n^*}}|\vec x_{k_1,\ldots,k_{n^*}}) & \coloneqq \Tr[(M_{\vec a_{k_1,\ldots,k_{n^*}}|\vec x_{k_1,\ldots,k_{n^*}}})^T\big(W_{(k_1,\ldots,k_{n^*})}^{(k_{n^*+1},\ldots,k_N)}\otimes\id^{A_{k_{n^*}}^O}\big)] \ge 0, \notag \\
    \text{and} \quad \tilde{p}_{\pi,k_n}(a_{k_n}|\vec x_{k_1,\ldots,k_n},\vec a_{k_1,\ldots,k_{n-1}}) & \coloneqq \frac{s_{(\N\backslash\{k_n,\ldots,k_N\},k_n)}^{(k_{n+1},\ldots,k_N)}(\cdot|\cdot)}{f_{\N\backslash\{k_n,\ldots,k_N\}}^{(k_n,\ldots,k_N)}(\cdot|\cdot)} \ge 0 \quad \text{for} \ n=n^*+1,\ldots,N.
\end{align}
From the definition of $W_{({\cal K}_{n^*} \backslash k_{n^*},k_{n^*})}^{(k_{n^*+1},\ldots,k_N)}$ just before Eq.~\eqref{eq:W_NIOn_decomp}, it follows that $s_{({\cal K}_{n^*} \backslash k_{n^*},k_{n^*})}^{(k_{n^*+1},\ldots,k_N)}(\cdot|\cdot) = \sum_{(k_1,\ldots,k_{n^*-1})\in{\cal K}_{n^*} \backslash k_{n^*}} \tilde{p}(\pi,\vec a_{k_1,\ldots,k_{n^*}}|\vec x_{k_1,\ldots,k_{n^*}})$, so that from Eq.~\eqref{eq:proof_QC-NIO_1} we can write
\begin{align}
    p(\vec a|\vec x) & = \sum_{\pi=(k_1,\ldots,k_{n^*},\ldots,k_N)} \tilde{p}(\pi,\vec a_{k_1,\ldots,k_{n^*}}|\vec x_{k_1,\ldots,k_{n^*}}) \prod_{n=n^*+1}^N \tilde{p}_{\pi,k_n}(a_{k_n}|\vec x_{k_1,\ldots,k_n},\vec a_{k_1,\ldots,k_{n-1}}) \ = \ \sum_\pi p(\pi,\vec a|\vec x) \notag \\
    & \hspace{5mm} \text{with} \qquad p\big(\pi=(k_1,\ldots,k_{n^*},\ldots,k_N),\vec a\big|\vec x\big) \coloneqq \tilde{p}(\pi,\vec a_{k_1,\ldots,k_{n^*}}|\vec x_{k_1,\ldots,k_{n^*}}) \prod_{n=n^*+1}^N \tilde{p}_{\pi,k_n}(a_{k_n}|\vec x_{k_1,\ldots,k_n},\vec a_{k_1,\ldots,k_{n-1}}). \label{eq:proof_QC-NIO_2}
\end{align}

Now, it can be shown as in Eq.~\eqref{eq:sum_akn_tilde_p_supFO} that $\sum_{a_{k_n}} \tilde p_{\pi,k_n}(a_{k_n}|\vec x_{k_1,\ldots,k_n}, \vec a_{k_1,\ldots,k_{n-1}}) = 1$. Hence for $n\ge  n^*$ and any $(k_1,\ldots,k_n,k_{n+1})$, $p\big((k_1,\ldots,k_{n+1}),\vec a_{k_1,\ldots,k_n}\big|\vec x\big) = \sum_{(k_{n+2},\ldots,k_N),\vec a_{k_{n+1},\ldots,k_N}} p\big(\pi{=}(k_1,\ldots,k_N),\vec a\big|\vec x\big) = \sum_{(k_{n+2},\ldots,k_N)} \tilde{p}\big(\pi{=}(k_1,\ldots,k_N),\vec a_{k_1,\ldots,k_{n^*}}\big|\vec x_{k_1,\ldots,k_{n^*}}\big) \prod_{j=n^*+1}^n \tilde{p}_{\pi=(k_1,\ldots,k_N),k_j}(a_{k_j}|\vec x_{k_1,\ldots,k_j},\vec a_{k_1,\ldots,k_{j-1}})$ does not depend on $\vec x_{\N\backslash\{k_1,\ldots,k_n\}}$.
For $n< n^*$ on the other hand, and still any $(k_1,\ldots,k_n,k_{n+1})$, it can be shown recursively from the second to last constraint in Eq.~\eqref{eq:W_NIOn_decomp} (in a similar way to Eqs.~\eqref{eq:proof_QCCC_causal_decomp2}--\eqref{eq:proof_QCCC_causal_decomp4}, and using the definition of $W_{(k_1,\ldots,k_{n^*})}$ just before Eq.~\eqref{eq:W_NIOn_decomp} for the case $n=n^*-1$) that $p\big((k_1,\ldots,k_n,k_{n+1}),\vec a_{k_1,\ldots,k_n}\big|\vec x\big) = \sum_{(k_{n+2},\ldots,k_N),\vec a_{k_{n+1},\ldots,k_{n^*}}} \tilde{p}\big(\pi{=}(k_1,\ldots,k_N),\vec a_{k_1,\ldots,k_{n^*}}\big|\vec x_{k_1,\ldots,k_{n^*}}\big) = \Tr\big[\big(M_{\vec a_{k_1,\ldots,k_n}|\vec x_{k_1,\ldots,k_n}}\big)^T\big(\Tr_{A_{k_{n+1}}^I}W_{(k_1,\ldots,k_{n+1})}\big)\big]$, which also does not depend on $\vec x_{\N\backslash\{k_1,\ldots,k_n\}}$.

All in all, we thus find that Eq.~\eqref{eq:proof_QC-NIO_2} provides a causal decomposition of $p(\vec a|\vec x) = \sum_\pi p(\pi,\vec a|\vec x)$ that satisfies Eq.~\eqref{eq:def_causal_p}.
Recalling that each $W_{(k_1,\ldots,k_{n^*})}^{(k_{n^*+1},\ldots,k_N)}$ is assumed to be proportional to a valid process matrix, we can write $W_{(k_1,\ldots,k_{n^*})}^{(k_{n^*+1},\ldots,k_N)} = q_{(k_1,\ldots,k_{n^*},\ldots,k_N)} \tilde{W}_{(k_1,\ldots,k_{n^*})}^{(k_{n^*+1},\ldots,k_N)}$, for some fixed weight $q_{(k_1,\ldots,k_{n^*},\ldots,k_N)}$ and some now properly normalised process matrix $\tilde{W}_{(k_1,\ldots,k_{n^*})}^{(k_{n^*+1},\ldots,k_N)}$. We then have, for all $(k_1,\ldots,k_{n^*},\ldots,k_N)$,
\begin{align}
    & p\big(\pi=(k_1,\ldots,k_{n^*},\ldots,k_N)\big|\vec x\big) \, = \, \sum_{\vec  a} p\big(\pi=(k_1,\ldots,k_{n^*},\ldots,k_N),\vec a\big|\vec x\big) \notag \\
    & \hspace{10mm} = \sum_{\vec a_{k_1,\ldots,k_{n^*}}} \tilde{p}\big(\pi=(k_1,\ldots,k_{n^*},\ldots,k_N),\vec a_{k_1,\ldots,k_{n^*}}\big|\vec x_{k_1,\ldots,k_{n^*}}\big) \notag \\
    & \hspace{10mm} = q_{(k_1,\ldots,k_{n^*},\ldots,k_N)} \underbrace{{\textstyle\sum_{\vec a_{k_1,\ldots,k_{n^*}}}} \Tr\big[(M_{\vec a_{k_1,\ldots,k_{n^*}}|\vec x_{k_1,\ldots,k_{n^*}}})^T\big(\tilde{W}_{(k_1,\ldots,k_{n^*})}^{(k_{n^*+1},\ldots,k_N)}\otimes\id^{A_{k_{n^*}}^O}\big)\big]}_{=1\ (\text{validity and normalisation of }\tilde{W}_{(k_1,\ldots,k_{n^*})}^{(k_{n^*+1},\ldots,k_N)})} = q_{(k_1,\ldots,k_{n^*},\ldots,k_N)},
\end{align}
which does not depend on $\vec x$ and concludes the proof that the correlations are in ${\cal P}_{\textup{NIO}}$.

\bigskip

Beyond the \textup{\textsf{QC-NIO}} class introduced here, one may also construct even larger classes whose correlations are restricted to be in ${\cal P}_\textup{NIO}$. E.g., one can conceive classes of a similar type as $\textup{\textsf{QC-NIO}}_{n^*}$, but allowing for different critical values $n^*$ attached to different orders $(k_1,\ldots,k_N)$ (and then considering convex combinations of such classes). Formalising this properly becomes rather tedious---we only illustrate it with a specific example, depicted in Fig.~\ref{fig:graph_QCNIO_N4_diff_nstar} using the graphical representation of Appendix~\ref{app:subsec:graphical_rep}.

\begin{figure}[hbtp]
    \centering
    \includegraphics[width=.65\columnwidth]{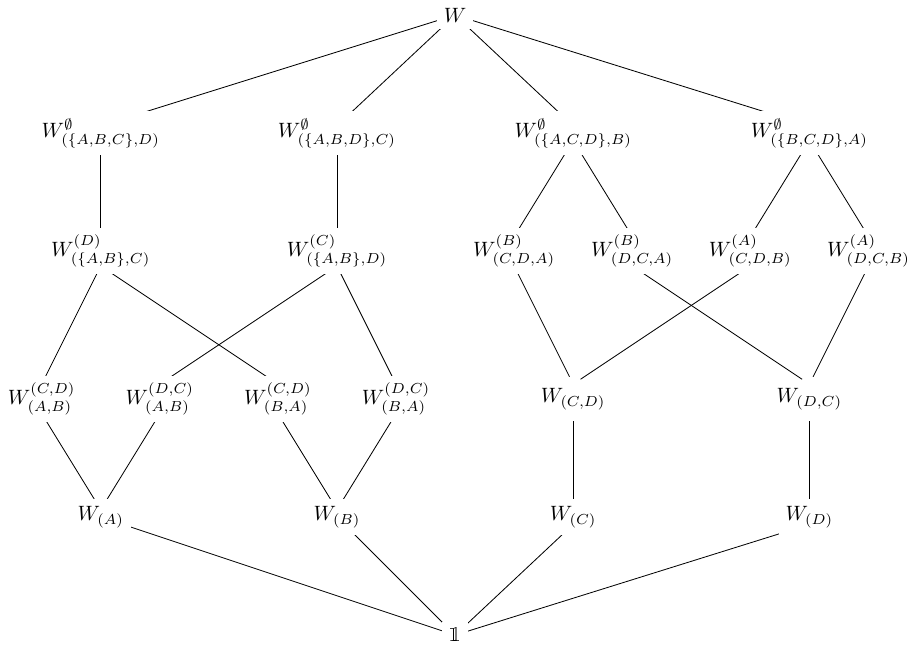}
    \caption{An example of a class of quantum circuits with $N=4$, defined through the graphical representation of Appendix~\ref{app:subsec:graphical_rep}. Assuming that the matrices of the form $W_{(k_1,k_2)}^{(k_3,k_4)}$ (in the left part of the graph) and of the form $W_{(k_1,k_2,k_3)}^{(k_4)}$ (in the right part) are valid process matrices (up to normalisation), one can show that this class only generates correlations in ${\cal P}_\textup{NIO}$.}
    \label{fig:graph_QCNIO_N4_diff_nstar}
\end{figure}

\subsection{Obtaining any correlation in ${\cal P}_{\textnormal{convFO}}$ or ${\cal P}_{\textnormal{causal}}$ with a QC-convFO or a QC-CC}
\label{app:obtaining_any_correl}

Here we show, for any correlation in ${\cal P}_{\text{convFO}}$, how to construct a convex mixture of quantum---and even classical---circuits with fixed causal order that generates it; and similarly for any correlation in ${\cal P}_{\text{causal}}$, how it can be obtained from a quantum---and even classical---circuit with classical control of causal order. We first propose a general construction that achieves this; this construction is of course not unique, and indeed for the specific correlations saturating the bounds~\eqref{eq:bnd_I4_FO}--\eqref{eq:bnd_I4_causal} on $I_4$, which we exhibited in our paper, we provide further below simpler ways to obtain them from classical circuits.

\medskip

The basic idea of the general construction is that the correlation one wishes to reproduce is established ``inside'' the process matrix. The process just indicates to each party what they should output, for each of their possible inputs; the party then does as instructed and indicates to the process which was their input.

To formalise this, we attribute to each party $A_{k}$ an input Hilbert space $\HS^{A_k^I}$ whose basis states describe the list of all possible outputs for the party, for all of its possible inputs: these are taken to be of the form $\ket{\vec\alpha_k}^{A_k^I}$, with $\vec\alpha_k = (\alpha_{k,x_k})_{x_k}$ and with the understanding that each $\alpha_{k,x_k}$ corresponds to the output that the party should return when their input is $x_k$. As for the output Hilbert space $\HS^{A_k^O}$, its basis states are taken to be of the form $\ket{x_k}^{A_k^O}$, so as to encode the party's input. The (classical) instrument consisting in reading off from $\HS^{A_k^I}$ the value $a_k$ to be outputted for input $x_k$, and sending out in $\HS^{A_k^O}$ the value of that input, is then described in the Choi picture by
\begin{align}
    M_{a_k|x_k} = \sum_{\vec\alpha_k} \, \delta_{a_k,\alpha_{k,x_k}} \,[\![\vec\alpha_k]\!]^{A_k^I}\otimes[\![x_k]\!]^{A_k^O}, \label{eq:classical_instr}
\end{align}
where we again used the notation $[\![\vec\alpha_k]\!]^{A_k^I} \coloneqq \ketbra{\vec\alpha_k}{\vec\alpha_k}^{A_k^I}$ and $[\![x_k]\!]^{A_k^O} \coloneqq \ketbra{x_k}{x_k}^{A_k^O}$ for effectively classical states.

\subsubsection{Any correlation in ${\cal P}_{\textup{convFO}}$ can be obtained from a QC-convFO}
\label{app:obtaining_any_convFO_correl}

Consider a correlation in ${\cal P}_{\text{convFO}}$, i.e.\ a mixture of correlations $p_\pi(\vec a|\vec x)$ compatible with the fixed causal order $\pi=(k_1,\ldots,k_N)$.
Recalling from Eq.~\eqref{eq:def_convFO_corr} that (for $n = 1,\ldots,N-1$) $p_\pi(\vec a_{k_1,\ldots,k_n}|\vec x)$ does not depend on $\vec x_{\mathcal{N}\backslash\{k_1,\ldots,k_n\}}$, one can thus decompose $p(\vec a|\vec x)$ as
\begin{align}
    p(\vec a|\vec x) = \sum_\pi q_\pi\,p_\pi(\vec a|\vec x) \quad & \text{with } q_\pi\ge 0, \ {\textstyle\sum}_\pi q_\pi = 1, \notag \\
    \text{and} \quad p_{\pi=(k_1,\ldots,k_N)}(\vec a|\vec x) & = p_\pi(a_{k_1}|x_{k_1})\,p_\pi(a_{k_2}|x_{k_1},a_{k_1},x_{k_2}) \cdots p_\pi(a_{k_N}|\vec x_{k_1,\ldots,k_{N-1}},\vec a_{k_1,\ldots,k_{N-1}},x_{k_N}) \notag \\[-1mm]
    & = \prod_{n=1}^N p_\pi(a_{k_n}|\vec x_{k_1,\ldots,k_{n-1}},\vec a_{k_1,\ldots,k_{n-1}},x_{k_n})\,.
\end{align}

Let us then define, for any $n=1,\ldots,N$, with $\vec\alpha_{k_n} = (\alpha_{k_n,x_{k_n}})_{x_{k_n}}$ as above, and with the shorthand notation $\overrightarrow{[\alpha_x]}_{k_1,\ldots,k_{n-1}}\coloneqq(\alpha_{k_1,x_{k_1}},\ldots,\alpha_{k_{n-1},x_{k_{n-1}}})$,
\begin{align}
    & \mathfrak{p}_\pi(\vec \alpha_{k_n}|\vec x_{k_1,\ldots,k_{n-1}},\overrightarrow{[\alpha_x]}_{k_1,\ldots,k_{n-1}}) \coloneqq \prod_{x_{k_n}'} p_\pi(\alpha_{k_n,x_{k_n}'}|\vec x_{k_1,\ldots,k_{n-1}},\overrightarrow{[\alpha_x]}_{k_1,\ldots,k_{n-1}},x_{k_n}'),
\end{align}
such that (for some $a_{k_n}, x_{k_n}$)
\begin{align}
    & \sum_{\vec \alpha_{k_n}} \delta_{a_{k_n},\alpha_{k_n,x_{k_n}}} \mathfrak{p}_\pi(\vec \alpha_{k_n}|\vec x_{k_1,\ldots,k_{n-1}},\overrightarrow{[\alpha_x]}_{k_1,\ldots,k_{n-1}}) = p_\pi(a_{k_n}|\vec x_{k_1,\ldots,k_{n-1}},\overrightarrow{[\alpha_x]}_{k_1,\ldots,k_{n-1}},x_{k_n}),
\end{align}
and
\begin{align}
    W \coloneqq & \sum_{\pi=(k_1,\ldots,k_N)} q_\pi\ \sum_{\substack{\vec\alpha_{k_1},\ldots,\vec\alpha_{k_N},\\ x_{k_1},\ldots,x_{k_N}}} \ \bigotimes_{n=1}^N\Big(\mathfrak{p}_\pi(\vec \alpha_{k_n}|\vec x_{k_1,\ldots,k_{n-1}},\overrightarrow{[\alpha_x]}_{k_1,\ldots,k_{n-1}}) \, [\![\vec\alpha_{k_n}]\!]^{A_{k_n}^I}\otimes[\![x_{k_n}]\!]^{A_{k_n}^O}\Big).
\end{align}
It can be verified that $W$ thus defined is the process matrix of a QC-convFO,%
\footnote{Indeed, it has a decomposition as in Eq.~\eqref{eq:charact_W_QC_convFO_decomp}, with 
\begin{align}
    W_{(k_1,\ldots,k_N,F)} & = q_{\pi=(k_1,\ldots,k_N)}\,{\textstyle\sum_{\substack{\vec\alpha_{k_1},\ldots,\vec\alpha_{k_N},\\ x_{k_1},\ldots,x_{k_N}}}} \ {\textstyle\bigotimes_{n=1}^N}\big(\mathfrak{p}_{\pi=(k_1,\ldots,k_N)}(\vec \alpha_{k_n}|\vec x_{k_1,\ldots,k_{n-1}},\overrightarrow{[\alpha_x]}_{k_1,\ldots,k_{n-1}}) \, [\![\vec\alpha_{k_n}]\!]^{A_{k_n}^I}\otimes[\![x_{k_n}]\!]^{A_{k_n}^O}\big), \notag \\
    W_{(k_1,\ldots,k_n)}^{(k_{n+1}\ldots,k_N)} & = q_{\pi=(k_1,\ldots,k_n,k_{n+1}\ldots,k_N)}  \,{\textstyle\sum_{\substack{\vec\alpha_{k_1},\ldots,\vec\alpha_{k_n},\\ x_{k_1},\ldots,x_{k_{n-1}}}}} \big({\textstyle\bigotimes_{j=1}^{n-1}}\mathfrak{p}_{\pi=(k_1,\ldots,k_N)}(\vec \alpha_{k_j}|\vec x_{k_1,\ldots,k_{j-1}},\overrightarrow{[\alpha_x]}_{k_1,\ldots,k_{j-1}}) \, [\![\vec\alpha_{k_j}]\!]^{A_{k_j}^I}\otimes[\![x_{k_j}]\!]^{A_{k_j}^O}\big) \notag \\[-3mm]
    & \hspace{75mm}\otimes\big(\mathfrak{p}_{\pi=(k_1,\ldots,k_N)}(\vec \alpha_{k_n}|\vec x_{k_1,\ldots,k_{n-1}},\overrightarrow{[\alpha_x]}_{k_1,\ldots,k_{n-1}}) \, [\![\vec\alpha_{k_n}]\!]^{A_{k_n}^I}\big). \notag
\end{align}
}
and that with the choice of instruments of Eq.~\eqref{eq:classical_instr} for each of the parties, it reproduces the correlation $p(\vec a|\vec x)$ under consideration, as desired.

\subsubsection{Any correlation in ${\cal P}_{\textup{causal}}$ can be obtained from a QC-CC}
\label{app:obtaining_any_causal_correl}

Consider now a correlation in ${\cal P}_{\text{causal}}$, which can be decomposed as in Eq.~\eqref{eq:def_causal_p_v4}:
\begin{align}
    & p(\vec a|\vec x) = \sum_\pi p(\pi,\vec a|\vec x) \notag \\[-4mm]
    & \hspace{10mm} \text{with} \quad p\big(\pi=(k_1,\ldots,k_N),\vec a\big|\vec x\big) = \prod_{n=1}^N \tilde{p}\big((k_n),a_{k_n}\big|(k_1,\ldots,k_{n-1}),\vec x_{k_1,\ldots,k_n},\vec a_{k_1,\ldots,k_{n-1}}\big),
\end{align}
such that (for all $n$) $\sum_{a_{k_n}} \tilde{p}\big((k_n),a_{k_n}\big|(k_1,\ldots,k_{n-1}),\vec x_{k_1,\ldots,k_n},\vec a_{k_1,\ldots,k_{n-1}}\big)$ does not depend on $x_{k_n}$.

The latter requirement allows us to write $\tilde{p}\big((k_n),a_{k_n}\big|(k_1,\ldots,k_{n-1}),\vec x_{k_1,\ldots,k_n},\vec a_{k_1,\ldots,k_{n-1}}\big) = \tilde{p}\big((k_n)\big|(k_1,\ldots,k_{n-1}),\vec x_{k_1,\ldots,k_{n-1}},\vec a_{k_1,\ldots,k_{n-1}}\big) \, \tilde{p}\big(a_{k_n}\big|(k_1,\ldots,k_{n-1},k_n),\vec x_{k_1,\ldots,k_{n-1}},\vec a_{k_1,\ldots,k_{n-1}},x_{k_n}\big)$.
Following a similar construction as above, let us then define
\begin{align}
    \mathfrak{p}\big((k_n),\vec \alpha_{k_n}\big|\vec x_{k_1,\ldots,k_{n-1}},\overrightarrow{[\alpha_x]}_{k_1,\ldots,k_{n-1}}\big) & \coloneqq \tilde{p}\big((k_n)\big|(k_1,\ldots,k_{n-1}),\vec x_{k_1,\ldots,k_{n-1}},\overrightarrow{[\alpha_x]}_{k_1,\ldots,k_{n-1}}\big) \notag \\
    & \hspace{5mm} \times \prod_{x_{k_n}'} \tilde{p}\big(\alpha_{k_n,x_{k_n}'}\big|(k_1,\ldots,k_{n-1},k_n),\vec x_{k_1,\ldots,k_{n-1}},\overrightarrow{[\alpha_x]}_{k_1,\ldots,k_{n-1}},x_{k_n}'\big),
\end{align}
such that (for some $a_{k_n}, x_{k_n}$)
\begin{align}
    & \sum_{\vec \alpha_{k_n}} \delta_{a_{k_n},\alpha_{k_n,x_{k_n}}} \mathfrak{p}\big((k_n),\vec \alpha_{k_n}\big|\vec x_{k_1,\ldots,k_{n-1}},\overrightarrow{[\alpha_x]}_{k_1,\ldots,k_{n-1}}\big) = \tilde{p}\big((k_n),a_{k_n}\big|(k_1,\ldots,k_{n-1}),\vec x_{k_1,\ldots,k_n},\overrightarrow{[\alpha_x]}_{k_1,\ldots,k_{n-1}}\big),
\end{align}
and
\begin{align}
    W \coloneqq & \sum_{\pi=(k_1,\ldots,k_N)} \ \sum_{\substack{\vec\alpha_{k_1},\ldots,\vec\alpha_{k_N},\\ x_{k_1},\ldots,x_{k_N}}} \ \bigotimes_{n=1}^N\Big(\mathfrak{p}\big((k_n),\vec \alpha_{k_n}\big|\vec x_{k_1,\ldots,k_{n-1}},\overrightarrow{[\alpha_x]}_{k_1,\ldots,k_{n-1}}\big) \, [\![\vec\alpha_{k_n}]\!]^{A_{k_n}^I}\otimes[\![x_{k_n}]\!]^{A_{k_n}^O}\Big).
\end{align}
It can be verified here that $W$ thus defined is the process matrix of a QC-CC,%
\footnote{Indeed, it has a decomposition as in Eq.~\eqref{eq:charact_W_QCCC_decomp}, with 
\begin{align}
    W_{(k_1,\ldots,k_N,F)} & = {\textstyle\sum_{\substack{\vec\alpha_{k_1},\ldots,\vec\alpha_{k_N},\\ x_{k_1},\ldots,x_{k_N}}}} \ {\textstyle\bigotimes_{n=1}^N}\big(\mathfrak{p}((k_n),\vec \alpha_{k_n}|\vec x_{k_1,\ldots,k_{n-1}},\overrightarrow{[\alpha_x]}_{k_1,\ldots,k_{n-1}}) \, [\![\vec\alpha_{k_n}]\!]^{A_{k_n}^I}\otimes[\![x_{k_n}]\!]^{A_{k_n}^O}\big), \notag \\
    W_{(k_1,\ldots,k_n)} & = {\textstyle\sum_{\substack{\vec\alpha_{k_1},\ldots,\vec\alpha_{k_n},\\ x_{k_1},\ldots,x_{k_{n-1}}}}} \ \big({\textstyle\bigotimes_{j=1}^{n-1}}\mathfrak{p}((k_j),\vec \alpha_{k_j}|\vec x_{k_1,\ldots,k_{j-1}},\overrightarrow{[\alpha_x]}_{k_1,\ldots,k_{j-1}}) \, [\![\vec\alpha_{k_j}]\!]^{A_{k_j}^I}\otimes[\![x_{k_j}]\!]^{A_{k_j}^O}\big) \notag \\[-3mm]
    & \hspace{70mm}\otimes\big(\mathfrak{p}((k_n),\vec \alpha_{k_n}|\vec x_{k_1,\ldots,k_{n-1}},\overrightarrow{[\alpha_x]}_{k_1,\ldots,k_{n-1}}) \, [\![\vec\alpha_{k_n}]\!]^{A_{k_n}^I}\big). \notag
\end{align}}
and that with the choice of instruments of Eq.~\eqref{eq:classical_instr} for each of the parties, it reproduces the correlation $p(\vec a|\vec x)$ under consideration, as desired.

\subsubsection{Saturating the bounds~\eqref{eq:bnd_I4_FO}--\eqref{eq:bnd_I4_causal} on $I_4$ with quantum circuits}

While the constructions above are quite general and allow one to find process matrices reproducing any desired correlation in ${\cal P}_{\text{convFO}}$ or ${\cal P}_{\text{causal}}$, in some cases (in particular, if some party does not signal to any other party, or if only limited communication is required) there also exist simpler process matrix realisations.

\bigskip

For instance, the correlation $p(a,b,c,d|x,y,z,t) = \delta_{a,0}\,\delta_{b,0}\,\delta_{c,0}\,\delta_{d,zt}$ (given below Eq.~\eqref{eq:bnd_I4_FO}), compatible with any causal order where $C\prec D$ and which saturates $I_4 \leq_{\text{convFO}} \frac{15}{16}$, can be rather trivially obtained from the QC-FO $W = [\![\id]\!]^{C^OD^I} \coloneqq \sum_{z=0,1}[\![z]\!]^{C^O}\otimes[\![z]\!]^{D^I} \in \textup{\textsf{QC-convFO}}\subset\textup{\textsf{QC-supFO}}$, representing a classical identity (or quantum dephasing) channel between $\HS^{C^O}$ and $\HS^{D^I}$ (with trivial systems for $A$, $B$, $C^I$ and $D^O$), and with the instruments $M_{a|x}^A=\delta_{a,0}$, $M_{b|y}^B=\delta_{b,0}$, $M_{c|z}^C=\delta_{c,0}[\![z]\!]^{C^O}$, $M_{0|t=0}^D=\id^{D^I}$ and $M_{d|t=1}^D=[\![d]\!]^{D^I}$.

\bigskip

Similarly, one can obtain the correlation $p(a,b,c,d|x,y,z,t) = \delta_{a,0}\,\delta_{b,0} \left\{\begin{array}{ll}
    \delta_{c,0}\,\delta_{d,zt} & \text{if}\ xy=0 \\
    \delta_{d,0}\,\delta_{c,zt} & \text{if}\ xy=1
\end{array}\right.$ of Eq.~\eqref{eq:correl_I4_1}, which saturates the trivial bound $I_4 \leq_{\text{causal}} 1$, with $W = ([\![0,0]\!]+[\![0,1]\!]+[\![1,0]\!])^{A^OB^O} \otimes [\![0]\!]^{C^I}\otimes[\![\id]\!]^{C^OD^I}\otimes\id^{D^O} + [\![1,1]\!]^{A^OB^O} \otimes [\![0]\!]^{D^I}\otimes[\![\id]\!]^{D^OC^I}\otimes\id^{C^O} \in \textup{\textsf{QC-CC}}\subset\textup{\textsf{QC-QC}}$ (with trivial systems for $A^O,B^I$) and the instruments 
\begin{align}
    & M_{a|x}^A=\delta_{a,0}\ketbra{x}{x}^{A^O}, \quad 
    M_{b|y}^B=\delta_{b,0}\ketbra{y}{y}^{B^O}, \notag \\
    & M_{0|z=0}^C=\id^{C^I}\!\otimes[\![0]\!]^{C^O}, \quad
    M_{c|z=1}^C=[\![c]\!]^{C^I}\!\otimes[\![1]\!]^{C^O}, \quad
    M_{0|t=0}^D=\id^{D^I}\!\otimes[\![0]\!]^{D^O} \quad \text{and} \quad 
    M_{d|t=1}^D=[\![d]\!]^{D^I}\!\otimes[\![1]\!]^{D^O}. \label{eq:instr_ABCD}
\end{align}

\bigskip

Let us now turn to the correlation with non-influenceable causal order of Eq.~\eqref{eq:saturate_NIO}, which saturates the $I_4 \leq_{\text{NIO}} \frac{31}{32}$ bound. Considering the local decomposition of the marginal distribution $p(a,b|x,y)$ given in Footnote~\ref{ftn:decomp_saturate_I4NIO}, one can see that the correlation of Eq.~\eqref{eq:saturate_NIO} can be obtained from the process matrix
\begin{align}
    W & = W_\rightarrow \otimes [\![0]\!]^{C^I}\otimes[\![\id]\!]^{C^OD^I}\otimes\id^{D^O} + W_\leftarrow \otimes [\![0]\!]^{D^I}\otimes[\![\id]\!]^{D^OC^I}\otimes\id^{C^O} \label{eq:W_saturate_NIO}
\end{align}
with
\begin{align}
   W_\rightarrow & = {\textstyle\frac12}{\textstyle\sum_{\lambda,x,y=0,1}} \big({\textstyle\frac12} \delta_{x=y=0} + \delta_{\lambda x=\lambda y=xy} \delta_{\neg(x=y=0)}\big) [\![\lambda,x,\lambda,y]\!]^{A^IA^OB^IB^O} , \notag \\[1mm]
   W_\leftarrow & = {\textstyle\frac12}{\textstyle\sum_{\lambda,x,y=0,1}} \big({\textstyle\frac12} \delta_{x=y=0} + \delta_{\neg(\lambda x=\lambda y=xy)} \delta_{\neg(x=y=0)}\big) [\![\lambda,x,\lambda,y]\!]^{A^IA^OB^IB^O},
\end{align}
together with Alice's instruments $M_{0|x=0}^A=\id^{A^I}\otimes[\![0]\!]^{A^O}$ and $M_{a|x=1}^A=[\![a]\!]^{A^I}\otimes[\![1]\!]^{A^O}$, and with the three other parties' instruments defined in the same way.
Noting in particular that $W_\rightarrow + W_\leftarrow = \frac12\sum_\lambda [\![\lambda,\lambda]\!]^{A^IB^I}\otimes\id^{A^OB^O}$ defines a (non-signalling) process matrix compatible with any fixed order between $A$ and $B$, it can be seen that $W \in \textup{\textsf{QC-CC}}$, with Eq.~\eqref{eq:W_saturate_NIO} readily providing a QC-CC decomposition as in Eq.~\eqref{eq:charact_W_QCCC_decomp}. However, $W_\rightarrow$ and $W_\leftarrow$, individually, are \emph{not} valid process matrices, and Eq.~\eqref{eq:charact_W_QCCC_decomp} does not give a QC-NICC decomposition. In fact it can be verified via SDP that $W \notin \textup{\textsf{QC-NICC}}$.
(Unsurprisingly, $W \notin \textup{\textsf{QC-NIQC}}$ either.)
As we explain in the main text, we were not able to saturate the $I_4 \leq_{\text{NIO}} \frac{31}{32}$ bound with process matrices in \textup{\textsf{QC-NICC}}.

The same holds for the $I_4 \leq_{\text{NIO}'} \frac{47}{48}$ bound: using an analogous approach (which we omit detailing explicitly here) one can construct the process matrix of a QC-QC that reproduces the correlation with non-influenceable coarse-grained order of Eq.~\eqref{eq:saturate_NIO'}, which saturates this bound. However that process matrix is not a QC-NIQC; we were also not able to saturate the $I_4 \leq_{\text{NIO}'} \frac{47}{48}$ bound with process matrices in \textup{\textsf{QC-NIQC}}.

Hence, contrary to the previous cases of ${\cal P}_{\textup{convFO}}$ and ${\cal P}_{\textup{causal}}$ in which any correlation could be obtained from a circuit in the corresponding classes of \textup{\textsf{QC-convFO}} (or \textup{\textsf{QC-supFO}}) and \textup{\textsf{QC-CC}} (or \textup{\textsf{QC-QC}}), respectively, it seems unlikely that all correlations in ${\cal P}_{\textup{NIO}}$ and ${\cal P}_{\textup{NIO}'}$ can be reached by processes in \textup{\textsf{QC-NICC}} and \textup{\textsf{QC-NIQC}}.

\subsection{Dynamical correlations from quantum circuits with non-influenceable control of causal order}
\label{app:violations}

In the main text, we calculated numerically the optimal value of $I_4$ (Eq.~\eqref{eq:def_I4}) obtainable with QC-NICCs and QC-NIQCs by using the techniques of Ref.~\cite{liu24} to bound single-trigger inequalities.
This tight bound is obtained by restricting to the so-called ``canonical instruments'' which (in the lazy-scenario) act on a qubit input space $\HS^{A_k^I}$ and two-qubit output space $\HS^{A_k^O A_k^{O'}}$ and have the form
\begin{equation}
    M_{0|0}^{A_k} = \kketbra{\id}{\id}^{A_k^I A_k^O}\otimes \ketbra{0}{0}^{A_k^{O'}}, \quad M_{a_k|1}^{A_k} = \ketbra{a_k}{a_k}^{A_k^I}\otimes\ketbra{a_k}{a_k}^{A_k^O}\otimes\ketbra{1}{1}^{A_k^{O'}},
    \label{eq:canonical_instruments}
\end{equation}
where here $A_k = A,B,C,D$ and $a_k=a,b,c,d$, respectively.
These ``labelled projective measurements'' have a simple interpretation: the system sent out in the auxiliary space $\HS^{A_k^{O'}}$ is precisely the classical input $x_k$, while, when $x_k=0$ the measurement outputs $a_k=0$ and transmits the system in $\HS^{A_k^I}$ to $\HS^{A_k^O}$ via an identity channel, while if $x_k=1$ a projective measurement in the computational basis is performed to determine $a_k$, which is then sent back to the process in $\HS^{A_k^O}$.

As we mentioned in the main text, the optimal numerical values we obtained for $I_4$, for both classes of circuits, can in fact be obtained with the simpler qubit instruments for each party $A$, $B$, $C$ and $D$:
\begin{align}
    M^A_{0|0} = M^B_{0|0} = \kketbra{\id}{\id}, \quad  &M^A_{0|1} = M^B_{0|1} = \ketbra{0}{0}\otimes \ketbra{0}{0}, \quad M^A_{1|1} = M^B_{1|1} = \ketbra{1}{1}\otimes \ketbra{0}{0}, \notag \\
    M^C_{0|0} = M^D_{0|0} = \id\otimes\ketbra{0}{0}, \quad &M^C_{0|1} = M^D_{0|1} = \ketbra{0}{0}\otimes \ketbra{1}{1}, \quad M^C_{1|1} = M^D_{1|1} = \ketbra{1}{1}\otimes \ketbra{1}{1},
    \label{eq:instruments_violating_I_4_QC-NICC-NIQC}
\end{align}
and processes compatible with the coarse-grained order $\{A,B\}\prec \{C,D\}$ (and even, for QC-NICC, $A\prec B$). 
The instruments for parties $A$ and $B$ correspond to the ones chosen in Ref.~\cite{branciard15} to show that $W_{+\alpha}$ violates the LGYNI inequality: on input $0$, $A$ and $B$ output $0$ (as required here in the lazy scenario), and simply transmit the physical system they received, untouched; while on input $1$, they perform a measurement in the $Z$ basis, output its classical outcome, and send out a physical system in the state $\ketbra{0}{0}$. 
The instruments for $C$ and $D$ are such that on input $0$, they output $0$, while on input $1$ they measure their input in the computational basis and output the classical result of that measurement; in both cases, they then transmit their classical input to the process in the corresponding state, $\ketbra{0}{0}$ or $\ketbra{1}{1}$. (Notice that these instruments for $C$ and $D$ are the same as those in Eq.~\eqref{eq:instr_ABCD} of the previous appendix.)

As far as the processes reaching the optimal values are concerned, for \textsf{QC-NICC} one can take $W$ of the form of Eq.~\eqref{eq:QC-NICC-example},%
\footnote{In fact since, with these instruments, information is only sent between $C$ and $D$ in the computational basis, one can replace the identity channels between them in Eq.~\eqref{eq:QC-NICC-example} by their classical versions $[\![\id]\!]$ and still obtain the same correlations. One can readily check that both $W_\textup{NICC}$ and the process $W$ obtained by replacing the indicated terms by those in Eq.~\eqref{eq:W0_W1_alternative} and~\eqref{eq:W_AB_alternative} remain in \textsf{QC-NICC}.}
with $\ket{\psi}=\ket{0}$ and replacing $\frac12 W_{+\alpha}$ and $\frac12 W_{-\alpha}$ by 
$W_0$ and $W_1$, defined as 
\begin{align}
W_1 &= (1-\tfrac{1}{\sqrt{2}})\ketbra{1}{1}^{B^I}\otimes[\![\id]\!]^{B^OA^I}\otimes\id^{A^O}, \quad
W_0 = W_{\{A\},B}\otimes\id^{B^O} - W_1,
\label{eq:W0_W1_alternative}
\end{align}
with 
\begin{equation}
    W_{\{A\},B} = \tfrac12\id^{A^I}\otimes\id^{A^O}\otimes\ketbra{1}{1}^{B^I} + \tfrac12(1-\tfrac{1}{\sqrt{2}})\kketbra{\id}{\id}^{A^IA^O}\otimes Z^{B^I}
    \label{eq:W_AB_alternative}
\end{equation}
(and with $[\![\id]\!] = \sum_{j=0,1} \ketbra{jj}{jj}$, as in the previous appendix).
By direct calculation, one then finds $I_4=1-\frac{\sqrt{2}-1}{8}\simeq 0.9482$, which agrees (up to numerical precision) with the optimal value obtained numerically using the canonical instruments of Eq.~\eqref{eq:canonical_instruments}.
This process has a realisation as a circuit with similar structure to Fig.~\ref{fig:QC-NICC-example} except that the left-hand part, between $A$ and $B$, is a sequential quantum comb (compatible with $A\prec B$) rather than a parallel circuit whose output is measured to determine the order between $C$ and $D$. The precise details of the circuit realisation, however, are not so simple and not particularly insightful, so we omit its full description here.

For \textsf{QC-NIQC} it proved more complicated to obtain a clean expression for an optimal process.
One can take, for instance, $W$ of the form of Eq.~\eqref{eq:QC-NIQC-example},%
\footnote{Once again, one can even take the channel between $C$ and $D$ to be a classical, rather than quantum, identity channel. Both $W_\textup{NIQC}$ and the process of a similar form described here are easily seen to remain in \textsf{QC-NIQC}. The same remarks also hold for the process $W'_\textup{NIQC}$ defined below in Eq.~\eqref{eq:QC-NIQC-example-2}.}
with $\ket{\psi}=\ket{0}$ and replacing $\frac12 W_{0/1}$ by $W_{0/1}$, defined as 
\begin{equation}
    W_1 = \omega \left(\tfrac12 \ketbra{1}{1}^{A^I}\otimes [\![\id]\!]^{A^OB^I}\otimes \id^{B^O} + \tfrac12 \ketbra{1}{1}^{B^I}\otimes [\![\id]\!]^{B^OA^I}\otimes\id^{A^O}\right), \quad W_0 = W_{AB} - W_1,
\end{equation}
with 
\begin{align}
    W_{AB} &= W_{\{A\},B}^{A^IA^OB^I}\otimes\id^{B^O} + W_{\{B\},A}^{B^IB^OA^I}\otimes\id^{A^O}\\
    W_{\{A\},B}^{A^IA^OB^I} &= \frac18\left(\id^{A^IA^OB^I} - \alpha\id\id Z+(1-\alpha)ZZZ+\beta Z\id Z + \sqrt{\alpha^2-\beta^2}(XX-YY)Z - (\beta+\gamma)Z\id\id - \gamma \id ZZ\right),
\end{align}
and with $W_{\{B\},A}^{B^IB^OA^I}$ defined in a similar way as $W_{\{A\},B}^{A^IA^OB^I}$, with $A$ and $B$ exchanged. (In the definition of $W_{\{A\},B}^{A^IA^OB^I}$ above, we left the tensor products implicit and kept the order $A^IA^OB^I$ for all terms.)
Here the 4 remaining parameters $\alpha, \beta, \gamma, \omega$ should satisfy $1+2 \alpha +\beta +\gamma-\sqrt{1+4 \alpha ^2+6 \beta+5 \beta ^2 +2 \gamma+6 \beta  \gamma +\gamma ^2} = 4 \omega$ and $(\alpha+\beta)(1-2 \alpha +\beta ^2 -\gamma ^2) = 2  (1-\alpha -\alpha ^2-\beta+\alpha  \beta +\beta ^2-\alpha  \gamma +\beta  \gamma -\gamma ^2) \omega$, which make certain eigenvalues of $W_0$ be equal to 0, rather than strictly positive. 
The value of $I_4$ is then $I_4=\frac{1}{16}(15-\alpha+\sqrt{\alpha^2-\beta^2}+2\omega)$, which one can numerically optimize under the above constraints to obtain $I_4 \simeq 0.9527$, in agreement with the value obtained numerically using the canonical instruments of Eq.~\eqref{eq:canonical_instruments}. 
Numerically, we find the optimal values $\alpha\simeq 0.3690, \beta\simeq 0.0338, \gamma\simeq 0.3081, \omega\simeq 0.1226$. 
This process has a realisation as a circuit with similar structure to Fig.~\ref{fig:QC-NIQC-example}, with just a different input state from $\ket{\Phi^+}$ (a partially entangled state, instead), a different internal isometry from the Swap and a Controlled-$Z$ gates, and a different measurement $E_\pm$. 
All these elements could, in principle, be given in terms of the parameters $\alpha, \beta, \gamma, \omega$, but little insight is gained by giving further details here.

\bigskip

While the processes described above give the optimal violation of our inequality~\eqref{eq:bnd_I4_FO}, $I_4 \leq_{\text{convFO}} \frac{15}{16}$, using QC-NICCs and QC-NIQCs, we were not able, as mentioned in the main text, to find any violation with the explicit examples of process matrices $W_{\textup{NICC}}$ and $W_{\textup{NIQC}}$ that we presented in the main text (Eqs.~\eqref{eq:QC-NICC-example} and \eqref{eq:QC-NIQC-example}). 
Nonetheless, we were able to come up with another inequality that also bounds correlations in $\mathcal{P}_\textup{convFO}$, and find explicit analytical violations with $W_{\textup{NICC}}$ as well as with another example $W_{\textup{NIQC}}'\in \text{\textup{\textsf{QC-NIQC}}}\setminus(\text{\textup{\textsf{QC-supFO}}}\cup\text{\textup{\textsf{QC-NICC}}})$ that is defined below.
\medskip

We recall that given a convex polytope $\mathcal{P}$ characterised in terms of its $n_{\mathcal{P}}$ vertices $\{\vec{p}_i\}_{i=1}^{n_{\mathcal{P}}}$, one can use the following linear program to test whether a correlation vector $\vec p$ is inside or outside the polytope:
   \begin{align}
\min_{r,\vec{\lambda}} r \quad
\text{s.t.} \quad &M_\mathcal{P}\vec{\lambda} = \vec{p} + r \vec{u}, \ \vec{\lambda} \geq 0, \ r \geq 0 , \label{eq:lp_primal}
\end{align}
where $\vec{u} = \frac{1}{n_{\mathcal{P}}} \sum_{i=1}^{n_{\mathcal{P}}} \vec{p}_i$, and $M_\mathcal{P}$ is the matrix formed by the vectors $\{\vec{p}_i\}_i$ in column. If the optimal value $r^*$ is zero, then the correlation vector $\vec p$ belongs to the polytope, while if $r^*>0$, then the correlation vector $\vec p$ is outside of the polytope, and one can consider the dual optimisation problem to Eq.~\eqref{eq:lp_primal} to derive an inequality satisfied by any point in $\mathcal{P}$ and violated by the correlation of interest, $\vec p$.

To obtain such an inequality and find a correlation created by $W_{\textup{NICC}}$ of Eq.~\eqref{eq:QC-NICC-example}
in the fourpartite lazy scenario that is outside of $\mathcal{P}_\textup{convFO}$, we fixed the initial state as $\ket{\psi}=\ket{0}$ and considered the instruments of Eq.~\eqref{eq:instruments_violating_I_4_QC-NICC-NIQC} above. Solving the optimisation problem of Eq.~\eqref{eq:lp_primal} for the correlation vector $\vec p_{\textup{NICC}} = (p_{\textup{NICC}}(\vec a|\vec x))_{\vec a, \vec x}$, 
defined with the process $W_{\textup{NICC}}$ (with $\ket{\psi}=\ket{0}$ and the instruments of Eq.~\eqref{eq:instruments_violating_I_4_QC-NICC-NIQC} using generalised Born rule (see Eq.~\eqref{eq:Born_rule}), we obtained $r^*_{\textup{NICC}} \simeq 0.063>0$, proving that $\vec p_{\textup{NICC}}$ is outside of $\mathcal{P}_{\text{convFO}}$. Solving the corresponding dual optimisation problem, we could derive the following inequality, which is satisfied by any correlation in $\mathcal{P}_{\text{convFO}}$ but violated by $\vec p_{\textup{NICC}}$:
\begin{align}
    I_4' \coloneqq & \ p(a{=}1,c{=}0|x{=}1,y{=}0,z{=}1,t{=}1) + p(a{=}1,c{=}1|x{=}1,y{=}0,z{=}1,t{=}0) \notag \\ 
    & \ + p(b{=}1,c{=}0|x{=}0,y{=}1,z{=}1,t{=}1) + p(b{=}1,c{=}1|x{=}0,y{=}1,z{=}1,t{=}0) \notag \\
    & \ + p(a{=}1,b{=}1,d{=}1|x{=}1,y{=}1,z{=}0,t{=}1) - p(a{=}1,b{=}1,c{=}0,d{=}1|x{=}1,y{=}1,z{=}1,t{=}1) \underset{\text{convFO}}{\ge} 0.
    \label{app:ineq_WNICC}
\end{align}
This inequality can indeed be verified to hold for all the 481\,776 (deterministic) vertices of $\mathcal{P}_{\text{convFO}}$
(and thus for any point in $\mathcal{P}_{\text{convFO}}$, since any correlation in this polytope can be expressed as a convex combination of its vertices). 
One can then see that the correlation vector $\vec p_{\textup{NICC}}$ violates this inequality, reaching $I_4' = \frac{6-5\sqrt{2}}{16} \simeq - 0.067 < 0$.

We could not find any analogous violation with $W_{\text{NIQC}}$, or any correlation outside of $\mathcal{P}_\text{convFO}$. Nevertheless, we could come up with another example of a fourpartite quantum circuit with dynamical but non-influenceable quantum control of causal order in $\text{\textup{\textsf{QC-NIQC}}}\setminus(\text{\textup{\textsf{QC-supFO}}}~\cup~\text{\textup{\textsf{QC-NICC}}})$, and show that this process violates Eq.~\eqref{app:ineq_WNICC} with the same instruments as those considered in Eq.~\eqref{eq:instruments_violating_I_4_QC-NICC-NIQC} above. 
This process is defined analogously to $W_{\text{NICC}}$ from Eq.~\eqref{eq:QC-NICC-example}, with $\ket{\psi}=\ket{0}$ and replacing $W_{-\alpha}$ by $W_{+\beta} \coloneqq \frac{1}{4} (\id^{\otimes 4} + \beta (Z Z Z \id + Z \id X X)) \in \mathcal{L}(\mathcal{H}^{{A^{IO}B^{IO}}})$ with $\beta = 1 - \frac{1}{\sqrt{2}}$:%
\footnote{As already mentioned in the main text, one can construct an explicit ``witness'' to prove that this process is indeed not in \textup{\textsf{QC-NICC}}, using the technique of causal witnesses~\cite{Araujo15,branciard16}.}
\begin{align}
    W'_{\textup{NIQC}} &= \frac{1}{2}W_{+\alpha}\otimes \ketbra{0}{0}^{C^I} \otimes \kketbra{\id}{\id}^{C^OD^I} \otimes \id^{D^O} + \frac{1}{2}W_{+\beta}\otimes \ketbra{0}{0}^{D^I} \otimes \kketbra{\id}{\id}^{D^OC^I} \otimes \id^{C^O}. \label{eq:QC-NIQC-example-2}
\end{align}
One also obtains $I_4' = \frac{6-5\sqrt{2}}{16} \simeq - 0.067 < 0$ for the correlation thus generated, showing that it is indeed outside of $\mathcal{P}_\text{convFO}$.

\bigskip

Let us provide, for completeness, the other bounds on the expression $I_4'$ for the other polytopes considered in this work: we find $I_4' \ge_{\text{NIO}} -\frac12$, $I_4' \ge_{\text{NIO}'} -\frac23$, $I_4' \ge_{\text{causal}} -1$. While, by construction, $I_4'\ge_{\text{convFO}} 0$ is a facet of $\mathcal{P}_{\text{convFO}}$, we do not expect these other bounds to define facets of the corresponding polytopes.
Likewise, using again the techniques of Ref.~\cite{liu24} used in Sec.~\ref{sec:QCQC_correlations} of the main text to find the maximal value of $I_4$ using QC-NICCs and QC-NIQCs, we find the minimal values of $I_4'$ obtainable with these classes to be $-0.1827$ and $-0.2436$, respectively.
As for $I_4$, these are obtainable using processes with qubit instruments and compatible with $\{A,B\}\prec\{C,D\}.$
Once again, similar to the case for $I_4$, we note that neither of these values reaches the NIO bound $I_4' \ge_{\text{NIO}} -\frac12$, and $I_4'$ does not allow one to certify QC-NIQC correlations outside of $\mathcal{P}_{\textup{NIO}}$.

The technique we used above, consisting of cooking up some particular correlation and finding the facet of the $\mathcal{P}_\textup{convFO}$ polytope ``right below'' it, could be exploited further to find other inequalities, possibly with some particular form, or some nice interpretation in terms of games. As explained in the main text, it is much more challenging to obtain facet inequalities for the $\mathcal{P}_\textup{NIO}$ and  $\mathcal{P}_{\textup{NIO}'}$ polytopes. Still, for some specific expression obtained as a facet for $\mathcal{P}_\textup{convFO}$, one can easily obtain the corresponding bound for $\mathcal{P}_\textup{NIO}$ and  $\mathcal{P}_{\textup{NIO}'}$ (and potentially test if, by any chance, it also corresponds to a facet of these polytopes).

\end{document}